\pdfoutput=1
\PassOptionsToPackage{svgnames}{xcolor}
\PassOptionsToPackage{hyphens}{url}

\RequirePackage[stable, hang, flushmargin, bottom]{footmisc}

\documentclass[usenatbib]{mnras}
\usepackage[T1]{fontenc}
\DeclareRobustCommand{\VAN}[3]{#2}
\let\VANthebibliography\thebibliography
\def\thebibliography{\DeclareRobustCommand{\VAN}[3]{##3}\VANthebibliography}

\usepackage[tbtags]{amsmath}
\usepackage{amssymb}

\usepackage{xparse}
\usepackage{ifthen}
\usepackage{etoolbox} 

\usepackage{fancyvrb}
\usepackage{xspace}
\usepackage{csquotes}

\usepackage{bm}      

\usepackage{dsfont}  
\usepackage[mathscr]{euscript}  
\usepackage[normalem]{ulem}    
\usepackage[scr=boondoxo]{mathalfa}

\usepackage{booktabs}
\usepackage{multirow}
\usepackage{varwidth}
\usepackage{subcaption}

\usepackage{cleveref}

\usepackage[allowspaces]{mathtools}
\usepackage{esvect}
\usepackage{physics}  

\usepackage{siunitx}[=v2]
\input{lib/utils}
\usepackage{dowith}

\begingroup\lccode`\|=`\\
\lowercase{\endgroup\def\removebs#1{\if#1|\else#1\fi}}
\newcommand{\macroname}[1]{\expandafter\removebs\string#1}

\NewDocumentCommand{\newmathcommand}{m O{} m}{\newcommand{#1}[#2]{\ensuremath{#3}}}
\NewDocumentCommand{\renewmathcommand}{m O{} m}{\renewcommand{#1}[#2]{\ensuremath{#3}}}

\makeatletter

\def\@upgreekone #1{\csdef{#1}{\csuse{up#1}}}
\newcommand{\upgreek}[1]{{%
	\forcsvlist\@upgreekone{alpha, beta, gamma, Gamma, delta, Delta, epsilon, varepsilon, zeta, eta, theta, vartheta, Theta, iota, kappa, lambda, Lambda, mu, nu, xi, Xi, pi, Pi, rho, varrho, sigma, Sigma, tau, upsilon, Upsilon, phi, varphi, Phi, chi, psi, Psi, omega, Omega}%
	#1
}}

\newcommand{\clearextraction}{
	\providecommand{\extractedmain}{} \renewcommand{\extractedmain}{}
	\providecommand{\extractedsub}{} \renewcommand{\extractedsub}{}
	\providecommand{\extractedsuper}{} \renewcommand{\extractedsuper}{}
}
\clearextraction

\newcommand{\splitargs}[4]{%
	\def#3{#1}%
	\def#4{#2}%
}
\NewDocumentCommand{\extractsubsuper}{> {\SplitArgument{1}{_}} u{\relax}}{%
	\splitargs#1{\gfirst}{\gsecond}%
	\expandafter\SplitArgument\expandafter1\expandafter^\expandafter{\gfirst}%
	\expandafter\splitargs\ProcessedArgument\first\second%
	\expandafter\SplitArgument\expandafter1\expandafter^\expandafter{\gsecond}%
	\expandafter\splitargs\ProcessedArgument\third\fourth%
	\edef\extractedmain{\expandonce{\first}}%
	\def\extractedsub{}%
	\def\extractedsuper{}%
	\expandafter\IfNoValueTF\expandafter{\second}{%
		\expandafter\IfNoValueF\expandafter{\third}{%
			\edef\extractedsub{\expandonce{\third}}%
			\expandafter\IfNoValueF\expandafter{\fourth}{%
				\edef\extractedsuper{\expandonce{\fourth}}%
			}%
		}%
	}{\global\edef\extractedsuper{\expandonce{\second}}}%
}

\NewDocumentCommand{\baseextensive}{o m}{%
    \ensuremath{\opbraces{\IfNoValueTF{#1}{#2}{#1{#2}}}}
}

\newcommand{\variabling}[1]{\bm{\mathrm{\upgreek{#1}}}}
\newcommand{\variable}{\baseextensive[\variabling]}

\newcommand{\starredbe}[3]{%
	\expandafter\NewDocumentCommand\csname\macroname{#1}@handler\endcsname{s m}{%
		\IfBooleanTF{##1}{}{\dosmash}{#2{##2}} \vphantom{##2}
	}
	\providecommand{#1}{}%
	\renewcommand{#1}{\@ifstar{\csuse{\macroname{#1}@@}}{\csuse{\macroname{#1}@}}}%
	\csdef{\macroname{#1}@}{\baseextensive[\csuse{\macroname{#1}@handler}]}%
	\csdef{\macroname{#1}@@}{\baseextensive[\csuse{\macroname{#1}@handler}*]}%
}

\newcommand{\vectoring}[1]{\mathmbox{\vec{#1}}}
\starredbe{\vector}{\vectoring}

\newcommand{\vectmatrixing}[1]{\mathmbox{\uuline{#1}}}
\starredbe{\vectmatrix}{\vectmatrixing}

\newcommand{\varmatrix}{\variable}

\newcommand{\comp}[1]{{\renewcommand{\vectoring}[1]{##1}#1}}
\newcommand{\varcomp}[1]{{\renewcommand{\variabling}[1]{##1}#1}}

\makeatother

\newmathcommand{\vectdet}[1]{\abs{#1}}
\newmathcommand{\vardet}[1]{\abs{#1}}

\renewcommand{\i}{\mathrm{i}}
\newcommand{\e}{\mathrm{e}}
\newcommand{\speedoflight}{c}
\newcommand{\gravG}{G}
\DeclareSIUnit{\Msun}{M_{\odot}}
\DeclareSIUnit{\parsec}{pc}
\DeclareSIUnit{\Mpc}{\Mega\parsec}
\DeclareSIUnit{\dex}{dex}

\newcommand{\vectgrad}[1][]{\opbraces{\vector{\nabla}_{\mkern-5mu #1}}}
\newcommand{\vectlap}[1][]{\renewcommand{\vnabla}{\nabla}\laplacian_{\mkern-5mu #1}}
\newcommand{\vargrad}[1][]{\opbraces{\vnabla\ifthenelse{\equal{#1}{}}{}{_{\mkern-4mu #1}}}}

\NewDocumentCommand{\defineverb}{m m}{
	\DefineShortVerb{\|}\SaveVerb{#1}|#2|\UndefineShortVerb{\|}
    \expandafter\newcommand\csname #1\endcsname{\UseVerb{#1}\xspace}
}
\defineverb{python}{Python}
\defineverb{numpy}{NumPy}
\defineverb{scipy}{SciPy}
\defineverb{astropy}{AstroPy}
\defineverb{matplotlib}{matplotlib}
\defineverb{pyro}{Pyro}
\defineverb{torch}{PyTorch}
\defineverb{keops}{PyKeOps}
\defineverb{gpytorch}{GPyTorch}
\defineverb{tqdm}{tqdm}

\def\Sersic{S\'ersic\xspace}
\newcommand{\stscirelease}[2]{(\href{https://hubblesite.org/contents/news-releases/#1/news-#1-#2.html}{STScI~#1-#2})}

\newcommand{\given}{\,|\,}

\NewDocumentCommand{\contract}{ O{} m G{} }{\ensuremath{#2^\transpose \ifblank{#1}{}{#1} \ifblank{#3}{#2}{#3}}}

\newcommand{\transpose}{\top}

\usepackage{environ}
\NewEnviron{eq}{\begin{equation}\begin{split}\BODY\end{split}\end{equation}}

\shortcut*{\dmmass}{\baseextensive{m}_X}[mass of a hypothetical dark matter particle]

\shortcut{\hmmass}{\baseextensive{M}_{\text{hm}}}[half-mode mass for a warm dark matter model]
\def\shmass{\baseextensive{M}}

\shortcut{\subn}{n}
\shortcut*{\shmfraction}{\baseextensive{f}_{\text{sub}}}
\shortcut{\shmfslope}{\alpha}[slope of the (sub)halo mass function]
\shortcut{\shmfb}{\beta}

\vectorshortcut{\ximg}{\xi}
\vectorshortcut{\x}{x}

\newmathcommand{\distance}[1]{D_{#1}}
\shortcut{\dl}{\distance{L}}
\shortcut{\ds}{\distance{S}}
\shortcut{\dls}{\distance{LS}}

\shortcut{\redshift}{z}
\shortcut{\zlens}{\redshift_{\text{lens}}}
\shortcut{\zsrc}{\redshift_{\text{src}}}
\shortcut{\dc}{\distance{C}}

\shortcut{\potfull}{\psi}
\shortcut{\pot}{\Psi}
\shortcut{\sdens}{\Sigma}[mass surface density at location $\ximg$ in the image plane][$\sdens(\ximg)$]
\shortcut{\sbr}{\beta}[surface brightness at location $\x$ in the source plane][$\sbr(\x)$]
\shortcut{\sdenscrit}{\sdens_{\text{cr}}}
\vectorshortcut{\disp}{\alpha}[displacement field][$\disp(\ximg)$]
\shortcut{\convergence}{\kappa}
\shortcut{\shear}{\gamma}
\shortcut{\magnification}{\mathcal{M}}

\shortcut{\spleslope}{\gamma}
\shortcut{\splerein}{\theta_E}
\shortcut{\ellq}{q}
\shortcut{\ellangle}{\varphi}
\shortcut{\ellR}{R}
\shortcut{\ellarg}{\phi}

\shortcut{\extshear}{\shear}

\shortcut{\rhocrit}{\rho_{\text{cr}}}
\shortcut{\nsub}{N_{\text{sub}}}
\shortcut*{\psub}{\P_{\text{sub}}}
\shortcut*{\msub}{\shmass_{200}}
\shortcut{\nfwr}{r}
\shortcut{\nfwrs}{\nfwr_s}
\shortcut{\nfwrt}{\nfwr_t}
\shortcut{\nfwtau}{\tau}
\shortcut{\nfwrho}{\rho}
\shortcut{\nfwrhos}{\nfwrho_s}
\shortcut{\nfws}{s}
\shortcut{\nfwc}{c_{200}}

\shortcut{\smoothmcut}{m_{\text{s}}}
\shortcut{\smoothsigma}{\sigma_{\text{s}}}
\shortcut{\smoothn}{n_{\text{s}}}
\shortcut{\smoothfunc}{S}

\shortcut{\sersicn}{n}

\def\anydata{\variable{d}}
\shortcut*{\data}{\anydata}
\shortcut*{\datavar}{\varmatrix{S}}
\shortcut{\dataerr}{\varepsilon}

\shortcut{\n}{N}[number of data (unmasked pixels)]
\shortcut{\PSF}{\operatorname{PSF}}

\vectorshortcut{\P}{\variable{\mathfrak{p}}}
\vectorshortcut{\p}{\variable{p}}
\shortcut*{\flux}{\variable{f}}[pixel values as predicted by a source model]

\shortcut*{\senssigma}{\sigma_{\text{det}}}[Detection significance for a single subhalo]

\newcommand{\mean}[1]{\hat{#1}}

\shortcut{\param}{\variable{\Theta}}
\shortcut{\gparam}{\variable{\Phi}}
\shortcut{\hparam}{\variable{\Xi}}

\newcommand{\params}[1]{\param_{\text{#1}}}
\newcommand{\hparams}[1]{\hparam_{\text{#1}}}
\newcommand{\meanparams}[1]{\mean{\param}_{\text{#1}}}

\shortcut{\prob}{p}
\def\posterior{\prob}
\shortcut{\proposal}{q}[proposal distribution (guide) for the model parameters $\param$, itself parameterised by $\gparam$][$\proposal_{\gparam}(\param)$]

\shortcut{\expectation}{\mathds{E}}

\DeclareMathOperator{\cov}{cov}
\DeclareMathOperator{\variance}{Var}

\DeclareMathOperator{\elbo}{ELBO}
\DeclareMathOperator{\KL}{KL}
\newcommand{\kl}[2]{\ensuremath{\KL[#1 \:\Vert\: #2]}}

\shortcut{\uniform}{\mathcal{U}}
\shortcut{\Normal}{\mathcal{N}}
\shortcut{\Gaussian}{\mathcal{G}}
\shortcut{\Loss}{\mathcal{L}}

\shortcut{\paramsGP}{\params{GP}}
\shortcut{\hparamsGP}{\hparams{GP}}

\shortcut*{\kfunc}{k}[GP covariance function][$k_{\hparamsGP}$]

\shortcut{\ovar}{\alpha}
\shortcut{\Ovar}{\varmatrix{A}}
\shortcut{\ovarvi}{\mean{\ovar}}
\shortcut{\ovartrue}{\ovar_0}

\newcommand{\KernelSymbol}{\Sigma}
\newcommand{\varKernelSymbol}{Z}
\vectmatrixshortcut{\Kernel}{\KernelSymbol}
\shortcut{\kernelsize}{\sigma}
\shortcut*{\krbf}{\operatorname{\mathscr{k}}}[Gaussian radial basis function ($\hparamsGP \rightarrow \qty{\ovar, \Kernel}$)][$\mathscr{k}$]

\shortcut{\mpred}{\variable{\mu}}
\shortcut{\epred}{\variable{\sigma}}
\shortcut{\vpred}{\variable{V}}
\shortcut{\meanpred}{\mpred_{\flux}}
\shortcut{\varpred}{\vpred_{\flux}}
\shortcut{\errpred}{\epred_{\flux}}
\shortcut{\errpredhat}{\hat{\variable{\sigma}}_{\flux}}

\shortcut{\Err}{\varmatrix{\Sigma}}
\shortcut{\err}{\variable{\sigma}}

\shortcut{\K}{\varmatrix{K}}[GP covariance matrix of the training data ($\K \equiv \K_{\hparamsGP}\qty(\p, \p)$)]
\shortcut{\Ktrue}{\K_0}
\shortcut{\T}{\varmatrix{T}}[GP transmission matrix ($\K \rightarrow \T \T^\transpose$) and rescaled version ($\T_* \equiv \T / \ovar$)][$\T,\ \T_*$]

\shortcut{\priorCov}{\ensuremath{\ovar^2 \varmatrix{\Identity}}}

\vectmatrixshortcut{\varKernel}{\variable{\varKernelSymbol}}[kernels associated with pixels]
\shortcut{\gpwindow}{w}
\shortcut{\gpwindowimg}{\mathfrak{w}_i}
\vectorshortcut{\q}{\variable{q}}
\shortcut{\density}{n}

\shortcut{\y}{\variable{\theta}}[Source parameters and rescaled version ($\mpred \equiv \ovar \y$)][$\y,\ \y_*$]
\shortcut{\meany}{\mean{\y}}
\shortcut{\meanyi}{\varcomp{\meany}_i}
\shortcut{\erry}{\err_{\y}}
\shortcut{\erryi}{\sigma_{\y, i}}
\shortcut{\Erry}{\Err_{\y}}
\shortcut{\meanpredy}{\mpred_{\y}}
\shortcut{\varpredy}{\vpred_{\y}}

\shortcut{\nlayers}{N_l}
\newcommand{\layerindex}[3][]{\ensuremath{#2_{\qty(#3)#1}}}
\newcommand{\spindex}[2][]{\ensuremath{#2_{#1\qty(\text{sp})}}}

\shortcut*{\Identity}{\mathds{I}}[the identity matrix (of size $m$ that could be evident from context)][$\Identity_m$]

\shortcut{\nc}{N_c}
\shortcut{\u}{\variable{u}}
\newcommand{\block}[1]{\ensuremath{\bar{#1}}}
\shortcut{\blockdatavar}{\block{\varmatrix{S}}_0}

\usepackage{xcolor}
\usepackage{graphbox}

\usepackage{pgfplots}
\usepackage{pgfplotstable}
\pgfplotsset{compat=newest, filter discard warning=false}
\usepgfplotslibrary{groupplots,fillbetween,patchplots}

\usepackage{tikz}
\usetikzlibrary{external, calc, patterns, decorations, shapes, arrows, fit}
\usetikzlibrary{
	graphs, graphs.standard,  
	backgrounds, intersections
}
\usepackage{tkz-euclide}

\definecolor{viridis0}{rgb}{0.267004, 0.004874, 0.329415}
\definecolor{viridiss0}{rgb}{0.283072, 0.130895, 0.449241}
\def\globalblackcolor{black}
\def\globalwhitecolor{white}

\pgfplotsset{
	invert cmap/.code={\pgfplotsset{colormap={#1 reverse}{indices of colormap={\pgfplotscolormaplastindexof{#1},...,0 of #1}}}},
	colormap={traffic}{rgb=(0.8,0,0), rgb=(0.9,0.45,0.07), rgb=(1, 0.9, 0.14), rgb=(0,0.7,0)},
	invert cmap=traffic,
	colormap={seismic}{rgb=(0,0,0.3) rgb=(0,0,1) rgb=(1,1,1) rgb=(1,0,0) rgb=(0.5,0,0)},
	invert cmap=seismic,
	colormap={bwr}{rgb=(0,0,1) rgb=(1,1,1) rgb=(1,0,0)},
	invert cmap=bwr,
	colormap={inferno}{rgb=(0.001,0.000,0.014) rgb=(0.002,0.001,0.019) rgb=(0.003,0.002,0.024) rgb=(0.005,0.003,0.031) rgb=(0.006,0.005,0.039) rgb=(0.008,0.006,0.047) rgb=(0.010,0.008,0.055) rgb=(0.012,0.009,0.063) rgb=(0.014,0.011,0.072) rgb=(0.017,0.013,0.080) rgb=(0.019,0.015,0.089) rgb=(0.022,0.017,0.097) rgb=(0.026,0.019,0.106) rgb=(0.029,0.022,0.115) rgb=(0.033,0.024,0.123) rgb=(0.038,0.026,0.132) rgb=(0.042,0.028,0.141) rgb=(0.047,0.030,0.150) rgb=(0.052,0.032,0.159) rgb=(0.056,0.035,0.168) rgb=(0.061,0.037,0.178) rgb=(0.066,0.039,0.187) rgb=(0.071,0.040,0.196) rgb=(0.077,0.042,0.206) rgb=(0.082,0.043,0.215) rgb=(0.087,0.045,0.225) rgb=(0.093,0.046,0.234) rgb=(0.099,0.046,0.244) rgb=(0.105,0.047,0.253) rgb=(0.111,0.047,0.263) rgb=(0.117,0.048,0.272) rgb=(0.123,0.048,0.282) rgb=(0.129,0.047,0.291) rgb=(0.136,0.047,0.300) rgb=(0.142,0.046,0.309) rgb=(0.149,0.045,0.317) rgb=(0.156,0.045,0.325) rgb=(0.163,0.044,0.333) rgb=(0.170,0.042,0.341) rgb=(0.176,0.041,0.348) rgb=(0.183,0.040,0.355) rgb=(0.190,0.039,0.361) rgb=(0.197,0.038,0.368) rgb=(0.204,0.038,0.373) rgb=(0.211,0.037,0.379) rgb=(0.218,0.037,0.384) rgb=(0.225,0.036,0.388) rgb=(0.232,0.036,0.392) rgb=(0.238,0.037,0.396) rgb=(0.245,0.037,0.400) rgb=(0.252,0.038,0.403) rgb=(0.258,0.039,0.406) rgb=(0.265,0.040,0.409) rgb=(0.271,0.041,0.412) rgb=(0.278,0.042,0.414) rgb=(0.284,0.044,0.417) rgb=(0.291,0.046,0.419) rgb=(0.297,0.047,0.420) rgb=(0.304,0.049,0.422) rgb=(0.310,0.051,0.424) rgb=(0.316,0.053,0.425) rgb=(0.323,0.056,0.426) rgb=(0.329,0.058,0.428) rgb=(0.335,0.060,0.429) rgb=(0.342,0.062,0.429) rgb=(0.348,0.065,0.430) rgb=(0.354,0.067,0.431) rgb=(0.360,0.069,0.431) rgb=(0.367,0.072,0.432) rgb=(0.373,0.074,0.432) rgb=(0.379,0.076,0.433) rgb=(0.385,0.079,0.433) rgb=(0.391,0.081,0.433) rgb=(0.398,0.083,0.433) rgb=(0.404,0.086,0.433) rgb=(0.410,0.088,0.433) rgb=(0.416,0.090,0.433) rgb=(0.423,0.093,0.433) rgb=(0.429,0.095,0.432) rgb=(0.435,0.097,0.432) rgb=(0.441,0.099,0.432) rgb=(0.447,0.102,0.431) rgb=(0.454,0.104,0.430) rgb=(0.460,0.106,0.430) rgb=(0.466,0.108,0.429) rgb=(0.472,0.111,0.428) rgb=(0.479,0.113,0.427) rgb=(0.485,0.115,0.427) rgb=(0.491,0.117,0.426) rgb=(0.497,0.119,0.424) rgb=(0.503,0.122,0.423) rgb=(0.510,0.124,0.422) rgb=(0.516,0.126,0.421) rgb=(0.522,0.128,0.420) rgb=(0.528,0.130,0.418) rgb=(0.535,0.133,0.417) rgb=(0.541,0.135,0.415) rgb=(0.547,0.137,0.414) rgb=(0.553,0.139,0.412) rgb=(0.560,0.141,0.410) rgb=(0.566,0.144,0.408) rgb=(0.572,0.146,0.406) rgb=(0.578,0.148,0.404) rgb=(0.585,0.150,0.402) rgb=(0.591,0.153,0.400) rgb=(0.597,0.155,0.398) rgb=(0.603,0.157,0.396) rgb=(0.609,0.159,0.394) rgb=(0.616,0.162,0.391) rgb=(0.622,0.164,0.389) rgb=(0.628,0.167,0.386) rgb=(0.634,0.169,0.384) rgb=(0.640,0.171,0.381) rgb=(0.646,0.174,0.378) rgb=(0.652,0.176,0.376) rgb=(0.658,0.179,0.373) rgb=(0.665,0.182,0.370) rgb=(0.671,0.184,0.367) rgb=(0.677,0.187,0.364) rgb=(0.683,0.190,0.361) rgb=(0.689,0.192,0.358) rgb=(0.695,0.195,0.354) rgb=(0.701,0.198,0.351) rgb=(0.707,0.201,0.348) rgb=(0.712,0.204,0.344) rgb=(0.718,0.207,0.341) rgb=(0.724,0.210,0.337) rgb=(0.730,0.213,0.334) rgb=(0.736,0.216,0.330) rgb=(0.741,0.219,0.327) rgb=(0.747,0.222,0.323) rgb=(0.753,0.226,0.319) rgb=(0.758,0.229,0.315) rgb=(0.764,0.233,0.311) rgb=(0.770,0.236,0.307) rgb=(0.775,0.240,0.304) rgb=(0.781,0.243,0.300) rgb=(0.786,0.247,0.295) rgb=(0.791,0.251,0.291) rgb=(0.797,0.255,0.287) rgb=(0.802,0.259,0.283) rgb=(0.807,0.263,0.279) rgb=(0.812,0.267,0.275) rgb=(0.817,0.271,0.270) rgb=(0.822,0.275,0.266) rgb=(0.827,0.280,0.262) rgb=(0.832,0.284,0.257) rgb=(0.837,0.288,0.253) rgb=(0.842,0.293,0.249) rgb=(0.847,0.298,0.244) rgb=(0.851,0.302,0.240) rgb=(0.856,0.307,0.235) rgb=(0.861,0.312,0.231) rgb=(0.865,0.317,0.226) rgb=(0.869,0.322,0.221) rgb=(0.874,0.327,0.217) rgb=(0.878,0.332,0.212) rgb=(0.882,0.337,0.208) rgb=(0.886,0.343,0.203) rgb=(0.890,0.348,0.198) rgb=(0.894,0.353,0.194) rgb=(0.898,0.359,0.189) rgb=(0.902,0.364,0.184) rgb=(0.906,0.370,0.179) rgb=(0.909,0.376,0.175) rgb=(0.913,0.382,0.170) rgb=(0.916,0.387,0.165) rgb=(0.920,0.393,0.160) rgb=(0.923,0.399,0.155) rgb=(0.926,0.405,0.150) rgb=(0.930,0.411,0.145) rgb=(0.933,0.418,0.140) rgb=(0.936,0.424,0.135) rgb=(0.939,0.430,0.130) rgb=(0.942,0.436,0.125) rgb=(0.944,0.443,0.120) rgb=(0.947,0.449,0.115) rgb=(0.950,0.456,0.110) rgb=(0.952,0.462,0.105) rgb=(0.955,0.469,0.100) rgb=(0.957,0.475,0.095) rgb=(0.959,0.482,0.089) rgb=(0.961,0.489,0.084) rgb=(0.963,0.495,0.079) rgb=(0.965,0.502,0.074) rgb=(0.967,0.509,0.069) rgb=(0.969,0.516,0.063) rgb=(0.971,0.523,0.058) rgb=(0.973,0.530,0.053) rgb=(0.974,0.537,0.048) rgb=(0.976,0.544,0.044) rgb=(0.977,0.551,0.039) rgb=(0.978,0.558,0.035) rgb=(0.980,0.565,0.031) rgb=(0.981,0.572,0.029) rgb=(0.982,0.579,0.026) rgb=(0.983,0.587,0.025) rgb=(0.984,0.594,0.024) rgb=(0.985,0.601,0.024) rgb=(0.985,0.608,0.024) rgb=(0.986,0.616,0.026) rgb=(0.987,0.623,0.028) rgb=(0.987,0.630,0.031) rgb=(0.987,0.638,0.035) rgb=(0.988,0.645,0.040) rgb=(0.988,0.653,0.046) rgb=(0.988,0.660,0.052) rgb=(0.988,0.668,0.058) rgb=(0.988,0.675,0.065) rgb=(0.988,0.683,0.072) rgb=(0.987,0.690,0.080) rgb=(0.987,0.698,0.088) rgb=(0.987,0.706,0.096) rgb=(0.986,0.713,0.104) rgb=(0.986,0.721,0.112) rgb=(0.985,0.728,0.121) rgb=(0.984,0.736,0.130) rgb=(0.983,0.744,0.138) rgb=(0.982,0.751,0.148) rgb=(0.981,0.759,0.157) rgb=(0.980,0.767,0.166) rgb=(0.979,0.775,0.176) rgb=(0.977,0.782,0.186) rgb=(0.976,0.790,0.196) rgb=(0.975,0.798,0.206) rgb=(0.973,0.805,0.217) rgb=(0.971,0.813,0.228) rgb=(0.970,0.821,0.239) rgb=(0.968,0.829,0.250) rgb=(0.966,0.836,0.262) rgb=(0.964,0.844,0.273) rgb=(0.963,0.851,0.286) rgb=(0.961,0.859,0.298) rgb=(0.959,0.867,0.311) rgb=(0.957,0.874,0.324) rgb=(0.955,0.882,0.337) rgb=(0.953,0.889,0.351) rgb=(0.952,0.896,0.366) rgb=(0.950,0.903,0.380) rgb=(0.949,0.910,0.395) rgb=(0.948,0.917,0.411) rgb=(0.947,0.924,0.426) rgb=(0.946,0.931,0.442) rgb=(0.946,0.937,0.459) rgb=(0.947,0.943,0.475) rgb=(0.948,0.949,0.491) rgb=(0.950,0.955,0.508) rgb=(0.952,0.961,0.524) rgb=(0.955,0.966,0.540) rgb=(0.958,0.971,0.556) rgb=(0.962,0.976,0.572) rgb=(0.966,0.981,0.587) rgb=(0.971,0.985,0.602) rgb=(0.977,0.990,0.617) rgb=(0.982,0.994,0.631) rgb=(0.988,0.998,0.645)},
	invert cmap=inferno,
}

\newcounter{marknumber}
\pgfplotsset{
    error bars/every nth mark/.style={
		/pgfplots/error bars/draw error bar/.prefix code={
			\pgfmathtruncatemacro\marknumbercheck{mod(floor(\themarknumber/2),#1)}
			\ifnum\marknumbercheck=0
			\else
			\begin{scope}[opacity=0]
				\fi
			},
			/pgfplots/error bars/draw error bar/.append code={
				\ifnum\marknumbercheck=0
				\else
			\end{scope}
			\fi
			\stepcounter{marknumber}
		}
	}
}

\newif\ifcolorbarishorizontal
\pgfplotsset{
    log xticks with fixed point/.style={
		xticklabel={
			\pgfkeys{/pgf/fpu=true}
			\pgfmathparse{exp(\tick)}%
			\pgfmathprintnumber[fixed relative, precision=3]{\pgfmathresult}
			\pgfkeys{/pgf/fpu=false}
		}
	},
	log yticks with fixed point/.style={
		yticklabel={
			\pgfkeys{/pgf/fpu=true}
			\pgfmathparse{exp(\tick)}%
			\pgfmathprintnumber[fixed relative, precision=3]{\pgfmathresult}
			\pgfkeys{/pgf/fpu=false}
		}
	},
	tight layout/.code={
		\pgfmathparse{#1-1}\edef\res{\pgfmathresult}
		\pgfplotsset{
			scale only axis,
			width={1/#1*\linewidth - \res/#1*\pgfkeysvalueof{/pgfplots/group/horizontal sep}},
		}
	},
	histogram/.style={
		hist/data filter/.code={\pgfmathparse{and(##1 > \pgfkeysvalueof{/pgfplots/hist/data min}, ##1 < \pgfkeysvalueof{/pgfplots/hist/data max}) ? ##1 : inf}}
	},
	extent/.code={
		\pgfplotsset{xmin={-#1}, xmax={#1}, ymin={-#1}, ymax={#1}}
	},
	colorbar is horizontal/.is if=colorbarishorizontal,
	colorbar label/.code={
		\ifcolorbarishorizontal
			\pgfplotsset{colorbar style={xlabel=#1}}
		\else
			\pgfplotsset{colorbar style={ylabel=#1}}
		\fi
	},
	plot graphics/!src/.code={
		\pgfplotsset{
			plot graphics/lowlevel draw/.code 2 args={\includegraphics[\pgfkeysvalueof{/pgfplots/plot graphics/includegraphics},width=##1,height=##2] {#1}}
		}
	}
}

\NewDocumentCommand{\axhline}{O{} m O{0} O{1} O{0}}{
	\draw[#1] ({{rel axis cs:#3,0}} |- {{axis cs:#5,#2}}) -| (rel axis cs:#4,0);}
\NewDocumentCommand{\axvline}{O{} m O{0} O{1} O{0}}{
	\draw[#1] ({{rel axis cs:0,#3}} -| {{axis cs:#2,#5}}) |- (rel axis cs:0,#4);}

\newcommand{\pgfplotsforeach}[3]{\foreach #1 in #2 {\edef\temp{#3}\temp}}

\newcommand{\drawnode}[2][]{\draw[#1] (#2.south west) rectangle (#2.north east);}
\NewDocumentCommand{\labelPoint}{O{} r() m}{\tkzDrawPoint[\globalblackcolor](#2)\tkzLabelPoint[\globalblackcolor, #1](#2){#3}}

\tikzset{
	filled/.style={fill=black!20},
	graph edge/.style={shorten >=1.5pt},
	graphnode/.style={minimum size=1.8em, draw, >graph edge, <graph edge},
	input/.style={graphnode, circle, draw=none, minimum size=0pt, inner sep=2pt},
	param/.style={graphnode, filled},
	calc/.style={graphnode, circle},
	dist/.style={graphnode, densely dash dot, >shorten >=0pt},
	guide edge/.style={graph edge, densely dotted},
	guide*/.style={draw, densely dash dot},
	guide/.style={dist, <guide edge},
	output/.style={calc, filled},
	sum/.style={"+"},
 	>={Stealth}
}

\pgfplotsset{
	corner/.style={
		group/x descriptions at = edge bottom,
		group/y descriptions at = edge left,
		group/horizontal sep=0.1cm, group/vertical sep=0.1cm,
		height={\pgfkeysvalueof{/pgfplots/width}},
		domain={\pgfkeysvalueof{/pgfplots/xmin}:\pgfkeysvalueof{/pgfplots/xmax}},
		scaled ticks=false,
		tickwidth=2pt,
		tick label style={
			rotate=45, anchor=east, font=\scriptsize,
			/pgf/number format/.cd, fixed, zerofill, precision=4
		},
		xticklabel={\minimumprecision{3}{\tick}},
		yticklabel={\minimumprecision{3}{\tick}},
		label style={at={(ticklabel* cs:0.5,3em)}, anchor=base, inner sep=0pt,},
		ylabel style={at={(ticklabel* cs:0.5,2.5em)}},
	},
}

\tikzset{
    cntr/.style={ultra thin,},
	cntr 1/.style={cntr, fill=green},
	cntr 2/.style={cntr, fill=yellow},
	cntr 3/.style={cntr, fill=red},
	hist1d/.style={no marks, samples=51},
	truth line/.style={black, no marks, thin},
	truth/.style={black, mark=*, mark size=1pt},
}

\usepgfplotslibrary{colorbrewer}

\pgfplotsset{
	every axis/.append style={
		axis on top,
		label style={font=\footnotesize},
		tick label style={font=\footnotesize},
		tick style={thin, black}, tickwidth=5pt,
		tick pos=left, tick align=outside,
	},
	every axis plot/.append style={
		ultra thick, line join=round, line cap=butt,
		domain={\pgfkeysvalueof{/pgfplots/xmin}:\pgfkeysvalueof{/pgfplots/xmax}},
	},
	every axis legend/.append style={
		font=\footnotesize, cells={anchor=west}, fill=none,
	},
	filled legend/.style={legend style={fill=white, fill opacity=0.9}},
	colorbar base/.style={
		axis equal image=false, tickwidth=3pt,
		xlabel style={at={(ticklabel cs:0.5)}, inner sep=4pt, anchor=near ticklabel, text depth=0pt},
		ticklabel pos=right, tick pos=right,
	},
	colorbar top/.style={
		colorbar horizontal,
		colorbar style={
			colorbar base, height=0.5em,
			at={(parent axis.north west)}, anchor=south west,
		},
		colorbar shift/.style={yshift=0.1cm},
		colorbar is horizontal=true,
	},
	colorbar bottom/.style={
		colorbar horizontal,
		colorbar style={
			colorbar base, height=0.5em,
			at={(parent axis.south west)}, anchor=north west,
			xtick pos=bottom, xticklabel pos=bottom,
		},
		colorbar shift/.style={yshift=-0.1cm},
		colorbar is horizontal=true,
	},
	colorbar right/.append style={
		colorbar style={colorbar base, width=0.5em, at={(parent axis.north east)}},
		colorbar shift/.style={xshift=0.1cm},
		colorbar is horizontal=false,
	},
	colorbar left/.append style={
		colorbar style={ytick pos=left},
		colorbar is horizontal=false
	}
}

\pgfplotsset{
	imageplane/.style={
		xlabel={$\comp{\ximg}_{x} / \si{arcsec}$}, ylabel={$\comp{\ximg}_{y} / \si{arcsec}$},
		extent=2.5,
		plot graphics/.cd, xmin=-2.5, xmax=2.5, ymin=-2.5, ymax=2.5
	},
	sourceplane/.style={
		xlabel={$x / \si{arcsec}$}, ylabel={$y / \si{arcsec}$},
		extent=0.5,
		/tikz/lbl/.style={above right, at={(axis description cs:0,0)}, white},
		/tikz/kernel/.style={Yellow, thin},
		plot graphics/.cd, xmin=-0.5, xmax=0.5, ymin=-0.5, ymax=0.5
	},
	/tikz/critical/.style={no markers, color=Red, thin, dashed, opacity=0.8},
	/tikz/caustic/.style={critical}
}

\tikzset{
	vigp defs/.style={
		exact/.style={black},
		full/.style={green!80!black},
		diag/.style={red!80!black},
		diag-fixed/.style={yellow!80!black},
	},
	ngpl-1s/.style={/pgfplots/table/y={ngpl-1s}, color=Paired-L},
	ngpl-1b/.style={/pgfplots/table/y={ngpl-1b}, color=Paired-E},
	ngpl-2/.style={/pgfplots/table/y={ngpl-2}, color=Paired-F},
	ngpl-3/.style={/pgfplots/table/y={ngpl-3}, color=Paired-I},
	ngpl-4/.style={/pgfplots/table/y={ngpl-4}, color=Paired-J},
	ngpl-5/.style={/pgfplots/table/y={ngpl-5}, color=Paired-A},
	ngpl-6/.style={/pgfplots/table/y={ngpl-6}, color=Paired-B},
	ngpl-7/.style={/pgfplots/table/y={ngpl-7}, color=Paired-C},
	ngpl-8/.style={/pgfplots/table/y={ngpl-8}, color=Paired-D},
	core/.style={Green}, ring/.style={Red},
}

\usepackage{expl3}
\ExplSyntaxOn

\regex_const:Nn \l_ngpl_regex {(\d+)([^\d]*)}
\newcommand{\ngplprocess}[2]{\regex_extract_once:NnNTF \l_ngpl_regex {#1} \l_ngpl_match_seq {#2} {}}
\newcommand{\ngplformat}[1]{\ngplprocess{#1}{\seq_item:Nn \l_ngpl_match_seq 2 \tl_if_blank:eTF {\seq_item:Nn \l_ngpl_match_seq 3} {} {\c_math_subscript_token{\seq_item:Nn \l_ngpl_match_seq 3}}}}

\ExplSyntaxOff

\usepackage{stringstrings}   
\newcommand{\ngplnum}[1]{\Treatments{0}{0}{0}{1}{0}{0}\substring[v]{#1}{1}{$}}
\newcommand{\ngplnumm}[1]{\Treatments{0}{0}{0}{1}{0}{0}\substring[q]{#1}{1}{$}}
\newcommand{\ngplnumtomacro}[2]{\ngplnumm{#2}\edef#1{\thestring}}

\newcommand{\getdata}[1]{\pgfkeysvalueof{/private/#1}}

\newcommand{\kernelcirclepos}{0.4, 0.4}

\ExplSyntaxOn
\newcommand{\formatuncertain}[2]{\exp_last_unbraced:Ne \prg_do_nothing: {
	\fp_eval:n {round(#1, -logb(sqrt(#2)) - ((sqrt(#2) / 10^logb(sqrt(#2)) > 5) ? 1 : 0))}
	+-
	\fp_eval:n {round(sqrt(#2), -logb(sqrt(#2)) - ((sqrt(#2) / 10^logb(sqrt(#2)) > 5) ? 1 : 0))}
}}
\ExplSyntaxOff

\newcommand{\afitresult}[2]{\formatuncertain{\getdata{#1/mu/#2}}{\getdata{#1/S/#2/#2}}}
\newcommand{\fitresult}[1]{\afitresult{hoags_object/ngpl-3}{#1}}
\newcommand{\atruth}[2]{\getdata{#1/truth/#2}}
\newcommand{\truth}[1]{\atruth{hoags_object}{#1}}

\pgfplotsset{,
	hoags_object-ngpl-1b-img-mean/.style={plot graphics/xmin={-2.5}, plot graphics/xmax={2.5}, plot graphics/ymin={-2.5}, plot graphics/ymax={2.5}, point meta min={0.0}, point meta max={30.0}, plot graphics/!src={tikz/gallery/hoags_object/hoags_object-ngpl-1b-img-mean-viridis-0-30}, colormap name={viridis}},
	hoags_object-ngpl-1b-img-std/.style={plot graphics/xmin={-2.5}, plot graphics/xmax={2.5}, plot graphics/ymin={-2.5}, plot graphics/ymax={2.5}, point meta min={0.0}, point meta max={0.7}, plot graphics/!src={tikz/gallery/hoags_object/hoags_object-ngpl-1b-img-std-viridis-0-1}, colormap name={viridis}},
	hoags_object-ngpl-1b-simg-deproj-mean/.style={plot graphics/xmin={-0.5}, plot graphics/xmax={0.5}, plot graphics/ymin={-0.5}, plot graphics/ymax={0.5}, point meta min={0.0}, point meta max={30.0}, plot graphics/!src={tikz/gallery/hoags_object/hoags_object-ngpl-1b-simg-deproj-mean-viridis-0-30}, colormap name={viridis}},
	hoags_object-ngpl-1b-simg-deproj-std/.style={plot graphics/xmin={-0.5}, plot graphics/xmax={0.5}, plot graphics/ymin={-0.5}, plot graphics/ymax={0.5}, point meta min={0.0}, point meta max={0.7}, plot graphics/!src={tikz/gallery/hoags_object/hoags_object-ngpl-1b-simg-deproj-std-viridis-0-1}, colormap name={viridis}},
	hoags_object-ngpl-1b-simg-mean/.style={plot graphics/xmin={-0.5}, plot graphics/xmax={0.5}, plot graphics/ymin={-0.5}, plot graphics/ymax={0.5}, point meta min={0.0}, point meta max={30.0}, plot graphics/!src={tikz/gallery/hoags_object/hoags_object-ngpl-1b-simg-mean-viridis-0-30}, colormap name={viridis}},
	hoags_object-ngpl-1b-simg-std/.style={plot graphics/xmin={-0.5}, plot graphics/xmax={0.5}, plot graphics/ymin={-0.5}, plot graphics/ymax={0.5}, point meta min={0.0}, point meta max={0.7}, plot graphics/!src={tikz/gallery/hoags_object/hoags_object-ngpl-1b-simg-std-viridis-0-1}, colormap name={viridis}},
	hoags_object-ngpl-1b-simg-std-interp/.style={plot graphics/xmin={-0.5}, plot graphics/xmax={0.5}, plot graphics/ymin={-0.5}, plot graphics/ymax={0.5}, point meta min={0.0}, point meta max={0.7}, plot graphics/!src={tikz/gallery/hoags_object/hoags_object-ngpl-1b-simg-std-interp-viridis-0-1}, colormap name={viridis}},
	hoags_object-ngpl-1b-simg-std-winterp/.style={plot graphics/xmin={-0.5}, plot graphics/xmax={0.5}, plot graphics/ymin={-0.5}, plot graphics/ymax={0.5}, point meta min={0.0}, point meta max={0.7}, plot graphics/!src={tikz/gallery/hoags_object/hoags_object-ngpl-1b-simg-std-winterp-viridis-0-1}, colormap name={viridis}},
	hoags_object-ngpl-1b-simg-std-interp-img/.style={plot graphics/xmin={-0.5}, plot graphics/xmax={0.5}, plot graphics/ymin={-0.5}, plot graphics/ymax={0.5}, point meta min={0.0}, point meta max={0.7}, plot graphics/!src={tikz/gallery/hoags_object/hoags_object-ngpl-1b-simg-std-interp-img-viridis-0-1}, colormap name={viridis}},
	hoags_object-ngpl-1b-srcs-src-mean/.style={plot graphics/xmin={-0.5}, plot graphics/xmax={0.5}, plot graphics/ymin={-0.5}, plot graphics/ymax={0.5}, point meta min={-0.6136965751647949}, point meta max={29.295188903808594}, plot graphics/!src={tikz/gallery/hoags_object/hoags_object-ngpl-1b-srcs-src-mean-viridis--1-29}, colormap name={viridis}},
	hoags_object-ngpl-1b-srcs-src-std/.style={plot graphics/xmin={-0.5}, plot graphics/xmax={0.5}, plot graphics/ymin={-0.5}, plot graphics/ymax={0.5}, point meta min={0.0}, point meta max={0.7}, plot graphics/!src={tikz/gallery/hoags_object/hoags_object-ngpl-1b-srcs-src-std-viridis-0-1}, colormap name={viridis}},
	hoags_object-ngpl-1b-srcs-gp-1-mean/.style={plot graphics/xmin={-0.5}, plot graphics/xmax={0.5}, plot graphics/ymin={-0.5}, plot graphics/ymax={0.5}, point meta min={-2.140028715133667}, point meta max={5.631145477294922}, plot graphics/!src={tikz/gallery/hoags_object/hoags_object-ngpl-1b-srcs-gp-1-mean-viridis--2-6}, colormap name={viridis}},
	hoags_object-ngpl-1b-srcs-gp-1-std/.style={plot graphics/xmin={-0.5}, plot graphics/xmax={0.5}, plot graphics/ymin={-0.5}, plot graphics/ymax={0.5}, point meta min={0.0}, point meta max={0.7}, plot graphics/!src={tikz/gallery/hoags_object/hoags_object-ngpl-1b-srcs-gp-1-std-viridis-0-1}, colormap name={viridis}},
	hoags_object-ngpl-1b-srcs-gp-1-std-interp/.style={plot graphics/xmin={-0.5}, plot graphics/xmax={0.5}, plot graphics/ymin={-0.5}, plot graphics/ymax={0.5}, point meta min={0.0}, point meta max={0.7}, plot graphics/!src={tikz/gallery/hoags_object/hoags_object-ngpl-1b-srcs-gp-1-std-interp-viridis-0-1}, colormap name={viridis}},
	hoags_object-ngpl-1b-srcs-gp-1-std-winterp/.style={plot graphics/xmin={-0.5}, plot graphics/xmax={0.5}, plot graphics/ymin={-0.5}, plot graphics/ymax={0.5}, point meta min={0.0}, point meta max={0.7}, plot graphics/!src={tikz/gallery/hoags_object/hoags_object-ngpl-1b-srcs-gp-1-std-winterp-viridis-0-1}, colormap name={viridis}},
	hoags_object-ngpl-1b-srcs-gp-1-std-interp-img/.style={plot graphics/xmin={-0.5}, plot graphics/xmax={0.5}, plot graphics/ymin={-0.5}, plot graphics/ymax={0.5}, point meta min={0.0}, point meta max={0.7}, plot graphics/!src={tikz/gallery/hoags_object/hoags_object-ngpl-1b-srcs-gp-1-std-interp-img-viridis-0-1}, colormap name={viridis}},
	hoags_object-ngpl-1s-img-mean/.style={plot graphics/xmin={-2.5}, plot graphics/xmax={2.5}, plot graphics/ymin={-2.5}, plot graphics/ymax={2.5}, point meta min={0.0}, point meta max={30.0}, plot graphics/!src={tikz/gallery/hoags_object/hoags_object-ngpl-1s-img-mean-viridis-0-30}, colormap name={viridis}},
	hoags_object-ngpl-1s-img-std/.style={plot graphics/xmin={-2.5}, plot graphics/xmax={2.5}, plot graphics/ymin={-2.5}, plot graphics/ymax={2.5}, point meta min={0.0}, point meta max={0.7}, plot graphics/!src={tikz/gallery/hoags_object/hoags_object-ngpl-1s-img-std-viridis-0-1}, colormap name={viridis}},
	hoags_object-ngpl-1s-simg-deproj-mean/.style={plot graphics/xmin={-0.5}, plot graphics/xmax={0.5}, plot graphics/ymin={-0.5}, plot graphics/ymax={0.5}, point meta min={0.0}, point meta max={30.0}, plot graphics/!src={tikz/gallery/hoags_object/hoags_object-ngpl-1s-simg-deproj-mean-viridis-0-30}, colormap name={viridis}},
	hoags_object-ngpl-1s-simg-deproj-std/.style={plot graphics/xmin={-0.5}, plot graphics/xmax={0.5}, plot graphics/ymin={-0.5}, plot graphics/ymax={0.5}, point meta min={0.0}, point meta max={0.7}, plot graphics/!src={tikz/gallery/hoags_object/hoags_object-ngpl-1s-simg-deproj-std-viridis-0-1}, colormap name={viridis}},
	hoags_object-ngpl-1s-simg-mean/.style={plot graphics/xmin={-0.5}, plot graphics/xmax={0.5}, plot graphics/ymin={-0.5}, plot graphics/ymax={0.5}, point meta min={0.0}, point meta max={30.0}, plot graphics/!src={tikz/gallery/hoags_object/hoags_object-ngpl-1s-simg-mean-viridis-0-30}, colormap name={viridis}},
	hoags_object-ngpl-1s-simg-std/.style={plot graphics/xmin={-0.5}, plot graphics/xmax={0.5}, plot graphics/ymin={-0.5}, plot graphics/ymax={0.5}, point meta min={0.0}, point meta max={0.7}, plot graphics/!src={tikz/gallery/hoags_object/hoags_object-ngpl-1s-simg-std-viridis-0-1}, colormap name={viridis}},
	hoags_object-ngpl-1s-simg-std-interp/.style={plot graphics/xmin={-0.5}, plot graphics/xmax={0.5}, plot graphics/ymin={-0.5}, plot graphics/ymax={0.5}, point meta min={0.0}, point meta max={0.7}, plot graphics/!src={tikz/gallery/hoags_object/hoags_object-ngpl-1s-simg-std-interp-viridis-0-1}, colormap name={viridis}},
	hoags_object-ngpl-1s-simg-std-winterp/.style={plot graphics/xmin={-0.5}, plot graphics/xmax={0.5}, plot graphics/ymin={-0.5}, plot graphics/ymax={0.5}, point meta min={0.0}, point meta max={0.7}, plot graphics/!src={tikz/gallery/hoags_object/hoags_object-ngpl-1s-simg-std-winterp-viridis-0-1}, colormap name={viridis}},
	hoags_object-ngpl-1s-simg-std-interp-img/.style={plot graphics/xmin={-0.5}, plot graphics/xmax={0.5}, plot graphics/ymin={-0.5}, plot graphics/ymax={0.5}, point meta min={0.0}, point meta max={0.7}, plot graphics/!src={tikz/gallery/hoags_object/hoags_object-ngpl-1s-simg-std-interp-img-viridis-0-1}, colormap name={viridis}},
	hoags_object-ngpl-1s-srcs-src-mean/.style={plot graphics/xmin={-0.5}, plot graphics/xmax={0.5}, plot graphics/ymin={-0.5}, plot graphics/ymax={0.5}, point meta min={-0.6212714910507202}, point meta max={29.8585262298584}, plot graphics/!src={tikz/gallery/hoags_object/hoags_object-ngpl-1s-srcs-src-mean-viridis--1-30}, colormap name={viridis}},
	hoags_object-ngpl-1s-srcs-src-std/.style={plot graphics/xmin={-0.5}, plot graphics/xmax={0.5}, plot graphics/ymin={-0.5}, plot graphics/ymax={0.5}, point meta min={0.0}, point meta max={0.7}, plot graphics/!src={tikz/gallery/hoags_object/hoags_object-ngpl-1s-srcs-src-std-viridis-0-1}, colormap name={viridis}},
	hoags_object-ngpl-1s-srcs-gp-1-mean/.style={plot graphics/xmin={-0.5}, plot graphics/xmax={0.5}, plot graphics/ymin={-0.5}, plot graphics/ymax={0.5}, point meta min={-0.4717896282672882}, point meta max={1.1919282674789429}, plot graphics/!src={tikz/gallery/hoags_object/hoags_object-ngpl-1s-srcs-gp-1-mean-viridis--0-1}, colormap name={viridis}},
	hoags_object-ngpl-1s-srcs-gp-1-std/.style={plot graphics/xmin={-0.5}, plot graphics/xmax={0.5}, plot graphics/ymin={-0.5}, plot graphics/ymax={0.5}, point meta min={0.0}, point meta max={0.7}, plot graphics/!src={tikz/gallery/hoags_object/hoags_object-ngpl-1s-srcs-gp-1-std-viridis-0-1}, colormap name={viridis}},
	hoags_object-ngpl-1s-srcs-gp-1-std-interp/.style={plot graphics/xmin={-0.5}, plot graphics/xmax={0.5}, plot graphics/ymin={-0.5}, plot graphics/ymax={0.5}, point meta min={0.0}, point meta max={0.7}, plot graphics/!src={tikz/gallery/hoags_object/hoags_object-ngpl-1s-srcs-gp-1-std-interp-viridis-0-1}, colormap name={viridis}},
	hoags_object-ngpl-1s-srcs-gp-1-std-winterp/.style={plot graphics/xmin={-0.5}, plot graphics/xmax={0.5}, plot graphics/ymin={-0.5}, plot graphics/ymax={0.5}, point meta min={0.0}, point meta max={0.7}, plot graphics/!src={tikz/gallery/hoags_object/hoags_object-ngpl-1s-srcs-gp-1-std-winterp-viridis-0-1}, colormap name={viridis}},
	hoags_object-ngpl-1s-srcs-gp-1-std-interp-img/.style={plot graphics/xmin={-0.5}, plot graphics/xmax={0.5}, plot graphics/ymin={-0.5}, plot graphics/ymax={0.5}, point meta min={0.0}, point meta max={0.7}, plot graphics/!src={tikz/gallery/hoags_object/hoags_object-ngpl-1s-srcs-gp-1-std-interp-img-viridis-0-1}, colormap name={viridis}},
	hoags_object-ngpl-2-img-mean/.style={plot graphics/xmin={-2.5}, plot graphics/xmax={2.5}, plot graphics/ymin={-2.5}, plot graphics/ymax={2.5}, point meta min={0.0}, point meta max={30.0}, plot graphics/!src={tikz/gallery/hoags_object/hoags_object-ngpl-2-img-mean-viridis-0-30}, colormap name={viridis}},
	hoags_object-ngpl-2-img-std/.style={plot graphics/xmin={-2.5}, plot graphics/xmax={2.5}, plot graphics/ymin={-2.5}, plot graphics/ymax={2.5}, point meta min={0.0}, point meta max={0.7}, plot graphics/!src={tikz/gallery/hoags_object/hoags_object-ngpl-2-img-std-viridis-0-1}, colormap name={viridis}},
	hoags_object-ngpl-2-simg-deproj-mean/.style={plot graphics/xmin={-0.5}, plot graphics/xmax={0.5}, plot graphics/ymin={-0.5}, plot graphics/ymax={0.5}, point meta min={0.0}, point meta max={30.0}, plot graphics/!src={tikz/gallery/hoags_object/hoags_object-ngpl-2-simg-deproj-mean-viridis-0-30}, colormap name={viridis}},
	hoags_object-ngpl-2-simg-deproj-std/.style={plot graphics/xmin={-0.5}, plot graphics/xmax={0.5}, plot graphics/ymin={-0.5}, plot graphics/ymax={0.5}, point meta min={0.0}, point meta max={0.7}, plot graphics/!src={tikz/gallery/hoags_object/hoags_object-ngpl-2-simg-deproj-std-viridis-0-1}, colormap name={viridis}},
	hoags_object-ngpl-2-simg-mean/.style={plot graphics/xmin={-0.5}, plot graphics/xmax={0.5}, plot graphics/ymin={-0.5}, plot graphics/ymax={0.5}, point meta min={0.0}, point meta max={30.0}, plot graphics/!src={tikz/gallery/hoags_object/hoags_object-ngpl-2-simg-mean-viridis-0-30}, colormap name={viridis}},
	hoags_object-ngpl-2-simg-std/.style={plot graphics/xmin={-0.5}, plot graphics/xmax={0.5}, plot graphics/ymin={-0.5}, plot graphics/ymax={0.5}, point meta min={0.0}, point meta max={0.7}, plot graphics/!src={tikz/gallery/hoags_object/hoags_object-ngpl-2-simg-std-viridis-0-1}, colormap name={viridis}},
	hoags_object-ngpl-2-simg-std-interp/.style={plot graphics/xmin={-0.5}, plot graphics/xmax={0.5}, plot graphics/ymin={-0.5}, plot graphics/ymax={0.5}, point meta min={0.0}, point meta max={0.7}, plot graphics/!src={tikz/gallery/hoags_object/hoags_object-ngpl-2-simg-std-interp-viridis-0-1}, colormap name={viridis}},
	hoags_object-ngpl-2-simg-std-winterp/.style={plot graphics/xmin={-0.5}, plot graphics/xmax={0.5}, plot graphics/ymin={-0.5}, plot graphics/ymax={0.5}, point meta min={0.0}, point meta max={0.7}, plot graphics/!src={tikz/gallery/hoags_object/hoags_object-ngpl-2-simg-std-winterp-viridis-0-1}, colormap name={viridis}},
	hoags_object-ngpl-2-simg-std-interp-img/.style={plot graphics/xmin={-0.5}, plot graphics/xmax={0.5}, plot graphics/ymin={-0.5}, plot graphics/ymax={0.5}, point meta min={0.0}, point meta max={0.7}, plot graphics/!src={tikz/gallery/hoags_object/hoags_object-ngpl-2-simg-std-interp-img-viridis-0-1}, colormap name={viridis}},
	hoags_object-ngpl-2-srcs-src-mean/.style={plot graphics/xmin={-0.5}, plot graphics/xmax={0.5}, plot graphics/ymin={-0.5}, plot graphics/ymax={0.5}, point meta min={-0.6335113644599915}, point meta max={29.2791748046875}, plot graphics/!src={tikz/gallery/hoags_object/hoags_object-ngpl-2-srcs-src-mean-viridis--1-29}, colormap name={viridis}},
	hoags_object-ngpl-2-srcs-src-std/.style={plot graphics/xmin={-0.5}, plot graphics/xmax={0.5}, plot graphics/ymin={-0.5}, plot graphics/ymax={0.5}, point meta min={0.0}, point meta max={0.7}, plot graphics/!src={tikz/gallery/hoags_object/hoags_object-ngpl-2-srcs-src-std-viridis-0-1}, colormap name={viridis}},
	hoags_object-ngpl-2-srcs-gp-1-mean/.style={plot graphics/xmin={-0.5}, plot graphics/xmax={0.5}, plot graphics/ymin={-0.5}, plot graphics/ymax={0.5}, point meta min={-2.054121971130371}, point meta max={5.433931350708008}, plot graphics/!src={tikz/gallery/hoags_object/hoags_object-ngpl-2-srcs-gp-1-mean-viridis--2-5}, colormap name={viridis}},
	hoags_object-ngpl-2-srcs-gp-1-std/.style={plot graphics/xmin={-0.5}, plot graphics/xmax={0.5}, plot graphics/ymin={-0.5}, plot graphics/ymax={0.5}, point meta min={0.0}, point meta max={0.7}, plot graphics/!src={tikz/gallery/hoags_object/hoags_object-ngpl-2-srcs-gp-1-std-viridis-0-1}, colormap name={viridis}},
	hoags_object-ngpl-2-srcs-gp-1-std-interp/.style={plot graphics/xmin={-0.5}, plot graphics/xmax={0.5}, plot graphics/ymin={-0.5}, plot graphics/ymax={0.5}, point meta min={0.0}, point meta max={0.7}, plot graphics/!src={tikz/gallery/hoags_object/hoags_object-ngpl-2-srcs-gp-1-std-interp-viridis-0-1}, colormap name={viridis}},
	hoags_object-ngpl-2-srcs-gp-1-std-winterp/.style={plot graphics/xmin={-0.5}, plot graphics/xmax={0.5}, plot graphics/ymin={-0.5}, plot graphics/ymax={0.5}, point meta min={0.0}, point meta max={0.7}, plot graphics/!src={tikz/gallery/hoags_object/hoags_object-ngpl-2-srcs-gp-1-std-winterp-viridis-0-1}, colormap name={viridis}},
	hoags_object-ngpl-2-srcs-gp-1-std-interp-img/.style={plot graphics/xmin={-0.5}, plot graphics/xmax={0.5}, plot graphics/ymin={-0.5}, plot graphics/ymax={0.5}, point meta min={0.0}, point meta max={0.7}, plot graphics/!src={tikz/gallery/hoags_object/hoags_object-ngpl-2-srcs-gp-1-std-interp-img-viridis-0-1}, colormap name={viridis}},
	hoags_object-ngpl-2-srcs-gp-2-mean/.style={plot graphics/xmin={-0.5}, plot graphics/xmax={0.5}, plot graphics/ymin={-0.5}, plot graphics/ymax={0.5}, point meta min={-0.2697353959083557}, point meta max={0.48380300402641296}, plot graphics/!src={tikz/gallery/hoags_object/hoags_object-ngpl-2-srcs-gp-2-mean-viridis--0-0}, colormap name={viridis}},
	hoags_object-ngpl-2-srcs-gp-2-std/.style={plot graphics/xmin={-0.5}, plot graphics/xmax={0.5}, plot graphics/ymin={-0.5}, plot graphics/ymax={0.5}, point meta min={0.0}, point meta max={0.7}, plot graphics/!src={tikz/gallery/hoags_object/hoags_object-ngpl-2-srcs-gp-2-std-viridis-0-1}, colormap name={viridis}},
	hoags_object-ngpl-2-srcs-gp-2-std-interp/.style={plot graphics/xmin={-0.5}, plot graphics/xmax={0.5}, plot graphics/ymin={-0.5}, plot graphics/ymax={0.5}, point meta min={0.0}, point meta max={0.7}, plot graphics/!src={tikz/gallery/hoags_object/hoags_object-ngpl-2-srcs-gp-2-std-interp-viridis-0-1}, colormap name={viridis}},
	hoags_object-ngpl-2-srcs-gp-2-std-winterp/.style={plot graphics/xmin={-0.5}, plot graphics/xmax={0.5}, plot graphics/ymin={-0.5}, plot graphics/ymax={0.5}, point meta min={0.0}, point meta max={0.7}, plot graphics/!src={tikz/gallery/hoags_object/hoags_object-ngpl-2-srcs-gp-2-std-winterp-viridis-0-1}, colormap name={viridis}},
	hoags_object-ngpl-2-srcs-gp-2-std-interp-img/.style={plot graphics/xmin={-0.5}, plot graphics/xmax={0.5}, plot graphics/ymin={-0.5}, plot graphics/ymax={0.5}, point meta min={0.0}, point meta max={0.7}, plot graphics/!src={tikz/gallery/hoags_object/hoags_object-ngpl-2-srcs-gp-2-std-interp-img-viridis-0-1}, colormap name={viridis}},
	hoags_object-ngpl-3-img-mean/.style={plot graphics/xmin={-2.5}, plot graphics/xmax={2.5}, plot graphics/ymin={-2.5}, plot graphics/ymax={2.5}, point meta min={0.0}, point meta max={30.0}, plot graphics/!src={tikz/gallery/hoags_object/hoags_object-ngpl-3-img-mean-viridis-0-30}, colormap name={viridis}},
	hoags_object-ngpl-3-img-std/.style={plot graphics/xmin={-2.5}, plot graphics/xmax={2.5}, plot graphics/ymin={-2.5}, plot graphics/ymax={2.5}, point meta min={0.0}, point meta max={0.7}, plot graphics/!src={tikz/gallery/hoags_object/hoags_object-ngpl-3-img-std-viridis-0-1}, colormap name={viridis}},
	hoags_object-ngpl-3-simg-deproj-mean/.style={plot graphics/xmin={-0.5}, plot graphics/xmax={0.5}, plot graphics/ymin={-0.5}, plot graphics/ymax={0.5}, point meta min={0.0}, point meta max={30.0}, plot graphics/!src={tikz/gallery/hoags_object/hoags_object-ngpl-3-simg-deproj-mean-viridis-0-30}, colormap name={viridis}},
	hoags_object-ngpl-3-simg-deproj-std/.style={plot graphics/xmin={-0.5}, plot graphics/xmax={0.5}, plot graphics/ymin={-0.5}, plot graphics/ymax={0.5}, point meta min={0.0}, point meta max={0.7}, plot graphics/!src={tikz/gallery/hoags_object/hoags_object-ngpl-3-simg-deproj-std-viridis-0-1}, colormap name={viridis}},
	hoags_object-ngpl-3-simg-mean/.style={plot graphics/xmin={-0.5}, plot graphics/xmax={0.5}, plot graphics/ymin={-0.5}, plot graphics/ymax={0.5}, point meta min={0.0}, point meta max={30.0}, plot graphics/!src={tikz/gallery/hoags_object/hoags_object-ngpl-3-simg-mean-viridis-0-30}, colormap name={viridis}},
	hoags_object-ngpl-3-simg-std/.style={plot graphics/xmin={-0.5}, plot graphics/xmax={0.5}, plot graphics/ymin={-0.5}, plot graphics/ymax={0.5}, point meta min={0.0}, point meta max={0.7}, plot graphics/!src={tikz/gallery/hoags_object/hoags_object-ngpl-3-simg-std-viridis-0-1}, colormap name={viridis}},
	hoags_object-ngpl-3-simg-std-interp/.style={plot graphics/xmin={-0.5}, plot graphics/xmax={0.5}, plot graphics/ymin={-0.5}, plot graphics/ymax={0.5}, point meta min={0.0}, point meta max={0.7}, plot graphics/!src={tikz/gallery/hoags_object/hoags_object-ngpl-3-simg-std-interp-viridis-0-1}, colormap name={viridis}},
	hoags_object-ngpl-3-simg-std-winterp/.style={plot graphics/xmin={-0.5}, plot graphics/xmax={0.5}, plot graphics/ymin={-0.5}, plot graphics/ymax={0.5}, point meta min={0.0}, point meta max={0.7}, plot graphics/!src={tikz/gallery/hoags_object/hoags_object-ngpl-3-simg-std-winterp-viridis-0-1}, colormap name={viridis}},
	hoags_object-ngpl-3-simg-std-interp-img/.style={plot graphics/xmin={-0.5}, plot graphics/xmax={0.5}, plot graphics/ymin={-0.5}, plot graphics/ymax={0.5}, point meta min={0.0}, point meta max={0.7}, plot graphics/!src={tikz/gallery/hoags_object/hoags_object-ngpl-3-simg-std-interp-img-viridis-0-1}, colormap name={viridis}},
	hoags_object-ngpl-3-srcs-src-mean/.style={plot graphics/xmin={-0.5}, plot graphics/xmax={0.5}, plot graphics/ymin={-0.5}, plot graphics/ymax={0.5}, point meta min={-0.5899906158447266}, point meta max={29.335567474365234}, plot graphics/!src={tikz/gallery/hoags_object/hoags_object-ngpl-3-srcs-src-mean-viridis--1-29}, colormap name={viridis}},
	hoags_object-ngpl-3-srcs-src-std/.style={plot graphics/xmin={-0.5}, plot graphics/xmax={0.5}, plot graphics/ymin={-0.5}, plot graphics/ymax={0.5}, point meta min={0.0}, point meta max={0.7}, plot graphics/!src={tikz/gallery/hoags_object/hoags_object-ngpl-3-srcs-src-std-viridis-0-1}, colormap name={viridis}},
	hoags_object-ngpl-3-srcs-gp-1-mean/.style={plot graphics/xmin={-0.5}, plot graphics/xmax={0.5}, plot graphics/ymin={-0.5}, plot graphics/ymax={0.5}, point meta min={-1.80923330783844}, point meta max={4.679297924041748}, plot graphics/!src={tikz/gallery/hoags_object/hoags_object-ngpl-3-srcs-gp-1-mean-viridis--2-5}, colormap name={viridis}},
	hoags_object-ngpl-3-srcs-gp-1-std/.style={plot graphics/xmin={-0.5}, plot graphics/xmax={0.5}, plot graphics/ymin={-0.5}, plot graphics/ymax={0.5}, point meta min={0.0}, point meta max={0.7}, plot graphics/!src={tikz/gallery/hoags_object/hoags_object-ngpl-3-srcs-gp-1-std-viridis-0-1}, colormap name={viridis}},
	hoags_object-ngpl-3-srcs-gp-1-std-interp/.style={plot graphics/xmin={-0.5}, plot graphics/xmax={0.5}, plot graphics/ymin={-0.5}, plot graphics/ymax={0.5}, point meta min={0.0}, point meta max={0.7}, plot graphics/!src={tikz/gallery/hoags_object/hoags_object-ngpl-3-srcs-gp-1-std-interp-viridis-0-1}, colormap name={viridis}},
	hoags_object-ngpl-3-srcs-gp-1-std-winterp/.style={plot graphics/xmin={-0.5}, plot graphics/xmax={0.5}, plot graphics/ymin={-0.5}, plot graphics/ymax={0.5}, point meta min={0.0}, point meta max={0.7}, plot graphics/!src={tikz/gallery/hoags_object/hoags_object-ngpl-3-srcs-gp-1-std-winterp-viridis-0-1}, colormap name={viridis}},
	hoags_object-ngpl-3-srcs-gp-1-std-interp-img/.style={plot graphics/xmin={-0.5}, plot graphics/xmax={0.5}, plot graphics/ymin={-0.5}, plot graphics/ymax={0.5}, point meta min={0.0}, point meta max={0.7}, plot graphics/!src={tikz/gallery/hoags_object/hoags_object-ngpl-3-srcs-gp-1-std-interp-img-viridis-0-1}, colormap name={viridis}},
	hoags_object-ngpl-3-srcs-gp-2-mean/.style={plot graphics/xmin={-0.5}, plot graphics/xmax={0.5}, plot graphics/ymin={-0.5}, plot graphics/ymax={0.5}, point meta min={-0.4605604410171509}, point meta max={1.274808645248413}, plot graphics/!src={tikz/gallery/hoags_object/hoags_object-ngpl-3-srcs-gp-2-mean-viridis--0-1}, colormap name={viridis}},
	hoags_object-ngpl-3-srcs-gp-2-std/.style={plot graphics/xmin={-0.5}, plot graphics/xmax={0.5}, plot graphics/ymin={-0.5}, plot graphics/ymax={0.5}, point meta min={0.0}, point meta max={0.7}, plot graphics/!src={tikz/gallery/hoags_object/hoags_object-ngpl-3-srcs-gp-2-std-viridis-0-1}, colormap name={viridis}},
	hoags_object-ngpl-3-srcs-gp-2-std-interp/.style={plot graphics/xmin={-0.5}, plot graphics/xmax={0.5}, plot graphics/ymin={-0.5}, plot graphics/ymax={0.5}, point meta min={0.0}, point meta max={0.7}, plot graphics/!src={tikz/gallery/hoags_object/hoags_object-ngpl-3-srcs-gp-2-std-interp-viridis-0-1}, colormap name={viridis}},
	hoags_object-ngpl-3-srcs-gp-2-std-winterp/.style={plot graphics/xmin={-0.5}, plot graphics/xmax={0.5}, plot graphics/ymin={-0.5}, plot graphics/ymax={0.5}, point meta min={0.0}, point meta max={0.7}, plot graphics/!src={tikz/gallery/hoags_object/hoags_object-ngpl-3-srcs-gp-2-std-winterp-viridis-0-1}, colormap name={viridis}},
	hoags_object-ngpl-3-srcs-gp-2-std-interp-img/.style={plot graphics/xmin={-0.5}, plot graphics/xmax={0.5}, plot graphics/ymin={-0.5}, plot graphics/ymax={0.5}, point meta min={0.0}, point meta max={0.7}, plot graphics/!src={tikz/gallery/hoags_object/hoags_object-ngpl-3-srcs-gp-2-std-interp-img-viridis-0-1}, colormap name={viridis}},
	hoags_object-ngpl-3-srcs-gp-3-mean/.style={plot graphics/xmin={-0.5}, plot graphics/xmax={0.5}, plot graphics/ymin={-0.5}, plot graphics/ymax={0.5}, point meta min={-0.3167674243450165}, point meta max={0.5735495686531067}, plot graphics/!src={tikz/gallery/hoags_object/hoags_object-ngpl-3-srcs-gp-3-mean-viridis--0-1}, colormap name={viridis}},
	hoags_object-ngpl-3-srcs-gp-3-std/.style={plot graphics/xmin={-0.5}, plot graphics/xmax={0.5}, plot graphics/ymin={-0.5}, plot graphics/ymax={0.5}, point meta min={0.0}, point meta max={0.7}, plot graphics/!src={tikz/gallery/hoags_object/hoags_object-ngpl-3-srcs-gp-3-std-viridis-0-1}, colormap name={viridis}},
	hoags_object-ngpl-3-srcs-gp-3-std-interp/.style={plot graphics/xmin={-0.5}, plot graphics/xmax={0.5}, plot graphics/ymin={-0.5}, plot graphics/ymax={0.5}, point meta min={0.0}, point meta max={0.7}, plot graphics/!src={tikz/gallery/hoags_object/hoags_object-ngpl-3-srcs-gp-3-std-interp-viridis-0-1}, colormap name={viridis}},
	hoags_object-ngpl-3-srcs-gp-3-std-winterp/.style={plot graphics/xmin={-0.5}, plot graphics/xmax={0.5}, plot graphics/ymin={-0.5}, plot graphics/ymax={0.5}, point meta min={0.0}, point meta max={0.7}, plot graphics/!src={tikz/gallery/hoags_object/hoags_object-ngpl-3-srcs-gp-3-std-winterp-viridis-0-1}, colormap name={viridis}},
	hoags_object-ngpl-3-srcs-gp-3-std-interp-img/.style={plot graphics/xmin={-0.5}, plot graphics/xmax={0.5}, plot graphics/ymin={-0.5}, plot graphics/ymax={0.5}, point meta min={0.0}, point meta max={0.7}, plot graphics/!src={tikz/gallery/hoags_object/hoags_object-ngpl-3-srcs-gp-3-std-interp-img-viridis-0-1}, colormap name={viridis}},
	hoags_object-ngpl-4-img-mean/.style={plot graphics/xmin={-2.5}, plot graphics/xmax={2.5}, plot graphics/ymin={-2.5}, plot graphics/ymax={2.5}, point meta min={0.0}, point meta max={30.0}, plot graphics/!src={tikz/gallery/hoags_object/hoags_object-ngpl-4-img-mean-viridis-0-30}, colormap name={viridis}},
	hoags_object-ngpl-4-img-std/.style={plot graphics/xmin={-2.5}, plot graphics/xmax={2.5}, plot graphics/ymin={-2.5}, plot graphics/ymax={2.5}, point meta min={0.0}, point meta max={0.7}, plot graphics/!src={tikz/gallery/hoags_object/hoags_object-ngpl-4-img-std-viridis-0-1}, colormap name={viridis}},
	hoags_object-ngpl-4-simg-deproj-mean/.style={plot graphics/xmin={-0.5}, plot graphics/xmax={0.5}, plot graphics/ymin={-0.5}, plot graphics/ymax={0.5}, point meta min={0.0}, point meta max={30.0}, plot graphics/!src={tikz/gallery/hoags_object/hoags_object-ngpl-4-simg-deproj-mean-viridis-0-30}, colormap name={viridis}},
	hoags_object-ngpl-4-simg-deproj-std/.style={plot graphics/xmin={-0.5}, plot graphics/xmax={0.5}, plot graphics/ymin={-0.5}, plot graphics/ymax={0.5}, point meta min={0.0}, point meta max={0.7}, plot graphics/!src={tikz/gallery/hoags_object/hoags_object-ngpl-4-simg-deproj-std-viridis-0-1}, colormap name={viridis}},
	hoags_object-ngpl-4-simg-mean/.style={plot graphics/xmin={-0.5}, plot graphics/xmax={0.5}, plot graphics/ymin={-0.5}, plot graphics/ymax={0.5}, point meta min={0.0}, point meta max={30.0}, plot graphics/!src={tikz/gallery/hoags_object/hoags_object-ngpl-4-simg-mean-viridis-0-30}, colormap name={viridis}},
	hoags_object-ngpl-4-simg-std/.style={plot graphics/xmin={-0.5}, plot graphics/xmax={0.5}, plot graphics/ymin={-0.5}, plot graphics/ymax={0.5}, point meta min={0.0}, point meta max={0.7}, plot graphics/!src={tikz/gallery/hoags_object/hoags_object-ngpl-4-simg-std-viridis-0-1}, colormap name={viridis}},
	hoags_object-ngpl-4-simg-std-interp/.style={plot graphics/xmin={-0.5}, plot graphics/xmax={0.5}, plot graphics/ymin={-0.5}, plot graphics/ymax={0.5}, point meta min={0.0}, point meta max={0.7}, plot graphics/!src={tikz/gallery/hoags_object/hoags_object-ngpl-4-simg-std-interp-viridis-0-1}, colormap name={viridis}},
	hoags_object-ngpl-4-simg-std-winterp/.style={plot graphics/xmin={-0.5}, plot graphics/xmax={0.5}, plot graphics/ymin={-0.5}, plot graphics/ymax={0.5}, point meta min={0.0}, point meta max={0.7}, plot graphics/!src={tikz/gallery/hoags_object/hoags_object-ngpl-4-simg-std-winterp-viridis-0-1}, colormap name={viridis}},
	hoags_object-ngpl-4-simg-std-interp-img/.style={plot graphics/xmin={-0.5}, plot graphics/xmax={0.5}, plot graphics/ymin={-0.5}, plot graphics/ymax={0.5}, point meta min={0.0}, point meta max={0.7}, plot graphics/!src={tikz/gallery/hoags_object/hoags_object-ngpl-4-simg-std-interp-img-viridis-0-1}, colormap name={viridis}},
	hoags_object-ngpl-4-srcs-src-mean/.style={plot graphics/xmin={-0.5}, plot graphics/xmax={0.5}, plot graphics/ymin={-0.5}, plot graphics/ymax={0.5}, point meta min={-0.6443514823913574}, point meta max={29.44578742980957}, plot graphics/!src={tikz/gallery/hoags_object/hoags_object-ngpl-4-srcs-src-mean-viridis--1-29}, colormap name={viridis}},
	hoags_object-ngpl-4-srcs-src-std/.style={plot graphics/xmin={-0.5}, plot graphics/xmax={0.5}, plot graphics/ymin={-0.5}, plot graphics/ymax={0.5}, point meta min={0.0}, point meta max={0.7}, plot graphics/!src={tikz/gallery/hoags_object/hoags_object-ngpl-4-srcs-src-std-viridis-0-1}, colormap name={viridis}},
	hoags_object-ngpl-4-srcs-gp-1-mean/.style={plot graphics/xmin={-0.5}, plot graphics/xmax={0.5}, plot graphics/ymin={-0.5}, plot graphics/ymax={0.5}, point meta min={-1.2262486219406128}, point meta max={3.127842664718628}, plot graphics/!src={tikz/gallery/hoags_object/hoags_object-ngpl-4-srcs-gp-1-mean-viridis--1-3}, colormap name={viridis}},
	hoags_object-ngpl-4-srcs-gp-1-std/.style={plot graphics/xmin={-0.5}, plot graphics/xmax={0.5}, plot graphics/ymin={-0.5}, plot graphics/ymax={0.5}, point meta min={0.0}, point meta max={0.7}, plot graphics/!src={tikz/gallery/hoags_object/hoags_object-ngpl-4-srcs-gp-1-std-viridis-0-1}, colormap name={viridis}},
	hoags_object-ngpl-4-srcs-gp-1-std-interp/.style={plot graphics/xmin={-0.5}, plot graphics/xmax={0.5}, plot graphics/ymin={-0.5}, plot graphics/ymax={0.5}, point meta min={0.0}, point meta max={0.7}, plot graphics/!src={tikz/gallery/hoags_object/hoags_object-ngpl-4-srcs-gp-1-std-interp-viridis-0-1}, colormap name={viridis}},
	hoags_object-ngpl-4-srcs-gp-1-std-winterp/.style={plot graphics/xmin={-0.5}, plot graphics/xmax={0.5}, plot graphics/ymin={-0.5}, plot graphics/ymax={0.5}, point meta min={0.0}, point meta max={0.7}, plot graphics/!src={tikz/gallery/hoags_object/hoags_object-ngpl-4-srcs-gp-1-std-winterp-viridis-0-1}, colormap name={viridis}},
	hoags_object-ngpl-4-srcs-gp-1-std-interp-img/.style={plot graphics/xmin={-0.5}, plot graphics/xmax={0.5}, plot graphics/ymin={-0.5}, plot graphics/ymax={0.5}, point meta min={0.0}, point meta max={0.7}, plot graphics/!src={tikz/gallery/hoags_object/hoags_object-ngpl-4-srcs-gp-1-std-interp-img-viridis-0-1}, colormap name={viridis}},
	hoags_object-ngpl-4-srcs-gp-2-mean/.style={plot graphics/xmin={-0.5}, plot graphics/xmax={0.5}, plot graphics/ymin={-0.5}, plot graphics/ymax={0.5}, point meta min={-0.6314948797225952}, point meta max={2.2603256702423096}, plot graphics/!src={tikz/gallery/hoags_object/hoags_object-ngpl-4-srcs-gp-2-mean-viridis--1-2}, colormap name={viridis}},
	hoags_object-ngpl-4-srcs-gp-2-std/.style={plot graphics/xmin={-0.5}, plot graphics/xmax={0.5}, plot graphics/ymin={-0.5}, plot graphics/ymax={0.5}, point meta min={0.0}, point meta max={0.7}, plot graphics/!src={tikz/gallery/hoags_object/hoags_object-ngpl-4-srcs-gp-2-std-viridis-0-1}, colormap name={viridis}},
	hoags_object-ngpl-4-srcs-gp-2-std-interp/.style={plot graphics/xmin={-0.5}, plot graphics/xmax={0.5}, plot graphics/ymin={-0.5}, plot graphics/ymax={0.5}, point meta min={0.0}, point meta max={0.7}, plot graphics/!src={tikz/gallery/hoags_object/hoags_object-ngpl-4-srcs-gp-2-std-interp-viridis-0-1}, colormap name={viridis}},
	hoags_object-ngpl-4-srcs-gp-2-std-winterp/.style={plot graphics/xmin={-0.5}, plot graphics/xmax={0.5}, plot graphics/ymin={-0.5}, plot graphics/ymax={0.5}, point meta min={0.0}, point meta max={0.7}, plot graphics/!src={tikz/gallery/hoags_object/hoags_object-ngpl-4-srcs-gp-2-std-winterp-viridis-0-1}, colormap name={viridis}},
	hoags_object-ngpl-4-srcs-gp-2-std-interp-img/.style={plot graphics/xmin={-0.5}, plot graphics/xmax={0.5}, plot graphics/ymin={-0.5}, plot graphics/ymax={0.5}, point meta min={0.0}, point meta max={0.7}, plot graphics/!src={tikz/gallery/hoags_object/hoags_object-ngpl-4-srcs-gp-2-std-interp-img-viridis-0-1}, colormap name={viridis}},
	hoags_object-ngpl-4-srcs-gp-3-mean/.style={plot graphics/xmin={-0.5}, plot graphics/xmax={0.5}, plot graphics/ymin={-0.5}, plot graphics/ymax={0.5}, point meta min={-0.3881877064704895}, point meta max={1.0376837253570557}, plot graphics/!src={tikz/gallery/hoags_object/hoags_object-ngpl-4-srcs-gp-3-mean-viridis--0-1}, colormap name={viridis}},
	hoags_object-ngpl-4-srcs-gp-3-std/.style={plot graphics/xmin={-0.5}, plot graphics/xmax={0.5}, plot graphics/ymin={-0.5}, plot graphics/ymax={0.5}, point meta min={0.0}, point meta max={0.7}, plot graphics/!src={tikz/gallery/hoags_object/hoags_object-ngpl-4-srcs-gp-3-std-viridis-0-1}, colormap name={viridis}},
	hoags_object-ngpl-4-srcs-gp-3-std-interp/.style={plot graphics/xmin={-0.5}, plot graphics/xmax={0.5}, plot graphics/ymin={-0.5}, plot graphics/ymax={0.5}, point meta min={0.0}, point meta max={0.7}, plot graphics/!src={tikz/gallery/hoags_object/hoags_object-ngpl-4-srcs-gp-3-std-interp-viridis-0-1}, colormap name={viridis}},
	hoags_object-ngpl-4-srcs-gp-3-std-winterp/.style={plot graphics/xmin={-0.5}, plot graphics/xmax={0.5}, plot graphics/ymin={-0.5}, plot graphics/ymax={0.5}, point meta min={0.0}, point meta max={0.7}, plot graphics/!src={tikz/gallery/hoags_object/hoags_object-ngpl-4-srcs-gp-3-std-winterp-viridis-0-1}, colormap name={viridis}},
	hoags_object-ngpl-4-srcs-gp-3-std-interp-img/.style={plot graphics/xmin={-0.5}, plot graphics/xmax={0.5}, plot graphics/ymin={-0.5}, plot graphics/ymax={0.5}, point meta min={0.0}, point meta max={0.7}, plot graphics/!src={tikz/gallery/hoags_object/hoags_object-ngpl-4-srcs-gp-3-std-interp-img-viridis-0-1}, colormap name={viridis}},
	hoags_object-ngpl-4-srcs-gp-4-mean/.style={plot graphics/xmin={-0.5}, plot graphics/xmax={0.5}, plot graphics/ymin={-0.5}, plot graphics/ymax={0.5}, point meta min={-0.3465268015861511}, point meta max={0.7156921029090881}, plot graphics/!src={tikz/gallery/hoags_object/hoags_object-ngpl-4-srcs-gp-4-mean-viridis--0-1}, colormap name={viridis}},
	hoags_object-ngpl-4-srcs-gp-4-std/.style={plot graphics/xmin={-0.5}, plot graphics/xmax={0.5}, plot graphics/ymin={-0.5}, plot graphics/ymax={0.5}, point meta min={0.0}, point meta max={0.7}, plot graphics/!src={tikz/gallery/hoags_object/hoags_object-ngpl-4-srcs-gp-4-std-viridis-0-1}, colormap name={viridis}},
	hoags_object-ngpl-4-srcs-gp-4-std-interp/.style={plot graphics/xmin={-0.5}, plot graphics/xmax={0.5}, plot graphics/ymin={-0.5}, plot graphics/ymax={0.5}, point meta min={0.0}, point meta max={0.7}, plot graphics/!src={tikz/gallery/hoags_object/hoags_object-ngpl-4-srcs-gp-4-std-interp-viridis-0-1}, colormap name={viridis}},
	hoags_object-ngpl-4-srcs-gp-4-std-winterp/.style={plot graphics/xmin={-0.5}, plot graphics/xmax={0.5}, plot graphics/ymin={-0.5}, plot graphics/ymax={0.5}, point meta min={0.0}, point meta max={0.7}, plot graphics/!src={tikz/gallery/hoags_object/hoags_object-ngpl-4-srcs-gp-4-std-winterp-viridis-0-1}, colormap name={viridis}},
	hoags_object-ngpl-4-srcs-gp-4-std-interp-img/.style={plot graphics/xmin={-0.5}, plot graphics/xmax={0.5}, plot graphics/ymin={-0.5}, plot graphics/ymax={0.5}, point meta min={0.0}, point meta max={0.7}, plot graphics/!src={tikz/gallery/hoags_object/hoags_object-ngpl-4-srcs-gp-4-std-interp-img-viridis-0-1}, colormap name={viridis}},
	hoags_object-ngpl-5-img-mean/.style={plot graphics/xmin={-2.5}, plot graphics/xmax={2.5}, plot graphics/ymin={-2.5}, plot graphics/ymax={2.5}, point meta min={0.0}, point meta max={30.0}, plot graphics/!src={tikz/gallery/hoags_object/hoags_object-ngpl-5-img-mean-viridis-0-30}, colormap name={viridis}},
	hoags_object-ngpl-5-img-std/.style={plot graphics/xmin={-2.5}, plot graphics/xmax={2.5}, plot graphics/ymin={-2.5}, plot graphics/ymax={2.5}, point meta min={0.0}, point meta max={0.7}, plot graphics/!src={tikz/gallery/hoags_object/hoags_object-ngpl-5-img-std-viridis-0-1}, colormap name={viridis}},
	hoags_object-ngpl-5-simg-deproj-mean/.style={plot graphics/xmin={-0.5}, plot graphics/xmax={0.5}, plot graphics/ymin={-0.5}, plot graphics/ymax={0.5}, point meta min={0.0}, point meta max={30.0}, plot graphics/!src={tikz/gallery/hoags_object/hoags_object-ngpl-5-simg-deproj-mean-viridis-0-30}, colormap name={viridis}},
	hoags_object-ngpl-5-simg-deproj-std/.style={plot graphics/xmin={-0.5}, plot graphics/xmax={0.5}, plot graphics/ymin={-0.5}, plot graphics/ymax={0.5}, point meta min={0.0}, point meta max={0.7}, plot graphics/!src={tikz/gallery/hoags_object/hoags_object-ngpl-5-simg-deproj-std-viridis-0-1}, colormap name={viridis}},
	hoags_object-ngpl-5-simg-mean/.style={plot graphics/xmin={-0.5}, plot graphics/xmax={0.5}, plot graphics/ymin={-0.5}, plot graphics/ymax={0.5}, point meta min={0.0}, point meta max={30.0}, plot graphics/!src={tikz/gallery/hoags_object/hoags_object-ngpl-5-simg-mean-viridis-0-30}, colormap name={viridis}},
	hoags_object-ngpl-5-simg-std/.style={plot graphics/xmin={-0.5}, plot graphics/xmax={0.5}, plot graphics/ymin={-0.5}, plot graphics/ymax={0.5}, point meta min={0.0}, point meta max={0.7}, plot graphics/!src={tikz/gallery/hoags_object/hoags_object-ngpl-5-simg-std-viridis-0-1}, colormap name={viridis}},
	hoags_object-ngpl-5-simg-std-interp/.style={plot graphics/xmin={-0.5}, plot graphics/xmax={0.5}, plot graphics/ymin={-0.5}, plot graphics/ymax={0.5}, point meta min={0.0}, point meta max={0.7}, plot graphics/!src={tikz/gallery/hoags_object/hoags_object-ngpl-5-simg-std-interp-viridis-0-1}, colormap name={viridis}},
	hoags_object-ngpl-5-simg-std-winterp/.style={plot graphics/xmin={-0.5}, plot graphics/xmax={0.5}, plot graphics/ymin={-0.5}, plot graphics/ymax={0.5}, point meta min={0.0}, point meta max={0.7}, plot graphics/!src={tikz/gallery/hoags_object/hoags_object-ngpl-5-simg-std-winterp-viridis-0-1}, colormap name={viridis}},
	hoags_object-ngpl-5-simg-std-interp-img/.style={plot graphics/xmin={-0.5}, plot graphics/xmax={0.5}, plot graphics/ymin={-0.5}, plot graphics/ymax={0.5}, point meta min={0.0}, point meta max={0.7}, plot graphics/!src={tikz/gallery/hoags_object/hoags_object-ngpl-5-simg-std-interp-img-viridis-0-1}, colormap name={viridis}},
	hoags_object-ngpl-5-srcs-src-mean/.style={plot graphics/xmin={-0.5}, plot graphics/xmax={0.5}, plot graphics/ymin={-0.5}, plot graphics/ymax={0.5}, point meta min={-0.6083585023880005}, point meta max={29.380146026611328}, plot graphics/!src={tikz/gallery/hoags_object/hoags_object-ngpl-5-srcs-src-mean-viridis--1-29}, colormap name={viridis}},
	hoags_object-ngpl-5-srcs-src-std/.style={plot graphics/xmin={-0.5}, plot graphics/xmax={0.5}, plot graphics/ymin={-0.5}, plot graphics/ymax={0.5}, point meta min={0.0}, point meta max={0.7}, plot graphics/!src={tikz/gallery/hoags_object/hoags_object-ngpl-5-srcs-src-std-viridis-0-1}, colormap name={viridis}},
	hoags_object-ngpl-5-srcs-gp-1-mean/.style={plot graphics/xmin={-0.5}, plot graphics/xmax={0.5}, plot graphics/ymin={-0.5}, plot graphics/ymax={0.5}, point meta min={-0.8965653777122498}, point meta max={2.0961127281188965}, plot graphics/!src={tikz/gallery/hoags_object/hoags_object-ngpl-5-srcs-gp-1-mean-viridis--1-2}, colormap name={viridis}},
	hoags_object-ngpl-5-srcs-gp-1-std/.style={plot graphics/xmin={-0.5}, plot graphics/xmax={0.5}, plot graphics/ymin={-0.5}, plot graphics/ymax={0.5}, point meta min={0.0}, point meta max={0.7}, plot graphics/!src={tikz/gallery/hoags_object/hoags_object-ngpl-5-srcs-gp-1-std-viridis-0-1}, colormap name={viridis}},
	hoags_object-ngpl-5-srcs-gp-1-std-interp/.style={plot graphics/xmin={-0.5}, plot graphics/xmax={0.5}, plot graphics/ymin={-0.5}, plot graphics/ymax={0.5}, point meta min={0.0}, point meta max={0.7}, plot graphics/!src={tikz/gallery/hoags_object/hoags_object-ngpl-5-srcs-gp-1-std-interp-viridis-0-1}, colormap name={viridis}},
	hoags_object-ngpl-5-srcs-gp-1-std-winterp/.style={plot graphics/xmin={-0.5}, plot graphics/xmax={0.5}, plot graphics/ymin={-0.5}, plot graphics/ymax={0.5}, point meta min={0.0}, point meta max={0.7}, plot graphics/!src={tikz/gallery/hoags_object/hoags_object-ngpl-5-srcs-gp-1-std-winterp-viridis-0-1}, colormap name={viridis}},
	hoags_object-ngpl-5-srcs-gp-1-std-interp-img/.style={plot graphics/xmin={-0.5}, plot graphics/xmax={0.5}, plot graphics/ymin={-0.5}, plot graphics/ymax={0.5}, point meta min={0.0}, point meta max={0.7}, plot graphics/!src={tikz/gallery/hoags_object/hoags_object-ngpl-5-srcs-gp-1-std-interp-img-viridis-0-1}, colormap name={viridis}},
	hoags_object-ngpl-5-srcs-gp-2-mean/.style={plot graphics/xmin={-0.5}, plot graphics/xmax={0.5}, plot graphics/ymin={-0.5}, plot graphics/ymax={0.5}, point meta min={-0.7454653978347778}, point meta max={2.4723470211029053}, plot graphics/!src={tikz/gallery/hoags_object/hoags_object-ngpl-5-srcs-gp-2-mean-viridis--1-2}, colormap name={viridis}},
	hoags_object-ngpl-5-srcs-gp-2-std/.style={plot graphics/xmin={-0.5}, plot graphics/xmax={0.5}, plot graphics/ymin={-0.5}, plot graphics/ymax={0.5}, point meta min={0.0}, point meta max={0.7}, plot graphics/!src={tikz/gallery/hoags_object/hoags_object-ngpl-5-srcs-gp-2-std-viridis-0-1}, colormap name={viridis}},
	hoags_object-ngpl-5-srcs-gp-2-std-interp/.style={plot graphics/xmin={-0.5}, plot graphics/xmax={0.5}, plot graphics/ymin={-0.5}, plot graphics/ymax={0.5}, point meta min={0.0}, point meta max={0.7}, plot graphics/!src={tikz/gallery/hoags_object/hoags_object-ngpl-5-srcs-gp-2-std-interp-viridis-0-1}, colormap name={viridis}},
	hoags_object-ngpl-5-srcs-gp-2-std-winterp/.style={plot graphics/xmin={-0.5}, plot graphics/xmax={0.5}, plot graphics/ymin={-0.5}, plot graphics/ymax={0.5}, point meta min={0.0}, point meta max={0.7}, plot graphics/!src={tikz/gallery/hoags_object/hoags_object-ngpl-5-srcs-gp-2-std-winterp-viridis-0-1}, colormap name={viridis}},
	hoags_object-ngpl-5-srcs-gp-2-std-interp-img/.style={plot graphics/xmin={-0.5}, plot graphics/xmax={0.5}, plot graphics/ymin={-0.5}, plot graphics/ymax={0.5}, point meta min={0.0}, point meta max={0.7}, plot graphics/!src={tikz/gallery/hoags_object/hoags_object-ngpl-5-srcs-gp-2-std-interp-img-viridis-0-1}, colormap name={viridis}},
	hoags_object-ngpl-5-srcs-gp-3-mean/.style={plot graphics/xmin={-0.5}, plot graphics/xmax={0.5}, plot graphics/ymin={-0.5}, plot graphics/ymax={0.5}, point meta min={-0.44970884919166565}, point meta max={1.459811806678772}, plot graphics/!src={tikz/gallery/hoags_object/hoags_object-ngpl-5-srcs-gp-3-mean-viridis--0-1}, colormap name={viridis}},
	hoags_object-ngpl-5-srcs-gp-3-std/.style={plot graphics/xmin={-0.5}, plot graphics/xmax={0.5}, plot graphics/ymin={-0.5}, plot graphics/ymax={0.5}, point meta min={0.0}, point meta max={0.7}, plot graphics/!src={tikz/gallery/hoags_object/hoags_object-ngpl-5-srcs-gp-3-std-viridis-0-1}, colormap name={viridis}},
	hoags_object-ngpl-5-srcs-gp-3-std-interp/.style={plot graphics/xmin={-0.5}, plot graphics/xmax={0.5}, plot graphics/ymin={-0.5}, plot graphics/ymax={0.5}, point meta min={0.0}, point meta max={0.7}, plot graphics/!src={tikz/gallery/hoags_object/hoags_object-ngpl-5-srcs-gp-3-std-interp-viridis-0-1}, colormap name={viridis}},
	hoags_object-ngpl-5-srcs-gp-3-std-winterp/.style={plot graphics/xmin={-0.5}, plot graphics/xmax={0.5}, plot graphics/ymin={-0.5}, plot graphics/ymax={0.5}, point meta min={0.0}, point meta max={0.7}, plot graphics/!src={tikz/gallery/hoags_object/hoags_object-ngpl-5-srcs-gp-3-std-winterp-viridis-0-1}, colormap name={viridis}},
	hoags_object-ngpl-5-srcs-gp-3-std-interp-img/.style={plot graphics/xmin={-0.5}, plot graphics/xmax={0.5}, plot graphics/ymin={-0.5}, plot graphics/ymax={0.5}, point meta min={0.0}, point meta max={0.7}, plot graphics/!src={tikz/gallery/hoags_object/hoags_object-ngpl-5-srcs-gp-3-std-interp-img-viridis-0-1}, colormap name={viridis}},
	hoags_object-ngpl-5-srcs-gp-4-mean/.style={plot graphics/xmin={-0.5}, plot graphics/xmax={0.5}, plot graphics/ymin={-0.5}, plot graphics/ymax={0.5}, point meta min={-0.4279448986053467}, point meta max={0.9171560406684875}, plot graphics/!src={tikz/gallery/hoags_object/hoags_object-ngpl-5-srcs-gp-4-mean-viridis--0-1}, colormap name={viridis}},
	hoags_object-ngpl-5-srcs-gp-4-std/.style={plot graphics/xmin={-0.5}, plot graphics/xmax={0.5}, plot graphics/ymin={-0.5}, plot graphics/ymax={0.5}, point meta min={0.0}, point meta max={0.7}, plot graphics/!src={tikz/gallery/hoags_object/hoags_object-ngpl-5-srcs-gp-4-std-viridis-0-1}, colormap name={viridis}},
	hoags_object-ngpl-5-srcs-gp-4-std-interp/.style={plot graphics/xmin={-0.5}, plot graphics/xmax={0.5}, plot graphics/ymin={-0.5}, plot graphics/ymax={0.5}, point meta min={0.0}, point meta max={0.7}, plot graphics/!src={tikz/gallery/hoags_object/hoags_object-ngpl-5-srcs-gp-4-std-interp-viridis-0-1}, colormap name={viridis}},
	hoags_object-ngpl-5-srcs-gp-4-std-winterp/.style={plot graphics/xmin={-0.5}, plot graphics/xmax={0.5}, plot graphics/ymin={-0.5}, plot graphics/ymax={0.5}, point meta min={0.0}, point meta max={0.7}, plot graphics/!src={tikz/gallery/hoags_object/hoags_object-ngpl-5-srcs-gp-4-std-winterp-viridis-0-1}, colormap name={viridis}},
	hoags_object-ngpl-5-srcs-gp-4-std-interp-img/.style={plot graphics/xmin={-0.5}, plot graphics/xmax={0.5}, plot graphics/ymin={-0.5}, plot graphics/ymax={0.5}, point meta min={0.0}, point meta max={0.7}, plot graphics/!src={tikz/gallery/hoags_object/hoags_object-ngpl-5-srcs-gp-4-std-interp-img-viridis-0-1}, colormap name={viridis}},
	hoags_object-ngpl-5-srcs-gp-5-mean/.style={plot graphics/xmin={-0.5}, plot graphics/xmax={0.5}, plot graphics/ymin={-0.5}, plot graphics/ymax={0.5}, point meta min={-0.4417813718318939}, point meta max={0.6601653695106506}, plot graphics/!src={tikz/gallery/hoags_object/hoags_object-ngpl-5-srcs-gp-5-mean-viridis--0-1}, colormap name={viridis}},
	hoags_object-ngpl-5-srcs-gp-5-std/.style={plot graphics/xmin={-0.5}, plot graphics/xmax={0.5}, plot graphics/ymin={-0.5}, plot graphics/ymax={0.5}, point meta min={0.0}, point meta max={0.7}, plot graphics/!src={tikz/gallery/hoags_object/hoags_object-ngpl-5-srcs-gp-5-std-viridis-0-1}, colormap name={viridis}},
	hoags_object-ngpl-5-srcs-gp-5-std-interp/.style={plot graphics/xmin={-0.5}, plot graphics/xmax={0.5}, plot graphics/ymin={-0.5}, plot graphics/ymax={0.5}, point meta min={0.0}, point meta max={0.7}, plot graphics/!src={tikz/gallery/hoags_object/hoags_object-ngpl-5-srcs-gp-5-std-interp-viridis-0-1}, colormap name={viridis}},
	hoags_object-ngpl-5-srcs-gp-5-std-winterp/.style={plot graphics/xmin={-0.5}, plot graphics/xmax={0.5}, plot graphics/ymin={-0.5}, plot graphics/ymax={0.5}, point meta min={0.0}, point meta max={0.7}, plot graphics/!src={tikz/gallery/hoags_object/hoags_object-ngpl-5-srcs-gp-5-std-winterp-viridis-0-1}, colormap name={viridis}},
	hoags_object-ngpl-5-srcs-gp-5-std-interp-img/.style={plot graphics/xmin={-0.5}, plot graphics/xmax={0.5}, plot graphics/ymin={-0.5}, plot graphics/ymax={0.5}, point meta min={0.0}, point meta max={0.7}, plot graphics/!src={tikz/gallery/hoags_object/hoags_object-ngpl-5-srcs-gp-5-std-interp-img-viridis-0-1}, colormap name={viridis}},
	hoags_object-ngpl-6-img-mean/.style={plot graphics/xmin={-2.5}, plot graphics/xmax={2.5}, plot graphics/ymin={-2.5}, plot graphics/ymax={2.5}, point meta min={0.0}, point meta max={30.0}, plot graphics/!src={tikz/gallery/hoags_object/hoags_object-ngpl-6-img-mean-viridis-0-30}, colormap name={viridis}},
	hoags_object-ngpl-6-img-std/.style={plot graphics/xmin={-2.5}, plot graphics/xmax={2.5}, plot graphics/ymin={-2.5}, plot graphics/ymax={2.5}, point meta min={0.0}, point meta max={0.7}, plot graphics/!src={tikz/gallery/hoags_object/hoags_object-ngpl-6-img-std-viridis-0-1}, colormap name={viridis}},
	hoags_object-ngpl-6-simg-deproj-mean/.style={plot graphics/xmin={-0.5}, plot graphics/xmax={0.5}, plot graphics/ymin={-0.5}, plot graphics/ymax={0.5}, point meta min={0.0}, point meta max={30.0}, plot graphics/!src={tikz/gallery/hoags_object/hoags_object-ngpl-6-simg-deproj-mean-viridis-0-30}, colormap name={viridis}},
	hoags_object-ngpl-6-simg-deproj-std/.style={plot graphics/xmin={-0.5}, plot graphics/xmax={0.5}, plot graphics/ymin={-0.5}, plot graphics/ymax={0.5}, point meta min={0.0}, point meta max={0.7}, plot graphics/!src={tikz/gallery/hoags_object/hoags_object-ngpl-6-simg-deproj-std-viridis-0-1}, colormap name={viridis}},
	hoags_object-ngpl-6-simg-mean/.style={plot graphics/xmin={-0.5}, plot graphics/xmax={0.5}, plot graphics/ymin={-0.5}, plot graphics/ymax={0.5}, point meta min={0.0}, point meta max={30.0}, plot graphics/!src={tikz/gallery/hoags_object/hoags_object-ngpl-6-simg-mean-viridis-0-30}, colormap name={viridis}},
	hoags_object-ngpl-6-simg-std/.style={plot graphics/xmin={-0.5}, plot graphics/xmax={0.5}, plot graphics/ymin={-0.5}, plot graphics/ymax={0.5}, point meta min={0.0}, point meta max={0.7}, plot graphics/!src={tikz/gallery/hoags_object/hoags_object-ngpl-6-simg-std-viridis-0-1}, colormap name={viridis}},
	hoags_object-ngpl-6-simg-std-interp/.style={plot graphics/xmin={-0.5}, plot graphics/xmax={0.5}, plot graphics/ymin={-0.5}, plot graphics/ymax={0.5}, point meta min={0.0}, point meta max={0.7}, plot graphics/!src={tikz/gallery/hoags_object/hoags_object-ngpl-6-simg-std-interp-viridis-0-1}, colormap name={viridis}},
	hoags_object-ngpl-6-simg-std-winterp/.style={plot graphics/xmin={-0.5}, plot graphics/xmax={0.5}, plot graphics/ymin={-0.5}, plot graphics/ymax={0.5}, point meta min={0.0}, point meta max={0.7}, plot graphics/!src={tikz/gallery/hoags_object/hoags_object-ngpl-6-simg-std-winterp-viridis-0-1}, colormap name={viridis}},
	hoags_object-ngpl-6-simg-std-interp-img/.style={plot graphics/xmin={-0.5}, plot graphics/xmax={0.5}, plot graphics/ymin={-0.5}, plot graphics/ymax={0.5}, point meta min={0.0}, point meta max={0.7}, plot graphics/!src={tikz/gallery/hoags_object/hoags_object-ngpl-6-simg-std-interp-img-viridis-0-1}, colormap name={viridis}},
	hoags_object-ngpl-6-srcs-src-mean/.style={plot graphics/xmin={-0.5}, plot graphics/xmax={0.5}, plot graphics/ymin={-0.5}, plot graphics/ymax={0.5}, point meta min={-0.598825991153717}, point meta max={29.346805572509766}, plot graphics/!src={tikz/gallery/hoags_object/hoags_object-ngpl-6-srcs-src-mean-viridis--1-29}, colormap name={viridis}},
	hoags_object-ngpl-6-srcs-src-std/.style={plot graphics/xmin={-0.5}, plot graphics/xmax={0.5}, plot graphics/ymin={-0.5}, plot graphics/ymax={0.5}, point meta min={0.0}, point meta max={0.7}, plot graphics/!src={tikz/gallery/hoags_object/hoags_object-ngpl-6-srcs-src-std-viridis-0-1}, colormap name={viridis}},
	hoags_object-ngpl-6-srcs-gp-1-mean/.style={plot graphics/xmin={-0.5}, plot graphics/xmax={0.5}, plot graphics/ymin={-0.5}, plot graphics/ymax={0.5}, point meta min={-0.6791581511497498}, point meta max={1.7026429176330566}, plot graphics/!src={tikz/gallery/hoags_object/hoags_object-ngpl-6-srcs-gp-1-mean-viridis--1-2}, colormap name={viridis}},
	hoags_object-ngpl-6-srcs-gp-1-std/.style={plot graphics/xmin={-0.5}, plot graphics/xmax={0.5}, plot graphics/ymin={-0.5}, plot graphics/ymax={0.5}, point meta min={0.0}, point meta max={0.7}, plot graphics/!src={tikz/gallery/hoags_object/hoags_object-ngpl-6-srcs-gp-1-std-viridis-0-1}, colormap name={viridis}},
	hoags_object-ngpl-6-srcs-gp-1-std-interp/.style={plot graphics/xmin={-0.5}, plot graphics/xmax={0.5}, plot graphics/ymin={-0.5}, plot graphics/ymax={0.5}, point meta min={0.0}, point meta max={0.7}, plot graphics/!src={tikz/gallery/hoags_object/hoags_object-ngpl-6-srcs-gp-1-std-interp-viridis-0-1}, colormap name={viridis}},
	hoags_object-ngpl-6-srcs-gp-1-std-winterp/.style={plot graphics/xmin={-0.5}, plot graphics/xmax={0.5}, plot graphics/ymin={-0.5}, plot graphics/ymax={0.5}, point meta min={0.0}, point meta max={0.7}, plot graphics/!src={tikz/gallery/hoags_object/hoags_object-ngpl-6-srcs-gp-1-std-winterp-viridis-0-1}, colormap name={viridis}},
	hoags_object-ngpl-6-srcs-gp-1-std-interp-img/.style={plot graphics/xmin={-0.5}, plot graphics/xmax={0.5}, plot graphics/ymin={-0.5}, plot graphics/ymax={0.5}, point meta min={0.0}, point meta max={0.7}, plot graphics/!src={tikz/gallery/hoags_object/hoags_object-ngpl-6-srcs-gp-1-std-interp-img-viridis-0-1}, colormap name={viridis}},
	hoags_object-ngpl-6-srcs-gp-2-mean/.style={plot graphics/xmin={-0.5}, plot graphics/xmax={0.5}, plot graphics/ymin={-0.5}, plot graphics/ymax={0.5}, point meta min={-0.6321155428886414}, point meta max={1.9880651235580444}, plot graphics/!src={tikz/gallery/hoags_object/hoags_object-ngpl-6-srcs-gp-2-mean-viridis--1-2}, colormap name={viridis}},
	hoags_object-ngpl-6-srcs-gp-2-std/.style={plot graphics/xmin={-0.5}, plot graphics/xmax={0.5}, plot graphics/ymin={-0.5}, plot graphics/ymax={0.5}, point meta min={0.0}, point meta max={0.7}, plot graphics/!src={tikz/gallery/hoags_object/hoags_object-ngpl-6-srcs-gp-2-std-viridis-0-1}, colormap name={viridis}},
	hoags_object-ngpl-6-srcs-gp-2-std-interp/.style={plot graphics/xmin={-0.5}, plot graphics/xmax={0.5}, plot graphics/ymin={-0.5}, plot graphics/ymax={0.5}, point meta min={0.0}, point meta max={0.7}, plot graphics/!src={tikz/gallery/hoags_object/hoags_object-ngpl-6-srcs-gp-2-std-interp-viridis-0-1}, colormap name={viridis}},
	hoags_object-ngpl-6-srcs-gp-2-std-winterp/.style={plot graphics/xmin={-0.5}, plot graphics/xmax={0.5}, plot graphics/ymin={-0.5}, plot graphics/ymax={0.5}, point meta min={0.0}, point meta max={0.7}, plot graphics/!src={tikz/gallery/hoags_object/hoags_object-ngpl-6-srcs-gp-2-std-winterp-viridis-0-1}, colormap name={viridis}},
	hoags_object-ngpl-6-srcs-gp-2-std-interp-img/.style={plot graphics/xmin={-0.5}, plot graphics/xmax={0.5}, plot graphics/ymin={-0.5}, plot graphics/ymax={0.5}, point meta min={0.0}, point meta max={0.7}, plot graphics/!src={tikz/gallery/hoags_object/hoags_object-ngpl-6-srcs-gp-2-std-interp-img-viridis-0-1}, colormap name={viridis}},
	hoags_object-ngpl-6-srcs-gp-3-mean/.style={plot graphics/xmin={-0.5}, plot graphics/xmax={0.5}, plot graphics/ymin={-0.5}, plot graphics/ymax={0.5}, point meta min={-0.42354220151901245}, point meta max={1.514434576034546}, plot graphics/!src={tikz/gallery/hoags_object/hoags_object-ngpl-6-srcs-gp-3-mean-viridis--0-2}, colormap name={viridis}},
	hoags_object-ngpl-6-srcs-gp-3-std/.style={plot graphics/xmin={-0.5}, plot graphics/xmax={0.5}, plot graphics/ymin={-0.5}, plot graphics/ymax={0.5}, point meta min={0.0}, point meta max={0.7}, plot graphics/!src={tikz/gallery/hoags_object/hoags_object-ngpl-6-srcs-gp-3-std-viridis-0-1}, colormap name={viridis}},
	hoags_object-ngpl-6-srcs-gp-3-std-interp/.style={plot graphics/xmin={-0.5}, plot graphics/xmax={0.5}, plot graphics/ymin={-0.5}, plot graphics/ymax={0.5}, point meta min={0.0}, point meta max={0.7}, plot graphics/!src={tikz/gallery/hoags_object/hoags_object-ngpl-6-srcs-gp-3-std-interp-viridis-0-1}, colormap name={viridis}},
	hoags_object-ngpl-6-srcs-gp-3-std-winterp/.style={plot graphics/xmin={-0.5}, plot graphics/xmax={0.5}, plot graphics/ymin={-0.5}, plot graphics/ymax={0.5}, point meta min={0.0}, point meta max={0.7}, plot graphics/!src={tikz/gallery/hoags_object/hoags_object-ngpl-6-srcs-gp-3-std-winterp-viridis-0-1}, colormap name={viridis}},
	hoags_object-ngpl-6-srcs-gp-3-std-interp-img/.style={plot graphics/xmin={-0.5}, plot graphics/xmax={0.5}, plot graphics/ymin={-0.5}, plot graphics/ymax={0.5}, point meta min={0.0}, point meta max={0.7}, plot graphics/!src={tikz/gallery/hoags_object/hoags_object-ngpl-6-srcs-gp-3-std-interp-img-viridis-0-1}, colormap name={viridis}},
	hoags_object-ngpl-6-srcs-gp-4-mean/.style={plot graphics/xmin={-0.5}, plot graphics/xmax={0.5}, plot graphics/ymin={-0.5}, plot graphics/ymax={0.5}, point meta min={-0.397795706987381}, point meta max={1.0642422437667847}, plot graphics/!src={tikz/gallery/hoags_object/hoags_object-ngpl-6-srcs-gp-4-mean-viridis--0-1}, colormap name={viridis}},
	hoags_object-ngpl-6-srcs-gp-4-std/.style={plot graphics/xmin={-0.5}, plot graphics/xmax={0.5}, plot graphics/ymin={-0.5}, plot graphics/ymax={0.5}, point meta min={0.0}, point meta max={0.7}, plot graphics/!src={tikz/gallery/hoags_object/hoags_object-ngpl-6-srcs-gp-4-std-viridis-0-1}, colormap name={viridis}},
	hoags_object-ngpl-6-srcs-gp-4-std-interp/.style={plot graphics/xmin={-0.5}, plot graphics/xmax={0.5}, plot graphics/ymin={-0.5}, plot graphics/ymax={0.5}, point meta min={0.0}, point meta max={0.7}, plot graphics/!src={tikz/gallery/hoags_object/hoags_object-ngpl-6-srcs-gp-4-std-interp-viridis-0-1}, colormap name={viridis}},
	hoags_object-ngpl-6-srcs-gp-4-std-winterp/.style={plot graphics/xmin={-0.5}, plot graphics/xmax={0.5}, plot graphics/ymin={-0.5}, plot graphics/ymax={0.5}, point meta min={0.0}, point meta max={0.7}, plot graphics/!src={tikz/gallery/hoags_object/hoags_object-ngpl-6-srcs-gp-4-std-winterp-viridis-0-1}, colormap name={viridis}},
	hoags_object-ngpl-6-srcs-gp-4-std-interp-img/.style={plot graphics/xmin={-0.5}, plot graphics/xmax={0.5}, plot graphics/ymin={-0.5}, plot graphics/ymax={0.5}, point meta min={0.0}, point meta max={0.7}, plot graphics/!src={tikz/gallery/hoags_object/hoags_object-ngpl-6-srcs-gp-4-std-interp-img-viridis-0-1}, colormap name={viridis}},
	hoags_object-ngpl-6-srcs-gp-5-mean/.style={plot graphics/xmin={-0.5}, plot graphics/xmax={0.5}, plot graphics/ymin={-0.5}, plot graphics/ymax={0.5}, point meta min={-0.33005291223526}, point meta max={0.8093322515487671}, plot graphics/!src={tikz/gallery/hoags_object/hoags_object-ngpl-6-srcs-gp-5-mean-viridis--0-1}, colormap name={viridis}},
	hoags_object-ngpl-6-srcs-gp-5-std/.style={plot graphics/xmin={-0.5}, plot graphics/xmax={0.5}, plot graphics/ymin={-0.5}, plot graphics/ymax={0.5}, point meta min={0.0}, point meta max={0.7}, plot graphics/!src={tikz/gallery/hoags_object/hoags_object-ngpl-6-srcs-gp-5-std-viridis-0-1}, colormap name={viridis}},
	hoags_object-ngpl-6-srcs-gp-5-std-interp/.style={plot graphics/xmin={-0.5}, plot graphics/xmax={0.5}, plot graphics/ymin={-0.5}, plot graphics/ymax={0.5}, point meta min={0.0}, point meta max={0.7}, plot graphics/!src={tikz/gallery/hoags_object/hoags_object-ngpl-6-srcs-gp-5-std-interp-viridis-0-1}, colormap name={viridis}},
	hoags_object-ngpl-6-srcs-gp-5-std-winterp/.style={plot graphics/xmin={-0.5}, plot graphics/xmax={0.5}, plot graphics/ymin={-0.5}, plot graphics/ymax={0.5}, point meta min={0.0}, point meta max={0.7}, plot graphics/!src={tikz/gallery/hoags_object/hoags_object-ngpl-6-srcs-gp-5-std-winterp-viridis-0-1}, colormap name={viridis}},
	hoags_object-ngpl-6-srcs-gp-5-std-interp-img/.style={plot graphics/xmin={-0.5}, plot graphics/xmax={0.5}, plot graphics/ymin={-0.5}, plot graphics/ymax={0.5}, point meta min={0.0}, point meta max={0.7}, plot graphics/!src={tikz/gallery/hoags_object/hoags_object-ngpl-6-srcs-gp-5-std-interp-img-viridis-0-1}, colormap name={viridis}},
	hoags_object-ngpl-6-srcs-gp-6-mean/.style={plot graphics/xmin={-0.5}, plot graphics/xmax={0.5}, plot graphics/ymin={-0.5}, plot graphics/ymax={0.5}, point meta min={-0.326918363571167}, point meta max={0.6808406710624695}, plot graphics/!src={tikz/gallery/hoags_object/hoags_object-ngpl-6-srcs-gp-6-mean-viridis--0-1}, colormap name={viridis}},
	hoags_object-ngpl-6-srcs-gp-6-std/.style={plot graphics/xmin={-0.5}, plot graphics/xmax={0.5}, plot graphics/ymin={-0.5}, plot graphics/ymax={0.5}, point meta min={0.0}, point meta max={0.7}, plot graphics/!src={tikz/gallery/hoags_object/hoags_object-ngpl-6-srcs-gp-6-std-viridis-0-1}, colormap name={viridis}},
	hoags_object-ngpl-6-srcs-gp-6-std-interp/.style={plot graphics/xmin={-0.5}, plot graphics/xmax={0.5}, plot graphics/ymin={-0.5}, plot graphics/ymax={0.5}, point meta min={0.0}, point meta max={0.7}, plot graphics/!src={tikz/gallery/hoags_object/hoags_object-ngpl-6-srcs-gp-6-std-interp-viridis-0-1}, colormap name={viridis}},
	hoags_object-ngpl-6-srcs-gp-6-std-winterp/.style={plot graphics/xmin={-0.5}, plot graphics/xmax={0.5}, plot graphics/ymin={-0.5}, plot graphics/ymax={0.5}, point meta min={0.0}, point meta max={0.7}, plot graphics/!src={tikz/gallery/hoags_object/hoags_object-ngpl-6-srcs-gp-6-std-winterp-viridis-0-1}, colormap name={viridis}},
	hoags_object-ngpl-6-srcs-gp-6-std-interp-img/.style={plot graphics/xmin={-0.5}, plot graphics/xmax={0.5}, plot graphics/ymin={-0.5}, plot graphics/ymax={0.5}, point meta min={0.0}, point meta max={0.7}, plot graphics/!src={tikz/gallery/hoags_object/hoags_object-ngpl-6-srcs-gp-6-std-interp-img-viridis-0-1}, colormap name={viridis}},
	hoags_object-ngpl-7-img-mean/.style={plot graphics/xmin={-2.5}, plot graphics/xmax={2.5}, plot graphics/ymin={-2.5}, plot graphics/ymax={2.5}, point meta min={0.0}, point meta max={30.0}, plot graphics/!src={tikz/gallery/hoags_object/hoags_object-ngpl-7-img-mean-viridis-0-30}, colormap name={viridis}},
	hoags_object-ngpl-7-img-std/.style={plot graphics/xmin={-2.5}, plot graphics/xmax={2.5}, plot graphics/ymin={-2.5}, plot graphics/ymax={2.5}, point meta min={0.0}, point meta max={0.7}, plot graphics/!src={tikz/gallery/hoags_object/hoags_object-ngpl-7-img-std-viridis-0-1}, colormap name={viridis}},
	hoags_object-ngpl-7-simg-deproj-mean/.style={plot graphics/xmin={-0.5}, plot graphics/xmax={0.5}, plot graphics/ymin={-0.5}, plot graphics/ymax={0.5}, point meta min={0.0}, point meta max={30.0}, plot graphics/!src={tikz/gallery/hoags_object/hoags_object-ngpl-7-simg-deproj-mean-viridis-0-30}, colormap name={viridis}},
	hoags_object-ngpl-7-simg-deproj-std/.style={plot graphics/xmin={-0.5}, plot graphics/xmax={0.5}, plot graphics/ymin={-0.5}, plot graphics/ymax={0.5}, point meta min={0.0}, point meta max={0.7}, plot graphics/!src={tikz/gallery/hoags_object/hoags_object-ngpl-7-simg-deproj-std-viridis-0-1}, colormap name={viridis}},
	hoags_object-ngpl-7-simg-mean/.style={plot graphics/xmin={-0.5}, plot graphics/xmax={0.5}, plot graphics/ymin={-0.5}, plot graphics/ymax={0.5}, point meta min={0.0}, point meta max={30.0}, plot graphics/!src={tikz/gallery/hoags_object/hoags_object-ngpl-7-simg-mean-viridis-0-30}, colormap name={viridis}},
	hoags_object-ngpl-7-simg-std/.style={plot graphics/xmin={-0.5}, plot graphics/xmax={0.5}, plot graphics/ymin={-0.5}, plot graphics/ymax={0.5}, point meta min={0.0}, point meta max={0.7}, plot graphics/!src={tikz/gallery/hoags_object/hoags_object-ngpl-7-simg-std-viridis-0-1}, colormap name={viridis}},
	hoags_object-ngpl-7-simg-std-interp/.style={plot graphics/xmin={-0.5}, plot graphics/xmax={0.5}, plot graphics/ymin={-0.5}, plot graphics/ymax={0.5}, point meta min={0.0}, point meta max={0.7}, plot graphics/!src={tikz/gallery/hoags_object/hoags_object-ngpl-7-simg-std-interp-viridis-0-1}, colormap name={viridis}},
	hoags_object-ngpl-7-simg-std-winterp/.style={plot graphics/xmin={-0.5}, plot graphics/xmax={0.5}, plot graphics/ymin={-0.5}, plot graphics/ymax={0.5}, point meta min={0.0}, point meta max={0.7}, plot graphics/!src={tikz/gallery/hoags_object/hoags_object-ngpl-7-simg-std-winterp-viridis-0-1}, colormap name={viridis}},
	hoags_object-ngpl-7-simg-std-interp-img/.style={plot graphics/xmin={-0.5}, plot graphics/xmax={0.5}, plot graphics/ymin={-0.5}, plot graphics/ymax={0.5}, point meta min={0.0}, point meta max={0.7}, plot graphics/!src={tikz/gallery/hoags_object/hoags_object-ngpl-7-simg-std-interp-img-viridis-0-1}, colormap name={viridis}},
	hoags_object-ngpl-7-srcs-src-mean/.style={plot graphics/xmin={-0.5}, plot graphics/xmax={0.5}, plot graphics/ymin={-0.5}, plot graphics/ymax={0.5}, point meta min={-0.5757520794868469}, point meta max={29.338680267333984}, plot graphics/!src={tikz/gallery/hoags_object/hoags_object-ngpl-7-srcs-src-mean-viridis--1-29}, colormap name={viridis}},
	hoags_object-ngpl-7-srcs-src-std/.style={plot graphics/xmin={-0.5}, plot graphics/xmax={0.5}, plot graphics/ymin={-0.5}, plot graphics/ymax={0.5}, point meta min={0.0}, point meta max={0.7}, plot graphics/!src={tikz/gallery/hoags_object/hoags_object-ngpl-7-srcs-src-std-viridis-0-1}, colormap name={viridis}},
	hoags_object-ngpl-7-srcs-gp-1-mean/.style={plot graphics/xmin={-0.5}, plot graphics/xmax={0.5}, plot graphics/ymin={-0.5}, plot graphics/ymax={0.5}, point meta min={-0.5896649956703186}, point meta max={1.4431936740875244}, plot graphics/!src={tikz/gallery/hoags_object/hoags_object-ngpl-7-srcs-gp-1-mean-viridis--1-1}, colormap name={viridis}},
	hoags_object-ngpl-7-srcs-gp-1-std/.style={plot graphics/xmin={-0.5}, plot graphics/xmax={0.5}, plot graphics/ymin={-0.5}, plot graphics/ymax={0.5}, point meta min={0.0}, point meta max={0.7}, plot graphics/!src={tikz/gallery/hoags_object/hoags_object-ngpl-7-srcs-gp-1-std-viridis-0-1}, colormap name={viridis}},
	hoags_object-ngpl-7-srcs-gp-1-std-interp/.style={plot graphics/xmin={-0.5}, plot graphics/xmax={0.5}, plot graphics/ymin={-0.5}, plot graphics/ymax={0.5}, point meta min={0.0}, point meta max={0.7}, plot graphics/!src={tikz/gallery/hoags_object/hoags_object-ngpl-7-srcs-gp-1-std-interp-viridis-0-1}, colormap name={viridis}},
	hoags_object-ngpl-7-srcs-gp-1-std-winterp/.style={plot graphics/xmin={-0.5}, plot graphics/xmax={0.5}, plot graphics/ymin={-0.5}, plot graphics/ymax={0.5}, point meta min={0.0}, point meta max={0.7}, plot graphics/!src={tikz/gallery/hoags_object/hoags_object-ngpl-7-srcs-gp-1-std-winterp-viridis-0-1}, colormap name={viridis}},
	hoags_object-ngpl-7-srcs-gp-1-std-interp-img/.style={plot graphics/xmin={-0.5}, plot graphics/xmax={0.5}, plot graphics/ymin={-0.5}, plot graphics/ymax={0.5}, point meta min={0.0}, point meta max={0.7}, plot graphics/!src={tikz/gallery/hoags_object/hoags_object-ngpl-7-srcs-gp-1-std-interp-img-viridis-0-1}, colormap name={viridis}},
	hoags_object-ngpl-7-srcs-gp-2-mean/.style={plot graphics/xmin={-0.5}, plot graphics/xmax={0.5}, plot graphics/ymin={-0.5}, plot graphics/ymax={0.5}, point meta min={-0.5741198658943176}, point meta max={1.6962424516677856}, plot graphics/!src={tikz/gallery/hoags_object/hoags_object-ngpl-7-srcs-gp-2-mean-viridis--1-2}, colormap name={viridis}},
	hoags_object-ngpl-7-srcs-gp-2-std/.style={plot graphics/xmin={-0.5}, plot graphics/xmax={0.5}, plot graphics/ymin={-0.5}, plot graphics/ymax={0.5}, point meta min={0.0}, point meta max={0.7}, plot graphics/!src={tikz/gallery/hoags_object/hoags_object-ngpl-7-srcs-gp-2-std-viridis-0-1}, colormap name={viridis}},
	hoags_object-ngpl-7-srcs-gp-2-std-interp/.style={plot graphics/xmin={-0.5}, plot graphics/xmax={0.5}, plot graphics/ymin={-0.5}, plot graphics/ymax={0.5}, point meta min={0.0}, point meta max={0.7}, plot graphics/!src={tikz/gallery/hoags_object/hoags_object-ngpl-7-srcs-gp-2-std-interp-viridis-0-1}, colormap name={viridis}},
	hoags_object-ngpl-7-srcs-gp-2-std-winterp/.style={plot graphics/xmin={-0.5}, plot graphics/xmax={0.5}, plot graphics/ymin={-0.5}, plot graphics/ymax={0.5}, point meta min={0.0}, point meta max={0.7}, plot graphics/!src={tikz/gallery/hoags_object/hoags_object-ngpl-7-srcs-gp-2-std-winterp-viridis-0-1}, colormap name={viridis}},
	hoags_object-ngpl-7-srcs-gp-2-std-interp-img/.style={plot graphics/xmin={-0.5}, plot graphics/xmax={0.5}, plot graphics/ymin={-0.5}, plot graphics/ymax={0.5}, point meta min={0.0}, point meta max={0.7}, plot graphics/!src={tikz/gallery/hoags_object/hoags_object-ngpl-7-srcs-gp-2-std-interp-img-viridis-0-1}, colormap name={viridis}},
	hoags_object-ngpl-7-srcs-gp-3-mean/.style={plot graphics/xmin={-0.5}, plot graphics/xmax={0.5}, plot graphics/ymin={-0.5}, plot graphics/ymax={0.5}, point meta min={-0.4206708073616028}, point meta max={1.5018329620361328}, plot graphics/!src={tikz/gallery/hoags_object/hoags_object-ngpl-7-srcs-gp-3-mean-viridis--0-2}, colormap name={viridis}},
	hoags_object-ngpl-7-srcs-gp-3-std/.style={plot graphics/xmin={-0.5}, plot graphics/xmax={0.5}, plot graphics/ymin={-0.5}, plot graphics/ymax={0.5}, point meta min={0.0}, point meta max={0.7}, plot graphics/!src={tikz/gallery/hoags_object/hoags_object-ngpl-7-srcs-gp-3-std-viridis-0-1}, colormap name={viridis}},
	hoags_object-ngpl-7-srcs-gp-3-std-interp/.style={plot graphics/xmin={-0.5}, plot graphics/xmax={0.5}, plot graphics/ymin={-0.5}, plot graphics/ymax={0.5}, point meta min={0.0}, point meta max={0.7}, plot graphics/!src={tikz/gallery/hoags_object/hoags_object-ngpl-7-srcs-gp-3-std-interp-viridis-0-1}, colormap name={viridis}},
	hoags_object-ngpl-7-srcs-gp-3-std-winterp/.style={plot graphics/xmin={-0.5}, plot graphics/xmax={0.5}, plot graphics/ymin={-0.5}, plot graphics/ymax={0.5}, point meta min={0.0}, point meta max={0.7}, plot graphics/!src={tikz/gallery/hoags_object/hoags_object-ngpl-7-srcs-gp-3-std-winterp-viridis-0-1}, colormap name={viridis}},
	hoags_object-ngpl-7-srcs-gp-3-std-interp-img/.style={plot graphics/xmin={-0.5}, plot graphics/xmax={0.5}, plot graphics/ymin={-0.5}, plot graphics/ymax={0.5}, point meta min={0.0}, point meta max={0.7}, plot graphics/!src={tikz/gallery/hoags_object/hoags_object-ngpl-7-srcs-gp-3-std-interp-img-viridis-0-1}, colormap name={viridis}},
	hoags_object-ngpl-7-srcs-gp-4-mean/.style={plot graphics/xmin={-0.5}, plot graphics/xmax={0.5}, plot graphics/ymin={-0.5}, plot graphics/ymax={0.5}, point meta min={-0.3420698046684265}, point meta max={1.2173362970352173}, plot graphics/!src={tikz/gallery/hoags_object/hoags_object-ngpl-7-srcs-gp-4-mean-viridis--0-1}, colormap name={viridis}},
	hoags_object-ngpl-7-srcs-gp-4-std/.style={plot graphics/xmin={-0.5}, plot graphics/xmax={0.5}, plot graphics/ymin={-0.5}, plot graphics/ymax={0.5}, point meta min={0.0}, point meta max={0.7}, plot graphics/!src={tikz/gallery/hoags_object/hoags_object-ngpl-7-srcs-gp-4-std-viridis-0-1}, colormap name={viridis}},
	hoags_object-ngpl-7-srcs-gp-4-std-interp/.style={plot graphics/xmin={-0.5}, plot graphics/xmax={0.5}, plot graphics/ymin={-0.5}, plot graphics/ymax={0.5}, point meta min={0.0}, point meta max={0.7}, plot graphics/!src={tikz/gallery/hoags_object/hoags_object-ngpl-7-srcs-gp-4-std-interp-viridis-0-1}, colormap name={viridis}},
	hoags_object-ngpl-7-srcs-gp-4-std-winterp/.style={plot graphics/xmin={-0.5}, plot graphics/xmax={0.5}, plot graphics/ymin={-0.5}, plot graphics/ymax={0.5}, point meta min={0.0}, point meta max={0.7}, plot graphics/!src={tikz/gallery/hoags_object/hoags_object-ngpl-7-srcs-gp-4-std-winterp-viridis-0-1}, colormap name={viridis}},
	hoags_object-ngpl-7-srcs-gp-4-std-interp-img/.style={plot graphics/xmin={-0.5}, plot graphics/xmax={0.5}, plot graphics/ymin={-0.5}, plot graphics/ymax={0.5}, point meta min={0.0}, point meta max={0.7}, plot graphics/!src={tikz/gallery/hoags_object/hoags_object-ngpl-7-srcs-gp-4-std-interp-img-viridis-0-1}, colormap name={viridis}},
	hoags_object-ngpl-7-srcs-gp-5-mean/.style={plot graphics/xmin={-0.5}, plot graphics/xmax={0.5}, plot graphics/ymin={-0.5}, plot graphics/ymax={0.5}, point meta min={-0.3337073028087616}, point meta max={0.8932861089706421}, plot graphics/!src={tikz/gallery/hoags_object/hoags_object-ngpl-7-srcs-gp-5-mean-viridis--0-1}, colormap name={viridis}},
	hoags_object-ngpl-7-srcs-gp-5-std/.style={plot graphics/xmin={-0.5}, plot graphics/xmax={0.5}, plot graphics/ymin={-0.5}, plot graphics/ymax={0.5}, point meta min={0.0}, point meta max={0.7}, plot graphics/!src={tikz/gallery/hoags_object/hoags_object-ngpl-7-srcs-gp-5-std-viridis-0-1}, colormap name={viridis}},
	hoags_object-ngpl-7-srcs-gp-5-std-interp/.style={plot graphics/xmin={-0.5}, plot graphics/xmax={0.5}, plot graphics/ymin={-0.5}, plot graphics/ymax={0.5}, point meta min={0.0}, point meta max={0.7}, plot graphics/!src={tikz/gallery/hoags_object/hoags_object-ngpl-7-srcs-gp-5-std-interp-viridis-0-1}, colormap name={viridis}},
	hoags_object-ngpl-7-srcs-gp-5-std-winterp/.style={plot graphics/xmin={-0.5}, plot graphics/xmax={0.5}, plot graphics/ymin={-0.5}, plot graphics/ymax={0.5}, point meta min={0.0}, point meta max={0.7}, plot graphics/!src={tikz/gallery/hoags_object/hoags_object-ngpl-7-srcs-gp-5-std-winterp-viridis-0-1}, colormap name={viridis}},
	hoags_object-ngpl-7-srcs-gp-5-std-interp-img/.style={plot graphics/xmin={-0.5}, plot graphics/xmax={0.5}, plot graphics/ymin={-0.5}, plot graphics/ymax={0.5}, point meta min={0.0}, point meta max={0.7}, plot graphics/!src={tikz/gallery/hoags_object/hoags_object-ngpl-7-srcs-gp-5-std-interp-img-viridis-0-1}, colormap name={viridis}},
	hoags_object-ngpl-7-srcs-gp-6-mean/.style={plot graphics/xmin={-0.5}, plot graphics/xmax={0.5}, plot graphics/ymin={-0.5}, plot graphics/ymax={0.5}, point meta min={-0.33746200799942017}, point meta max={0.7328528165817261}, plot graphics/!src={tikz/gallery/hoags_object/hoags_object-ngpl-7-srcs-gp-6-mean-viridis--0-1}, colormap name={viridis}},
	hoags_object-ngpl-7-srcs-gp-6-std/.style={plot graphics/xmin={-0.5}, plot graphics/xmax={0.5}, plot graphics/ymin={-0.5}, plot graphics/ymax={0.5}, point meta min={0.0}, point meta max={0.7}, plot graphics/!src={tikz/gallery/hoags_object/hoags_object-ngpl-7-srcs-gp-6-std-viridis-0-1}, colormap name={viridis}},
	hoags_object-ngpl-7-srcs-gp-6-std-interp/.style={plot graphics/xmin={-0.5}, plot graphics/xmax={0.5}, plot graphics/ymin={-0.5}, plot graphics/ymax={0.5}, point meta min={0.0}, point meta max={0.7}, plot graphics/!src={tikz/gallery/hoags_object/hoags_object-ngpl-7-srcs-gp-6-std-interp-viridis-0-1}, colormap name={viridis}},
	hoags_object-ngpl-7-srcs-gp-6-std-winterp/.style={plot graphics/xmin={-0.5}, plot graphics/xmax={0.5}, plot graphics/ymin={-0.5}, plot graphics/ymax={0.5}, point meta min={0.0}, point meta max={0.7}, plot graphics/!src={tikz/gallery/hoags_object/hoags_object-ngpl-7-srcs-gp-6-std-winterp-viridis-0-1}, colormap name={viridis}},
	hoags_object-ngpl-7-srcs-gp-6-std-interp-img/.style={plot graphics/xmin={-0.5}, plot graphics/xmax={0.5}, plot graphics/ymin={-0.5}, plot graphics/ymax={0.5}, point meta min={0.0}, point meta max={0.7}, plot graphics/!src={tikz/gallery/hoags_object/hoags_object-ngpl-7-srcs-gp-6-std-interp-img-viridis-0-1}, colormap name={viridis}},
	hoags_object-ngpl-7-srcs-gp-7-mean/.style={plot graphics/xmin={-0.5}, plot graphics/xmax={0.5}, plot graphics/ymin={-0.5}, plot graphics/ymax={0.5}, point meta min={-0.37413331866264343}, point meta max={0.5985175371170044}, plot graphics/!src={tikz/gallery/hoags_object/hoags_object-ngpl-7-srcs-gp-7-mean-viridis--0-1}, colormap name={viridis}},
	hoags_object-ngpl-7-srcs-gp-7-std/.style={plot graphics/xmin={-0.5}, plot graphics/xmax={0.5}, plot graphics/ymin={-0.5}, plot graphics/ymax={0.5}, point meta min={0.0}, point meta max={0.7}, plot graphics/!src={tikz/gallery/hoags_object/hoags_object-ngpl-7-srcs-gp-7-std-viridis-0-1}, colormap name={viridis}},
	hoags_object-ngpl-7-srcs-gp-7-std-interp/.style={plot graphics/xmin={-0.5}, plot graphics/xmax={0.5}, plot graphics/ymin={-0.5}, plot graphics/ymax={0.5}, point meta min={0.0}, point meta max={0.7}, plot graphics/!src={tikz/gallery/hoags_object/hoags_object-ngpl-7-srcs-gp-7-std-interp-viridis-0-1}, colormap name={viridis}},
	hoags_object-ngpl-7-srcs-gp-7-std-winterp/.style={plot graphics/xmin={-0.5}, plot graphics/xmax={0.5}, plot graphics/ymin={-0.5}, plot graphics/ymax={0.5}, point meta min={0.0}, point meta max={0.7}, plot graphics/!src={tikz/gallery/hoags_object/hoags_object-ngpl-7-srcs-gp-7-std-winterp-viridis-0-1}, colormap name={viridis}},
	hoags_object-ngpl-7-srcs-gp-7-std-interp-img/.style={plot graphics/xmin={-0.5}, plot graphics/xmax={0.5}, plot graphics/ymin={-0.5}, plot graphics/ymax={0.5}, point meta min={0.0}, point meta max={0.7}, plot graphics/!src={tikz/gallery/hoags_object/hoags_object-ngpl-7-srcs-gp-7-std-interp-img-viridis-0-1}, colormap name={viridis}},
	hoags_object-ngpl-8-img-mean/.style={plot graphics/xmin={-2.5}, plot graphics/xmax={2.5}, plot graphics/ymin={-2.5}, plot graphics/ymax={2.5}, point meta min={0.0}, point meta max={30.0}, plot graphics/!src={tikz/gallery/hoags_object/hoags_object-ngpl-8-img-mean-viridis-0-30}, colormap name={viridis}},
	hoags_object-ngpl-8-img-std/.style={plot graphics/xmin={-2.5}, plot graphics/xmax={2.5}, plot graphics/ymin={-2.5}, plot graphics/ymax={2.5}, point meta min={0.0}, point meta max={0.7}, plot graphics/!src={tikz/gallery/hoags_object/hoags_object-ngpl-8-img-std-viridis-0-1}, colormap name={viridis}},
	hoags_object-ngpl-8-simg-deproj-mean/.style={plot graphics/xmin={-0.5}, plot graphics/xmax={0.5}, plot graphics/ymin={-0.5}, plot graphics/ymax={0.5}, point meta min={0.0}, point meta max={30.0}, plot graphics/!src={tikz/gallery/hoags_object/hoags_object-ngpl-8-simg-deproj-mean-viridis-0-30}, colormap name={viridis}},
	hoags_object-ngpl-8-simg-deproj-std/.style={plot graphics/xmin={-0.5}, plot graphics/xmax={0.5}, plot graphics/ymin={-0.5}, plot graphics/ymax={0.5}, point meta min={0.0}, point meta max={0.7}, plot graphics/!src={tikz/gallery/hoags_object/hoags_object-ngpl-8-simg-deproj-std-viridis-0-1}, colormap name={viridis}},
	hoags_object-ngpl-8-simg-mean/.style={plot graphics/xmin={-0.5}, plot graphics/xmax={0.5}, plot graphics/ymin={-0.5}, plot graphics/ymax={0.5}, point meta min={0.0}, point meta max={30.0}, plot graphics/!src={tikz/gallery/hoags_object/hoags_object-ngpl-8-simg-mean-viridis-0-30}, colormap name={viridis}},
	hoags_object-ngpl-8-simg-std/.style={plot graphics/xmin={-0.5}, plot graphics/xmax={0.5}, plot graphics/ymin={-0.5}, plot graphics/ymax={0.5}, point meta min={0.0}, point meta max={0.7}, plot graphics/!src={tikz/gallery/hoags_object/hoags_object-ngpl-8-simg-std-viridis-0-1}, colormap name={viridis}},
	hoags_object-ngpl-8-simg-std-interp/.style={plot graphics/xmin={-0.5}, plot graphics/xmax={0.5}, plot graphics/ymin={-0.5}, plot graphics/ymax={0.5}, point meta min={0.0}, point meta max={0.7}, plot graphics/!src={tikz/gallery/hoags_object/hoags_object-ngpl-8-simg-std-interp-viridis-0-1}, colormap name={viridis}},
	hoags_object-ngpl-8-simg-std-winterp/.style={plot graphics/xmin={-0.5}, plot graphics/xmax={0.5}, plot graphics/ymin={-0.5}, plot graphics/ymax={0.5}, point meta min={0.0}, point meta max={0.7}, plot graphics/!src={tikz/gallery/hoags_object/hoags_object-ngpl-8-simg-std-winterp-viridis-0-1}, colormap name={viridis}},
	hoags_object-ngpl-8-simg-std-interp-img/.style={plot graphics/xmin={-0.5}, plot graphics/xmax={0.5}, plot graphics/ymin={-0.5}, plot graphics/ymax={0.5}, point meta min={0.0}, point meta max={0.7}, plot graphics/!src={tikz/gallery/hoags_object/hoags_object-ngpl-8-simg-std-interp-img-viridis-0-1}, colormap name={viridis}},
	hoags_object-ngpl-8-srcs-src-mean/.style={plot graphics/xmin={-0.5}, plot graphics/xmax={0.5}, plot graphics/ymin={-0.5}, plot graphics/ymax={0.5}, point meta min={-0.5852934122085571}, point meta max={29.424766540527344}, plot graphics/!src={tikz/gallery/hoags_object/hoags_object-ngpl-8-srcs-src-mean-viridis--1-29}, colormap name={viridis}},
	hoags_object-ngpl-8-srcs-src-std/.style={plot graphics/xmin={-0.5}, plot graphics/xmax={0.5}, plot graphics/ymin={-0.5}, plot graphics/ymax={0.5}, point meta min={0.0}, point meta max={0.7}, plot graphics/!src={tikz/gallery/hoags_object/hoags_object-ngpl-8-srcs-src-std-viridis-0-1}, colormap name={viridis}},
	hoags_object-ngpl-8-srcs-gp-1-mean/.style={plot graphics/xmin={-0.5}, plot graphics/xmax={0.5}, plot graphics/ymin={-0.5}, plot graphics/ymax={0.5}, point meta min={-0.5069890022277832}, point meta max={1.2279870510101318}, plot graphics/!src={tikz/gallery/hoags_object/hoags_object-ngpl-8-srcs-gp-1-mean-viridis--1-1}, colormap name={viridis}},
	hoags_object-ngpl-8-srcs-gp-1-std/.style={plot graphics/xmin={-0.5}, plot graphics/xmax={0.5}, plot graphics/ymin={-0.5}, plot graphics/ymax={0.5}, point meta min={0.0}, point meta max={0.7}, plot graphics/!src={tikz/gallery/hoags_object/hoags_object-ngpl-8-srcs-gp-1-std-viridis-0-1}, colormap name={viridis}},
	hoags_object-ngpl-8-srcs-gp-1-std-interp/.style={plot graphics/xmin={-0.5}, plot graphics/xmax={0.5}, plot graphics/ymin={-0.5}, plot graphics/ymax={0.5}, point meta min={0.0}, point meta max={0.7}, plot graphics/!src={tikz/gallery/hoags_object/hoags_object-ngpl-8-srcs-gp-1-std-interp-viridis-0-1}, colormap name={viridis}},
	hoags_object-ngpl-8-srcs-gp-1-std-winterp/.style={plot graphics/xmin={-0.5}, plot graphics/xmax={0.5}, plot graphics/ymin={-0.5}, plot graphics/ymax={0.5}, point meta min={0.0}, point meta max={0.7}, plot graphics/!src={tikz/gallery/hoags_object/hoags_object-ngpl-8-srcs-gp-1-std-winterp-viridis-0-1}, colormap name={viridis}},
	hoags_object-ngpl-8-srcs-gp-1-std-interp-img/.style={plot graphics/xmin={-0.5}, plot graphics/xmax={0.5}, plot graphics/ymin={-0.5}, plot graphics/ymax={0.5}, point meta min={0.0}, point meta max={0.7}, plot graphics/!src={tikz/gallery/hoags_object/hoags_object-ngpl-8-srcs-gp-1-std-interp-img-viridis-0-1}, colormap name={viridis}},
	hoags_object-ngpl-8-srcs-gp-2-mean/.style={plot graphics/xmin={-0.5}, plot graphics/xmax={0.5}, plot graphics/ymin={-0.5}, plot graphics/ymax={0.5}, point meta min={-0.4956877827644348}, point meta max={1.4369945526123047}, plot graphics/!src={tikz/gallery/hoags_object/hoags_object-ngpl-8-srcs-gp-2-mean-viridis--0-1}, colormap name={viridis}},
	hoags_object-ngpl-8-srcs-gp-2-std/.style={plot graphics/xmin={-0.5}, plot graphics/xmax={0.5}, plot graphics/ymin={-0.5}, plot graphics/ymax={0.5}, point meta min={0.0}, point meta max={0.7}, plot graphics/!src={tikz/gallery/hoags_object/hoags_object-ngpl-8-srcs-gp-2-std-viridis-0-1}, colormap name={viridis}},
	hoags_object-ngpl-8-srcs-gp-2-std-interp/.style={plot graphics/xmin={-0.5}, plot graphics/xmax={0.5}, plot graphics/ymin={-0.5}, plot graphics/ymax={0.5}, point meta min={0.0}, point meta max={0.7}, plot graphics/!src={tikz/gallery/hoags_object/hoags_object-ngpl-8-srcs-gp-2-std-interp-viridis-0-1}, colormap name={viridis}},
	hoags_object-ngpl-8-srcs-gp-2-std-winterp/.style={plot graphics/xmin={-0.5}, plot graphics/xmax={0.5}, plot graphics/ymin={-0.5}, plot graphics/ymax={0.5}, point meta min={0.0}, point meta max={0.7}, plot graphics/!src={tikz/gallery/hoags_object/hoags_object-ngpl-8-srcs-gp-2-std-winterp-viridis-0-1}, colormap name={viridis}},
	hoags_object-ngpl-8-srcs-gp-2-std-interp-img/.style={plot graphics/xmin={-0.5}, plot graphics/xmax={0.5}, plot graphics/ymin={-0.5}, plot graphics/ymax={0.5}, point meta min={0.0}, point meta max={0.7}, plot graphics/!src={tikz/gallery/hoags_object/hoags_object-ngpl-8-srcs-gp-2-std-interp-img-viridis-0-1}, colormap name={viridis}},
	hoags_object-ngpl-8-srcs-gp-3-mean/.style={plot graphics/xmin={-0.5}, plot graphics/xmax={0.5}, plot graphics/ymin={-0.5}, plot graphics/ymax={0.5}, point meta min={-0.39823469519615173}, point meta max={1.385358452796936}, plot graphics/!src={tikz/gallery/hoags_object/hoags_object-ngpl-8-srcs-gp-3-mean-viridis--0-1}, colormap name={viridis}},
	hoags_object-ngpl-8-srcs-gp-3-std/.style={plot graphics/xmin={-0.5}, plot graphics/xmax={0.5}, plot graphics/ymin={-0.5}, plot graphics/ymax={0.5}, point meta min={0.0}, point meta max={0.7}, plot graphics/!src={tikz/gallery/hoags_object/hoags_object-ngpl-8-srcs-gp-3-std-viridis-0-1}, colormap name={viridis}},
	hoags_object-ngpl-8-srcs-gp-3-std-interp/.style={plot graphics/xmin={-0.5}, plot graphics/xmax={0.5}, plot graphics/ymin={-0.5}, plot graphics/ymax={0.5}, point meta min={0.0}, point meta max={0.7}, plot graphics/!src={tikz/gallery/hoags_object/hoags_object-ngpl-8-srcs-gp-3-std-interp-viridis-0-1}, colormap name={viridis}},
	hoags_object-ngpl-8-srcs-gp-3-std-winterp/.style={plot graphics/xmin={-0.5}, plot graphics/xmax={0.5}, plot graphics/ymin={-0.5}, plot graphics/ymax={0.5}, point meta min={0.0}, point meta max={0.7}, plot graphics/!src={tikz/gallery/hoags_object/hoags_object-ngpl-8-srcs-gp-3-std-winterp-viridis-0-1}, colormap name={viridis}},
	hoags_object-ngpl-8-srcs-gp-3-std-interp-img/.style={plot graphics/xmin={-0.5}, plot graphics/xmax={0.5}, plot graphics/ymin={-0.5}, plot graphics/ymax={0.5}, point meta min={0.0}, point meta max={0.7}, plot graphics/!src={tikz/gallery/hoags_object/hoags_object-ngpl-8-srcs-gp-3-std-interp-img-viridis-0-1}, colormap name={viridis}},
	hoags_object-ngpl-8-srcs-gp-4-mean/.style={plot graphics/xmin={-0.5}, plot graphics/xmax={0.5}, plot graphics/ymin={-0.5}, plot graphics/ymax={0.5}, point meta min={-0.31279483437538147}, point meta max={1.2035950422286987}, plot graphics/!src={tikz/gallery/hoags_object/hoags_object-ngpl-8-srcs-gp-4-mean-viridis--0-1}, colormap name={viridis}},
	hoags_object-ngpl-8-srcs-gp-4-std/.style={plot graphics/xmin={-0.5}, plot graphics/xmax={0.5}, plot graphics/ymin={-0.5}, plot graphics/ymax={0.5}, point meta min={0.0}, point meta max={0.7}, plot graphics/!src={tikz/gallery/hoags_object/hoags_object-ngpl-8-srcs-gp-4-std-viridis-0-1}, colormap name={viridis}},
	hoags_object-ngpl-8-srcs-gp-4-std-interp/.style={plot graphics/xmin={-0.5}, plot graphics/xmax={0.5}, plot graphics/ymin={-0.5}, plot graphics/ymax={0.5}, point meta min={0.0}, point meta max={0.7}, plot graphics/!src={tikz/gallery/hoags_object/hoags_object-ngpl-8-srcs-gp-4-std-interp-viridis-0-1}, colormap name={viridis}},
	hoags_object-ngpl-8-srcs-gp-4-std-winterp/.style={plot graphics/xmin={-0.5}, plot graphics/xmax={0.5}, plot graphics/ymin={-0.5}, plot graphics/ymax={0.5}, point meta min={0.0}, point meta max={0.7}, plot graphics/!src={tikz/gallery/hoags_object/hoags_object-ngpl-8-srcs-gp-4-std-winterp-viridis-0-1}, colormap name={viridis}},
	hoags_object-ngpl-8-srcs-gp-4-std-interp-img/.style={plot graphics/xmin={-0.5}, plot graphics/xmax={0.5}, plot graphics/ymin={-0.5}, plot graphics/ymax={0.5}, point meta min={0.0}, point meta max={0.7}, plot graphics/!src={tikz/gallery/hoags_object/hoags_object-ngpl-8-srcs-gp-4-std-interp-img-viridis-0-1}, colormap name={viridis}},
	hoags_object-ngpl-8-srcs-gp-5-mean/.style={plot graphics/xmin={-0.5}, plot graphics/xmax={0.5}, plot graphics/ymin={-0.5}, plot graphics/ymax={0.5}, point meta min={-0.3053920269012451}, point meta max={0.9488711953163147}, plot graphics/!src={tikz/gallery/hoags_object/hoags_object-ngpl-8-srcs-gp-5-mean-viridis--0-1}, colormap name={viridis}},
	hoags_object-ngpl-8-srcs-gp-5-std/.style={plot graphics/xmin={-0.5}, plot graphics/xmax={0.5}, plot graphics/ymin={-0.5}, plot graphics/ymax={0.5}, point meta min={0.0}, point meta max={0.7}, plot graphics/!src={tikz/gallery/hoags_object/hoags_object-ngpl-8-srcs-gp-5-std-viridis-0-1}, colormap name={viridis}},
	hoags_object-ngpl-8-srcs-gp-5-std-interp/.style={plot graphics/xmin={-0.5}, plot graphics/xmax={0.5}, plot graphics/ymin={-0.5}, plot graphics/ymax={0.5}, point meta min={0.0}, point meta max={0.7}, plot graphics/!src={tikz/gallery/hoags_object/hoags_object-ngpl-8-srcs-gp-5-std-interp-viridis-0-1}, colormap name={viridis}},
	hoags_object-ngpl-8-srcs-gp-5-std-winterp/.style={plot graphics/xmin={-0.5}, plot graphics/xmax={0.5}, plot graphics/ymin={-0.5}, plot graphics/ymax={0.5}, point meta min={0.0}, point meta max={0.7}, plot graphics/!src={tikz/gallery/hoags_object/hoags_object-ngpl-8-srcs-gp-5-std-winterp-viridis-0-1}, colormap name={viridis}},
	hoags_object-ngpl-8-srcs-gp-5-std-interp-img/.style={plot graphics/xmin={-0.5}, plot graphics/xmax={0.5}, plot graphics/ymin={-0.5}, plot graphics/ymax={0.5}, point meta min={0.0}, point meta max={0.7}, plot graphics/!src={tikz/gallery/hoags_object/hoags_object-ngpl-8-srcs-gp-5-std-interp-img-viridis-0-1}, colormap name={viridis}},
	hoags_object-ngpl-8-srcs-gp-6-mean/.style={plot graphics/xmin={-0.5}, plot graphics/xmax={0.5}, plot graphics/ymin={-0.5}, plot graphics/ymax={0.5}, point meta min={-0.31989219784736633}, point meta max={0.7742252349853516}, plot graphics/!src={tikz/gallery/hoags_object/hoags_object-ngpl-8-srcs-gp-6-mean-viridis--0-1}, colormap name={viridis}},
	hoags_object-ngpl-8-srcs-gp-6-std/.style={plot graphics/xmin={-0.5}, plot graphics/xmax={0.5}, plot graphics/ymin={-0.5}, plot graphics/ymax={0.5}, point meta min={0.0}, point meta max={0.7}, plot graphics/!src={tikz/gallery/hoags_object/hoags_object-ngpl-8-srcs-gp-6-std-viridis-0-1}, colormap name={viridis}},
	hoags_object-ngpl-8-srcs-gp-6-std-interp/.style={plot graphics/xmin={-0.5}, plot graphics/xmax={0.5}, plot graphics/ymin={-0.5}, plot graphics/ymax={0.5}, point meta min={0.0}, point meta max={0.7}, plot graphics/!src={tikz/gallery/hoags_object/hoags_object-ngpl-8-srcs-gp-6-std-interp-viridis-0-1}, colormap name={viridis}},
	hoags_object-ngpl-8-srcs-gp-6-std-winterp/.style={plot graphics/xmin={-0.5}, plot graphics/xmax={0.5}, plot graphics/ymin={-0.5}, plot graphics/ymax={0.5}, point meta min={0.0}, point meta max={0.7}, plot graphics/!src={tikz/gallery/hoags_object/hoags_object-ngpl-8-srcs-gp-6-std-winterp-viridis-0-1}, colormap name={viridis}},
	hoags_object-ngpl-8-srcs-gp-6-std-interp-img/.style={plot graphics/xmin={-0.5}, plot graphics/xmax={0.5}, plot graphics/ymin={-0.5}, plot graphics/ymax={0.5}, point meta min={0.0}, point meta max={0.7}, plot graphics/!src={tikz/gallery/hoags_object/hoags_object-ngpl-8-srcs-gp-6-std-interp-img-viridis-0-1}, colormap name={viridis}},
	hoags_object-ngpl-8-srcs-gp-7-mean/.style={plot graphics/xmin={-0.5}, plot graphics/xmax={0.5}, plot graphics/ymin={-0.5}, plot graphics/ymax={0.5}, point meta min={-0.29311132431030273}, point meta max={0.6789448261260986}, plot graphics/!src={tikz/gallery/hoags_object/hoags_object-ngpl-8-srcs-gp-7-mean-viridis--0-1}, colormap name={viridis}},
	hoags_object-ngpl-8-srcs-gp-7-std/.style={plot graphics/xmin={-0.5}, plot graphics/xmax={0.5}, plot graphics/ymin={-0.5}, plot graphics/ymax={0.5}, point meta min={0.0}, point meta max={0.7}, plot graphics/!src={tikz/gallery/hoags_object/hoags_object-ngpl-8-srcs-gp-7-std-viridis-0-1}, colormap name={viridis}},
	hoags_object-ngpl-8-srcs-gp-7-std-interp/.style={plot graphics/xmin={-0.5}, plot graphics/xmax={0.5}, plot graphics/ymin={-0.5}, plot graphics/ymax={0.5}, point meta min={0.0}, point meta max={0.7}, plot graphics/!src={tikz/gallery/hoags_object/hoags_object-ngpl-8-srcs-gp-7-std-interp-viridis-0-1}, colormap name={viridis}},
	hoags_object-ngpl-8-srcs-gp-7-std-winterp/.style={plot graphics/xmin={-0.5}, plot graphics/xmax={0.5}, plot graphics/ymin={-0.5}, plot graphics/ymax={0.5}, point meta min={0.0}, point meta max={0.7}, plot graphics/!src={tikz/gallery/hoags_object/hoags_object-ngpl-8-srcs-gp-7-std-winterp-viridis-0-1}, colormap name={viridis}},
	hoags_object-ngpl-8-srcs-gp-7-std-interp-img/.style={plot graphics/xmin={-0.5}, plot graphics/xmax={0.5}, plot graphics/ymin={-0.5}, plot graphics/ymax={0.5}, point meta min={0.0}, point meta max={0.7}, plot graphics/!src={tikz/gallery/hoags_object/hoags_object-ngpl-8-srcs-gp-7-std-interp-img-viridis-0-1}, colormap name={viridis}},
	hoags_object-ngpl-8-srcs-gp-8-mean/.style={plot graphics/xmin={-0.5}, plot graphics/xmax={0.5}, plot graphics/ymin={-0.5}, plot graphics/ymax={0.5}, point meta min={-0.3349716067314148}, point meta max={0.6098037958145142}, plot graphics/!src={tikz/gallery/hoags_object/hoags_object-ngpl-8-srcs-gp-8-mean-viridis--0-1}, colormap name={viridis}},
	hoags_object-ngpl-8-srcs-gp-8-std/.style={plot graphics/xmin={-0.5}, plot graphics/xmax={0.5}, plot graphics/ymin={-0.5}, plot graphics/ymax={0.5}, point meta min={0.0}, point meta max={0.7}, plot graphics/!src={tikz/gallery/hoags_object/hoags_object-ngpl-8-srcs-gp-8-std-viridis-0-1}, colormap name={viridis}},
	hoags_object-ngpl-8-srcs-gp-8-std-interp/.style={plot graphics/xmin={-0.5}, plot graphics/xmax={0.5}, plot graphics/ymin={-0.5}, plot graphics/ymax={0.5}, point meta min={0.0}, point meta max={0.7}, plot graphics/!src={tikz/gallery/hoags_object/hoags_object-ngpl-8-srcs-gp-8-std-interp-viridis-0-1}, colormap name={viridis}},
	hoags_object-ngpl-8-srcs-gp-8-std-winterp/.style={plot graphics/xmin={-0.5}, plot graphics/xmax={0.5}, plot graphics/ymin={-0.5}, plot graphics/ymax={0.5}, point meta min={0.0}, point meta max={0.7}, plot graphics/!src={tikz/gallery/hoags_object/hoags_object-ngpl-8-srcs-gp-8-std-winterp-viridis-0-1}, colormap name={viridis}},
	hoags_object-ngpl-8-srcs-gp-8-std-interp-img/.style={plot graphics/xmin={-0.5}, plot graphics/xmax={0.5}, plot graphics/ymin={-0.5}, plot graphics/ymax={0.5}, point meta min={0.0}, point meta max={0.7}, plot graphics/!src={tikz/gallery/hoags_object/hoags_object-ngpl-8-srcs-gp-8-std-interp-img-viridis-0-1}, colormap name={viridis}},
	hoags_object-obs/.style={plot graphics/xmin={-2.5}, plot graphics/xmax={2.5}, plot graphics/ymin={-2.5}, plot graphics/ymax={2.5}, point meta min={0.0}, point meta max={30.0}, plot graphics/!src={tikz/gallery/hoags_object/hoags_object-obs-viridis-0-30}, colormap name={viridis}},
	hoags_object-simg/.style={plot graphics/xmin={-0.5}, plot graphics/xmax={0.5}, plot graphics/ymin={-0.5}, plot graphics/ymax={0.5}, point meta min={0.0}, point meta max={30.0}, plot graphics/!src={tikz/gallery/hoags_object/hoags_object-simg-viridis-0-30}, colormap name={viridis}},
	hoags_object-simg-deproj/.style={plot graphics/xmin={-0.5}, plot graphics/xmax={0.5}, plot graphics/ymin={-0.5}, plot graphics/ymax={0.5}, point meta min={0.0}, point meta max={30.0}, plot graphics/!src={tikz/gallery/hoags_object/hoags_object-simg-deproj-viridis-0-30}, colormap name={viridis}},
	hoags_object-simg-wdeproj/.style={plot graphics/xmin={-0.5}, plot graphics/xmax={0.5}, plot graphics/ymin={-0.5}, plot graphics/ymax={0.5}, point meta min={0.0}, point meta max={30.0}, plot graphics/!src={tikz/gallery/hoags_object/hoags_object-simg-wdeproj-viridis-0-30}, colormap name={viridis}},
	hoags_object-ngpl-1b-img-resid-obs-noise/.style={plot graphics/xmin={-2.5}, plot graphics/xmax={2.5}, plot graphics/ymin={-2.5}, plot graphics/ymax={2.5}, point meta min={-3.0}, point meta max={3.0}, plot graphics/!src={tikz/gallery/hoags_object/hoags_object-ngpl-1b-img-resid-obs-noise-bwr--3-3}, colormap name={bwr}},
	hoags_object-ngpl-1b-img-resid-obs-noise+std/.style={plot graphics/xmin={-2.5}, plot graphics/xmax={2.5}, plot graphics/ymin={-2.5}, plot graphics/ymax={2.5}, point meta min={-3.0}, point meta max={3.0}, plot graphics/!src={tikz/gallery/hoags_object/hoags_object-ngpl-1b-img-resid-obs-noise+std-bwr--3-3}, colormap name={bwr}},
	hoags_object-ngpl-1b-img-resid-mu/.style={plot graphics/xmin={-2.5}, plot graphics/xmax={2.5}, plot graphics/ymin={-2.5}, plot graphics/ymax={2.5}, point meta min={-3.0}, point meta max={3.0}, plot graphics/!src={tikz/gallery/hoags_object/hoags_object-ngpl-1b-img-resid-mu-bwr--3-3}, colormap name={bwr}},
	hoags_object-ngpl-1b-simg-resid-noise/.style={plot graphics/xmin={-0.5}, plot graphics/xmax={0.5}, plot graphics/ymin={-0.5}, plot graphics/ymax={0.5}, point meta min={-3.0}, point meta max={3.0}, plot graphics/!src={tikz/gallery/hoags_object/hoags_object-ngpl-1b-simg-resid-noise-bwr--3-3}, colormap name={bwr}},
	hoags_object-ngpl-1b-simg-resid-std/.style={plot graphics/xmin={-0.5}, plot graphics/xmax={0.5}, plot graphics/ymin={-0.5}, plot graphics/ymax={0.5}, point meta min={-3.0}, point meta max={3.0}, plot graphics/!src={tikz/gallery/hoags_object/hoags_object-ngpl-1b-simg-resid-std-bwr--3-3}, colormap name={bwr}},
	hoags_object-ngpl-1b-simg-resid-deproj-noise/.style={plot graphics/xmin={-0.5}, plot graphics/xmax={0.5}, plot graphics/ymin={-0.5}, plot graphics/ymax={0.5}, point meta min={-3.0}, point meta max={3.0}, plot graphics/!src={tikz/gallery/hoags_object/hoags_object-ngpl-1b-simg-resid-deproj-noise-bwr--3-3}, colormap name={bwr}},
	hoags_object-ngpl-1b-simg-resid-deproj-std/.style={plot graphics/xmin={-0.5}, plot graphics/xmax={0.5}, plot graphics/ymin={-0.5}, plot graphics/ymax={0.5}, point meta min={-3.0}, point meta max={3.0}, plot graphics/!src={tikz/gallery/hoags_object/hoags_object-ngpl-1b-simg-resid-deproj-std-bwr--3-3}, colormap name={bwr}},
	hoags_object-ngpl-1b-simg-resid-wdeproj-noise/.style={plot graphics/xmin={-0.5}, plot graphics/xmax={0.5}, plot graphics/ymin={-0.5}, plot graphics/ymax={0.5}, point meta min={-3.0}, point meta max={3.0}, plot graphics/!src={tikz/gallery/hoags_object/hoags_object-ngpl-1b-simg-resid-wdeproj-noise-bwr--3-3}, colormap name={bwr}},
	hoags_object-ngpl-1b-simg-resid-wdeproj-std/.style={plot graphics/xmin={-0.5}, plot graphics/xmax={0.5}, plot graphics/ymin={-0.5}, plot graphics/ymax={0.5}, point meta min={-3.0}, point meta max={3.0}, plot graphics/!src={tikz/gallery/hoags_object/hoags_object-ngpl-1b-simg-resid-wdeproj-std-bwr--3-3}, colormap name={bwr}},
	hoags_object-ngpl-1s-img-resid-obs-noise/.style={plot graphics/xmin={-2.5}, plot graphics/xmax={2.5}, plot graphics/ymin={-2.5}, plot graphics/ymax={2.5}, point meta min={-3.0}, point meta max={3.0}, plot graphics/!src={tikz/gallery/hoags_object/hoags_object-ngpl-1s-img-resid-obs-noise-bwr--3-3}, colormap name={bwr}},
	hoags_object-ngpl-1s-img-resid-obs-noise+std/.style={plot graphics/xmin={-2.5}, plot graphics/xmax={2.5}, plot graphics/ymin={-2.5}, plot graphics/ymax={2.5}, point meta min={-3.0}, point meta max={3.0}, plot graphics/!src={tikz/gallery/hoags_object/hoags_object-ngpl-1s-img-resid-obs-noise+std-bwr--3-3}, colormap name={bwr}},
	hoags_object-ngpl-1s-img-resid-mu/.style={plot graphics/xmin={-2.5}, plot graphics/xmax={2.5}, plot graphics/ymin={-2.5}, plot graphics/ymax={2.5}, point meta min={-3.0}, point meta max={3.0}, plot graphics/!src={tikz/gallery/hoags_object/hoags_object-ngpl-1s-img-resid-mu-bwr--3-3}, colormap name={bwr}},
	hoags_object-ngpl-1s-simg-resid-noise/.style={plot graphics/xmin={-0.5}, plot graphics/xmax={0.5}, plot graphics/ymin={-0.5}, plot graphics/ymax={0.5}, point meta min={-3.0}, point meta max={3.0}, plot graphics/!src={tikz/gallery/hoags_object/hoags_object-ngpl-1s-simg-resid-noise-bwr--3-3}, colormap name={bwr}},
	hoags_object-ngpl-1s-simg-resid-std/.style={plot graphics/xmin={-0.5}, plot graphics/xmax={0.5}, plot graphics/ymin={-0.5}, plot graphics/ymax={0.5}, point meta min={-3.0}, point meta max={3.0}, plot graphics/!src={tikz/gallery/hoags_object/hoags_object-ngpl-1s-simg-resid-std-bwr--3-3}, colormap name={bwr}},
	hoags_object-ngpl-1s-simg-resid-deproj-noise/.style={plot graphics/xmin={-0.5}, plot graphics/xmax={0.5}, plot graphics/ymin={-0.5}, plot graphics/ymax={0.5}, point meta min={-3.0}, point meta max={3.0}, plot graphics/!src={tikz/gallery/hoags_object/hoags_object-ngpl-1s-simg-resid-deproj-noise-bwr--3-3}, colormap name={bwr}},
	hoags_object-ngpl-1s-simg-resid-deproj-std/.style={plot graphics/xmin={-0.5}, plot graphics/xmax={0.5}, plot graphics/ymin={-0.5}, plot graphics/ymax={0.5}, point meta min={-3.0}, point meta max={3.0}, plot graphics/!src={tikz/gallery/hoags_object/hoags_object-ngpl-1s-simg-resid-deproj-std-bwr--3-3}, colormap name={bwr}},
	hoags_object-ngpl-1s-simg-resid-wdeproj-noise/.style={plot graphics/xmin={-0.5}, plot graphics/xmax={0.5}, plot graphics/ymin={-0.5}, plot graphics/ymax={0.5}, point meta min={-3.0}, point meta max={3.0}, plot graphics/!src={tikz/gallery/hoags_object/hoags_object-ngpl-1s-simg-resid-wdeproj-noise-bwr--3-3}, colormap name={bwr}},
	hoags_object-ngpl-1s-simg-resid-wdeproj-std/.style={plot graphics/xmin={-0.5}, plot graphics/xmax={0.5}, plot graphics/ymin={-0.5}, plot graphics/ymax={0.5}, point meta min={-3.0}, point meta max={3.0}, plot graphics/!src={tikz/gallery/hoags_object/hoags_object-ngpl-1s-simg-resid-wdeproj-std-bwr--3-3}, colormap name={bwr}},
	hoags_object-ngpl-2-img-resid-obs-noise/.style={plot graphics/xmin={-2.5}, plot graphics/xmax={2.5}, plot graphics/ymin={-2.5}, plot graphics/ymax={2.5}, point meta min={-3.0}, point meta max={3.0}, plot graphics/!src={tikz/gallery/hoags_object/hoags_object-ngpl-2-img-resid-obs-noise-bwr--3-3}, colormap name={bwr}},
	hoags_object-ngpl-2-img-resid-obs-noise+std/.style={plot graphics/xmin={-2.5}, plot graphics/xmax={2.5}, plot graphics/ymin={-2.5}, plot graphics/ymax={2.5}, point meta min={-3.0}, point meta max={3.0}, plot graphics/!src={tikz/gallery/hoags_object/hoags_object-ngpl-2-img-resid-obs-noise+std-bwr--3-3}, colormap name={bwr}},
	hoags_object-ngpl-2-img-resid-mu/.style={plot graphics/xmin={-2.5}, plot graphics/xmax={2.5}, plot graphics/ymin={-2.5}, plot graphics/ymax={2.5}, point meta min={-3.0}, point meta max={3.0}, plot graphics/!src={tikz/gallery/hoags_object/hoags_object-ngpl-2-img-resid-mu-bwr--3-3}, colormap name={bwr}},
	hoags_object-ngpl-2-simg-resid-noise/.style={plot graphics/xmin={-0.5}, plot graphics/xmax={0.5}, plot graphics/ymin={-0.5}, plot graphics/ymax={0.5}, point meta min={-3.0}, point meta max={3.0}, plot graphics/!src={tikz/gallery/hoags_object/hoags_object-ngpl-2-simg-resid-noise-bwr--3-3}, colormap name={bwr}},
	hoags_object-ngpl-2-simg-resid-std/.style={plot graphics/xmin={-0.5}, plot graphics/xmax={0.5}, plot graphics/ymin={-0.5}, plot graphics/ymax={0.5}, point meta min={-3.0}, point meta max={3.0}, plot graphics/!src={tikz/gallery/hoags_object/hoags_object-ngpl-2-simg-resid-std-bwr--3-3}, colormap name={bwr}},
	hoags_object-ngpl-2-simg-resid-deproj-noise/.style={plot graphics/xmin={-0.5}, plot graphics/xmax={0.5}, plot graphics/ymin={-0.5}, plot graphics/ymax={0.5}, point meta min={-3.0}, point meta max={3.0}, plot graphics/!src={tikz/gallery/hoags_object/hoags_object-ngpl-2-simg-resid-deproj-noise-bwr--3-3}, colormap name={bwr}},
	hoags_object-ngpl-2-simg-resid-deproj-std/.style={plot graphics/xmin={-0.5}, plot graphics/xmax={0.5}, plot graphics/ymin={-0.5}, plot graphics/ymax={0.5}, point meta min={-3.0}, point meta max={3.0}, plot graphics/!src={tikz/gallery/hoags_object/hoags_object-ngpl-2-simg-resid-deproj-std-bwr--3-3}, colormap name={bwr}},
	hoags_object-ngpl-2-simg-resid-wdeproj-noise/.style={plot graphics/xmin={-0.5}, plot graphics/xmax={0.5}, plot graphics/ymin={-0.5}, plot graphics/ymax={0.5}, point meta min={-3.0}, point meta max={3.0}, plot graphics/!src={tikz/gallery/hoags_object/hoags_object-ngpl-2-simg-resid-wdeproj-noise-bwr--3-3}, colormap name={bwr}},
	hoags_object-ngpl-2-simg-resid-wdeproj-std/.style={plot graphics/xmin={-0.5}, plot graphics/xmax={0.5}, plot graphics/ymin={-0.5}, plot graphics/ymax={0.5}, point meta min={-3.0}, point meta max={3.0}, plot graphics/!src={tikz/gallery/hoags_object/hoags_object-ngpl-2-simg-resid-wdeproj-std-bwr--3-3}, colormap name={bwr}},
	hoags_object-ngpl-3-img-resid-obs-noise/.style={plot graphics/xmin={-2.5}, plot graphics/xmax={2.5}, plot graphics/ymin={-2.5}, plot graphics/ymax={2.5}, point meta min={-3.0}, point meta max={3.0}, plot graphics/!src={tikz/gallery/hoags_object/hoags_object-ngpl-3-img-resid-obs-noise-bwr--3-3}, colormap name={bwr}},
	hoags_object-ngpl-3-img-resid-obs-noise+std/.style={plot graphics/xmin={-2.5}, plot graphics/xmax={2.5}, plot graphics/ymin={-2.5}, plot graphics/ymax={2.5}, point meta min={-3.0}, point meta max={3.0}, plot graphics/!src={tikz/gallery/hoags_object/hoags_object-ngpl-3-img-resid-obs-noise+std-bwr--3-3}, colormap name={bwr}},
	hoags_object-ngpl-3-img-resid-mu/.style={plot graphics/xmin={-2.5}, plot graphics/xmax={2.5}, plot graphics/ymin={-2.5}, plot graphics/ymax={2.5}, point meta min={-3.0}, point meta max={3.0}, plot graphics/!src={tikz/gallery/hoags_object/hoags_object-ngpl-3-img-resid-mu-bwr--3-3}, colormap name={bwr}},
	hoags_object-ngpl-3-simg-resid-noise/.style={plot graphics/xmin={-0.5}, plot graphics/xmax={0.5}, plot graphics/ymin={-0.5}, plot graphics/ymax={0.5}, point meta min={-3.0}, point meta max={3.0}, plot graphics/!src={tikz/gallery/hoags_object/hoags_object-ngpl-3-simg-resid-noise-bwr--3-3}, colormap name={bwr}},
	hoags_object-ngpl-3-simg-resid-std/.style={plot graphics/xmin={-0.5}, plot graphics/xmax={0.5}, plot graphics/ymin={-0.5}, plot graphics/ymax={0.5}, point meta min={-3.0}, point meta max={3.0}, plot graphics/!src={tikz/gallery/hoags_object/hoags_object-ngpl-3-simg-resid-std-bwr--3-3}, colormap name={bwr}},
	hoags_object-ngpl-3-simg-resid-deproj-noise/.style={plot graphics/xmin={-0.5}, plot graphics/xmax={0.5}, plot graphics/ymin={-0.5}, plot graphics/ymax={0.5}, point meta min={-3.0}, point meta max={3.0}, plot graphics/!src={tikz/gallery/hoags_object/hoags_object-ngpl-3-simg-resid-deproj-noise-bwr--3-3}, colormap name={bwr}},
	hoags_object-ngpl-3-simg-resid-deproj-std/.style={plot graphics/xmin={-0.5}, plot graphics/xmax={0.5}, plot graphics/ymin={-0.5}, plot graphics/ymax={0.5}, point meta min={-3.0}, point meta max={3.0}, plot graphics/!src={tikz/gallery/hoags_object/hoags_object-ngpl-3-simg-resid-deproj-std-bwr--3-3}, colormap name={bwr}},
	hoags_object-ngpl-3-simg-resid-wdeproj-noise/.style={plot graphics/xmin={-0.5}, plot graphics/xmax={0.5}, plot graphics/ymin={-0.5}, plot graphics/ymax={0.5}, point meta min={-3.0}, point meta max={3.0}, plot graphics/!src={tikz/gallery/hoags_object/hoags_object-ngpl-3-simg-resid-wdeproj-noise-bwr--3-3}, colormap name={bwr}},
	hoags_object-ngpl-3-simg-resid-wdeproj-std/.style={plot graphics/xmin={-0.5}, plot graphics/xmax={0.5}, plot graphics/ymin={-0.5}, plot graphics/ymax={0.5}, point meta min={-3.0}, point meta max={3.0}, plot graphics/!src={tikz/gallery/hoags_object/hoags_object-ngpl-3-simg-resid-wdeproj-std-bwr--3-3}, colormap name={bwr}},
	hoags_object-ngpl-4-img-resid-obs-noise/.style={plot graphics/xmin={-2.5}, plot graphics/xmax={2.5}, plot graphics/ymin={-2.5}, plot graphics/ymax={2.5}, point meta min={-3.0}, point meta max={3.0}, plot graphics/!src={tikz/gallery/hoags_object/hoags_object-ngpl-4-img-resid-obs-noise-bwr--3-3}, colormap name={bwr}},
	hoags_object-ngpl-4-img-resid-obs-noise+std/.style={plot graphics/xmin={-2.5}, plot graphics/xmax={2.5}, plot graphics/ymin={-2.5}, plot graphics/ymax={2.5}, point meta min={-3.0}, point meta max={3.0}, plot graphics/!src={tikz/gallery/hoags_object/hoags_object-ngpl-4-img-resid-obs-noise+std-bwr--3-3}, colormap name={bwr}},
	hoags_object-ngpl-4-img-resid-mu/.style={plot graphics/xmin={-2.5}, plot graphics/xmax={2.5}, plot graphics/ymin={-2.5}, plot graphics/ymax={2.5}, point meta min={-3.0}, point meta max={3.0}, plot graphics/!src={tikz/gallery/hoags_object/hoags_object-ngpl-4-img-resid-mu-bwr--3-3}, colormap name={bwr}},
	hoags_object-ngpl-4-simg-resid-noise/.style={plot graphics/xmin={-0.5}, plot graphics/xmax={0.5}, plot graphics/ymin={-0.5}, plot graphics/ymax={0.5}, point meta min={-3.0}, point meta max={3.0}, plot graphics/!src={tikz/gallery/hoags_object/hoags_object-ngpl-4-simg-resid-noise-bwr--3-3}, colormap name={bwr}},
	hoags_object-ngpl-4-simg-resid-std/.style={plot graphics/xmin={-0.5}, plot graphics/xmax={0.5}, plot graphics/ymin={-0.5}, plot graphics/ymax={0.5}, point meta min={-3.0}, point meta max={3.0}, plot graphics/!src={tikz/gallery/hoags_object/hoags_object-ngpl-4-simg-resid-std-bwr--3-3}, colormap name={bwr}},
	hoags_object-ngpl-4-simg-resid-deproj-noise/.style={plot graphics/xmin={-0.5}, plot graphics/xmax={0.5}, plot graphics/ymin={-0.5}, plot graphics/ymax={0.5}, point meta min={-3.0}, point meta max={3.0}, plot graphics/!src={tikz/gallery/hoags_object/hoags_object-ngpl-4-simg-resid-deproj-noise-bwr--3-3}, colormap name={bwr}},
	hoags_object-ngpl-4-simg-resid-deproj-std/.style={plot graphics/xmin={-0.5}, plot graphics/xmax={0.5}, plot graphics/ymin={-0.5}, plot graphics/ymax={0.5}, point meta min={-3.0}, point meta max={3.0}, plot graphics/!src={tikz/gallery/hoags_object/hoags_object-ngpl-4-simg-resid-deproj-std-bwr--3-3}, colormap name={bwr}},
	hoags_object-ngpl-4-simg-resid-wdeproj-noise/.style={plot graphics/xmin={-0.5}, plot graphics/xmax={0.5}, plot graphics/ymin={-0.5}, plot graphics/ymax={0.5}, point meta min={-3.0}, point meta max={3.0}, plot graphics/!src={tikz/gallery/hoags_object/hoags_object-ngpl-4-simg-resid-wdeproj-noise-bwr--3-3}, colormap name={bwr}},
	hoags_object-ngpl-4-simg-resid-wdeproj-std/.style={plot graphics/xmin={-0.5}, plot graphics/xmax={0.5}, plot graphics/ymin={-0.5}, plot graphics/ymax={0.5}, point meta min={-3.0}, point meta max={3.0}, plot graphics/!src={tikz/gallery/hoags_object/hoags_object-ngpl-4-simg-resid-wdeproj-std-bwr--3-3}, colormap name={bwr}},
	hoags_object-ngpl-5-img-resid-obs-noise/.style={plot graphics/xmin={-2.5}, plot graphics/xmax={2.5}, plot graphics/ymin={-2.5}, plot graphics/ymax={2.5}, point meta min={-3.0}, point meta max={3.0}, plot graphics/!src={tikz/gallery/hoags_object/hoags_object-ngpl-5-img-resid-obs-noise-bwr--3-3}, colormap name={bwr}},
	hoags_object-ngpl-5-img-resid-obs-noise+std/.style={plot graphics/xmin={-2.5}, plot graphics/xmax={2.5}, plot graphics/ymin={-2.5}, plot graphics/ymax={2.5}, point meta min={-3.0}, point meta max={3.0}, plot graphics/!src={tikz/gallery/hoags_object/hoags_object-ngpl-5-img-resid-obs-noise+std-bwr--3-3}, colormap name={bwr}},
	hoags_object-ngpl-5-img-resid-mu/.style={plot graphics/xmin={-2.5}, plot graphics/xmax={2.5}, plot graphics/ymin={-2.5}, plot graphics/ymax={2.5}, point meta min={-3.0}, point meta max={3.0}, plot graphics/!src={tikz/gallery/hoags_object/hoags_object-ngpl-5-img-resid-mu-bwr--3-3}, colormap name={bwr}},
	hoags_object-ngpl-5-simg-resid-noise/.style={plot graphics/xmin={-0.5}, plot graphics/xmax={0.5}, plot graphics/ymin={-0.5}, plot graphics/ymax={0.5}, point meta min={-3.0}, point meta max={3.0}, plot graphics/!src={tikz/gallery/hoags_object/hoags_object-ngpl-5-simg-resid-noise-bwr--3-3}, colormap name={bwr}},
	hoags_object-ngpl-5-simg-resid-std/.style={plot graphics/xmin={-0.5}, plot graphics/xmax={0.5}, plot graphics/ymin={-0.5}, plot graphics/ymax={0.5}, point meta min={-3.0}, point meta max={3.0}, plot graphics/!src={tikz/gallery/hoags_object/hoags_object-ngpl-5-simg-resid-std-bwr--3-3}, colormap name={bwr}},
	hoags_object-ngpl-5-simg-resid-deproj-noise/.style={plot graphics/xmin={-0.5}, plot graphics/xmax={0.5}, plot graphics/ymin={-0.5}, plot graphics/ymax={0.5}, point meta min={-3.0}, point meta max={3.0}, plot graphics/!src={tikz/gallery/hoags_object/hoags_object-ngpl-5-simg-resid-deproj-noise-bwr--3-3}, colormap name={bwr}},
	hoags_object-ngpl-5-simg-resid-deproj-std/.style={plot graphics/xmin={-0.5}, plot graphics/xmax={0.5}, plot graphics/ymin={-0.5}, plot graphics/ymax={0.5}, point meta min={-3.0}, point meta max={3.0}, plot graphics/!src={tikz/gallery/hoags_object/hoags_object-ngpl-5-simg-resid-deproj-std-bwr--3-3}, colormap name={bwr}},
	hoags_object-ngpl-5-simg-resid-wdeproj-noise/.style={plot graphics/xmin={-0.5}, plot graphics/xmax={0.5}, plot graphics/ymin={-0.5}, plot graphics/ymax={0.5}, point meta min={-3.0}, point meta max={3.0}, plot graphics/!src={tikz/gallery/hoags_object/hoags_object-ngpl-5-simg-resid-wdeproj-noise-bwr--3-3}, colormap name={bwr}},
	hoags_object-ngpl-5-simg-resid-wdeproj-std/.style={plot graphics/xmin={-0.5}, plot graphics/xmax={0.5}, plot graphics/ymin={-0.5}, plot graphics/ymax={0.5}, point meta min={-3.0}, point meta max={3.0}, plot graphics/!src={tikz/gallery/hoags_object/hoags_object-ngpl-5-simg-resid-wdeproj-std-bwr--3-3}, colormap name={bwr}},
	hoags_object-ngpl-6-img-resid-obs-noise/.style={plot graphics/xmin={-2.5}, plot graphics/xmax={2.5}, plot graphics/ymin={-2.5}, plot graphics/ymax={2.5}, point meta min={-3.0}, point meta max={3.0}, plot graphics/!src={tikz/gallery/hoags_object/hoags_object-ngpl-6-img-resid-obs-noise-bwr--3-3}, colormap name={bwr}},
	hoags_object-ngpl-6-img-resid-obs-noise+std/.style={plot graphics/xmin={-2.5}, plot graphics/xmax={2.5}, plot graphics/ymin={-2.5}, plot graphics/ymax={2.5}, point meta min={-3.0}, point meta max={3.0}, plot graphics/!src={tikz/gallery/hoags_object/hoags_object-ngpl-6-img-resid-obs-noise+std-bwr--3-3}, colormap name={bwr}},
	hoags_object-ngpl-6-img-resid-mu/.style={plot graphics/xmin={-2.5}, plot graphics/xmax={2.5}, plot graphics/ymin={-2.5}, plot graphics/ymax={2.5}, point meta min={-3.0}, point meta max={3.0}, plot graphics/!src={tikz/gallery/hoags_object/hoags_object-ngpl-6-img-resid-mu-bwr--3-3}, colormap name={bwr}},
	hoags_object-ngpl-6-simg-resid-noise/.style={plot graphics/xmin={-0.5}, plot graphics/xmax={0.5}, plot graphics/ymin={-0.5}, plot graphics/ymax={0.5}, point meta min={-3.0}, point meta max={3.0}, plot graphics/!src={tikz/gallery/hoags_object/hoags_object-ngpl-6-simg-resid-noise-bwr--3-3}, colormap name={bwr}},
	hoags_object-ngpl-6-simg-resid-std/.style={plot graphics/xmin={-0.5}, plot graphics/xmax={0.5}, plot graphics/ymin={-0.5}, plot graphics/ymax={0.5}, point meta min={-3.0}, point meta max={3.0}, plot graphics/!src={tikz/gallery/hoags_object/hoags_object-ngpl-6-simg-resid-std-bwr--3-3}, colormap name={bwr}},
	hoags_object-ngpl-6-simg-resid-deproj-noise/.style={plot graphics/xmin={-0.5}, plot graphics/xmax={0.5}, plot graphics/ymin={-0.5}, plot graphics/ymax={0.5}, point meta min={-3.0}, point meta max={3.0}, plot graphics/!src={tikz/gallery/hoags_object/hoags_object-ngpl-6-simg-resid-deproj-noise-bwr--3-3}, colormap name={bwr}},
	hoags_object-ngpl-6-simg-resid-deproj-std/.style={plot graphics/xmin={-0.5}, plot graphics/xmax={0.5}, plot graphics/ymin={-0.5}, plot graphics/ymax={0.5}, point meta min={-3.0}, point meta max={3.0}, plot graphics/!src={tikz/gallery/hoags_object/hoags_object-ngpl-6-simg-resid-deproj-std-bwr--3-3}, colormap name={bwr}},
	hoags_object-ngpl-6-simg-resid-wdeproj-noise/.style={plot graphics/xmin={-0.5}, plot graphics/xmax={0.5}, plot graphics/ymin={-0.5}, plot graphics/ymax={0.5}, point meta min={-3.0}, point meta max={3.0}, plot graphics/!src={tikz/gallery/hoags_object/hoags_object-ngpl-6-simg-resid-wdeproj-noise-bwr--3-3}, colormap name={bwr}},
	hoags_object-ngpl-6-simg-resid-wdeproj-std/.style={plot graphics/xmin={-0.5}, plot graphics/xmax={0.5}, plot graphics/ymin={-0.5}, plot graphics/ymax={0.5}, point meta min={-3.0}, point meta max={3.0}, plot graphics/!src={tikz/gallery/hoags_object/hoags_object-ngpl-6-simg-resid-wdeproj-std-bwr--3-3}, colormap name={bwr}},
	hoags_object-ngpl-7-img-resid-obs-noise/.style={plot graphics/xmin={-2.5}, plot graphics/xmax={2.5}, plot graphics/ymin={-2.5}, plot graphics/ymax={2.5}, point meta min={-3.0}, point meta max={3.0}, plot graphics/!src={tikz/gallery/hoags_object/hoags_object-ngpl-7-img-resid-obs-noise-bwr--3-3}, colormap name={bwr}},
	hoags_object-ngpl-7-img-resid-obs-noise+std/.style={plot graphics/xmin={-2.5}, plot graphics/xmax={2.5}, plot graphics/ymin={-2.5}, plot graphics/ymax={2.5}, point meta min={-3.0}, point meta max={3.0}, plot graphics/!src={tikz/gallery/hoags_object/hoags_object-ngpl-7-img-resid-obs-noise+std-bwr--3-3}, colormap name={bwr}},
	hoags_object-ngpl-7-img-resid-mu/.style={plot graphics/xmin={-2.5}, plot graphics/xmax={2.5}, plot graphics/ymin={-2.5}, plot graphics/ymax={2.5}, point meta min={-3.0}, point meta max={3.0}, plot graphics/!src={tikz/gallery/hoags_object/hoags_object-ngpl-7-img-resid-mu-bwr--3-3}, colormap name={bwr}},
	hoags_object-ngpl-7-simg-resid-noise/.style={plot graphics/xmin={-0.5}, plot graphics/xmax={0.5}, plot graphics/ymin={-0.5}, plot graphics/ymax={0.5}, point meta min={-3.0}, point meta max={3.0}, plot graphics/!src={tikz/gallery/hoags_object/hoags_object-ngpl-7-simg-resid-noise-bwr--3-3}, colormap name={bwr}},
	hoags_object-ngpl-7-simg-resid-std/.style={plot graphics/xmin={-0.5}, plot graphics/xmax={0.5}, plot graphics/ymin={-0.5}, plot graphics/ymax={0.5}, point meta min={-3.0}, point meta max={3.0}, plot graphics/!src={tikz/gallery/hoags_object/hoags_object-ngpl-7-simg-resid-std-bwr--3-3}, colormap name={bwr}},
	hoags_object-ngpl-7-simg-resid-deproj-noise/.style={plot graphics/xmin={-0.5}, plot graphics/xmax={0.5}, plot graphics/ymin={-0.5}, plot graphics/ymax={0.5}, point meta min={-3.0}, point meta max={3.0}, plot graphics/!src={tikz/gallery/hoags_object/hoags_object-ngpl-7-simg-resid-deproj-noise-bwr--3-3}, colormap name={bwr}},
	hoags_object-ngpl-7-simg-resid-deproj-std/.style={plot graphics/xmin={-0.5}, plot graphics/xmax={0.5}, plot graphics/ymin={-0.5}, plot graphics/ymax={0.5}, point meta min={-3.0}, point meta max={3.0}, plot graphics/!src={tikz/gallery/hoags_object/hoags_object-ngpl-7-simg-resid-deproj-std-bwr--3-3}, colormap name={bwr}},
	hoags_object-ngpl-7-simg-resid-wdeproj-noise/.style={plot graphics/xmin={-0.5}, plot graphics/xmax={0.5}, plot graphics/ymin={-0.5}, plot graphics/ymax={0.5}, point meta min={-3.0}, point meta max={3.0}, plot graphics/!src={tikz/gallery/hoags_object/hoags_object-ngpl-7-simg-resid-wdeproj-noise-bwr--3-3}, colormap name={bwr}},
	hoags_object-ngpl-7-simg-resid-wdeproj-std/.style={plot graphics/xmin={-0.5}, plot graphics/xmax={0.5}, plot graphics/ymin={-0.5}, plot graphics/ymax={0.5}, point meta min={-3.0}, point meta max={3.0}, plot graphics/!src={tikz/gallery/hoags_object/hoags_object-ngpl-7-simg-resid-wdeproj-std-bwr--3-3}, colormap name={bwr}},
	hoags_object-ngpl-8-img-resid-obs-noise/.style={plot graphics/xmin={-2.5}, plot graphics/xmax={2.5}, plot graphics/ymin={-2.5}, plot graphics/ymax={2.5}, point meta min={-3.0}, point meta max={3.0}, plot graphics/!src={tikz/gallery/hoags_object/hoags_object-ngpl-8-img-resid-obs-noise-bwr--3-3}, colormap name={bwr}},
	hoags_object-ngpl-8-img-resid-obs-noise+std/.style={plot graphics/xmin={-2.5}, plot graphics/xmax={2.5}, plot graphics/ymin={-2.5}, plot graphics/ymax={2.5}, point meta min={-3.0}, point meta max={3.0}, plot graphics/!src={tikz/gallery/hoags_object/hoags_object-ngpl-8-img-resid-obs-noise+std-bwr--3-3}, colormap name={bwr}},
	hoags_object-ngpl-8-img-resid-mu/.style={plot graphics/xmin={-2.5}, plot graphics/xmax={2.5}, plot graphics/ymin={-2.5}, plot graphics/ymax={2.5}, point meta min={-3.0}, point meta max={3.0}, plot graphics/!src={tikz/gallery/hoags_object/hoags_object-ngpl-8-img-resid-mu-bwr--3-3}, colormap name={bwr}},
	hoags_object-ngpl-8-simg-resid-noise/.style={plot graphics/xmin={-0.5}, plot graphics/xmax={0.5}, plot graphics/ymin={-0.5}, plot graphics/ymax={0.5}, point meta min={-3.0}, point meta max={3.0}, plot graphics/!src={tikz/gallery/hoags_object/hoags_object-ngpl-8-simg-resid-noise-bwr--3-3}, colormap name={bwr}},
	hoags_object-ngpl-8-simg-resid-std/.style={plot graphics/xmin={-0.5}, plot graphics/xmax={0.5}, plot graphics/ymin={-0.5}, plot graphics/ymax={0.5}, point meta min={-3.0}, point meta max={3.0}, plot graphics/!src={tikz/gallery/hoags_object/hoags_object-ngpl-8-simg-resid-std-bwr--3-3}, colormap name={bwr}},
	hoags_object-ngpl-8-simg-resid-deproj-noise/.style={plot graphics/xmin={-0.5}, plot graphics/xmax={0.5}, plot graphics/ymin={-0.5}, plot graphics/ymax={0.5}, point meta min={-3.0}, point meta max={3.0}, plot graphics/!src={tikz/gallery/hoags_object/hoags_object-ngpl-8-simg-resid-deproj-noise-bwr--3-3}, colormap name={bwr}},
	hoags_object-ngpl-8-simg-resid-deproj-std/.style={plot graphics/xmin={-0.5}, plot graphics/xmax={0.5}, plot graphics/ymin={-0.5}, plot graphics/ymax={0.5}, point meta min={-3.0}, point meta max={3.0}, plot graphics/!src={tikz/gallery/hoags_object/hoags_object-ngpl-8-simg-resid-deproj-std-bwr--3-3}, colormap name={bwr}},
	hoags_object-ngpl-8-simg-resid-wdeproj-noise/.style={plot graphics/xmin={-0.5}, plot graphics/xmax={0.5}, plot graphics/ymin={-0.5}, plot graphics/ymax={0.5}, point meta min={-3.0}, point meta max={3.0}, plot graphics/!src={tikz/gallery/hoags_object/hoags_object-ngpl-8-simg-resid-wdeproj-noise-bwr--3-3}, colormap name={bwr}},
	hoags_object-ngpl-8-simg-resid-wdeproj-std/.style={plot graphics/xmin={-0.5}, plot graphics/xmax={0.5}, plot graphics/ymin={-0.5}, plot graphics/ymax={0.5}, point meta min={-3.0}, point meta max={3.0}, plot graphics/!src={tikz/gallery/hoags_object/hoags_object-ngpl-8-simg-resid-wdeproj-std-bwr--3-3}, colormap name={bwr}},
}
\pgfplotsset{,
	antennae-ngpl-3-img-mean/.style={plot graphics/xmin={-2.5}, plot graphics/xmax={2.5}, plot graphics/ymin={-2.5}, plot graphics/ymax={2.5}, point meta min={0.0}, point meta max={30.0}, plot graphics/!src={tikz/gallery/antennae/antennae-ngpl-3-img-mean-viridis-0-30}, colormap name={viridis}},
	antennae-ngpl-3-img-std/.style={plot graphics/xmin={-2.5}, plot graphics/xmax={2.5}, plot graphics/ymin={-2.5}, plot graphics/ymax={2.5}, point meta min={0.0}, point meta max={0.7}, plot graphics/!src={tikz/gallery/antennae/antennae-ngpl-3-img-std-viridis-0-1}, colormap name={viridis}},
	antennae-ngpl-3-simg-deproj-mean/.style={plot graphics/xmin={-0.6}, plot graphics/xmax={0.6}, plot graphics/ymin={-0.6}, plot graphics/ymax={0.6}, point meta min={0.0}, point meta max={30.0}, plot graphics/!src={tikz/gallery/antennae/antennae-ngpl-3-simg-deproj-mean-viridis-0-30}, colormap name={viridis}},
	antennae-ngpl-3-simg-deproj-std/.style={plot graphics/xmin={-0.6}, plot graphics/xmax={0.6}, plot graphics/ymin={-0.6}, plot graphics/ymax={0.6}, point meta min={0.0}, point meta max={0.7}, plot graphics/!src={tikz/gallery/antennae/antennae-ngpl-3-simg-deproj-std-viridis-0-1}, colormap name={viridis}},
	antennae-ngpl-3-simg-mean/.style={plot graphics/xmin={-0.6}, plot graphics/xmax={0.6}, plot graphics/ymin={-0.6}, plot graphics/ymax={0.6}, point meta min={0.0}, point meta max={30.0}, plot graphics/!src={tikz/gallery/antennae/antennae-ngpl-3-simg-mean-viridis-0-30}, colormap name={viridis}},
	antennae-ngpl-3-simg-std/.style={plot graphics/xmin={-0.6}, plot graphics/xmax={0.6}, plot graphics/ymin={-0.6}, plot graphics/ymax={0.6}, point meta min={0.0}, point meta max={0.7}, plot graphics/!src={tikz/gallery/antennae/antennae-ngpl-3-simg-std-viridis-0-1}, colormap name={viridis}},
	antennae-ngpl-3-simg-std-interp/.style={plot graphics/xmin={-0.6}, plot graphics/xmax={0.6}, plot graphics/ymin={-0.6}, plot graphics/ymax={0.6}, point meta min={0.0}, point meta max={0.7}, plot graphics/!src={tikz/gallery/antennae/antennae-ngpl-3-simg-std-interp-viridis-0-1}, colormap name={viridis}},
	antennae-ngpl-3-simg-std-winterp/.style={plot graphics/xmin={-0.6}, plot graphics/xmax={0.6}, plot graphics/ymin={-0.6}, plot graphics/ymax={0.6}, point meta min={0.0}, point meta max={0.7}, plot graphics/!src={tikz/gallery/antennae/antennae-ngpl-3-simg-std-winterp-viridis-0-1}, colormap name={viridis}},
	antennae-ngpl-3-simg-std-interp-img/.style={plot graphics/xmin={-0.6}, plot graphics/xmax={0.6}, plot graphics/ymin={-0.6}, plot graphics/ymax={0.6}, point meta min={0.0}, point meta max={0.7}, plot graphics/!src={tikz/gallery/antennae/antennae-ngpl-3-simg-std-interp-img-viridis-0-1}, colormap name={viridis}},
	antennae-ngpl-3-srcs-src-mean/.style={plot graphics/xmin={-0.6}, plot graphics/xmax={0.6}, plot graphics/ymin={-0.6}, plot graphics/ymax={0.6}, point meta min={-0.6188250780105591}, point meta max={28.205341339111328}, plot graphics/!src={tikz/gallery/antennae/antennae-ngpl-3-srcs-src-mean-viridis--1-28}, colormap name={viridis}},
	antennae-ngpl-3-srcs-src-std/.style={plot graphics/xmin={-0.6}, plot graphics/xmax={0.6}, plot graphics/ymin={-0.6}, plot graphics/ymax={0.6}, point meta min={0.0}, point meta max={0.7}, plot graphics/!src={tikz/gallery/antennae/antennae-ngpl-3-srcs-src-std-viridis-0-1}, colormap name={viridis}},
	antennae-ngpl-3-srcs-gp-1-mean/.style={plot graphics/xmin={-0.6}, plot graphics/xmax={0.6}, plot graphics/ymin={-0.6}, plot graphics/ymax={0.6}, point meta min={-4.2482099533081055}, point meta max={4.94185209274292}, plot graphics/!src={tikz/gallery/antennae/antennae-ngpl-3-srcs-gp-1-mean-viridis--4-5}, colormap name={viridis}},
	antennae-ngpl-3-srcs-gp-1-std/.style={plot graphics/xmin={-0.6}, plot graphics/xmax={0.6}, plot graphics/ymin={-0.6}, plot graphics/ymax={0.6}, point meta min={0.0}, point meta max={0.7}, plot graphics/!src={tikz/gallery/antennae/antennae-ngpl-3-srcs-gp-1-std-viridis-0-1}, colormap name={viridis}},
	antennae-ngpl-3-srcs-gp-1-std-interp/.style={plot graphics/xmin={-0.6}, plot graphics/xmax={0.6}, plot graphics/ymin={-0.6}, plot graphics/ymax={0.6}, point meta min={0.0}, point meta max={0.7}, plot graphics/!src={tikz/gallery/antennae/antennae-ngpl-3-srcs-gp-1-std-interp-viridis-0-1}, colormap name={viridis}},
	antennae-ngpl-3-srcs-gp-1-std-winterp/.style={plot graphics/xmin={-0.6}, plot graphics/xmax={0.6}, plot graphics/ymin={-0.6}, plot graphics/ymax={0.6}, point meta min={0.0}, point meta max={0.7}, plot graphics/!src={tikz/gallery/antennae/antennae-ngpl-3-srcs-gp-1-std-winterp-viridis-0-1}, colormap name={viridis}},
	antennae-ngpl-3-srcs-gp-1-std-interp-img/.style={plot graphics/xmin={-0.6}, plot graphics/xmax={0.6}, plot graphics/ymin={-0.6}, plot graphics/ymax={0.6}, point meta min={0.0}, point meta max={0.7}, plot graphics/!src={tikz/gallery/antennae/antennae-ngpl-3-srcs-gp-1-std-interp-img-viridis-0-1}, colormap name={viridis}},
	antennae-ngpl-3-srcs-gp-2-mean/.style={plot graphics/xmin={-0.6}, plot graphics/xmax={0.6}, plot graphics/ymin={-0.6}, plot graphics/ymax={0.6}, point meta min={-1.1360419988632202}, point meta max={1.8180923461914062}, plot graphics/!src={tikz/gallery/antennae/antennae-ngpl-3-srcs-gp-2-mean-viridis--1-2}, colormap name={viridis}},
	antennae-ngpl-3-srcs-gp-2-std/.style={plot graphics/xmin={-0.6}, plot graphics/xmax={0.6}, plot graphics/ymin={-0.6}, plot graphics/ymax={0.6}, point meta min={0.0}, point meta max={0.7}, plot graphics/!src={tikz/gallery/antennae/antennae-ngpl-3-srcs-gp-2-std-viridis-0-1}, colormap name={viridis}},
	antennae-ngpl-3-srcs-gp-2-std-interp/.style={plot graphics/xmin={-0.6}, plot graphics/xmax={0.6}, plot graphics/ymin={-0.6}, plot graphics/ymax={0.6}, point meta min={0.0}, point meta max={0.7}, plot graphics/!src={tikz/gallery/antennae/antennae-ngpl-3-srcs-gp-2-std-interp-viridis-0-1}, colormap name={viridis}},
	antennae-ngpl-3-srcs-gp-2-std-winterp/.style={plot graphics/xmin={-0.6}, plot graphics/xmax={0.6}, plot graphics/ymin={-0.6}, plot graphics/ymax={0.6}, point meta min={0.0}, point meta max={0.7}, plot graphics/!src={tikz/gallery/antennae/antennae-ngpl-3-srcs-gp-2-std-winterp-viridis-0-1}, colormap name={viridis}},
	antennae-ngpl-3-srcs-gp-2-std-interp-img/.style={plot graphics/xmin={-0.6}, plot graphics/xmax={0.6}, plot graphics/ymin={-0.6}, plot graphics/ymax={0.6}, point meta min={0.0}, point meta max={0.7}, plot graphics/!src={tikz/gallery/antennae/antennae-ngpl-3-srcs-gp-2-std-interp-img-viridis-0-1}, colormap name={viridis}},
	antennae-ngpl-3-srcs-gp-3-mean/.style={plot graphics/xmin={-0.6}, plot graphics/xmax={0.6}, plot graphics/ymin={-0.6}, plot graphics/ymax={0.6}, point meta min={-0.46650344133377075}, point meta max={0.5891963243484497}, plot graphics/!src={tikz/gallery/antennae/antennae-ngpl-3-srcs-gp-3-mean-viridis--0-1}, colormap name={viridis}},
	antennae-ngpl-3-srcs-gp-3-std/.style={plot graphics/xmin={-0.6}, plot graphics/xmax={0.6}, plot graphics/ymin={-0.6}, plot graphics/ymax={0.6}, point meta min={0.0}, point meta max={0.7}, plot graphics/!src={tikz/gallery/antennae/antennae-ngpl-3-srcs-gp-3-std-viridis-0-1}, colormap name={viridis}},
	antennae-ngpl-3-srcs-gp-3-std-interp/.style={plot graphics/xmin={-0.6}, plot graphics/xmax={0.6}, plot graphics/ymin={-0.6}, plot graphics/ymax={0.6}, point meta min={0.0}, point meta max={0.7}, plot graphics/!src={tikz/gallery/antennae/antennae-ngpl-3-srcs-gp-3-std-interp-viridis-0-1}, colormap name={viridis}},
	antennae-ngpl-3-srcs-gp-3-std-winterp/.style={plot graphics/xmin={-0.6}, plot graphics/xmax={0.6}, plot graphics/ymin={-0.6}, plot graphics/ymax={0.6}, point meta min={0.0}, point meta max={0.7}, plot graphics/!src={tikz/gallery/antennae/antennae-ngpl-3-srcs-gp-3-std-winterp-viridis-0-1}, colormap name={viridis}},
	antennae-ngpl-3-srcs-gp-3-std-interp-img/.style={plot graphics/xmin={-0.6}, plot graphics/xmax={0.6}, plot graphics/ymin={-0.6}, plot graphics/ymax={0.6}, point meta min={0.0}, point meta max={0.7}, plot graphics/!src={tikz/gallery/antennae/antennae-ngpl-3-srcs-gp-3-std-interp-img-viridis-0-1}, colormap name={viridis}},
	antennae-obs/.style={plot graphics/xmin={-2.5}, plot graphics/xmax={2.5}, plot graphics/ymin={-2.5}, plot graphics/ymax={2.5}, point meta min={0.0}, point meta max={30.0}, plot graphics/!src={tikz/gallery/antennae/antennae-obs-viridis-0-30}, colormap name={viridis}},
	antennae-simg/.style={plot graphics/xmin={-0.6}, plot graphics/xmax={0.6}, plot graphics/ymin={-0.6}, plot graphics/ymax={0.6}, point meta min={0.0}, point meta max={30.0}, plot graphics/!src={tikz/gallery/antennae/antennae-simg-viridis-0-30}, colormap name={viridis}},
	antennae-simg-deproj/.style={plot graphics/xmin={-0.6}, plot graphics/xmax={0.6}, plot graphics/ymin={-0.6}, plot graphics/ymax={0.6}, point meta min={0.0}, point meta max={30.0}, plot graphics/!src={tikz/gallery/antennae/antennae-simg-deproj-viridis-0-30}, colormap name={viridis}},
	antennae-simg-wdeproj/.style={plot graphics/xmin={-0.6}, plot graphics/xmax={0.6}, plot graphics/ymin={-0.6}, plot graphics/ymax={0.6}, point meta min={0.0}, point meta max={30.0}, plot graphics/!src={tikz/gallery/antennae/antennae-simg-wdeproj-viridis-0-30}, colormap name={viridis}},
	antennae-ngpl-3-img-resid-obs-noise/.style={plot graphics/xmin={-2.5}, plot graphics/xmax={2.5}, plot graphics/ymin={-2.5}, plot graphics/ymax={2.5}, point meta min={-3.0}, point meta max={3.0}, plot graphics/!src={tikz/gallery/antennae/antennae-ngpl-3-img-resid-obs-noise-bwr--3-3}, colormap name={bwr}},
	antennae-ngpl-3-img-resid-obs-noise+std/.style={plot graphics/xmin={-2.5}, plot graphics/xmax={2.5}, plot graphics/ymin={-2.5}, plot graphics/ymax={2.5}, point meta min={-3.0}, point meta max={3.0}, plot graphics/!src={tikz/gallery/antennae/antennae-ngpl-3-img-resid-obs-noise+std-bwr--3-3}, colormap name={bwr}},
	antennae-ngpl-3-img-resid-mu/.style={plot graphics/xmin={-2.5}, plot graphics/xmax={2.5}, plot graphics/ymin={-2.5}, plot graphics/ymax={2.5}, point meta min={-3.0}, point meta max={3.0}, plot graphics/!src={tikz/gallery/antennae/antennae-ngpl-3-img-resid-mu-bwr--3-3}, colormap name={bwr}},
	antennae-ngpl-3-simg-resid-noise/.style={plot graphics/xmin={-0.6}, plot graphics/xmax={0.6}, plot graphics/ymin={-0.6}, plot graphics/ymax={0.6}, point meta min={-3.0}, point meta max={3.0}, plot graphics/!src={tikz/gallery/antennae/antennae-ngpl-3-simg-resid-noise-bwr--3-3}, colormap name={bwr}},
	antennae-ngpl-3-simg-resid-std/.style={plot graphics/xmin={-0.6}, plot graphics/xmax={0.6}, plot graphics/ymin={-0.6}, plot graphics/ymax={0.6}, point meta min={-3.0}, point meta max={3.0}, plot graphics/!src={tikz/gallery/antennae/antennae-ngpl-3-simg-resid-std-bwr--3-3}, colormap name={bwr}},
	antennae-ngpl-3-simg-resid-deproj-noise/.style={plot graphics/xmin={-0.6}, plot graphics/xmax={0.6}, plot graphics/ymin={-0.6}, plot graphics/ymax={0.6}, point meta min={-3.0}, point meta max={3.0}, plot graphics/!src={tikz/gallery/antennae/antennae-ngpl-3-simg-resid-deproj-noise-bwr--3-3}, colormap name={bwr}},
	antennae-ngpl-3-simg-resid-deproj-std/.style={plot graphics/xmin={-0.6}, plot graphics/xmax={0.6}, plot graphics/ymin={-0.6}, plot graphics/ymax={0.6}, point meta min={-3.0}, point meta max={3.0}, plot graphics/!src={tikz/gallery/antennae/antennae-ngpl-3-simg-resid-deproj-std-bwr--3-3}, colormap name={bwr}},
	antennae-ngpl-3-simg-resid-wdeproj-noise/.style={plot graphics/xmin={-0.6}, plot graphics/xmax={0.6}, plot graphics/ymin={-0.6}, plot graphics/ymax={0.6}, point meta min={-3.0}, point meta max={3.0}, plot graphics/!src={tikz/gallery/antennae/antennae-ngpl-3-simg-resid-wdeproj-noise-bwr--3-3}, colormap name={bwr}},
	antennae-ngpl-3-simg-resid-wdeproj-std/.style={plot graphics/xmin={-0.6}, plot graphics/xmax={0.6}, plot graphics/ymin={-0.6}, plot graphics/ymax={0.6}, point meta min={-3.0}, point meta max={3.0}, plot graphics/!src={tikz/gallery/antennae/antennae-ngpl-3-simg-resid-wdeproj-std-bwr--3-3}, colormap name={bwr}},
}
\pgfplotsset{,
	arp142-ngpl-3-img-mean/.style={plot graphics/xmin={-2.5}, plot graphics/xmax={2.5}, plot graphics/ymin={-2.5}, plot graphics/ymax={2.5}, point meta min={0.0}, point meta max={30.0}, plot graphics/!src={tikz/gallery/arp142/arp142-ngpl-3-img-mean-viridis-0-30}, colormap name={viridis}},
	arp142-ngpl-3-img-std/.style={plot graphics/xmin={-2.5}, plot graphics/xmax={2.5}, plot graphics/ymin={-2.5}, plot graphics/ymax={2.5}, point meta min={0.0}, point meta max={0.7}, plot graphics/!src={tikz/gallery/arp142/arp142-ngpl-3-img-std-viridis-0-1}, colormap name={viridis}},
	arp142-ngpl-3-simg-deproj-mean/.style={plot graphics/xmin={-0.5}, plot graphics/xmax={0.5}, plot graphics/ymin={-0.5}, plot graphics/ymax={0.5}, point meta min={0.0}, point meta max={30.0}, plot graphics/!src={tikz/gallery/arp142/arp142-ngpl-3-simg-deproj-mean-viridis-0-30}, colormap name={viridis}},
	arp142-ngpl-3-simg-deproj-std/.style={plot graphics/xmin={-0.5}, plot graphics/xmax={0.5}, plot graphics/ymin={-0.5}, plot graphics/ymax={0.5}, point meta min={0.0}, point meta max={0.7}, plot graphics/!src={tikz/gallery/arp142/arp142-ngpl-3-simg-deproj-std-viridis-0-1}, colormap name={viridis}},
	arp142-ngpl-3-simg-mean/.style={plot graphics/xmin={-0.5}, plot graphics/xmax={0.5}, plot graphics/ymin={-0.5}, plot graphics/ymax={0.5}, point meta min={0.0}, point meta max={30.0}, plot graphics/!src={tikz/gallery/arp142/arp142-ngpl-3-simg-mean-viridis-0-30}, colormap name={viridis}},
	arp142-ngpl-3-simg-std/.style={plot graphics/xmin={-0.5}, plot graphics/xmax={0.5}, plot graphics/ymin={-0.5}, plot graphics/ymax={0.5}, point meta min={0.0}, point meta max={0.7}, plot graphics/!src={tikz/gallery/arp142/arp142-ngpl-3-simg-std-viridis-0-1}, colormap name={viridis}},
	arp142-ngpl-3-simg-std-interp/.style={plot graphics/xmin={-0.5}, plot graphics/xmax={0.5}, plot graphics/ymin={-0.5}, plot graphics/ymax={0.5}, point meta min={0.0}, point meta max={0.7}, plot graphics/!src={tikz/gallery/arp142/arp142-ngpl-3-simg-std-interp-viridis-0-1}, colormap name={viridis}},
	arp142-ngpl-3-simg-std-winterp/.style={plot graphics/xmin={-0.5}, plot graphics/xmax={0.5}, plot graphics/ymin={-0.5}, plot graphics/ymax={0.5}, point meta min={0.0}, point meta max={0.7}, plot graphics/!src={tikz/gallery/arp142/arp142-ngpl-3-simg-std-winterp-viridis-0-1}, colormap name={viridis}},
	arp142-ngpl-3-simg-std-interp-img/.style={plot graphics/xmin={-0.5}, plot graphics/xmax={0.5}, plot graphics/ymin={-0.5}, plot graphics/ymax={0.5}, point meta min={0.0}, point meta max={0.7}, plot graphics/!src={tikz/gallery/arp142/arp142-ngpl-3-simg-std-interp-img-viridis-0-1}, colormap name={viridis}},
	arp142-ngpl-3-srcs-src-mean/.style={plot graphics/xmin={-0.5}, plot graphics/xmax={0.5}, plot graphics/ymin={-0.5}, plot graphics/ymax={0.5}, point meta min={-0.5017623901367188}, point meta max={26.487035751342773}, plot graphics/!src={tikz/gallery/arp142/arp142-ngpl-3-srcs-src-mean-viridis--1-26}, colormap name={viridis}},
	arp142-ngpl-3-srcs-src-std/.style={plot graphics/xmin={-0.5}, plot graphics/xmax={0.5}, plot graphics/ymin={-0.5}, plot graphics/ymax={0.5}, point meta min={0.0}, point meta max={0.7}, plot graphics/!src={tikz/gallery/arp142/arp142-ngpl-3-srcs-src-std-viridis-0-1}, colormap name={viridis}},
	arp142-ngpl-3-srcs-gp-1-mean/.style={plot graphics/xmin={-0.5}, plot graphics/xmax={0.5}, plot graphics/ymin={-0.5}, plot graphics/ymax={0.5}, point meta min={-2.1212663650512695}, point meta max={3.6515820026397705}, plot graphics/!src={tikz/gallery/arp142/arp142-ngpl-3-srcs-gp-1-mean-viridis--2-4}, colormap name={viridis}},
	arp142-ngpl-3-srcs-gp-1-std/.style={plot graphics/xmin={-0.5}, plot graphics/xmax={0.5}, plot graphics/ymin={-0.5}, plot graphics/ymax={0.5}, point meta min={0.0}, point meta max={0.7}, plot graphics/!src={tikz/gallery/arp142/arp142-ngpl-3-srcs-gp-1-std-viridis-0-1}, colormap name={viridis}},
	arp142-ngpl-3-srcs-gp-1-std-interp/.style={plot graphics/xmin={-0.5}, plot graphics/xmax={0.5}, plot graphics/ymin={-0.5}, plot graphics/ymax={0.5}, point meta min={0.0}, point meta max={0.7}, plot graphics/!src={tikz/gallery/arp142/arp142-ngpl-3-srcs-gp-1-std-interp-viridis-0-1}, colormap name={viridis}},
	arp142-ngpl-3-srcs-gp-1-std-winterp/.style={plot graphics/xmin={-0.5}, plot graphics/xmax={0.5}, plot graphics/ymin={-0.5}, plot graphics/ymax={0.5}, point meta min={0.0}, point meta max={0.7}, plot graphics/!src={tikz/gallery/arp142/arp142-ngpl-3-srcs-gp-1-std-winterp-viridis-0-1}, colormap name={viridis}},
	arp142-ngpl-3-srcs-gp-1-std-interp-img/.style={plot graphics/xmin={-0.5}, plot graphics/xmax={0.5}, plot graphics/ymin={-0.5}, plot graphics/ymax={0.5}, point meta min={0.0}, point meta max={0.7}, plot graphics/!src={tikz/gallery/arp142/arp142-ngpl-3-srcs-gp-1-std-interp-img-viridis-0-1}, colormap name={viridis}},
	arp142-ngpl-3-srcs-gp-2-mean/.style={plot graphics/xmin={-0.5}, plot graphics/xmax={0.5}, plot graphics/ymin={-0.5}, plot graphics/ymax={0.5}, point meta min={-0.8369684219360352}, point meta max={2.052353858947754}, plot graphics/!src={tikz/gallery/arp142/arp142-ngpl-3-srcs-gp-2-mean-viridis--1-2}, colormap name={viridis}},
	arp142-ngpl-3-srcs-gp-2-std/.style={plot graphics/xmin={-0.5}, plot graphics/xmax={0.5}, plot graphics/ymin={-0.5}, plot graphics/ymax={0.5}, point meta min={0.0}, point meta max={0.7}, plot graphics/!src={tikz/gallery/arp142/arp142-ngpl-3-srcs-gp-2-std-viridis-0-1}, colormap name={viridis}},
	arp142-ngpl-3-srcs-gp-2-std-interp/.style={plot graphics/xmin={-0.5}, plot graphics/xmax={0.5}, plot graphics/ymin={-0.5}, plot graphics/ymax={0.5}, point meta min={0.0}, point meta max={0.7}, plot graphics/!src={tikz/gallery/arp142/arp142-ngpl-3-srcs-gp-2-std-interp-viridis-0-1}, colormap name={viridis}},
	arp142-ngpl-3-srcs-gp-2-std-winterp/.style={plot graphics/xmin={-0.5}, plot graphics/xmax={0.5}, plot graphics/ymin={-0.5}, plot graphics/ymax={0.5}, point meta min={0.0}, point meta max={0.7}, plot graphics/!src={tikz/gallery/arp142/arp142-ngpl-3-srcs-gp-2-std-winterp-viridis-0-1}, colormap name={viridis}},
	arp142-ngpl-3-srcs-gp-2-std-interp-img/.style={plot graphics/xmin={-0.5}, plot graphics/xmax={0.5}, plot graphics/ymin={-0.5}, plot graphics/ymax={0.5}, point meta min={0.0}, point meta max={0.7}, plot graphics/!src={tikz/gallery/arp142/arp142-ngpl-3-srcs-gp-2-std-interp-img-viridis-0-1}, colormap name={viridis}},
	arp142-ngpl-3-srcs-gp-3-mean/.style={plot graphics/xmin={-0.5}, plot graphics/xmax={0.5}, plot graphics/ymin={-0.5}, plot graphics/ymax={0.5}, point meta min={-0.3223089575767517}, point meta max={0.7121658325195312}, plot graphics/!src={tikz/gallery/arp142/arp142-ngpl-3-srcs-gp-3-mean-viridis--0-1}, colormap name={viridis}},
	arp142-ngpl-3-srcs-gp-3-std/.style={plot graphics/xmin={-0.5}, plot graphics/xmax={0.5}, plot graphics/ymin={-0.5}, plot graphics/ymax={0.5}, point meta min={0.0}, point meta max={0.7}, plot graphics/!src={tikz/gallery/arp142/arp142-ngpl-3-srcs-gp-3-std-viridis-0-1}, colormap name={viridis}},
	arp142-ngpl-3-srcs-gp-3-std-interp/.style={plot graphics/xmin={-0.5}, plot graphics/xmax={0.5}, plot graphics/ymin={-0.5}, plot graphics/ymax={0.5}, point meta min={0.0}, point meta max={0.7}, plot graphics/!src={tikz/gallery/arp142/arp142-ngpl-3-srcs-gp-3-std-interp-viridis-0-1}, colormap name={viridis}},
	arp142-ngpl-3-srcs-gp-3-std-winterp/.style={plot graphics/xmin={-0.5}, plot graphics/xmax={0.5}, plot graphics/ymin={-0.5}, plot graphics/ymax={0.5}, point meta min={0.0}, point meta max={0.7}, plot graphics/!src={tikz/gallery/arp142/arp142-ngpl-3-srcs-gp-3-std-winterp-viridis-0-1}, colormap name={viridis}},
	arp142-ngpl-3-srcs-gp-3-std-interp-img/.style={plot graphics/xmin={-0.5}, plot graphics/xmax={0.5}, plot graphics/ymin={-0.5}, plot graphics/ymax={0.5}, point meta min={0.0}, point meta max={0.7}, plot graphics/!src={tikz/gallery/arp142/arp142-ngpl-3-srcs-gp-3-std-interp-img-viridis-0-1}, colormap name={viridis}},
	arp142-obs/.style={plot graphics/xmin={-2.5}, plot graphics/xmax={2.5}, plot graphics/ymin={-2.5}, plot graphics/ymax={2.5}, point meta min={0.0}, point meta max={30.0}, plot graphics/!src={tikz/gallery/arp142/arp142-obs-viridis-0-30}, colormap name={viridis}},
	arp142-simg/.style={plot graphics/xmin={-0.5}, plot graphics/xmax={0.5}, plot graphics/ymin={-0.5}, plot graphics/ymax={0.5}, point meta min={0.0}, point meta max={30.0}, plot graphics/!src={tikz/gallery/arp142/arp142-simg-viridis-0-30}, colormap name={viridis}},
	arp142-simg-deproj/.style={plot graphics/xmin={-0.5}, plot graphics/xmax={0.5}, plot graphics/ymin={-0.5}, plot graphics/ymax={0.5}, point meta min={0.0}, point meta max={30.0}, plot graphics/!src={tikz/gallery/arp142/arp142-simg-deproj-viridis-0-30}, colormap name={viridis}},
	arp142-simg-wdeproj/.style={plot graphics/xmin={-0.5}, plot graphics/xmax={0.5}, plot graphics/ymin={-0.5}, plot graphics/ymax={0.5}, point meta min={0.0}, point meta max={30.0}, plot graphics/!src={tikz/gallery/arp142/arp142-simg-wdeproj-viridis-0-30}, colormap name={viridis}},
	arp142-ngpl-3-img-resid-obs-noise/.style={plot graphics/xmin={-2.5}, plot graphics/xmax={2.5}, plot graphics/ymin={-2.5}, plot graphics/ymax={2.5}, point meta min={-3.0}, point meta max={3.0}, plot graphics/!src={tikz/gallery/arp142/arp142-ngpl-3-img-resid-obs-noise-bwr--3-3}, colormap name={bwr}},
	arp142-ngpl-3-img-resid-obs-noise+std/.style={plot graphics/xmin={-2.5}, plot graphics/xmax={2.5}, plot graphics/ymin={-2.5}, plot graphics/ymax={2.5}, point meta min={-3.0}, point meta max={3.0}, plot graphics/!src={tikz/gallery/arp142/arp142-ngpl-3-img-resid-obs-noise+std-bwr--3-3}, colormap name={bwr}},
	arp142-ngpl-3-img-resid-mu/.style={plot graphics/xmin={-2.5}, plot graphics/xmax={2.5}, plot graphics/ymin={-2.5}, plot graphics/ymax={2.5}, point meta min={-3.0}, point meta max={3.0}, plot graphics/!src={tikz/gallery/arp142/arp142-ngpl-3-img-resid-mu-bwr--3-3}, colormap name={bwr}},
	arp142-ngpl-3-simg-resid-noise/.style={plot graphics/xmin={-0.5}, plot graphics/xmax={0.5}, plot graphics/ymin={-0.5}, plot graphics/ymax={0.5}, point meta min={-3.0}, point meta max={3.0}, plot graphics/!src={tikz/gallery/arp142/arp142-ngpl-3-simg-resid-noise-bwr--3-3}, colormap name={bwr}},
	arp142-ngpl-3-simg-resid-std/.style={plot graphics/xmin={-0.5}, plot graphics/xmax={0.5}, plot graphics/ymin={-0.5}, plot graphics/ymax={0.5}, point meta min={-3.0}, point meta max={3.0}, plot graphics/!src={tikz/gallery/arp142/arp142-ngpl-3-simg-resid-std-bwr--3-3}, colormap name={bwr}},
	arp142-ngpl-3-simg-resid-deproj-noise/.style={plot graphics/xmin={-0.5}, plot graphics/xmax={0.5}, plot graphics/ymin={-0.5}, plot graphics/ymax={0.5}, point meta min={-3.0}, point meta max={3.0}, plot graphics/!src={tikz/gallery/arp142/arp142-ngpl-3-simg-resid-deproj-noise-bwr--3-3}, colormap name={bwr}},
	arp142-ngpl-3-simg-resid-deproj-std/.style={plot graphics/xmin={-0.5}, plot graphics/xmax={0.5}, plot graphics/ymin={-0.5}, plot graphics/ymax={0.5}, point meta min={-3.0}, point meta max={3.0}, plot graphics/!src={tikz/gallery/arp142/arp142-ngpl-3-simg-resid-deproj-std-bwr--3-3}, colormap name={bwr}},
	arp142-ngpl-3-simg-resid-wdeproj-noise/.style={plot graphics/xmin={-0.5}, plot graphics/xmax={0.5}, plot graphics/ymin={-0.5}, plot graphics/ymax={0.5}, point meta min={-3.0}, point meta max={3.0}, plot graphics/!src={tikz/gallery/arp142/arp142-ngpl-3-simg-resid-wdeproj-noise-bwr--3-3}, colormap name={bwr}},
	arp142-ngpl-3-simg-resid-wdeproj-std/.style={plot graphics/xmin={-0.5}, plot graphics/xmax={0.5}, plot graphics/ymin={-0.5}, plot graphics/ymax={0.5}, point meta min={-3.0}, point meta max={3.0}, plot graphics/!src={tikz/gallery/arp142/arp142-ngpl-3-simg-resid-wdeproj-std-bwr--3-3}, colormap name={bwr}},
}
\pgfplotsset{,
	arp148-ngpl-3-img-mean/.style={plot graphics/xmin={-2.5}, plot graphics/xmax={2.5}, plot graphics/ymin={-2.5}, plot graphics/ymax={2.5}, point meta min={0.0}, point meta max={30.0}, plot graphics/!src={tikz/gallery/arp148/arp148-ngpl-3-img-mean-viridis-0-30}, colormap name={viridis}},
	arp148-ngpl-3-img-std/.style={plot graphics/xmin={-2.5}, plot graphics/xmax={2.5}, plot graphics/ymin={-2.5}, plot graphics/ymax={2.5}, point meta min={0.0}, point meta max={1.0}, plot graphics/!src={tikz/gallery/arp148/arp148-ngpl-3-img-std-viridis-0-1}, colormap name={viridis}},
	arp148-ngpl-3-simg-deproj-mean/.style={plot graphics/xmin={-0.6}, plot graphics/xmax={0.6}, plot graphics/ymin={-0.6}, plot graphics/ymax={0.6}, point meta min={0.0}, point meta max={30.0}, plot graphics/!src={tikz/gallery/arp148/arp148-ngpl-3-simg-deproj-mean-viridis-0-30}, colormap name={viridis}},
	arp148-ngpl-3-simg-deproj-std/.style={plot graphics/xmin={-0.6}, plot graphics/xmax={0.6}, plot graphics/ymin={-0.6}, plot graphics/ymax={0.6}, point meta min={0.0}, point meta max={1.0}, plot graphics/!src={tikz/gallery/arp148/arp148-ngpl-3-simg-deproj-std-viridis-0-1}, colormap name={viridis}},
	arp148-ngpl-3-simg-mean/.style={plot graphics/xmin={-0.6}, plot graphics/xmax={0.6}, plot graphics/ymin={-0.6}, plot graphics/ymax={0.6}, point meta min={0.0}, point meta max={30.0}, plot graphics/!src={tikz/gallery/arp148/arp148-ngpl-3-simg-mean-viridis-0-30}, colormap name={viridis}},
	arp148-ngpl-3-simg-std/.style={plot graphics/xmin={-0.6}, plot graphics/xmax={0.6}, plot graphics/ymin={-0.6}, plot graphics/ymax={0.6}, point meta min={0.0}, point meta max={1.0}, plot graphics/!src={tikz/gallery/arp148/arp148-ngpl-3-simg-std-viridis-0-1}, colormap name={viridis}},
	arp148-ngpl-3-simg-std-interp/.style={plot graphics/xmin={-0.6}, plot graphics/xmax={0.6}, plot graphics/ymin={-0.6}, plot graphics/ymax={0.6}, point meta min={0.0}, point meta max={1.0}, plot graphics/!src={tikz/gallery/arp148/arp148-ngpl-3-simg-std-interp-viridis-0-1}, colormap name={viridis}},
	arp148-ngpl-3-simg-std-winterp/.style={plot graphics/xmin={-0.6}, plot graphics/xmax={0.6}, plot graphics/ymin={-0.6}, plot graphics/ymax={0.6}, point meta min={0.0}, point meta max={1.0}, plot graphics/!src={tikz/gallery/arp148/arp148-ngpl-3-simg-std-winterp-viridis-0-1}, colormap name={viridis}},
	arp148-ngpl-3-simg-std-interp-img/.style={plot graphics/xmin={-0.6}, plot graphics/xmax={0.6}, plot graphics/ymin={-0.6}, plot graphics/ymax={0.6}, point meta min={0.0}, point meta max={1.0}, plot graphics/!src={tikz/gallery/arp148/arp148-ngpl-3-simg-std-interp-img-viridis-0-1}, colormap name={viridis}},
	arp148-ngpl-3-srcs-src-mean/.style={plot graphics/xmin={-0.6}, plot graphics/xmax={0.6}, plot graphics/ymin={-0.6}, plot graphics/ymax={0.6}, point meta min={-0.9732298254966736}, point meta max={27.04579734802246}, plot graphics/!src={tikz/gallery/arp148/arp148-ngpl-3-srcs-src-mean-viridis--1-27}, colormap name={viridis}},
	arp148-ngpl-3-srcs-src-std/.style={plot graphics/xmin={-0.6}, plot graphics/xmax={0.6}, plot graphics/ymin={-0.6}, plot graphics/ymax={0.6}, point meta min={0.0}, point meta max={1.0}, plot graphics/!src={tikz/gallery/arp148/arp148-ngpl-3-srcs-src-std-viridis-0-1}, colormap name={viridis}},
	arp148-ngpl-3-srcs-gp-1-mean/.style={plot graphics/xmin={-0.6}, plot graphics/xmax={0.6}, plot graphics/ymin={-0.6}, plot graphics/ymax={0.6}, point meta min={-5.558814525604248}, point meta max={7.732621669769287}, plot graphics/!src={tikz/gallery/arp148/arp148-ngpl-3-srcs-gp-1-mean-viridis--6-8}, colormap name={viridis}},
	arp148-ngpl-3-srcs-gp-1-std/.style={plot graphics/xmin={-0.6}, plot graphics/xmax={0.6}, plot graphics/ymin={-0.6}, plot graphics/ymax={0.6}, point meta min={0.0}, point meta max={1.0}, plot graphics/!src={tikz/gallery/arp148/arp148-ngpl-3-srcs-gp-1-std-viridis-0-1}, colormap name={viridis}},
	arp148-ngpl-3-srcs-gp-1-std-interp/.style={plot graphics/xmin={-0.6}, plot graphics/xmax={0.6}, plot graphics/ymin={-0.6}, plot graphics/ymax={0.6}, point meta min={0.0}, point meta max={1.0}, plot graphics/!src={tikz/gallery/arp148/arp148-ngpl-3-srcs-gp-1-std-interp-viridis-0-1}, colormap name={viridis}},
	arp148-ngpl-3-srcs-gp-1-std-winterp/.style={plot graphics/xmin={-0.6}, plot graphics/xmax={0.6}, plot graphics/ymin={-0.6}, plot graphics/ymax={0.6}, point meta min={0.0}, point meta max={1.0}, plot graphics/!src={tikz/gallery/arp148/arp148-ngpl-3-srcs-gp-1-std-winterp-viridis-0-1}, colormap name={viridis}},
	arp148-ngpl-3-srcs-gp-1-std-interp-img/.style={plot graphics/xmin={-0.6}, plot graphics/xmax={0.6}, plot graphics/ymin={-0.6}, plot graphics/ymax={0.6}, point meta min={0.0}, point meta max={1.0}, plot graphics/!src={tikz/gallery/arp148/arp148-ngpl-3-srcs-gp-1-std-interp-img-viridis-0-1}, colormap name={viridis}},
	arp148-ngpl-3-srcs-gp-2-mean/.style={plot graphics/xmin={-0.6}, plot graphics/xmax={0.6}, plot graphics/ymin={-0.6}, plot graphics/ymax={0.6}, point meta min={-1.4316707849502563}, point meta max={2.4581496715545654}, plot graphics/!src={tikz/gallery/arp148/arp148-ngpl-3-srcs-gp-2-mean-viridis--1-2}, colormap name={viridis}},
	arp148-ngpl-3-srcs-gp-2-std/.style={plot graphics/xmin={-0.6}, plot graphics/xmax={0.6}, plot graphics/ymin={-0.6}, plot graphics/ymax={0.6}, point meta min={0.0}, point meta max={1.0}, plot graphics/!src={tikz/gallery/arp148/arp148-ngpl-3-srcs-gp-2-std-viridis-0-1}, colormap name={viridis}},
	arp148-ngpl-3-srcs-gp-2-std-interp/.style={plot graphics/xmin={-0.6}, plot graphics/xmax={0.6}, plot graphics/ymin={-0.6}, plot graphics/ymax={0.6}, point meta min={0.0}, point meta max={1.0}, plot graphics/!src={tikz/gallery/arp148/arp148-ngpl-3-srcs-gp-2-std-interp-viridis-0-1}, colormap name={viridis}},
	arp148-ngpl-3-srcs-gp-2-std-winterp/.style={plot graphics/xmin={-0.6}, plot graphics/xmax={0.6}, plot graphics/ymin={-0.6}, plot graphics/ymax={0.6}, point meta min={0.0}, point meta max={1.0}, plot graphics/!src={tikz/gallery/arp148/arp148-ngpl-3-srcs-gp-2-std-winterp-viridis-0-1}, colormap name={viridis}},
	arp148-ngpl-3-srcs-gp-2-std-interp-img/.style={plot graphics/xmin={-0.6}, plot graphics/xmax={0.6}, plot graphics/ymin={-0.6}, plot graphics/ymax={0.6}, point meta min={0.0}, point meta max={1.0}, plot graphics/!src={tikz/gallery/arp148/arp148-ngpl-3-srcs-gp-2-std-interp-img-viridis-0-1}, colormap name={viridis}},
	arp148-ngpl-3-srcs-gp-3-mean/.style={plot graphics/xmin={-0.6}, plot graphics/xmax={0.6}, plot graphics/ymin={-0.6}, plot graphics/ymax={0.6}, point meta min={-0.4059641659259796}, point meta max={0.7210601568222046}, plot graphics/!src={tikz/gallery/arp148/arp148-ngpl-3-srcs-gp-3-mean-viridis--0-1}, colormap name={viridis}},
	arp148-ngpl-3-srcs-gp-3-std/.style={plot graphics/xmin={-0.6}, plot graphics/xmax={0.6}, plot graphics/ymin={-0.6}, plot graphics/ymax={0.6}, point meta min={0.0}, point meta max={1.0}, plot graphics/!src={tikz/gallery/arp148/arp148-ngpl-3-srcs-gp-3-std-viridis-0-1}, colormap name={viridis}},
	arp148-ngpl-3-srcs-gp-3-std-interp/.style={plot graphics/xmin={-0.6}, plot graphics/xmax={0.6}, plot graphics/ymin={-0.6}, plot graphics/ymax={0.6}, point meta min={0.0}, point meta max={1.0}, plot graphics/!src={tikz/gallery/arp148/arp148-ngpl-3-srcs-gp-3-std-interp-viridis-0-1}, colormap name={viridis}},
	arp148-ngpl-3-srcs-gp-3-std-winterp/.style={plot graphics/xmin={-0.6}, plot graphics/xmax={0.6}, plot graphics/ymin={-0.6}, plot graphics/ymax={0.6}, point meta min={0.0}, point meta max={1.0}, plot graphics/!src={tikz/gallery/arp148/arp148-ngpl-3-srcs-gp-3-std-winterp-viridis-0-1}, colormap name={viridis}},
	arp148-ngpl-3-srcs-gp-3-std-interp-img/.style={plot graphics/xmin={-0.6}, plot graphics/xmax={0.6}, plot graphics/ymin={-0.6}, plot graphics/ymax={0.6}, point meta min={0.0}, point meta max={1.0}, plot graphics/!src={tikz/gallery/arp148/arp148-ngpl-3-srcs-gp-3-std-interp-img-viridis-0-1}, colormap name={viridis}},
	arp148-obs/.style={plot graphics/xmin={-2.5}, plot graphics/xmax={2.5}, plot graphics/ymin={-2.5}, plot graphics/ymax={2.5}, point meta min={0.0}, point meta max={30.0}, plot graphics/!src={tikz/gallery/arp148/arp148-obs-viridis-0-30}, colormap name={viridis}},
	arp148-simg/.style={plot graphics/xmin={-0.6}, plot graphics/xmax={0.6}, plot graphics/ymin={-0.6}, plot graphics/ymax={0.6}, point meta min={0.0}, point meta max={30.0}, plot graphics/!src={tikz/gallery/arp148/arp148-simg-viridis-0-30}, colormap name={viridis}},
	arp148-simg-deproj/.style={plot graphics/xmin={-0.6}, plot graphics/xmax={0.6}, plot graphics/ymin={-0.6}, plot graphics/ymax={0.6}, point meta min={0.0}, point meta max={30.0}, plot graphics/!src={tikz/gallery/arp148/arp148-simg-deproj-viridis-0-30}, colormap name={viridis}},
	arp148-simg-wdeproj/.style={plot graphics/xmin={-0.6}, plot graphics/xmax={0.6}, plot graphics/ymin={-0.6}, plot graphics/ymax={0.6}, point meta min={0.0}, point meta max={30.0}, plot graphics/!src={tikz/gallery/arp148/arp148-simg-wdeproj-viridis-0-30}, colormap name={viridis}},
	arp148-ngpl-3-img-resid-obs-noise/.style={plot graphics/xmin={-2.5}, plot graphics/xmax={2.5}, plot graphics/ymin={-2.5}, plot graphics/ymax={2.5}, point meta min={-3.0}, point meta max={3.0}, plot graphics/!src={tikz/gallery/arp148/arp148-ngpl-3-img-resid-obs-noise-bwr--3-3}, colormap name={bwr}},
	arp148-ngpl-3-img-resid-obs-noise+std/.style={plot graphics/xmin={-2.5}, plot graphics/xmax={2.5}, plot graphics/ymin={-2.5}, plot graphics/ymax={2.5}, point meta min={-3.0}, point meta max={3.0}, plot graphics/!src={tikz/gallery/arp148/arp148-ngpl-3-img-resid-obs-noise+std-bwr--3-3}, colormap name={bwr}},
	arp148-ngpl-3-img-resid-mu/.style={plot graphics/xmin={-2.5}, plot graphics/xmax={2.5}, plot graphics/ymin={-2.5}, plot graphics/ymax={2.5}, point meta min={-3.0}, point meta max={3.0}, plot graphics/!src={tikz/gallery/arp148/arp148-ngpl-3-img-resid-mu-bwr--3-3}, colormap name={bwr}},
	arp148-ngpl-3-simg-resid-noise/.style={plot graphics/xmin={-0.6}, plot graphics/xmax={0.6}, plot graphics/ymin={-0.6}, plot graphics/ymax={0.6}, point meta min={-3.0}, point meta max={3.0}, plot graphics/!src={tikz/gallery/arp148/arp148-ngpl-3-simg-resid-noise-bwr--3-3}, colormap name={bwr}},
	arp148-ngpl-3-simg-resid-std/.style={plot graphics/xmin={-0.6}, plot graphics/xmax={0.6}, plot graphics/ymin={-0.6}, plot graphics/ymax={0.6}, point meta min={-3.0}, point meta max={3.0}, plot graphics/!src={tikz/gallery/arp148/arp148-ngpl-3-simg-resid-std-bwr--3-3}, colormap name={bwr}},
	arp148-ngpl-3-simg-resid-deproj-noise/.style={plot graphics/xmin={-0.6}, plot graphics/xmax={0.6}, plot graphics/ymin={-0.6}, plot graphics/ymax={0.6}, point meta min={-3.0}, point meta max={3.0}, plot graphics/!src={tikz/gallery/arp148/arp148-ngpl-3-simg-resid-deproj-noise-bwr--3-3}, colormap name={bwr}},
	arp148-ngpl-3-simg-resid-deproj-std/.style={plot graphics/xmin={-0.6}, plot graphics/xmax={0.6}, plot graphics/ymin={-0.6}, plot graphics/ymax={0.6}, point meta min={-3.0}, point meta max={3.0}, plot graphics/!src={tikz/gallery/arp148/arp148-ngpl-3-simg-resid-deproj-std-bwr--3-3}, colormap name={bwr}},
	arp148-ngpl-3-simg-resid-wdeproj-noise/.style={plot graphics/xmin={-0.6}, plot graphics/xmax={0.6}, plot graphics/ymin={-0.6}, plot graphics/ymax={0.6}, point meta min={-3.0}, point meta max={3.0}, plot graphics/!src={tikz/gallery/arp148/arp148-ngpl-3-simg-resid-wdeproj-noise-bwr--3-3}, colormap name={bwr}},
	arp148-ngpl-3-simg-resid-wdeproj-std/.style={plot graphics/xmin={-0.6}, plot graphics/xmax={0.6}, plot graphics/ymin={-0.6}, plot graphics/ymax={0.6}, point meta min={-3.0}, point meta max={3.0}, plot graphics/!src={tikz/gallery/arp148/arp148-ngpl-3-simg-resid-wdeproj-std-bwr--3-3}, colormap name={bwr}},
}
\pgfplotsset{,
	ngc4414-ngpl-3-img-mean/.style={plot graphics/xmin={-2.5}, plot graphics/xmax={2.5}, plot graphics/ymin={-2.5}, plot graphics/ymax={2.5}, point meta min={0.0}, point meta max={30.0}, plot graphics/!src={tikz/gallery/ngc4414/ngc4414-ngpl-3-img-mean-viridis-0-30}, colormap name={viridis}},
	ngc4414-ngpl-3-img-std/.style={plot graphics/xmin={-2.5}, plot graphics/xmax={2.5}, plot graphics/ymin={-2.5}, plot graphics/ymax={2.5}, point meta min={0.0}, point meta max={1.0}, plot graphics/!src={tikz/gallery/ngc4414/ngc4414-ngpl-3-img-std-viridis-0-1}, colormap name={viridis}},
	ngc4414-ngpl-3-simg-deproj-mean/.style={plot graphics/xmin={-0.5}, plot graphics/xmax={0.5}, plot graphics/ymin={-0.5}, plot graphics/ymax={0.5}, point meta min={0.0}, point meta max={30.0}, plot graphics/!src={tikz/gallery/ngc4414/ngc4414-ngpl-3-simg-deproj-mean-viridis-0-30}, colormap name={viridis}},
	ngc4414-ngpl-3-simg-deproj-std/.style={plot graphics/xmin={-0.5}, plot graphics/xmax={0.5}, plot graphics/ymin={-0.5}, plot graphics/ymax={0.5}, point meta min={0.0}, point meta max={1.0}, plot graphics/!src={tikz/gallery/ngc4414/ngc4414-ngpl-3-simg-deproj-std-viridis-0-1}, colormap name={viridis}},
	ngc4414-ngpl-3-simg-mean/.style={plot graphics/xmin={-0.5}, plot graphics/xmax={0.5}, plot graphics/ymin={-0.5}, plot graphics/ymax={0.5}, point meta min={0.0}, point meta max={30.0}, plot graphics/!src={tikz/gallery/ngc4414/ngc4414-ngpl-3-simg-mean-viridis-0-30}, colormap name={viridis}},
	ngc4414-ngpl-3-simg-std/.style={plot graphics/xmin={-0.5}, plot graphics/xmax={0.5}, plot graphics/ymin={-0.5}, plot graphics/ymax={0.5}, point meta min={0.0}, point meta max={1.0}, plot graphics/!src={tikz/gallery/ngc4414/ngc4414-ngpl-3-simg-std-viridis-0-1}, colormap name={viridis}},
	ngc4414-ngpl-3-simg-std-interp/.style={plot graphics/xmin={-0.5}, plot graphics/xmax={0.5}, plot graphics/ymin={-0.5}, plot graphics/ymax={0.5}, point meta min={0.0}, point meta max={1.0}, plot graphics/!src={tikz/gallery/ngc4414/ngc4414-ngpl-3-simg-std-interp-viridis-0-1}, colormap name={viridis}},
	ngc4414-ngpl-3-simg-std-winterp/.style={plot graphics/xmin={-0.5}, plot graphics/xmax={0.5}, plot graphics/ymin={-0.5}, plot graphics/ymax={0.5}, point meta min={0.0}, point meta max={1.0}, plot graphics/!src={tikz/gallery/ngc4414/ngc4414-ngpl-3-simg-std-winterp-viridis-0-1}, colormap name={viridis}},
	ngc4414-ngpl-3-simg-std-interp-img/.style={plot graphics/xmin={-0.5}, plot graphics/xmax={0.5}, plot graphics/ymin={-0.5}, plot graphics/ymax={0.5}, point meta min={0.0}, point meta max={1.0}, plot graphics/!src={tikz/gallery/ngc4414/ngc4414-ngpl-3-simg-std-interp-img-viridis-0-1}, colormap name={viridis}},
	ngc4414-ngpl-3-srcs-src-mean/.style={plot graphics/xmin={-0.5}, plot graphics/xmax={0.5}, plot graphics/ymin={-0.5}, plot graphics/ymax={0.5}, point meta min={-0.526749312877655}, point meta max={28.77107810974121}, plot graphics/!src={tikz/gallery/ngc4414/ngc4414-ngpl-3-srcs-src-mean-viridis--1-29}, colormap name={viridis}},
	ngc4414-ngpl-3-srcs-src-std/.style={plot graphics/xmin={-0.5}, plot graphics/xmax={0.5}, plot graphics/ymin={-0.5}, plot graphics/ymax={0.5}, point meta min={0.0}, point meta max={1.0}, plot graphics/!src={tikz/gallery/ngc4414/ngc4414-ngpl-3-srcs-src-std-viridis-0-1}, colormap name={viridis}},
	ngc4414-ngpl-3-srcs-gp-1-mean/.style={plot graphics/xmin={-0.5}, plot graphics/xmax={0.5}, plot graphics/ymin={-0.5}, plot graphics/ymax={0.5}, point meta min={-0.6768194437026978}, point meta max={0.9986350536346436}, plot graphics/!src={tikz/gallery/ngc4414/ngc4414-ngpl-3-srcs-gp-1-mean-viridis--1-1}, colormap name={viridis}},
	ngc4414-ngpl-3-srcs-gp-1-std/.style={plot graphics/xmin={-0.5}, plot graphics/xmax={0.5}, plot graphics/ymin={-0.5}, plot graphics/ymax={0.5}, point meta min={0.0}, point meta max={1.0}, plot graphics/!src={tikz/gallery/ngc4414/ngc4414-ngpl-3-srcs-gp-1-std-viridis-0-1}, colormap name={viridis}},
	ngc4414-ngpl-3-srcs-gp-1-std-interp/.style={plot graphics/xmin={-0.5}, plot graphics/xmax={0.5}, plot graphics/ymin={-0.5}, plot graphics/ymax={0.5}, point meta min={0.0}, point meta max={1.0}, plot graphics/!src={tikz/gallery/ngc4414/ngc4414-ngpl-3-srcs-gp-1-std-interp-viridis-0-1}, colormap name={viridis}},
	ngc4414-ngpl-3-srcs-gp-1-std-winterp/.style={plot graphics/xmin={-0.5}, plot graphics/xmax={0.5}, plot graphics/ymin={-0.5}, plot graphics/ymax={0.5}, point meta min={0.0}, point meta max={1.0}, plot graphics/!src={tikz/gallery/ngc4414/ngc4414-ngpl-3-srcs-gp-1-std-winterp-viridis-0-1}, colormap name={viridis}},
	ngc4414-ngpl-3-srcs-gp-1-std-interp-img/.style={plot graphics/xmin={-0.5}, plot graphics/xmax={0.5}, plot graphics/ymin={-0.5}, plot graphics/ymax={0.5}, point meta min={0.0}, point meta max={1.0}, plot graphics/!src={tikz/gallery/ngc4414/ngc4414-ngpl-3-srcs-gp-1-std-interp-img-viridis-0-1}, colormap name={viridis}},
	ngc4414-ngpl-3-srcs-gp-2-mean/.style={plot graphics/xmin={-0.5}, plot graphics/xmax={0.5}, plot graphics/ymin={-0.5}, plot graphics/ymax={0.5}, point meta min={-4.69724178314209}, point meta max={6.20796012878418}, plot graphics/!src={tikz/gallery/ngc4414/ngc4414-ngpl-3-srcs-gp-2-mean-viridis--5-6}, colormap name={viridis}},
	ngc4414-ngpl-3-srcs-gp-2-std/.style={plot graphics/xmin={-0.5}, plot graphics/xmax={0.5}, plot graphics/ymin={-0.5}, plot graphics/ymax={0.5}, point meta min={0.0}, point meta max={1.0}, plot graphics/!src={tikz/gallery/ngc4414/ngc4414-ngpl-3-srcs-gp-2-std-viridis-0-1}, colormap name={viridis}},
	ngc4414-ngpl-3-srcs-gp-2-std-interp/.style={plot graphics/xmin={-0.5}, plot graphics/xmax={0.5}, plot graphics/ymin={-0.5}, plot graphics/ymax={0.5}, point meta min={0.0}, point meta max={1.0}, plot graphics/!src={tikz/gallery/ngc4414/ngc4414-ngpl-3-srcs-gp-2-std-interp-viridis-0-1}, colormap name={viridis}},
	ngc4414-ngpl-3-srcs-gp-2-std-winterp/.style={plot graphics/xmin={-0.5}, plot graphics/xmax={0.5}, plot graphics/ymin={-0.5}, plot graphics/ymax={0.5}, point meta min={0.0}, point meta max={1.0}, plot graphics/!src={tikz/gallery/ngc4414/ngc4414-ngpl-3-srcs-gp-2-std-winterp-viridis-0-1}, colormap name={viridis}},
	ngc4414-ngpl-3-srcs-gp-2-std-interp-img/.style={plot graphics/xmin={-0.5}, plot graphics/xmax={0.5}, plot graphics/ymin={-0.5}, plot graphics/ymax={0.5}, point meta min={0.0}, point meta max={1.0}, plot graphics/!src={tikz/gallery/ngc4414/ngc4414-ngpl-3-srcs-gp-2-std-interp-img-viridis-0-1}, colormap name={viridis}},
	ngc4414-ngpl-3-srcs-gp-3-mean/.style={plot graphics/xmin={-0.5}, plot graphics/xmax={0.5}, plot graphics/ymin={-0.5}, plot graphics/ymax={0.5}, point meta min={-0.6981844902038574}, point meta max={0.8448110818862915}, plot graphics/!src={tikz/gallery/ngc4414/ngc4414-ngpl-3-srcs-gp-3-mean-viridis--1-1}, colormap name={viridis}},
	ngc4414-ngpl-3-srcs-gp-3-std/.style={plot graphics/xmin={-0.5}, plot graphics/xmax={0.5}, plot graphics/ymin={-0.5}, plot graphics/ymax={0.5}, point meta min={0.0}, point meta max={1.0}, plot graphics/!src={tikz/gallery/ngc4414/ngc4414-ngpl-3-srcs-gp-3-std-viridis-0-1}, colormap name={viridis}},
	ngc4414-ngpl-3-srcs-gp-3-std-interp/.style={plot graphics/xmin={-0.5}, plot graphics/xmax={0.5}, plot graphics/ymin={-0.5}, plot graphics/ymax={0.5}, point meta min={0.0}, point meta max={1.0}, plot graphics/!src={tikz/gallery/ngc4414/ngc4414-ngpl-3-srcs-gp-3-std-interp-viridis-0-1}, colormap name={viridis}},
	ngc4414-ngpl-3-srcs-gp-3-std-winterp/.style={plot graphics/xmin={-0.5}, plot graphics/xmax={0.5}, plot graphics/ymin={-0.5}, plot graphics/ymax={0.5}, point meta min={0.0}, point meta max={1.0}, plot graphics/!src={tikz/gallery/ngc4414/ngc4414-ngpl-3-srcs-gp-3-std-winterp-viridis-0-1}, colormap name={viridis}},
	ngc4414-ngpl-3-srcs-gp-3-std-interp-img/.style={plot graphics/xmin={-0.5}, plot graphics/xmax={0.5}, plot graphics/ymin={-0.5}, plot graphics/ymax={0.5}, point meta min={0.0}, point meta max={1.0}, plot graphics/!src={tikz/gallery/ngc4414/ngc4414-ngpl-3-srcs-gp-3-std-interp-img-viridis-0-1}, colormap name={viridis}},
	ngc4414-obs/.style={plot graphics/xmin={-2.5}, plot graphics/xmax={2.5}, plot graphics/ymin={-2.5}, plot graphics/ymax={2.5}, point meta min={0.0}, point meta max={30.0}, plot graphics/!src={tikz/gallery/ngc4414/ngc4414-obs-viridis-0-30}, colormap name={viridis}},
	ngc4414-simg/.style={plot graphics/xmin={-0.5}, plot graphics/xmax={0.5}, plot graphics/ymin={-0.5}, plot graphics/ymax={0.5}, point meta min={0.0}, point meta max={30.0}, plot graphics/!src={tikz/gallery/ngc4414/ngc4414-simg-viridis-0-30}, colormap name={viridis}},
	ngc4414-simg-deproj/.style={plot graphics/xmin={-0.5}, plot graphics/xmax={0.5}, plot graphics/ymin={-0.5}, plot graphics/ymax={0.5}, point meta min={0.0}, point meta max={30.0}, plot graphics/!src={tikz/gallery/ngc4414/ngc4414-simg-deproj-viridis-0-30}, colormap name={viridis}},
	ngc4414-simg-wdeproj/.style={plot graphics/xmin={-0.5}, plot graphics/xmax={0.5}, plot graphics/ymin={-0.5}, plot graphics/ymax={0.5}, point meta min={0.0}, point meta max={30.0}, plot graphics/!src={tikz/gallery/ngc4414/ngc4414-simg-wdeproj-viridis-0-30}, colormap name={viridis}},
	ngc4414-ngpl-3-img-resid-obs-noise/.style={plot graphics/xmin={-2.5}, plot graphics/xmax={2.5}, plot graphics/ymin={-2.5}, plot graphics/ymax={2.5}, point meta min={-3.0}, point meta max={3.0}, plot graphics/!src={tikz/gallery/ngc4414/ngc4414-ngpl-3-img-resid-obs-noise-bwr--3-3}, colormap name={bwr}},
	ngc4414-ngpl-3-img-resid-obs-noise+std/.style={plot graphics/xmin={-2.5}, plot graphics/xmax={2.5}, plot graphics/ymin={-2.5}, plot graphics/ymax={2.5}, point meta min={-3.0}, point meta max={3.0}, plot graphics/!src={tikz/gallery/ngc4414/ngc4414-ngpl-3-img-resid-obs-noise+std-bwr--3-3}, colormap name={bwr}},
	ngc4414-ngpl-3-img-resid-mu/.style={plot graphics/xmin={-2.5}, plot graphics/xmax={2.5}, plot graphics/ymin={-2.5}, plot graphics/ymax={2.5}, point meta min={-3.0}, point meta max={3.0}, plot graphics/!src={tikz/gallery/ngc4414/ngc4414-ngpl-3-img-resid-mu-bwr--3-3}, colormap name={bwr}},
	ngc4414-ngpl-3-simg-resid-noise/.style={plot graphics/xmin={-0.5}, plot graphics/xmax={0.5}, plot graphics/ymin={-0.5}, plot graphics/ymax={0.5}, point meta min={-3.0}, point meta max={3.0}, plot graphics/!src={tikz/gallery/ngc4414/ngc4414-ngpl-3-simg-resid-noise-bwr--3-3}, colormap name={bwr}},
	ngc4414-ngpl-3-simg-resid-std/.style={plot graphics/xmin={-0.5}, plot graphics/xmax={0.5}, plot graphics/ymin={-0.5}, plot graphics/ymax={0.5}, point meta min={-3.0}, point meta max={3.0}, plot graphics/!src={tikz/gallery/ngc4414/ngc4414-ngpl-3-simg-resid-std-bwr--3-3}, colormap name={bwr}},
	ngc4414-ngpl-3-simg-resid-deproj-noise/.style={plot graphics/xmin={-0.5}, plot graphics/xmax={0.5}, plot graphics/ymin={-0.5}, plot graphics/ymax={0.5}, point meta min={-3.0}, point meta max={3.0}, plot graphics/!src={tikz/gallery/ngc4414/ngc4414-ngpl-3-simg-resid-deproj-noise-bwr--3-3}, colormap name={bwr}},
	ngc4414-ngpl-3-simg-resid-deproj-std/.style={plot graphics/xmin={-0.5}, plot graphics/xmax={0.5}, plot graphics/ymin={-0.5}, plot graphics/ymax={0.5}, point meta min={-3.0}, point meta max={3.0}, plot graphics/!src={tikz/gallery/ngc4414/ngc4414-ngpl-3-simg-resid-deproj-std-bwr--3-3}, colormap name={bwr}},
	ngc4414-ngpl-3-simg-resid-wdeproj-noise/.style={plot graphics/xmin={-0.5}, plot graphics/xmax={0.5}, plot graphics/ymin={-0.5}, plot graphics/ymax={0.5}, point meta min={-3.0}, point meta max={3.0}, plot graphics/!src={tikz/gallery/ngc4414/ngc4414-ngpl-3-simg-resid-wdeproj-noise-bwr--3-3}, colormap name={bwr}},
	ngc4414-ngpl-3-simg-resid-wdeproj-std/.style={plot graphics/xmin={-0.5}, plot graphics/xmax={0.5}, plot graphics/ymin={-0.5}, plot graphics/ymax={0.5}, point meta min={-3.0}, point meta max={3.0}, plot graphics/!src={tikz/gallery/ngc4414/ngc4414-ngpl-3-simg-resid-wdeproj-std-bwr--3-3}, colormap name={bwr}},
}
\pgfplotsset{
    flip y/.style={plot graphics/node/.append style={anchor=north west, yscale=-1}},
    plot graphics extent/.style={plot graphics/xmin={-#1}, plot graphics/xmax={#1}, plot graphics/ymin={-#1}, plot graphics/ymax={#1}},
    plot graphics extents/.style 2 args={plot graphics/xmin={-#1}, plot graphics/xmax={#1}, plot graphics/ymin={-#2}, plot graphics/ymax={#2}},
    hoags_object-img-color/.style={plot graphics extent=2.5, plot graphics/!src={tikz/gallery/hoags_object-img-color}},
    hoags_object-simg-color/.style={flip y, plot graphics extent=0.5, plot graphics/!src={tikz/gallery/hoags_object-simg-color}},
    antennae-img-color/.style={plot graphics extent=2.5, plot graphics/!src={tikz/gallery/antennae-img-color}},
    antennae-simg-color/.style={flip y, plot graphics extents={0.6}{0.596}, plot graphics/!src={tikz/gallery/antennae-simg-color}, extent=0.6},
    arp142-img-color/.style={plot graphics extent=2.5, plot graphics/!src={tikz/gallery/arp142-img-color}},
    arp142-simg-color/.style={flip y, plot graphics extent=0.5, plot graphics/!src={tikz/gallery/arp142-simg-color}},
    arp148-img-color/.style={plot graphics extent=2.5, plot graphics/!src={tikz/gallery/arp148-img-color}},
    arp148-simg-color/.style={flip y, plot graphics extent=0.6, plot graphics/!src={tikz/gallery/arp148-simg-color}},
    ngc4414-img-color/.style={plot graphics extent=2.5, plot graphics/!src={tikz/gallery/ngc4414-img-color}, extent=2.1},
    ngc4414-simg-color/.style={flip y, plot graphics extents={0.5}{0.4127}, plot graphics/!src={tikz/gallery/ngc4414-simg-color}},
}

\newcommand{\gallerysrcstd}{interp-img}

\newcommand{\showofffigure}[1]{
	\tikzsetnextfilename{showoff-#1}%
	\begin{scaletowidth}{\linewidth}
	\begin{tikzpicture}
		\begin{groupplot}[
			axis equal image, scale only axis, tight layout=2,
			group style={group size=2 by 1, horizontal sep=0.5em},
			xlabel style={at={(ticklabel* cs:0.5,0.9cm)}, anchor=base},
			ylabel style={at={(ticklabel* cs:0.5,0.7cm)}, anchor=base},
			every axis title shift=0pt,
			axis background/.style={fill=black},
			scale={\thescale}
		]
			\nextgroupplot[imageplane, extent=2, title={image plane}, #1-img-color]
				\addplot graphics {};
				\addplot[critical] table {tikz/object-data/#1/#1-critical.txt};
				
				\coordinate (ll) at (-0.5, -0.5);
				\coordinate (lr) at ( 0.5, -0.5);
				\coordinate (ul) at (-0.5,  0.5);
				\coordinate (ur) at ( 0.5,  0.5);
			\nextgroupplot[sourceplane, ytick pos=right, ylabel style={at={(ticklabel* cs:0.55,0.9cm)}, rotate=180}, title={source plane}, #1-simg-color]
				\addplot graphics {};
				\addplot[caustic] table {tikz/object-data/#1/#1-caustic.txt};
		\end{groupplot}
	
		\begin{scope}[ultra thin, black!20, opacity=0.5]
			\draw (ll) rectangle (ur);
			\draw (ll) -- (group c2r1.south west);
			\draw (lr) -- (intersection of {lr}--{group c2r1.south east} and {group c2r1.south west}--{group c2r1.north west});
			\draw (ul) -- (group c2r1.north west);
			\draw (ur) -- (intersection of {ur}--{group c2r1.north east} and {group c2r1.north west}--{group c2r1.south west});
		\end{scope}
	\end{tikzpicture}%
	\end{scaletowidth}
}

\pgfplotsset{imageplane/.append style={extent=1.8}}
\newcommand{\resultsfigwidth}{0.88\linewidth}
\newcommand{\resultsfigrowsep}{0.42em}

\newcommand{\resultsfigure}[2][hoags_object]{
	\setlength{\linewidth}{\resultsfigwidth}
	\newcommand{\prefix}{#1-ngpl-#2}
	\ngplnumtomacro{\ngpl}{#2}
	\pgfplotsset{
		axis equal image, tight layout=4, height={\pgfkeysvalueof{/pgfplots/width}},
		group style={
			group size=4 by 1,
			y descriptions at=edge left,
			horizontal sep={1.5em}
		},
		ylabel style={at={(ticklabel* cs:0.5,3em)}, anchor=base},
		observation/.style={colorbar label={observation / noise}},
		true source/.style={colorbar label={true source / noise}},
		rec mean/.style={colorbar label={reconstruction mean / noise}},
		rec stdev/.style={colorbar label={reconstruction st. dev. / noise}},
		residual/.style={colorbar label={residual / noise}}
	}
	\tikzset{
		label node/.style={font=\bfseries, color=white, inner xsep=0.5em},
		subcaption label node/.style={label node, yshift=0.5em, at={(axis description cs:0, 0)}, anchor=base west},
		plane label node/.style={label node, yshift=-0.2em, at={(axis description cs:0, 1)}, anchor=north west},
		trim axis group left, trim axis group right
	}
	\shortstack[l]{
		\tikzsetnextfilename{results-\prefix-img}%
		\begin{tikzpicture}
			\begin{groupplot}[imageplane, colorbar top]
				\nextgroupplot[#1-obs, observation]
					\addplot graphics {};
					\addplot[critical] table {tikz/object-data/#1/#1-critical.txt};
					\node[subcaption label node]{a.};
					\node[plane label node]{image plane};
				\nextgroupplot[\prefix-img-mean, rec mean]
					\addplot graphics {};
					\node[subcaption label node]{b.};
				\nextgroupplot[\prefix-img-std, rec stdev]
					\addplot graphics {};
					\node[subcaption label node]{c.};
				\nextgroupplot[\prefix-img-resid-obs-noise, residual]
					\addplot graphics {};
					\node[subcaption label node, color=black]{d.};
			\end{groupplot}
		\end{tikzpicture}%
		\\[\resultsfigrowsep]
		\tikzsetnextfilename{results-\prefix-simg}%
		\begin{tikzpicture}
			\begin{groupplot}[sourceplane]
				\nextgroupplot[#1-simg, true source]
					\addplot graphics {};
					\addplot[caustic] table {tikz/object-data/#1/#1-caustic.txt};
					\node[subcaption label node]{e.};
					\node[plane label node]{source plane};
				\nextgroupplot[\prefix-simg-mean, rec mean]
					\addplot graphics {};
					\kernelcircle{#1/ngpl-#2/src/sigma};
					\node[kernel, below, shift={(axis direction cs:0,-0.05)}] at (\kernelcirclepos) {$\layerindex{\kernelsize}{k}$};
					\pgfplotsforeachungrouped \i in {0,...,\ngpl-1} {\edef\tmp{
						\noexpand\kernelcircle{#1/ngpl-#2/gp/sigma/\i};
					}\tmp}
					\node[subcaption label node]{f.};
				\nextgroupplot[\prefix-simg-std-\gallerysrcstd, rec stdev]
					\addplot graphics {};
					\node[subcaption label node, color=black]{g.};
				\nextgroupplot[\prefix-simg-resid-deproj-noise, residual]
					\addplot graphics {};
					\node[subcaption label node, color=black]{h.};
			\end{groupplot}
		\end{tikzpicture}%
	}
}

\newcommand{\layersfigure}[2][hoags_object]{
	\newcommand{\prefix}{#1-ngpl-#2}
	\ngplnumtomacro{\ngpl}{#2}

	\tikzsetnextfilename{layers-\prefix}%
	\begin{tikzpicture}[trim axis group left, trim axis group right]
		\pgfmathtruncatemacro{\rows}{\ngpl + 2}
		\begin{groupplot}[
			axis equal image, scale only axis, tight layout=2, height={\pgfkeysvalueof{/pgfplots/width}},
			group style={
				columns=2, rows={\rows},
				horizontal sep=0.5em, vertical sep=0.5em,
			},
			tick align=inside, tick pos=both, xticklabels={}, yticklabels={}, tickwidth=3pt,
			sourceplane, xlabel={}, ylabel={},
			every axis post/.style={colorbar style={tick align=outside}},
		]
			\pgfplotsforeachungrouped \i in {\ngpl,...,1} {\edef\tmp{
				\noexpand\nextgroupplot[\prefix-srcs-gp-\i-mean, colorbar left]
					\noexpand\addplot graphics {};
					\noexpand\node[lbl] {image-plane layer \i};
					\noexpand\kernelcircle{#1/ngpl-#2/gp/sigma/\directmath{\i-1}};
				\noexpand\nextgroupplot[\prefix-srcs-gp-\i-std-\gallerysrcstd]
					\noexpand\addplot graphics {};
					\noexpand\ifthenelse{\noexpand\equal{\i}{\ngpl}}{\noexpand\pgfplotsset{colorbar top, every axis post}}{}
			}\tmp}
			\nextgroupplot[\prefix-srcs-src-mean, colorbar left]
				\addplot graphics {};
				\node[lbl] {source-plane source};
				\kernelcircle{#1/ngpl-#2/src/sigma};
			\nextgroupplot[\prefix-srcs-src-std]
				\addplot graphics {};
			\nextgroupplot[\prefix-simg-mean, colorbar left]
				\addplot graphics {};
				\node[lbl] {combined};
			\nextgroupplot[\prefix-simg-std-\gallerysrcstd]
				\addplot graphics {};
		\end{groupplot}
	\end{tikzpicture}%
}

\newcommand{\hlayersfigure}[2][hoags_object]{
	\newcommand{\prefix}{#1-ngpl-#2}
	\ngplnumtomacro{\ngpl}{#2}
	\pgfmathtruncatemacro{\cols}{\ngpl + 2}

	\pgfplotsset{meanimg/.style={colorbar bottom}}

	\setlength{\linewidth}{\resultsfigwidth}
	\tikzsetnextfilename{hlayers-\prefix}
	\begin{tikzpicture}[trim axis group left, trim axis group right]
		\begin{groupplot}[
			axis equal image, scale only axis, tight layout={\cols},
			height={\pgfkeysvalueof{/pgfplots/width}},
			group style={
				rows=2, columns={\cols},
				horizontal sep=0.5em, vertical sep=0.5em,
			},
			tick align=inside, tick pos=both, xticklabels={}, yticklabels={}, tickwidth=3pt,
			sourceplane, xlabel={}, ylabel={},
			every axis post/.style={colorbar style={tick align=outside}}
		]
			\nextgroupplot[\prefix-simg-std-\gallerysrcstd, ylabel={standard devation}]
				\addplot graphics {};
			\nextgroupplot[\prefix-srcs-src-std]
				\addplot graphics {};
			
			\pgfplotsforeachungrouped \i in {1,...,\ngpl} {\edef\tmp{
			\noexpand\nextgroupplot[\prefix-srcs-gp-\i-std-\gallerysrcstd,
				xlabel={image-plane layer \i}]
				\noexpand\addplot graphics {};
				\noexpand\ifthenelse{\noexpand\equal{\i}{\ngpl}}{\noexpand\pgfplotsset{colorbar right, every axis post}}{}
			}\tmp}
		
			\nextgroupplot[\prefix-simg-mean, ylabel={mean}, meanimg,
				colorbar style={x tick label style={xshift={(\tick==30)*(-0.5em)}}}]
				\addplot graphics {};
				\node[lbl] {combined};
			\nextgroupplot[\prefix-srcs-src-mean, meanimg,
				set layers=axis on top,
				minor tick={-0.615,-0.585,...,0.615}, every minor tick/.style={draw=none},
				grid=minor, grid style={ultra thin, opacity=0.2}]
				\addplot graphics {};
				\begin{pgfonlayer}{axis foreground}
					\kernelcircle{#1/ngpl-#2/src/sigma};
					\node[lbl, /pgfplots/on layer=axis background] {source-plane source};
				\end{pgfonlayer}
			
			\pgfplotsforeachungrouped \i in {1,...,\ngpl} {\edef\tmp{
			\noexpand\nextgroupplot[\prefix-srcs-gp-\i-mean, meanimg]
				\noexpand\addplot graphics {};
				\noexpand\kernelcircle{#1/ngpl-#2/gp/sigma/\directmath{\i-1}};
				\noexpand\node[lbl] {image-plane layer \i};
			}\tmp}
		
		\end{groupplot}
	\end{tikzpicture}%
}

\DeclareDocumentCommand{\resids}{O{hoags_object} O{3} O{true}}{
	\newcommand{\prefix}{tikz/resids/#1/resids-#1-ngpl-#2}
	
	\tikzsetnextfilename{resids-#1-ngpl-#2-#3}%
	\begin{tikzpicture}[
		trim axis group right,
		dots/.style={only marks, mark=*, mark size=1pt, mark options={line width=0pt, draw opacity=0, fill opacity=0.5}}
	]
		\newcommand{\doplot}[1][dots]{
			\addplot[ring, ##1] table {\prefix-ring.txt};
			\addplot[core, ##1] table {\prefix-core.txt};
		}
		\begin{groupplot}[
			group style={
				group size=1 by 3,
				vertical sep=0pt,
				x descriptions at=edge bottom
			},
			scale only axis, width=0.8\linewidth,
			xmin=-1, xmax=31, xlabel={true flux / noise},
			ylabel style={at={(ticklabel* cs:0.5,3.14em)}, anchor=base},
			table/x=#3,
		]
			\nextgroupplot[
				height={0.33*\pgfkeysvalueof{/pgfplots/width}},
				ymode=log,
				log base 10 number format code/.code={
					\ifthenelse{\lengthtest{##1pt > 3pt}}{\pgfkeys{/pgfplots/log number format basis={10}{##1}}}{\pgfmathparse{10^##1}\pgfmathprintnumber[precision=0]{\pgfmathresult}}
				},
				ylabel={number of pixels}, xtick align=center,
				legend image code/.code={
					\fill [##1] (-0.4em, -0.25em) rectangle (0.4em, 0.25em);},
				legend style={/tikz/every even column/.append style={column sep=0.3em}}
			]
			\doplot[hist={
				bins=30, data value=\pgfkeysvalueof{/data point/x}, data min=0, data max=30,
				handler/.style={const plot}
			}, forget plot]
		
			\addlegendimage{ring} \addlegendentry{ring}
			\addlegendimage{core} \addlegendentry{core}
			
			\nextgroupplot[
				height={0.66*\pgfkeysvalueof{/pgfplots/width}}, xtick align=center,
				table/y=mean, ylabel={reconstructed flux / noise},
				ymin=-1, ymax=31
			]
			\addplot[dashed, semithick, forget plot] coordinates {(0, 0) (30, 30)};
			\doplot
			
			\nextgroupplot[
				height={0.5*\pgfkeysvalueof{/pgfplots/width}}, table/y=resid-#3, ylabel={normalised residual}]
			\doplot
			\axhline[dashed]{0}{0}{1}\axhline[solid]{-1}{0}{1}\axhline[solid]{1}{0}{1}
		\end{groupplot}
	\end{tikzpicture}%
}

\pgfkeys{,
	/private/.cd,
	antennae/scale/.initial={1.2},
	antennae/truth/ext/gamma_1/.initial={-0.0146},
	antennae/truth/ext/gamma_2/.initial={-0.0512},
	antennae/truth/main/phi/.initial={-2.956},
	antennae/truth/main/q/.initial={0.3928},
	antennae/truth/main/r_ein/.initial={1.2161},
	antennae/truth/main/slope/.initial={2.4765},
	antennae/truth/main/x/.initial={-0.025},
	antennae/truth/main/y/.initial={-0.0642},
	antennae/init/ext/gamma_1/.initial={0.0},
	antennae/init/ext/gamma_2/.initial={0.0},
	antennae/init/main/phi/.initial={-2.5},
	antennae/init/main/q/.initial={0.5},
	antennae/init/main/r_ein/.initial={1.2},
	antennae/init/main/slope/.initial={2.0},
	antennae/init/main/x/.initial={0.0},
	antennae/init/main/y/.initial={0.0},
	arp142/scale/.initial={1.0},
	arp142/truth/ext/gamma_1/.initial={0.0443},
	arp142/truth/ext/gamma_2/.initial={0.0069},
	arp142/truth/main/phi/.initial={-3.0506},
	arp142/truth/main/q/.initial={0.8719},
	arp142/truth/main/r_ein/.initial={1.361},
	arp142/truth/main/slope/.initial={2.2335},
	arp142/truth/main/x/.initial={0.0102},
	arp142/truth/main/y/.initial={-0.0202},
	arp142/init/ext/gamma_1/.initial={0.0},
	arp142/init/ext/gamma_2/.initial={0.0},
	arp142/init/main/phi/.initial={-3.0},
	arp142/init/main/q/.initial={0.75},
	arp142/init/main/r_ein/.initial={1.3},
	arp142/init/main/slope/.initial={2.0},
	arp142/init/main/x/.initial={0.0},
	arp142/init/main/y/.initial={0.0},
	arp148/scale/.initial={1.2},
	arp148/truth/ext/gamma_1/.initial={-0.01},
	arp148/truth/ext/gamma_2/.initial={0.003},
	arp148/truth/main/phi/.initial={0.2},
	arp148/truth/main/q/.initial={0.5},
	arp148/truth/main/r_ein/.initial={1.3},
	arp148/truth/main/slope/.initial={2.4},
	arp148/truth/main/x/.initial={0.08},
	arp148/truth/main/y/.initial={-0.01},
	arp148/init/ext/gamma_1/.initial={0.0},
	arp148/init/ext/gamma_2/.initial={0.0},
	arp148/init/main/phi/.initial={0.3},
	arp148/init/main/q/.initial={0.42},
	arp148/init/main/r_ein/.initial={1.25},
	arp148/init/main/slope/.initial={2.0},
	arp148/init/main/x/.initial={0.0},
	arp148/init/main/y/.initial={0.0},
	hoags_object/scale/.initial={1.0},
	hoags_object/truth/ext/gamma_1/.initial={-0.0227},
	hoags_object/truth/ext/gamma_2/.initial={0.0047},
	hoags_object/truth/main/phi/.initial={1.742},
	hoags_object/truth/main/q/.initial={0.2753},
	hoags_object/truth/main/r_ein/.initial={1.1847},
	hoags_object/truth/main/slope/.initial={2.2193},
	hoags_object/truth/main/x/.initial={-0.0704},
	hoags_object/truth/main/y/.initial={-0.0193},
	hoags_object/init/ext/gamma_1/.initial={0.0},
	hoags_object/init/ext/gamma_2/.initial={0.0},
	hoags_object/init/main/phi/.initial={1.5},
	hoags_object/init/main/q/.initial={0.3},
	hoags_object/init/main/r_ein/.initial={1.2},
	hoags_object/init/main/slope/.initial={2.0},
	hoags_object/init/main/x/.initial={0.0},
	hoags_object/init/main/y/.initial={0.0},
	ngc4414/scale/.initial={1.0},
	ngc4414/truth/ext/gamma_1/.initial={0.007},
	ngc4414/truth/ext/gamma_2/.initial={0.01},
	ngc4414/truth/main/phi/.initial={1.0},
	ngc4414/truth/main/q/.initial={0.75},
	ngc4414/truth/main/r_ein/.initial={1.5},
	ngc4414/truth/main/slope/.initial={2.1},
	ngc4414/truth/main/x/.initial={-0.05},
	ngc4414/truth/main/y/.initial={0.1},
	ngc4414/init/ext/gamma_1/.initial={0.0},
	ngc4414/init/ext/gamma_2/.initial={0.0},
	ngc4414/init/main/phi/.initial={0.5},
	ngc4414/init/main/q/.initial={0.5},
	ngc4414/init/main/r_ein/.initial={1.42},
	ngc4414/init/main/slope/.initial={2.0},
	ngc4414/init/main/x/.initial={0.0},
	ngc4414/init/main/y/.initial={0.0},
}
\pgfkeys{,
	/private/hoags_object/.cd,
	ngpl-1b/src/sigma/.initial={0.04},
	ngpl-1b/src/alpha/.initial={1.9117287397384644},
	ngpl-1b/gp/sigma/0/.initial={0.01},
	ngpl-1b/gp/alpha/0/.initial={0.41702723503112793},
	ngpl-1b/recloss/obs/all/.initial={1.0341423749923706},
	ngpl-1b/recloss/obs/masked/.initial={1.1954715251922607},
	ngpl-1b/recloss/mu/all/.initial={inf},
	ngpl-1b/recloss/mu/masked/.initial={2.05889630317688},
	ngpl-1s/src/sigma/.initial={0.04},
	ngpl-1s/src/alpha/.initial={2.4210219383239746},
	ngpl-1s/gp/sigma/0/.initial={0.002},
	ngpl-1s/gp/alpha/0/.initial={0.196120947599411},
	ngpl-1s/recloss/obs/all/.initial={1.081154465675354},
	ngpl-1s/recloss/obs/masked/.initial={1.4617007970809937},
	ngpl-1s/recloss/mu/all/.initial={inf},
	ngpl-1s/recloss/mu/masked/.initial={9.311266899108887},
	ngpl-2/src/sigma/.initial={0.04},
	ngpl-2/src/alpha/.initial={1.9185967445373535},
	ngpl-2/gp/sigma/0/.initial={0.01},
	ngpl-2/gp/sigma/1/.initial={0.002},
	ngpl-2/gp/alpha/0/.initial={0.4151436388492584},
	ngpl-2/gp/alpha/1/.initial={0.12925708293914795},
	ngpl-2/recloss/obs/all/.initial={1.031891942024231},
	ngpl-2/recloss/obs/masked/.initial={1.18276047706604},
	ngpl-2/recloss/mu/all/.initial={inf},
	ngpl-2/recloss/mu/masked/.initial={1.977320671081543},
	ngpl-3/src/sigma/.initial={0.04},
	ngpl-3/src/alpha/.initial={1.934656023979187},
	ngpl-3/gp/sigma/0/.initial={0.01},
	ngpl-3/gp/sigma/1/.initial={0.00447213595499958},
	ngpl-3/gp/sigma/2/.initial={0.002},
	ngpl-3/gp/alpha/0/.initial={0.3813263773918152},
	ngpl-3/gp/alpha/1/.initial={0.19807367026805878},
	ngpl-3/gp/alpha/2/.initial={0.15156535804271698},
	ngpl-3/recloss/obs/all/.initial={1.0266326665878296},
	ngpl-3/recloss/obs/masked/.initial={1.1530756950378418},
	ngpl-3/recloss/mu/all/.initial={inf},
	ngpl-3/recloss/mu/masked/.initial={1.8073664903640747},
	ngpl-4/src/sigma/.initial={0.04},
	ngpl-4/src/alpha/.initial={1.9716228246688843},
	ngpl-4/gp/sigma/0/.initial={0.01},
	ngpl-4/gp/sigma/1/.initial={0.0058480354764257345},
	ngpl-4/gp/sigma/2/.initial={0.003419951893353393},
	ngpl-4/gp/sigma/3/.initial={0.002},
	ngpl-4/gp/alpha/0/.initial={0.30735164880752563},
	ngpl-4/gp/alpha/1/.initial={0.2561296224594116},
	ngpl-4/gp/alpha/2/.initial={0.19508086144924164},
	ngpl-4/gp/alpha/3/.initial={0.17328223586082458},
	ngpl-4/recloss/obs/all/.initial={1.0215572118759155},
	ngpl-4/recloss/obs/masked/.initial={1.1243634223937988},
	ngpl-4/recloss/mu/all/.initial={inf},
	ngpl-4/recloss/mu/masked/.initial={1.684205174446106},
	ngpl-5/src/sigma/.initial={0.04},
	ngpl-5/src/alpha/.initial={1.8991622924804688},
	ngpl-5/gp/sigma/0/.initial={0.01},
	ngpl-5/gp/sigma/1/.initial={0.00668740304976422},
	ngpl-5/gp/sigma/2/.initial={0.00447213595499958},
	ngpl-5/gp/sigma/3/.initial={0.002990697562442441},
	ngpl-5/gp/sigma/4/.initial={0.002},
	ngpl-5/gp/alpha/0/.initial={0.2549149692058563},
	ngpl-5/gp/alpha/1/.initial={0.26616623997688293},
	ngpl-5/gp/alpha/2/.initial={0.21969468891620636},
	ngpl-5/gp/alpha/3/.initial={0.191836416721344},
	ngpl-5/gp/alpha/4/.initial={0.17724300920963287},
	ngpl-5/recloss/obs/all/.initial={1.0180587768554688},
	ngpl-5/recloss/obs/masked/.initial={1.103907585144043},
	ngpl-5/recloss/mu/all/.initial={inf},
	ngpl-5/recloss/mu/masked/.initial={1.5651462078094482},
	ngpl-6/src/sigma/.initial={0.04},
	ngpl-6/src/alpha/.initial={1.9007112979888916},
	ngpl-6/gp/sigma/0/.initial={0.01},
	ngpl-6/gp/sigma/1/.initial={0.007247796636776954},
	ngpl-6/gp/sigma/2/.initial={0.0052530556088075326},
	ngpl-6/gp/sigma/3/.initial={0.003807307877431759},
	ngpl-6/gp/sigma/4/.initial={0.0027594593229224307},
	ngpl-6/gp/sigma/5/.initial={0.002},
	ngpl-6/gp/alpha/0/.initial={0.2275429666042328},
	ngpl-6/gp/alpha/1/.initial={0.24060247838497162},
	ngpl-6/gp/alpha/2/.initial={0.21803312003612518},
	ngpl-6/gp/alpha/3/.initial={0.19667643308639526},
	ngpl-6/gp/alpha/4/.initial={0.18275123834609985},
	ngpl-6/gp/alpha/5/.initial={0.17381423711776733},
	ngpl-6/recloss/obs/all/.initial={1.0170172452926636},
	ngpl-6/recloss/obs/masked/.initial={1.0979424715042114},
	ngpl-6/recloss/mu/all/.initial={inf},
	ngpl-6/recloss/mu/masked/.initial={1.5486544370651245},
	ngpl-7/src/sigma/.initial={0.04},
	ngpl-7/src/alpha/.initial={1.8988761901855469},
	ngpl-7/gp/sigma/0/.initial={0.01},
	ngpl-7/gp/sigma/1/.initial={0.007647244913317298},
	ngpl-7/gp/sigma/2/.initial={0.0058480354764257345},
	ngpl-7/gp/sigma/3/.initial={0.00447213595499958},
	ngpl-7/gp/sigma/4/.initial={0.003419951893353393},
	ngpl-7/gp/sigma/5/.initial={0.0026153209720236625},
	ngpl-7/gp/sigma/6/.initial={0.002},
	ngpl-7/gp/alpha/0/.initial={0.2089836597442627},
	ngpl-7/gp/alpha/1/.initial={0.22174657881259918},
	ngpl-7/gp/alpha/2/.initial={0.2122098058462143},
	ngpl-7/gp/alpha/3/.initial={0.19787251949310303},
	ngpl-7/gp/alpha/4/.initial={0.18635928630828857},
	ngpl-7/gp/alpha/5/.initial={0.17797501385211945},
	ngpl-7/gp/alpha/6/.initial={0.17256540060043335},
	ngpl-7/recloss/obs/all/.initial={1.0152239799499512},
	ngpl-7/recloss/obs/masked/.initial={1.087877631187439},
	ngpl-7/recloss/mu/all/.initial={inf},
	ngpl-7/recloss/mu/masked/.initial={1.4665985107421875},
	ngpl-8/src/sigma/.initial={0.04},
	ngpl-8/src/alpha/.initial={1.8966001272201538},
	ngpl-8/gp/sigma/0/.initial={0.01},
	ngpl-8/gp/sigma/1/.initial={0.007945974047018519},
	ngpl-8/gp/sigma/2/.initial={0.006313850355589193},
	ngpl-8/gp/sigma/3/.initial={0.005016969106227038},
	ngpl-8/gp/sigma/4/.initial={0.003986470631277378},
	ngpl-8/gp/sigma/5/.initial={0.003167639217533158},
	ngpl-8/gp/sigma/6/.initial={0.0025169979012836523},
	ngpl-8/gp/sigma/7/.initial={0.002},
	ngpl-8/gp/alpha/0/.initial={0.19525954127311707},
	ngpl-8/gp/alpha/1/.initial={0.20480625331401825},
	ngpl-8/gp/alpha/2/.initial={0.20181022584438324},
	ngpl-8/gp/alpha/3/.initial={0.19346685707569122},
	ngpl-8/gp/alpha/4/.initial={0.18520063161849976},
	ngpl-8/gp/alpha/5/.initial={0.17860080301761627},
	ngpl-8/gp/alpha/6/.initial={0.1735309362411499},
	ngpl-8/gp/alpha/7/.initial={0.16984523832798004},
	ngpl-8/recloss/obs/all/.initial={1.013848066329956},
	ngpl-8/recloss/obs/masked/.initial={1.080077886581421},
	ngpl-8/recloss/mu/all/.initial={inf},
	ngpl-8/recloss/mu/masked/.initial={1.4069641828536987},
}
\pgfkeys{,
	/private/antennae/.cd,
	ngpl-3/src/sigma/.initial={0.04},
	ngpl-3/src/alpha/.initial={3.6407997608184814},
	ngpl-3/gp/sigma/0/.initial={0.01},
	ngpl-3/gp/sigma/1/.initial={0.00447213595499958},
	ngpl-3/gp/sigma/2/.initial={0.002},
	ngpl-3/gp/alpha/0/.initial={0.4505695402622223},
	ngpl-3/gp/alpha/1/.initial={0.26779815554618835},
	ngpl-3/gp/alpha/2/.initial={0.18279710412025452},
	ngpl-3/recloss/obs/all/.initial={1.0226327180862427},
	ngpl-3/recloss/obs/masked/.initial={1.1104817390441895},
	ngpl-3/recloss/mu/all/.initial={inf},
	ngpl-3/recloss/mu/masked/.initial={1.406022548675537},
}
\pgfkeys{,
	/private/arp142/.cd,
	ngpl-3/src/sigma/.initial={0.04},
	ngpl-3/src/alpha/.initial={2.37630033493042},
	ngpl-3/gp/sigma/0/.initial={0.01},
	ngpl-3/gp/sigma/1/.initial={0.00447213595499958},
	ngpl-3/gp/sigma/2/.initial={0.002},
	ngpl-3/gp/alpha/0/.initial={0.3164946138858795},
	ngpl-3/gp/alpha/1/.initial={0.2462293952703476},
	ngpl-3/gp/alpha/2/.initial={0.17932790517807007},
	ngpl-3/recloss/obs/all/.initial={1.0104048252105713},
	ngpl-3/recloss/obs/masked/.initial={1.0645955801010132},
	ngpl-3/recloss/mu/all/.initial={inf},
	ngpl-3/recloss/mu/masked/.initial={1.33056640625},
}
\pgfkeys{,
	/private/arp148/.cd,
	ngpl-3/src/sigma/.initial={0.04},
	ngpl-3/src/alpha/.initial={2.9710936546325684},
	ngpl-3/gp/sigma/0/.initial={0.01},
	ngpl-3/gp/sigma/1/.initial={0.00447213595499958},
	ngpl-3/gp/sigma/2/.initial={0.002},
	ngpl-3/gp/alpha/0/.initial={0.6144081950187683},
	ngpl-3/gp/alpha/1/.initial={0.3262597918510437},
	ngpl-3/gp/alpha/2/.initial={0.1938321441411972},
	ngpl-3/recloss/obs/all/.initial={1.0381245613098145},
	ngpl-3/recloss/obs/masked/.initial={1.1606874465942383},
	ngpl-3/recloss/mu/all/.initial={inf},
	ngpl-3/recloss/mu/masked/.initial={1.092410683631897},
}
\pgfkeys{,
	/private/ngc4414/.cd,
	ngpl-3/src/sigma/.initial={0.04},
	ngpl-3/src/alpha/.initial={2.64845609664917},
	ngpl-3/gp/sigma/0/.initial={0.01},
	ngpl-3/gp/sigma/1/.initial={0.00447213595499958},
	ngpl-3/gp/sigma/2/.initial={0.002},
	ngpl-3/gp/alpha/0/.initial={0.20888687670230865},
	ngpl-3/gp/alpha/1/.initial={0.6650449633598328},
	ngpl-3/gp/alpha/2/.initial={0.2361769825220108},
}
\pgfkeys{,
	/private/hoags_object/.cd,
	ngpl-1b/init/ext/gamma_1/.initial={0.0},
	ngpl-1b/init/ext/gamma_2/.initial={0.0},
	ngpl-1b/init/main/phi/.initial={1.5},
	ngpl-1b/init/main/q/.initial={-1.252762794494629},
	ngpl-1b/init/main/r_ein/.initial={-0.9162903428077698},
	ngpl-1b/init/main/x/.initial={0.0},
	ngpl-1b/init/main/y/.initial={0.0},
	ngpl-1b/mu/ext/gamma_1/.initial={-0.022558968514204025},
	ngpl-1b/mu/ext/gamma_2/.initial={0.004485872574150562},
	ngpl-1b/mu/main/phi/.initial={1.7421149015426636},
	ngpl-1b/mu/main/q/.initial={0.27562445402145386},
	ngpl-1b/mu/main/r_ein/.initial={1.1846247911453247},
	ngpl-1b/mu/main/x/.initial={-0.07035234570503235},
	ngpl-1b/mu/main/y/.initial={-0.01939953863620758},
	ngpl-1b/S/ext/gamma_1/ext/gamma_1/.initial={2.6847169820598538e-08},
	ngpl-1b/S/ext/gamma_1/ext/gamma_2/.initial={5.848025175225757e-09},
	ngpl-1b/S/ext/gamma_1/main/phi/.initial={9.053888305743385e-09},
	ngpl-1b/S/ext/gamma_1/main/q/.initial={3.482035637603076e-08},
	ngpl-1b/S/ext/gamma_1/main/r_ein/.initial={-2.919426478342757e-08},
	ngpl-1b/S/ext/gamma_1/main/x/.initial={2.504215945720034e-09},
	ngpl-1b/S/ext/gamma_1/main/y/.initial={-3.994128139339637e-09},
	ngpl-1b/S/ext/gamma_2/ext/gamma_1/.initial={5.848025175225757e-09},
	ngpl-1b/S/ext/gamma_2/ext/gamma_2/.initial={1.9180179222644256e-08},
	ngpl-1b/S/ext/gamma_2/main/phi/.initial={-2.690157074880517e-08},
	ngpl-1b/S/ext/gamma_2/main/q/.initial={1.3726324965546155e-08},
	ngpl-1b/S/ext/gamma_2/main/r_ein/.initial={-1.3991980907235302e-08},
	ngpl-1b/S/ext/gamma_2/main/x/.initial={3.006867255184176e-10},
	ngpl-1b/S/ext/gamma_2/main/y/.initial={1.3005752030892381e-09},
	ngpl-1b/S/main/phi/ext/gamma_1/.initial={9.053888305743385e-09},
	ngpl-1b/S/main/phi/ext/gamma_2/.initial={-2.690157074880517e-08},
	ngpl-1b/S/main/phi/main/phi/.initial={6.255912410324527e-08},
	ngpl-1b/S/main/phi/main/q/.initial={4.0347840624122e-09},
	ngpl-1b/S/main/phi/main/r_ein/.initial={2.7542235159216943e-10},
	ngpl-1b/S/main/phi/main/x/.initial={-1.0059761912373233e-09},
	ngpl-1b/S/main/phi/main/y/.initial={-3.6785181567466907e-09},
	ngpl-1b/S/main/q/ext/gamma_1/.initial={3.482035637603076e-08},
	ngpl-1b/S/main/q/ext/gamma_2/.initial={1.3726324965546155e-08},
	ngpl-1b/S/main/q/main/phi/.initial={4.0347840624122e-09},
	ngpl-1b/S/main/q/main/q/.initial={7.647888367046107e-08},
	ngpl-1b/S/main/q/main/r_ein/.initial={-4.914534557087791e-08},
	ngpl-1b/S/main/q/main/x/.initial={6.943459140984487e-09},
	ngpl-1b/S/main/q/main/y/.initial={-1.1406827482574045e-08},
	ngpl-1b/S/main/r_ein/ext/gamma_1/.initial={-2.919426478342757e-08},
	ngpl-1b/S/main/r_ein/ext/gamma_2/.initial={-1.3991980907235302e-08},
	ngpl-1b/S/main/r_ein/main/phi/.initial={2.7542235159216943e-10},
	ngpl-1b/S/main/r_ein/main/q/.initial={-4.914534557087791e-08},
	ngpl-1b/S/main/r_ein/main/r_ein/.initial={4.198837544322487e-08},
	ngpl-1b/S/main/r_ein/main/x/.initial={-4.475874337828145e-09},
	ngpl-1b/S/main/r_ein/main/y/.initial={3.3625031647943615e-09},
	ngpl-1b/S/main/x/ext/gamma_1/.initial={2.504215945720034e-09},
	ngpl-1b/S/main/x/ext/gamma_2/.initial={3.006867255184176e-10},
	ngpl-1b/S/main/x/main/phi/.initial={-1.0059761912373233e-09},
	ngpl-1b/S/main/x/main/q/.initial={6.943459140984487e-09},
	ngpl-1b/S/main/x/main/r_ein/.initial={-4.475874337828145e-09},
	ngpl-1b/S/main/x/main/x/.initial={8.841916532276173e-09},
	ngpl-1b/S/main/x/main/y/.initial={-8.291398678750284e-09},
	ngpl-1b/S/main/y/ext/gamma_1/.initial={-3.994128139339637e-09},
	ngpl-1b/S/main/y/ext/gamma_2/.initial={1.3005752030892381e-09},
	ngpl-1b/S/main/y/main/phi/.initial={-3.6785181567466907e-09},
	ngpl-1b/S/main/y/main/q/.initial={-1.1406827482574045e-08},
	ngpl-1b/S/main/y/main/r_ein/.initial={3.3625031647943615e-09},
	ngpl-1b/S/main/y/main/x/.initial={-8.291398678750284e-09},
	ngpl-1b/S/main/y/main/y/.initial={3.503292234086075e-08},
	ngpl-5/init/ext/gamma_1/.initial={0.0},
	ngpl-5/init/ext/gamma_2/.initial={0.0},
	ngpl-5/init/main/phi/.initial={1.5},
	ngpl-5/init/main/q/.initial={-1.252762794494629},
	ngpl-5/init/main/r_ein/.initial={-0.9162903428077698},
	ngpl-5/init/main/x/.initial={0.0},
	ngpl-5/init/main/y/.initial={0.0},
	ngpl-5/mu/ext/gamma_1/.initial={-0.022629542276263237},
	ngpl-5/mu/ext/gamma_2/.initial={0.004648970440030098},
	ngpl-5/mu/main/phi/.initial={1.7418900728225708},
	ngpl-5/mu/main/q/.initial={0.27536505460739136},
	ngpl-5/mu/main/r_ein/.initial={1.184735894203186},
	ngpl-5/mu/main/x/.initial={-0.07038745284080505},
	ngpl-5/mu/main/y/.initial={-0.019591599702835083},
	ngpl-5/S/ext/gamma_1/ext/gamma_1/.initial={2.121186781778306e-08},
	ngpl-5/S/ext/gamma_1/ext/gamma_2/.initial={7.0388259665321584e-09},
	ngpl-5/S/ext/gamma_1/main/phi/.initial={-5.2547726170359965e-09},
	ngpl-5/S/ext/gamma_1/main/q/.initial={3.603333809110154e-08},
	ngpl-5/S/ext/gamma_1/main/r_ein/.initial={-2.2460637794097238e-08},
	ngpl-5/S/ext/gamma_1/main/x/.initial={1.8440295956878572e-09},
	ngpl-5/S/ext/gamma_1/main/y/.initial={1.1608757288783522e-09},
	ngpl-5/S/ext/gamma_2/ext/gamma_1/.initial={7.0388259665321584e-09},
	ngpl-5/S/ext/gamma_2/ext/gamma_2/.initial={1.4823102745253891e-08},
	ngpl-5/S/ext/gamma_2/main/phi/.initial={-1.3018862432545575e-08},
	ngpl-5/S/ext/gamma_2/main/q/.initial={1.6410774961173047e-08},
	ngpl-5/S/ext/gamma_2/main/r_ein/.initial={-9.013234603116871e-09},
	ngpl-5/S/ext/gamma_2/main/x/.initial={-2.8499846926877126e-09},
	ngpl-5/S/ext/gamma_2/main/y/.initial={3.6852454421421044e-09},
	ngpl-5/S/main/phi/ext/gamma_1/.initial={-5.2547726170359965e-09},
	ngpl-5/S/main/phi/ext/gamma_2/.initial={-1.3018862432545575e-08},
	ngpl-5/S/main/phi/main/phi/.initial={1.982181885296086e-08},
	ngpl-5/S/main/phi/main/q/.initial={-1.988354014770266e-08},
	ngpl-5/S/main/phi/main/r_ein/.initial={7.995780038072553e-09},
	ngpl-5/S/main/phi/main/x/.initial={1.9646870796918847e-09},
	ngpl-5/S/main/phi/main/y/.initial={-1.380564218500524e-09},
	ngpl-5/S/main/q/ext/gamma_1/.initial={3.603333809110154e-08},
	ngpl-5/S/main/q/ext/gamma_2/.initial={1.6410774961173047e-08},
	ngpl-5/S/main/q/main/phi/.initial={-1.988354014770266e-08},
	ngpl-5/S/main/q/main/q/.initial={8.900644843379268e-08},
	ngpl-5/S/main/q/main/r_ein/.initial={-5.278722881030262e-08},
	ngpl-5/S/main/q/main/x/.initial={2.070354998551238e-09},
	ngpl-5/S/main/q/main/y/.initial={1.2439246299678075e-09},
	ngpl-5/S/main/r_ein/ext/gamma_1/.initial={-2.2460637794097238e-08},
	ngpl-5/S/main/r_ein/ext/gamma_2/.initial={-9.013234603116871e-09},
	ngpl-5/S/main/r_ein/main/phi/.initial={7.995780038072553e-09},
	ngpl-5/S/main/r_ein/main/q/.initial={-5.278722881030262e-08},
	ngpl-5/S/main/r_ein/main/r_ein/.initial={3.565144979233992e-08},
	ngpl-5/S/main/r_ein/main/x/.initial={-1.0814367179534656e-09},
	ngpl-5/S/main/r_ein/main/y/.initial={-2.7948829917079365e-09},
	ngpl-5/S/main/x/ext/gamma_1/.initial={1.8440295956878572e-09},
	ngpl-5/S/main/x/ext/gamma_2/.initial={-2.8499846926877126e-09},
	ngpl-5/S/main/x/main/phi/.initial={1.9646870796918847e-09},
	ngpl-5/S/main/x/main/q/.initial={2.070354998551238e-09},
	ngpl-5/S/main/x/main/r_ein/.initial={-1.0814367179534656e-09},
	ngpl-5/S/main/x/main/x/.initial={7.025444670460956e-09},
	ngpl-5/S/main/x/main/y/.initial={-3.6145795245801082e-09},
	ngpl-5/S/main/y/ext/gamma_1/.initial={1.1608757288783522e-09},
	ngpl-5/S/main/y/ext/gamma_2/.initial={3.6852454421421044e-09},
	ngpl-5/S/main/y/main/phi/.initial={-1.380564218500524e-09},
	ngpl-5/S/main/y/main/q/.initial={1.2439246299678075e-09},
	ngpl-5/S/main/y/main/r_ein/.initial={-2.7948829917079365e-09},
	ngpl-5/S/main/y/main/x/.initial={-3.6145795245801082e-09},
	ngpl-5/S/main/y/main/y/.initial={1.9788453542446405e-08},
	ngpl-3/init/ext/gamma_1/.initial={0.0},
	ngpl-3/init/ext/gamma_2/.initial={0.0},
	ngpl-3/init/main/phi/.initial={1.5},
	ngpl-3/init/main/q/.initial={-1.252762794494629},
	ngpl-3/init/main/r_ein/.initial={-0.9162903428077698},
	ngpl-3/init/main/x/.initial={0.0},
	ngpl-3/init/main/y/.initial={0.0},
	ngpl-3/mu/ext/gamma_1/.initial={-0.022587629035115242},
	ngpl-3/mu/ext/gamma_2/.initial={0.004641156643629074},
	ngpl-3/mu/main/phi/.initial={1.7420324087142944},
	ngpl-3/mu/main/q/.initial={0.27551037073135376},
	ngpl-3/mu/main/r_ein/.initial={1.1846750974655151},
	ngpl-3/mu/main/x/.initial={-0.07039865851402283},
	ngpl-3/mu/main/y/.initial={-0.01946188509464264},
	ngpl-3/S/ext/gamma_1/ext/gamma_1/.initial={2.3289038253437866e-08},
	ngpl-3/S/ext/gamma_1/ext/gamma_2/.initial={6.327808943495938e-09},
	ngpl-3/S/ext/gamma_1/main/phi/.initial={1.928098924963706e-08},
	ngpl-3/S/ext/gamma_1/main/q/.initial={3.922232494346645e-08},
	ngpl-3/S/ext/gamma_1/main/r_ein/.initial={-2.893400541381652e-08},
	ngpl-3/S/ext/gamma_1/main/x/.initial={1.292405626962534e-09},
	ngpl-3/S/ext/gamma_1/main/y/.initial={6.5806333715556775e-09},
	ngpl-3/S/ext/gamma_2/ext/gamma_1/.initial={6.327808943495938e-09},
	ngpl-3/S/ext/gamma_2/ext/gamma_2/.initial={1.5511902873299732e-08},
	ngpl-3/S/ext/gamma_2/main/phi/.initial={-3.6778882162025184e-08},
	ngpl-3/S/ext/gamma_2/main/q/.initial={1.8223566655706236e-08},
	ngpl-3/S/ext/gamma_2/main/r_ein/.initial={-1.5043358558841646e-08},
	ngpl-3/S/ext/gamma_2/main/x/.initial={-1.786064185438363e-09},
	ngpl-3/S/ext/gamma_2/main/y/.initial={5.30025712208726e-09},
	ngpl-3/S/main/phi/ext/gamma_1/.initial={1.928098924963706e-08},
	ngpl-3/S/main/phi/ext/gamma_2/.initial={-3.6778882162025184e-08},
	ngpl-3/S/main/phi/main/phi/.initial={1.5417040799547976e-07},
	ngpl-3/S/main/phi/main/q/.initial={7.331880880201425e-09},
	ngpl-3/S/main/phi/main/r_ein/.initial={-3.0880809021027744e-09},
	ngpl-3/S/main/phi/main/x/.initial={6.55275478322892e-09},
	ngpl-3/S/main/phi/main/y/.initial={-4.020090926815101e-09},
	ngpl-3/S/main/q/ext/gamma_1/.initial={3.922232494346645e-08},
	ngpl-3/S/main/q/ext/gamma_2/.initial={1.8223566655706236e-08},
	ngpl-3/S/main/q/main/phi/.initial={7.331880880201425e-09},
	ngpl-3/S/main/q/main/q/.initial={9.431330738607357e-08},
	ngpl-3/S/main/q/main/r_ein/.initial={-6.052377443666046e-08},
	ngpl-3/S/main/q/main/x/.initial={-1.6969619043294415e-09},
	ngpl-3/S/main/q/main/y/.initial={5.391335378135409e-09},
	ngpl-3/S/main/r_ein/ext/gamma_1/.initial={-2.893400541381652e-08},
	ngpl-3/S/main/r_ein/ext/gamma_2/.initial={-1.5043358558841646e-08},
	ngpl-3/S/main/r_ein/main/phi/.initial={-3.0880809021027744e-09},
	ngpl-3/S/main/r_ein/main/q/.initial={-6.052377443666046e-08},
	ngpl-3/S/main/r_ein/main/r_ein/.initial={4.62253666455581e-08},
	ngpl-3/S/main/r_ein/main/x/.initial={1.5915319062642652e-09},
	ngpl-3/S/main/r_ein/main/y/.initial={-9.605586548389056e-09},
	ngpl-3/S/main/x/ext/gamma_1/.initial={1.292405626962534e-09},
	ngpl-3/S/main/x/ext/gamma_2/.initial={-1.786064185438363e-09},
	ngpl-3/S/main/x/main/phi/.initial={6.55275478322892e-09},
	ngpl-3/S/main/x/main/q/.initial={-1.6969619043294415e-09},
	ngpl-3/S/main/x/main/r_ein/.initial={1.5915319062642652e-09},
	ngpl-3/S/main/x/main/x/.initial={7.816117530978772e-09},
	ngpl-3/S/main/x/main/y/.initial={-4.874455505898823e-09},
	ngpl-3/S/main/y/ext/gamma_1/.initial={6.5806333715556775e-09},
	ngpl-3/S/main/y/ext/gamma_2/.initial={5.30025712208726e-09},
	ngpl-3/S/main/y/main/phi/.initial={-4.020090926815101e-09},
	ngpl-3/S/main/y/main/q/.initial={5.391335378135409e-09},
	ngpl-3/S/main/y/main/r_ein/.initial={-9.605586548389056e-09},
	ngpl-3/S/main/y/main/x/.initial={-4.874455505898823e-09},
	ngpl-3/S/main/y/main/y/.initial={2.982691782449365e-08},
	ngpl-7/init/ext/gamma_1/.initial={0.0},
	ngpl-7/init/ext/gamma_2/.initial={0.0},
	ngpl-7/init/main/phi/.initial={1.5},
	ngpl-7/init/main/q/.initial={-1.252762794494629},
	ngpl-7/init/main/r_ein/.initial={-0.9162903428077698},
	ngpl-7/init/main/x/.initial={0.0},
	ngpl-7/init/main/y/.initial={0.0},
	ngpl-7/mu/ext/gamma_1/.initial={-0.022618835791945457},
	ngpl-7/mu/ext/gamma_2/.initial={0.00471584452316165},
	ngpl-7/mu/main/phi/.initial={1.7418763637542725},
	ngpl-7/mu/main/q/.initial={0.27537956833839417},
	ngpl-7/mu/main/r_ein/.initial={1.1847457885742188},
	ngpl-7/mu/main/x/.initial={-0.0704125463962555},
	ngpl-7/mu/main/y/.initial={-0.019520223140716553},
	ngpl-7/S/ext/gamma_1/ext/gamma_1/.initial={1.9827789188298084e-08},
	ngpl-7/S/ext/gamma_1/ext/gamma_2/.initial={8.376881410754322e-09},
	ngpl-7/S/ext/gamma_1/main/phi/.initial={1.8560765369102228e-08},
	ngpl-7/S/ext/gamma_1/main/q/.initial={2.911234275870811e-08},
	ngpl-7/S/ext/gamma_1/main/r_ein/.initial={-1.9501364079133054e-08},
	ngpl-7/S/ext/gamma_1/main/x/.initial={6.373650496271921e-09},
	ngpl-7/S/ext/gamma_1/main/y/.initial={4.961189681296219e-09},
	ngpl-7/S/ext/gamma_2/ext/gamma_1/.initial={8.376881410754322e-09},
	ngpl-7/S/ext/gamma_2/ext/gamma_2/.initial={1.52861883151445e-08},
	ngpl-7/S/ext/gamma_2/main/phi/.initial={-7.6478618993292e-09},
	ngpl-7/S/ext/gamma_2/main/q/.initial={1.6963696225502645e-08},
	ngpl-7/S/ext/gamma_2/main/r_ein/.initial={-1.4076700693976818e-08},
	ngpl-7/S/ext/gamma_2/main/x/.initial={4.252751484301598e-09},
	ngpl-7/S/ext/gamma_2/main/y/.initial={1.0926287652424094e-09},
	ngpl-7/S/main/phi/ext/gamma_1/.initial={1.8560765369102228e-08},
	ngpl-7/S/main/phi/ext/gamma_2/.initial={-7.6478618993292e-09},
	ngpl-7/S/main/phi/main/phi/.initial={4.599962721840711e-08},
	ngpl-7/S/main/phi/main/q/.initial={1.5967776434422376e-08},
	ngpl-7/S/main/phi/main/r_ein/.initial={-7.951418190543791e-09},
	ngpl-7/S/main/phi/main/x/.initial={1.1340071104370963e-09},
	ngpl-7/S/main/phi/main/y/.initial={4.806224751519039e-09},
	ngpl-7/S/main/q/ext/gamma_1/.initial={2.911234275870811e-08},
	ngpl-7/S/main/q/ext/gamma_2/.initial={1.6963696225502645e-08},
	ngpl-7/S/main/q/main/phi/.initial={1.5967776434422376e-08},
	ngpl-7/S/main/q/main/q/.initial={6.664264162736799e-08},
	ngpl-7/S/main/q/main/r_ein/.initial={-3.5277196275274036e-08},
	ngpl-7/S/main/q/main/x/.initial={1.734545662657183e-08},
	ngpl-7/S/main/q/main/y/.initial={7.55477458369569e-09},
	ngpl-7/S/main/r_ein/ext/gamma_1/.initial={-1.9501364079133054e-08},
	ngpl-7/S/main/r_ein/ext/gamma_2/.initial={-1.4076700693976818e-08},
	ngpl-7/S/main/r_ein/main/phi/.initial={-7.951418190543791e-09},
	ngpl-7/S/main/r_ein/main/q/.initial={-3.5277196275274036e-08},
	ngpl-7/S/main/r_ein/main/r_ein/.initial={2.6158307520063317e-08},
	ngpl-7/S/main/r_ein/main/x/.initial={-9.474765860773005e-09},
	ngpl-7/S/main/r_ein/main/y/.initial={-4.391793595459603e-09},
	ngpl-7/S/main/x/ext/gamma_1/.initial={6.373650496271921e-09},
	ngpl-7/S/main/x/ext/gamma_2/.initial={4.252751484301598e-09},
	ngpl-7/S/main/x/main/phi/.initial={1.1340071104370963e-09},
	ngpl-7/S/main/x/main/q/.initial={1.734545662657183e-08},
	ngpl-7/S/main/x/main/r_ein/.initial={-9.474765860773005e-09},
	ngpl-7/S/main/x/main/x/.initial={1.0853376863906306e-08},
	ngpl-7/S/main/x/main/y/.initial={-3.343824772628068e-09},
	ngpl-7/S/main/y/ext/gamma_1/.initial={4.961189681296219e-09},
	ngpl-7/S/main/y/ext/gamma_2/.initial={1.0926287652424094e-09},
	ngpl-7/S/main/y/main/phi/.initial={4.806224751519039e-09},
	ngpl-7/S/main/y/main/q/.initial={7.55477458369569e-09},
	ngpl-7/S/main/y/main/r_ein/.initial={-4.391793595459603e-09},
	ngpl-7/S/main/y/main/x/.initial={-3.343824772628068e-09},
	ngpl-7/S/main/y/main/y/.initial={2.1701168861909537e-08},
	ngpl-4/init/ext/gamma_1/.initial={0.0},
	ngpl-4/init/ext/gamma_2/.initial={0.0},
	ngpl-4/init/main/phi/.initial={1.5},
	ngpl-4/init/main/q/.initial={-1.252762794494629},
	ngpl-4/init/main/r_ein/.initial={-0.9162903428077698},
	ngpl-4/init/main/x/.initial={0.0},
	ngpl-4/init/main/y/.initial={0.0},
	ngpl-4/mu/ext/gamma_1/.initial={-0.022696860134601593},
	ngpl-4/mu/ext/gamma_2/.initial={0.004686174914240837},
	ngpl-4/mu/main/phi/.initial={1.7419676780700684},
	ngpl-4/mu/main/q/.initial={0.27542632818222046},
	ngpl-4/mu/main/r_ein/.initial={1.1847052574157715},
	ngpl-4/mu/main/x/.initial={-0.07035115361213684},
	ngpl-4/mu/main/y/.initial={-0.019506633281707764},
	ngpl-4/S/ext/gamma_1/ext/gamma_1/.initial={2.1318177445550646e-08},
	ngpl-4/S/ext/gamma_1/ext/gamma_2/.initial={-2.8210396241235003e-09},
	ngpl-4/S/ext/gamma_1/main/phi/.initial={2.9954903002504807e-09},
	ngpl-4/S/ext/gamma_1/main/q/.initial={3.396994685544996e-08},
	ngpl-4/S/ext/gamma_1/main/r_ein/.initial={-2.298703449810091e-08},
	ngpl-4/S/ext/gamma_1/main/x/.initial={-1.6873377139958734e-09},
	ngpl-4/S/ext/gamma_1/main/y/.initial={-2.914007035670352e-09},
	ngpl-4/S/ext/gamma_2/ext/gamma_1/.initial={-2.8210396241235003e-09},
	ngpl-4/S/ext/gamma_2/ext/gamma_2/.initial={1.2748843758458861e-08},
	ngpl-4/S/ext/gamma_2/main/phi/.initial={-1.8584145777822414e-08},
	ngpl-4/S/ext/gamma_2/main/q/.initial={-9.748468698944635e-11},
	ngpl-4/S/ext/gamma_2/main/r_ein/.initial={5.916727108257192e-10},
	ngpl-4/S/ext/gamma_2/main/x/.initial={-1.3042861235490477e-09},
	ngpl-4/S/ext/gamma_2/main/y/.initial={-8.48362435856842e-10},
	ngpl-4/S/main/phi/ext/gamma_1/.initial={2.9954903002504807e-09},
	ngpl-4/S/main/phi/ext/gamma_2/.initial={-1.8584145777822414e-08},
	ngpl-4/S/main/phi/main/phi/.initial={3.611413390558482e-08},
	ngpl-4/S/main/phi/main/q/.initial={-2.4997359737710667e-09},
	ngpl-4/S/main/phi/main/r_ein/.initial={-9.40804989468802e-10},
	ngpl-4/S/main/phi/main/x/.initial={-1.0766823543839621e-10},
	ngpl-4/S/main/phi/main/y/.initial={2.967473378134855e-09},
	ngpl-4/S/main/q/ext/gamma_1/.initial={3.396994685544996e-08},
	ngpl-4/S/main/q/ext/gamma_2/.initial={-9.748468698944635e-11},
	ngpl-4/S/main/q/main/phi/.initial={-2.4997359737710667e-09},
	ngpl-4/S/main/q/main/q/.initial={7.757031994515273e-08},
	ngpl-4/S/main/q/main/r_ein/.initial={-4.544337528500364e-08},
	ngpl-4/S/main/q/main/x/.initial={-1.9648505045211095e-09},
	ngpl-4/S/main/q/main/y/.initial={-3.051453312252761e-09},
	ngpl-4/S/main/r_ein/ext/gamma_1/.initial={-2.298703449810091e-08},
	ngpl-4/S/main/r_ein/ext/gamma_2/.initial={5.916727108257192e-10},
	ngpl-4/S/main/r_ein/main/phi/.initial={-9.40804989468802e-10},
	ngpl-4/S/main/r_ein/main/q/.initial={-4.544337528500364e-08},
	ngpl-4/S/main/r_ein/main/r_ein/.initial={3.173794382860251e-08},
	ngpl-4/S/main/r_ein/main/x/.initial={1.5232834993383904e-09},
	ngpl-4/S/main/r_ein/main/y/.initial={-3.932434378128846e-10},
	ngpl-4/S/main/x/ext/gamma_1/.initial={-1.6873377139958734e-09},
	ngpl-4/S/main/x/ext/gamma_2/.initial={-1.3042861235490477e-09},
	ngpl-4/S/main/x/main/phi/.initial={-1.0766823543839621e-10},
	ngpl-4/S/main/x/main/q/.initial={-1.9648505045211095e-09},
	ngpl-4/S/main/x/main/r_ein/.initial={1.5232834993383904e-09},
	ngpl-4/S/main/x/main/x/.initial={7.339638674608295e-09},
	ngpl-4/S/main/x/main/y/.initial={-2.6672157815710307e-09},
	ngpl-4/S/main/y/ext/gamma_1/.initial={-2.914007035670352e-09},
	ngpl-4/S/main/y/ext/gamma_2/.initial={-8.48362435856842e-10},
	ngpl-4/S/main/y/main/phi/.initial={2.967473378134855e-09},
	ngpl-4/S/main/y/main/q/.initial={-3.051453312252761e-09},
	ngpl-4/S/main/y/main/r_ein/.initial={-3.932434378128846e-10},
	ngpl-4/S/main/y/main/x/.initial={-2.6672157815710307e-09},
	ngpl-4/S/main/y/main/y/.initial={2.283826106008746e-08},
	ngpl-1s/init/ext/gamma_1/.initial={0.0},
	ngpl-1s/init/ext/gamma_2/.initial={0.0},
	ngpl-1s/init/main/phi/.initial={1.5},
	ngpl-1s/init/main/q/.initial={-1.252762794494629},
	ngpl-1s/init/main/r_ein/.initial={-0.9162903428077698},
	ngpl-1s/init/main/x/.initial={0.0},
	ngpl-1s/init/main/y/.initial={0.0},
	ngpl-1s/mu/ext/gamma_1/.initial={-0.022615836933255196},
	ngpl-1s/mu/ext/gamma_2/.initial={0.004788591526448727},
	ngpl-1s/mu/main/phi/.initial={1.7420247793197632},
	ngpl-1s/mu/main/q/.initial={0.27560245990753174},
	ngpl-1s/mu/main/r_ein/.initial={1.1845775842666626},
	ngpl-1s/mu/main/x/.initial={-0.07056805491447449},
	ngpl-1s/mu/main/y/.initial={-0.0193987637758255},
	ngpl-1s/S/ext/gamma_1/ext/gamma_1/.initial={4.073562465123359e-08},
	ngpl-1s/S/ext/gamma_1/ext/gamma_2/.initial={-8.72330474521732e-09},
	ngpl-1s/S/ext/gamma_1/main/phi/.initial={1.4665533676350151e-08},
	ngpl-1s/S/ext/gamma_1/main/q/.initial={6.817307252049432e-08},
	ngpl-1s/S/ext/gamma_1/main/r_ein/.initial={-4.4387558517655634e-08},
	ngpl-1s/S/ext/gamma_1/main/x/.initial={-4.708610723547224e-10},
	ngpl-1s/S/ext/gamma_1/main/y/.initial={4.055143998371591e-10},
	ngpl-1s/S/ext/gamma_2/ext/gamma_1/.initial={-8.72330474521732e-09},
	ngpl-1s/S/ext/gamma_2/ext/gamma_2/.initial={3.0809186313263126e-08},
	ngpl-1s/S/ext/gamma_2/main/phi/.initial={-4.958881660854786e-08},
	ngpl-1s/S/ext/gamma_2/main/q/.initial={9.86467441066452e-09},
	ngpl-1s/S/ext/gamma_2/main/r_ein/.initial={-4.0796255262876e-09},
	ngpl-1s/S/ext/gamma_2/main/x/.initial={-2.1868904465094374e-09},
	ngpl-1s/S/ext/gamma_2/main/y/.initial={3.206582555037585e-09},
	ngpl-1s/S/main/phi/ext/gamma_1/.initial={1.4665533676350151e-08},
	ngpl-1s/S/main/phi/ext/gamma_2/.initial={-4.958881660854786e-08},
	ngpl-1s/S/main/phi/main/phi/.initial={9.808764644958501e-08},
	ngpl-1s/S/main/phi/main/q/.initial={-1.2692506601297282e-08},
	ngpl-1s/S/main/phi/main/r_ein/.initial={5.349535925347482e-09},
	ngpl-1s/S/main/phi/main/x/.initial={7.801331136647605e-09},
	ngpl-1s/S/main/phi/main/y/.initial={-4.957416255280123e-09},
	ngpl-1s/S/main/q/ext/gamma_1/.initial={6.817307252049432e-08},
	ngpl-1s/S/main/q/ext/gamma_2/.initial={9.86467441066452e-09},
	ngpl-1s/S/main/q/main/phi/.initial={-1.2692506601297282e-08},
	ngpl-1s/S/main/q/main/q/.initial={1.7879716551760794e-07},
	ngpl-1s/S/main/q/main/r_ein/.initial={-1.0190359489570255e-07},
	ngpl-1s/S/main/q/main/x/.initial={-3.662179670627097e-09},
	ngpl-1s/S/main/q/main/y/.initial={-5.148915960262457e-09},
	ngpl-1s/S/main/r_ein/ext/gamma_1/.initial={-4.4387558517655634e-08},
	ngpl-1s/S/main/r_ein/ext/gamma_2/.initial={-4.0796255262876e-09},
	ngpl-1s/S/main/r_ein/main/phi/.initial={5.349535925347482e-09},
	ngpl-1s/S/main/r_ein/main/q/.initial={-1.0190359489570255e-07},
	ngpl-1s/S/main/r_ein/main/r_ein/.initial={6.754693515631516e-08},
	ngpl-1s/S/main/r_ein/main/x/.initial={4.1087511171156166e-09},
	ngpl-1s/S/main/r_ein/main/y/.initial={-5.394005686554237e-09},
	ngpl-1s/S/main/x/ext/gamma_1/.initial={-4.708610723547224e-10},
	ngpl-1s/S/main/x/ext/gamma_2/.initial={-2.1868904465094374e-09},
	ngpl-1s/S/main/x/main/phi/.initial={7.801331136647605e-09},
	ngpl-1s/S/main/x/main/q/.initial={-3.662179670627097e-09},
	ngpl-1s/S/main/x/main/r_ein/.initial={4.1087511171156166e-09},
	ngpl-1s/S/main/x/main/x/.initial={1.2208126953794363e-08},
	ngpl-1s/S/main/x/main/y/.initial={-7.947773994487761e-09},
	ngpl-1s/S/main/y/ext/gamma_1/.initial={4.055143998371591e-10},
	ngpl-1s/S/main/y/ext/gamma_2/.initial={3.206582555037585e-09},
	ngpl-1s/S/main/y/main/phi/.initial={-4.957416255280123e-09},
	ngpl-1s/S/main/y/main/q/.initial={-5.148915960262457e-09},
	ngpl-1s/S/main/y/main/r_ein/.initial={-5.394005686554237e-09},
	ngpl-1s/S/main/y/main/x/.initial={-7.947773994487761e-09},
	ngpl-1s/S/main/y/main/y/.initial={4.785842122601025e-08},
	ngpl-2/init/ext/gamma_1/.initial={0.0},
	ngpl-2/init/ext/gamma_2/.initial={0.0},
	ngpl-2/init/main/phi/.initial={1.5},
	ngpl-2/init/main/q/.initial={-1.252762794494629},
	ngpl-2/init/main/r_ein/.initial={-0.9162903428077698},
	ngpl-2/init/main/x/.initial={0.0},
	ngpl-2/init/main/y/.initial={0.0},
	ngpl-2/mu/ext/gamma_1/.initial={-0.0224817655980587},
	ngpl-2/mu/ext/gamma_2/.initial={0.004687176086008549},
	ngpl-2/mu/main/phi/.initial={1.7420985698699951},
	ngpl-2/mu/main/q/.initial={0.2756154537200928},
	ngpl-2/mu/main/r_ein/.initial={1.1846134662628174},
	ngpl-2/mu/main/x/.initial={-0.07036803662776947},
	ngpl-2/mu/main/y/.initial={-0.019450992345809937},
	ngpl-2/S/ext/gamma_1/ext/gamma_1/.initial={2.522313025110634e-08},
	ngpl-2/S/ext/gamma_1/ext/gamma_2/.initial={9.154460300919709e-09},
	ngpl-2/S/ext/gamma_1/main/phi/.initial={1.4159648564771032e-08},
	ngpl-2/S/ext/gamma_1/main/q/.initial={4.491513294624383e-08},
	ngpl-2/S/ext/gamma_1/main/r_ein/.initial={-2.8714080002600895e-08},
	ngpl-2/S/ext/gamma_1/main/x/.initial={4.404030029547812e-09},
	ngpl-2/S/ext/gamma_1/main/y/.initial={-1.4344000431876225e-09},
	ngpl-2/S/ext/gamma_2/ext/gamma_1/.initial={9.154460300919709e-09},
	ngpl-2/S/ext/gamma_2/ext/gamma_2/.initial={1.9125360850580364e-08},
	ngpl-2/S/ext/gamma_2/main/phi/.initial={-1.5616663517903362e-08},
	ngpl-2/S/ext/gamma_2/main/q/.initial={2.867010806539838e-08},
	ngpl-2/S/ext/gamma_2/main/r_ein/.initial={-1.8721216576977895e-08},
	ngpl-2/S/ext/gamma_2/main/x/.initial={-1.1500426166932698e-09},
	ngpl-2/S/ext/gamma_2/main/y/.initial={-2.1523696158709527e-09},
	ngpl-2/S/main/phi/ext/gamma_1/.initial={1.4159648564771032e-08},
	ngpl-2/S/main/phi/ext/gamma_2/.initial={-1.5616663517903362e-08},
	ngpl-2/S/main/phi/main/phi/.initial={4.701553635300115e-08},
	ngpl-2/S/main/phi/main/q/.initial={4.9209063490707194e-09},
	ngpl-2/S/main/phi/main/r_ein/.initial={4.764704186754898e-10},
	ngpl-2/S/main/phi/main/x/.initial={8.292399655829286e-09},
	ngpl-2/S/main/phi/main/y/.initial={4.915970297503236e-09},
	ngpl-2/S/main/q/ext/gamma_1/.initial={4.491513294624383e-08},
	ngpl-2/S/main/q/ext/gamma_2/.initial={2.867010806539838e-08},
	ngpl-2/S/main/q/main/phi/.initial={4.9209063490707194e-09},
	ngpl-2/S/main/q/main/q/.initial={1.178872039986345e-07},
	ngpl-2/S/main/q/main/r_ein/.initial={-6.826788023772679e-08},
	ngpl-2/S/main/q/main/x/.initial={9.9604022807398e-09},
	ngpl-2/S/main/q/main/y/.initial={-1.628495027716781e-08},
	ngpl-2/S/main/r_ein/ext/gamma_1/.initial={-2.8714080002600895e-08},
	ngpl-2/S/main/r_ein/ext/gamma_2/.initial={-1.8721216576977895e-08},
	ngpl-2/S/main/r_ein/main/phi/.initial={4.764704186754898e-10},
	ngpl-2/S/main/r_ein/main/q/.initial={-6.826788023772679e-08},
	ngpl-2/S/main/r_ein/main/r_ein/.initial={4.665760755528936e-08},
	ngpl-2/S/main/r_ein/main/x/.initial={-2.0588850624392308e-09},
	ngpl-2/S/main/r_ein/main/y/.initial={7.151076619749119e-09},
	ngpl-2/S/main/x/ext/gamma_1/.initial={4.404030029547812e-09},
	ngpl-2/S/main/x/ext/gamma_2/.initial={-1.1500426166932698e-09},
	ngpl-2/S/main/x/main/phi/.initial={8.292399655829286e-09},
	ngpl-2/S/main/x/main/q/.initial={9.9604022807398e-09},
	ngpl-2/S/main/x/main/r_ein/.initial={-2.0588850624392308e-09},
	ngpl-2/S/main/x/main/x/.initial={1.0664940042204307e-08},
	ngpl-2/S/main/x/main/y/.initial={-1.7218465542256922e-09},
	ngpl-2/S/main/y/ext/gamma_1/.initial={-1.4344000431876225e-09},
	ngpl-2/S/main/y/ext/gamma_2/.initial={-2.1523696158709527e-09},
	ngpl-2/S/main/y/main/phi/.initial={4.915970297503236e-09},
	ngpl-2/S/main/y/main/q/.initial={-1.628495027716781e-08},
	ngpl-2/S/main/y/main/r_ein/.initial={7.151076619749119e-09},
	ngpl-2/S/main/y/main/x/.initial={-1.7218465542256922e-09},
	ngpl-2/S/main/y/main/y/.initial={2.8722674016989913e-08},
	ngpl-6/init/ext/gamma_1/.initial={0.0},
	ngpl-6/init/ext/gamma_2/.initial={0.0},
	ngpl-6/init/main/phi/.initial={1.5},
	ngpl-6/init/main/q/.initial={-1.252762794494629},
	ngpl-6/init/main/r_ein/.initial={-0.9162903428077698},
	ngpl-6/init/main/x/.initial={0.0},
	ngpl-6/init/main/y/.initial={0.0},
	ngpl-6/mu/ext/gamma_1/.initial={-0.02267424762248993},
	ngpl-6/mu/ext/gamma_2/.initial={0.004731396213173866},
	ngpl-6/mu/main/phi/.initial={1.7418831586837769},
	ngpl-6/mu/main/q/.initial={0.2753717601299286},
	ngpl-6/mu/main/r_ein/.initial={1.1847094297409058},
	ngpl-6/mu/main/x/.initial={-0.07039257884025574},
	ngpl-6/mu/main/y/.initial={-0.019573718309402466},
	ngpl-6/S/ext/gamma_1/ext/gamma_1/.initial={2.0894162844342645e-08},
	ngpl-6/S/ext/gamma_1/ext/gamma_2/.initial={3.967320250097828e-09},
	ngpl-6/S/ext/gamma_1/main/phi/.initial={3.836774009613464e-09},
	ngpl-6/S/ext/gamma_1/main/q/.initial={3.253151348303618e-08},
	ngpl-6/S/ext/gamma_1/main/r_ein/.initial={-2.3071027754895113e-08},
	ngpl-6/S/ext/gamma_1/main/x/.initial={2.022475076302044e-09},
	ngpl-6/S/ext/gamma_1/main/y/.initial={1.6672602187739471e-09},
	ngpl-6/S/ext/gamma_2/ext/gamma_1/.initial={3.967320250097828e-09},
	ngpl-6/S/ext/gamma_2/ext/gamma_2/.initial={1.3079837657414828e-08},
	ngpl-6/S/ext/gamma_2/main/phi/.initial={-1.0314054499360736e-08},
	ngpl-6/S/ext/gamma_2/main/q/.initial={7.193336148958451e-09},
	ngpl-6/S/ext/gamma_2/main/r_ein/.initial={-5.249553236552629e-09},
	ngpl-6/S/ext/gamma_2/main/x/.initial={5.673124747751501e-10},
	ngpl-6/S/ext/gamma_2/main/y/.initial={-6.104433070142079e-10},
	ngpl-6/S/main/phi/ext/gamma_1/.initial={3.836774009613464e-09},
	ngpl-6/S/main/phi/ext/gamma_2/.initial={-1.0314054499360736e-08},
	ngpl-6/S/main/phi/main/phi/.initial={1.8566630899385927e-08},
	ngpl-6/S/main/phi/main/q/.initial={4.5513912638739384e-09},
	ngpl-6/S/main/phi/main/r_ein/.initial={-6.913191796797946e-09},
	ngpl-6/S/main/phi/main/x/.initial={-2.0053306237777235e-10},
	ngpl-6/S/main/phi/main/y/.initial={2.1348096623796664e-09},
	ngpl-6/S/main/q/ext/gamma_1/.initial={3.253151348303618e-08},
	ngpl-6/S/main/q/ext/gamma_2/.initial={7.193336148958451e-09},
	ngpl-6/S/main/q/main/phi/.initial={4.5513912638739384e-09},
	ngpl-6/S/main/q/main/q/.initial={7.018498848765375e-08},
	ngpl-6/S/main/q/main/r_ein/.initial={-3.783048541095013e-08},
	ngpl-6/S/main/q/main/x/.initial={5.466054275871102e-09},
	ngpl-6/S/main/q/main/y/.initial={5.5548237121172406e-09},
	ngpl-6/S/main/r_ein/ext/gamma_1/.initial={-2.3071027754895113e-08},
	ngpl-6/S/main/r_ein/ext/gamma_2/.initial={-5.249553236552629e-09},
	ngpl-6/S/main/r_ein/main/phi/.initial={-6.913191796797946e-09},
	ngpl-6/S/main/r_ein/main/q/.initial={-3.783048541095013e-08},
	ngpl-6/S/main/r_ein/main/r_ein/.initial={3.0212554236186406e-08},
	ngpl-6/S/main/r_ein/main/x/.initial={-2.7656079648608056e-09},
	ngpl-6/S/main/r_ein/main/y/.initial={-3.6209559794997404e-09},
	ngpl-6/S/main/x/ext/gamma_1/.initial={2.022475076302044e-09},
	ngpl-6/S/main/x/ext/gamma_2/.initial={5.673124747751501e-10},
	ngpl-6/S/main/x/main/phi/.initial={-2.0053306237777235e-10},
	ngpl-6/S/main/x/main/q/.initial={5.466054275871102e-09},
	ngpl-6/S/main/x/main/r_ein/.initial={-2.7656079648608056e-09},
	ngpl-6/S/main/x/main/x/.initial={6.49188880430529e-09},
	ngpl-6/S/main/x/main/y/.initial={-1.2753978984036962e-09},
	ngpl-6/S/main/y/ext/gamma_1/.initial={1.6672602187739471e-09},
	ngpl-6/S/main/y/ext/gamma_2/.initial={-6.104433070142079e-10},
	ngpl-6/S/main/y/main/phi/.initial={2.1348096623796664e-09},
	ngpl-6/S/main/y/main/q/.initial={5.5548237121172406e-09},
	ngpl-6/S/main/y/main/r_ein/.initial={-3.6209559794997404e-09},
	ngpl-6/S/main/y/main/x/.initial={-1.2753978984036962e-09},
	ngpl-6/S/main/y/main/y/.initial={1.7921610862003945e-08},
	ngpl-8/init/ext/gamma_1/.initial={0.0},
	ngpl-8/init/ext/gamma_2/.initial={0.0},
	ngpl-8/init/main/phi/.initial={1.5},
	ngpl-8/init/main/q/.initial={-1.252762794494629},
	ngpl-8/init/main/r_ein/.initial={-0.9162903428077698},
	ngpl-8/init/main/x/.initial={0.0},
	ngpl-8/init/main/y/.initial={0.0},
	ngpl-8/mu/ext/gamma_1/.initial={-0.022703824564814568},
	ngpl-8/mu/ext/gamma_2/.initial={0.004800190683454275},
	ngpl-8/mu/main/phi/.initial={1.7418713569641113},
	ngpl-8/mu/main/q/.initial={0.27531930804252625},
	ngpl-8/mu/main/r_ein/.initial={1.184731125831604},
	ngpl-8/mu/main/x/.initial={-0.07040254771709442},
	ngpl-8/mu/main/y/.initial={-0.019537702202796936},
	ngpl-8/S/ext/gamma_1/ext/gamma_1/.initial={1.968497720383766e-08},
	ngpl-8/S/ext/gamma_1/ext/gamma_2/.initial={-5.081207676838062e-10},
	ngpl-8/S/ext/gamma_1/main/phi/.initial={1.0495442737123994e-08},
	ngpl-8/S/ext/gamma_1/main/q/.initial={2.5312392182286203e-08},
	ngpl-8/S/ext/gamma_1/main/r_ein/.initial={-2.054636638604279e-08},
	ngpl-8/S/ext/gamma_1/main/x/.initial={3.6752043075516383e-10},
	ngpl-8/S/ext/gamma_1/main/y/.initial={2.2915696007430597e-09},
	ngpl-8/S/ext/gamma_2/ext/gamma_1/.initial={-5.081207676838062e-10},
	ngpl-8/S/ext/gamma_2/ext/gamma_2/.initial={1.1639430752552471e-08},
	ngpl-8/S/ext/gamma_2/main/phi/.initial={-1.8209529883961295e-08},
	ngpl-8/S/ext/gamma_2/main/q/.initial={1.5112244788895168e-11},
	ngpl-8/S/ext/gamma_2/main/r_ein/.initial={-1.7533821061732624e-09},
	ngpl-8/S/ext/gamma_2/main/x/.initial={-7.429796222702123e-10},
	ngpl-8/S/ext/gamma_2/main/y/.initial={2.1034674002606835e-09},
	ngpl-8/S/main/phi/ext/gamma_1/.initial={1.0495442737123994e-08},
	ngpl-8/S/main/phi/ext/gamma_2/.initial={-1.8209529883961295e-08},
	ngpl-8/S/main/phi/main/phi/.initial={4.107691609078756e-08},
	ngpl-8/S/main/phi/main/q/.initial={9.090949326662212e-09},
	ngpl-8/S/main/phi/main/r_ein/.initial={-7.620483799541944e-09},
	ngpl-8/S/main/phi/main/x/.initial={1.459689591420954e-09},
	ngpl-8/S/main/phi/main/y/.initial={-3.1912175124659825e-09},
	ngpl-8/S/main/q/ext/gamma_1/.initial={2.5312392182286203e-08},
	ngpl-8/S/main/q/ext/gamma_2/.initial={1.5112244788895168e-11},
	ngpl-8/S/main/q/main/phi/.initial={9.090949326662212e-09},
	ngpl-8/S/main/q/main/q/.initial={5.259580859728885e-08},
	ngpl-8/S/main/q/main/r_ein/.initial={-3.146325866509869e-08},
	ngpl-8/S/main/q/main/x/.initial={1.9915284976690373e-09},
	ngpl-8/S/main/q/main/y/.initial={7.491815168236826e-09},
	ngpl-8/S/main/r_ein/ext/gamma_1/.initial={-2.054636638604279e-08},
	ngpl-8/S/main/r_ein/ext/gamma_2/.initial={-1.7533821061732624e-09},
	ngpl-8/S/main/r_ein/main/phi/.initial={-7.620483799541944e-09},
	ngpl-8/S/main/r_ein/main/q/.initial={-3.146325866509869e-08},
	ngpl-8/S/main/r_ein/main/r_ein/.initial={2.608852156527064e-08},
	ngpl-8/S/main/r_ein/main/x/.initial={-2.2564619617249093e-10},
	ngpl-8/S/main/r_ein/main/y/.initial={-5.0748178992421344e-09},
	ngpl-8/S/main/x/ext/gamma_1/.initial={3.6752043075516383e-10},
	ngpl-8/S/main/x/ext/gamma_2/.initial={-7.429796222702123e-10},
	ngpl-8/S/main/x/main/phi/.initial={1.459689591420954e-09},
	ngpl-8/S/main/x/main/q/.initial={1.9915284976690373e-09},
	ngpl-8/S/main/x/main/r_ein/.initial={-2.2564619617249093e-10},
	ngpl-8/S/main/x/main/x/.initial={5.882926146227874e-09},
	ngpl-8/S/main/x/main/y/.initial={-2.5864941299857946e-09},
	ngpl-8/S/main/y/ext/gamma_1/.initial={2.2915696007430597e-09},
	ngpl-8/S/main/y/ext/gamma_2/.initial={2.1034674002606835e-09},
	ngpl-8/S/main/y/main/phi/.initial={-3.1912175124659825e-09},
	ngpl-8/S/main/y/main/q/.initial={7.491815168236826e-09},
	ngpl-8/S/main/y/main/r_ein/.initial={-5.0748178992421344e-09},
	ngpl-8/S/main/y/main/x/.initial={-2.5864941299857946e-09},
	ngpl-8/S/main/y/main/y/.initial={1.8611718388683585e-08},
}

\graphicspath{{./}{figures/}}

\tikzset{
    external/system call={%
        lualatex %
        \tikzexternalcheckshellescape %
        -halt-on-error %
        -interaction=batchmode %
        -aux-directory=build %
        -jobname "\image" %
        "\texsource"%
    },
	/pgf/images/external info,
}
\DeclareGraphicsExtensions{.png,.pdf,.jpg}
\usepackage{currfile}  

\newlength{\rawwidth}
\NewEnviron{scaletowidth}[1]{
	\def\thescale{1}%
	\tikzifexternalizingnext{
		\settowidth\rawwidth{\bgroup\tikzexternaldisable\BODY\egroup}%
		\pgfmathsetmacro{\thescale}{#1 / \rawwidth * \thescale}%
		\settowidth\rawwidth{\bgroup\tikzexternaldisable\BODY\egroup}%
		\pgfmathsetmacro{\thescale}{#1 / \rawwidth * \thescale}%
		\settowidth\rawwidth{\bgroup\tikzexternaldisable\BODY\egroup}%
		\pgfmathsetmacro{\thescale}{#1 / \rawwidth * \thescale}%
		\settowidth\rawwidth{\bgroup\tikzexternaldisable\BODY\egroup}%
		\pgfmathsetmacro{\thescale}{#1 / \rawwidth * \thescale}%
	}{}
	\BODY
}

\usepackage{appendix}
\crefname{appendix}{appendix}{appendices}
\AtBeginEnvironment{appendices}{\crefalias{section}{appendix}\crefalias{subsection}{appendix}}

\usepackage{newtxtext,newtxmath}
\usepackage[shortcuts]{extdash}


\usepackage{tabularx}
\usepackage{makecell}
\renewcommand{\footnoterule}{%
	\kern -3pt
	\hrule width \linewidth height 1pt
	\kern 2pt
}

\newcommand{\captionnotes}[1]{{\itshape Notes}: #1}
\newcommand{\pcaption}[1]{\caption{\protecting{#1}}}

\newsavebox\commabox
\sbox{\commabox}{,}
\newlength\commalength
\newcommand{\commasup}{,\kern-\commalength\textsuperscript}

\newcommand{\GRAPPA}{%
	Gravitation Astroparticle Physics Amsterdam (GRAPPA),\\
	Institute for Theoretical Physics Amsterdam
	and Delta Institute for Theoretical Physics,\\
	University of Amsterdam, Science Park 904, 1098 XH Amsterdam, The Netherlands}
\newcommand{\SISSA}{%
	SISSA (Scuola Internazionale Superiore di Studi Avanzati)\\
	via Bonomea 265, I-34136 Trieste, Italy.}

\title[VIGP Strong Lensing]{Strong-lensing source reconstruction with variationally optimised Gaussian processes}

\author[Karchev et al.]{
	Konstantin Karchev\commasup{1,2}%
	\thanks{E-mail:
		\href{mailto:kkarchev@sissa.it}{kkarchev@sissa.it},
		\href{mailto:a.m.coogan@uva.nl}{a.m.coogan@uva.nl},
		\href{mailto:c.weniger@uva.nl}{c.weniger@uva.nl}
	}
	Adam Coogan\commasup{1}\textsuperscript{\color{blue}$\star$}
	and Christoph Weniger\textsuperscript{1}\textsuperscript{\color{blue}$\star$}
	\\
	\textsuperscript{1}\GRAPPA
	\\
	\textsuperscript{2}\SISSA
}

\pubyear{\the\year}

\begin{document}
\label{firstpage} 
\pagerange{\pageref{firstpage}--\pageref{lastpage}}

\maketitle

\begin{abstract}
    Strong-lensing images provide a wealth of information both about the magnified source and about the dark matter distribution in the lens. Precision analyses of these images can be used to constrain the nature of dark matter. However, this requires high-fidelity image reconstructions and careful treatment of the uncertainties of both lens mass distribution and source light, which are typically difficult to quantify. In anticipation of future high-resolution datasets, in this work we leverage a range of recent developments in machine learning to develop a new Bayesian strong-lensing image analysis pipeline. Its highlights are: (A)~a fast, GPU-enabled, end-to-end differentiable strong-lensing image simulator; (B)~a new, statistically principled source model based on a computationally highly efficient approximation to Gaussian processes that also takes into account pixellation; and (C)~a scalable variational inference framework that enables simultaneously deriving posteriors for tens of thousands of lens and source parameters and optimising hyperparameters via stochastic gradient descent. Besides efficient and accurate parameter estimation and lens model uncertainty quantification, the main aim of the pipeline is the generation of training data for targeted simulation-based inference of dark matter substructure, which we will exploit in a companion paper.
	\bigskip
\end{abstract}

\makeatletter
\@thanks
\makeatother

\section{Introduction}\label{section:intro}

\paragraph*{Gravitational lensing} has found application in a wide range of cosmological problems. As can be inferred from the name, it is of great aid in observations of the most distant objects in the Universe like the once most distant known galaxy EGSY8p7 \citep{egsy8p7}. It is an important component in the analysis of the cosmic microwave background (CMB) \citep{Planck-Collaboration_2018} and, perhaps, the only alternative to kinematics for measuring the mass of distant galaxies and clusters (proposed first by \citet{Zwicky_1937}). At the same time the effect known as microlensing has been used to detect objects as small as planets \citep{Tsapras_2018} and to look for mountain-mass primordial black holes \citep{Niikura_2019}. Finally, the time delay induced by lensing is being used to give a straightforward and independent of local and CMB measurements estimate of the Hubble constant \citep{holicow}.

\paragraph*{Dark matter} (DM) is a well established hypothetical component of the Universe that interacts with particles of the Standard Model predominantly gravitationally and only very weakly (if at all) through other means \citep{Bertone_2018}. Of particular interest in constraining models of warm dark matter is the distribution of DM structures on sub-galactic scales. In that context gravitational lensing is a powerful tool since it directly encodes the gravitational field of DM (which is its defining characteristic) into signatures in electromagnetic images.

Naturally, the objects of interest in substructure studies are galaxy-scale lenses, preferentially in one of two lens-source configurations: multiply (quadruply) imaged point sources and (again multiply imaged) extended arcs or Einstein rings (occurring near critical lines). In the former case, the ratios between fluxes of the source projections are used, together with strong assumptions about a \enquote{macroscopic} (smooth) lens mass distribution, to rate the \emph{marginalised statistical likelihood} of the presence of substructure with a given low-mass cutoff. Recent studies using this method \citep{Gilman_2020, Hsueh_2020} derive an upper limit $\hmmass \lesssim \SI{e7.5}{\Msun}$, roughly corresponding to constraints from other methods. On the other hand, techniques analysing extended sources aim, usually, at localising and characterising \emph{individual subhaloes}, whose (non\=/)detection allows conclusions on the mass function to be drawn based on the statistical power of the used technique \citep{Vegetti_2010, Vegetti_2012, Hezaveh_2016, Ritondale_2019}. An alternative is to add a \enquote{correction} potential on top of a smooth extended lens and from it extract information about the DM model \citep{Birrer_2017, Bayer_2018}. These studies employ a \enquote{classical} approach to statistical testing by utilising Monte Carlo (MC) simulations and comparing their outcome (possibly through the use of a test statistic) to the observation. Although statistically straightforward and robust to correlations, this approach is extremely time-consuming, necessitating the exploration of a very high-dimensional parameter space, especially if any sort of \enquote{pixellated} (i.e.\ highly resolved) lens or source model is used.

\paragraph*{Strong-lensing source reconstruction.} The main endeavour in modelling strong gravitational lenses is disentangling the effects of the lens from surface brightness inhomogeneities of the source. The two are largely (and, without priors, completely) degenerate. The degeneracy can be broken if multiple projections of the same source are identified. One can then expect that a macroscopic mass distribution accounts for mapping the multiple images to the same location in the source plane, while smaller dark matter structures only introduce local variations in one of the projections.  

Existing methods fall roughly into four categories: (i)~low-dimensional parametric models, (ii)~pixellation of the source plane and bi-linear regularisation, (iii)~source modelling through basis functions attached to the source plane, and (iv)~source regularisation through neural networks. The computationally fastest approach is to use simple parametric functions, including Gaussian and \Sersic profiles \citep[e.g.][]{Bussmann2013-uv, Spilker2016-iu}. In that case, the small number of parameters can be analysed using standard sampling techniques. This approach is, however, not suitable for capturing the complex light distribution of galaxies observed with high angular resolution.

In order to account for the complexity of observed galaxies, pixellation-based techniques treat the source as an unknown image to be reconstructed together with the lens parameters. Even though in the majority of cases the reconstruction of the source light from the observation, under fixed lens parameters, is a linear problem \citep{Warren2003-up}, the source image still needs to be derived, e.g.\ via conjugate gradient methods, at each step of a non-linear (MC) exploration of the lens parameters. In this semi-linear approach the source reconstruction is typically ill-defined, and its intensity needs to be regularised. This is commonly done using bilinear penalty terms in the likelihood function that restrict the source intensity, its gradient, and/or its curvature. The strength of the regularisation affects the reconstructed source image and is determined by hyperparameters. Optimal hyperparameters that maximise the marginal likelihood of a given image can be determined using Bayesian methods, as proposed by \citet{Suyu2006-lj}.

It was quickly realised that a simple Cartesian pixellation of the source plane is not optimal, since it does not account for variations of the source magnification \citep{Dye2005-es}. Pixellation of the source plane using Delaunay triangulation based on the projected image pixel positions was proposed by \citet{Vegetti_2009} and is now used in many works. This automatically increases the number of pixels in regions of the source that are magnified by the lens. However, there is no agreed-on optimal choice for source plane pixellation, and reconstruction results depend in general on the adopted regularisation and pixellation choices. The problem of selecting the right source pixellation and regularisation hyperparameters was recently studied and newly addressed in the automated approach of \citet{Nightingale2018-vz}, which employs 14 (!) hyperparameters that are all non-linearly optimised through nested sampling techniques.

Other approaches that address the problem of source regularisation make use of basis sets like shapelets, defined in the source plane \citep{Birrer2015-xb, Birrer2018-rq}. The number of regularisation parameters is reduced in this case, and, instead, choices about the maximum order of shapelet coefficients have to be made. Recently, source regularisation using neural networks trained on observed galaxy images has been proposed, based on recurrent inference machines \citep{Morningstar_2019} and variational auto-encoders \citep{Chianese_2020}. Although those approaches bake prior choices that might be difficult to control into the regularisation scheme/source prior, they have the advantage of removing the necessity of hyperparameter tuning for source regularisation.

\paragraph*{Computer science developments.} The recent widespread success of deep learning is due in large part to the advent of automatic differentiation (AD) frameworks that enable efficient gradient-based optimisation of large numbers of network weights; see \citet{Baydin2015-qw} for an overview. \emph{Differentiable programming} provides an abstraction layer for programming languages that treats programs---whether neural networks or not---as flexible building blocks for potential solutions, which are then optimised using gradient descent. Usually, AD is implemented through libraries and extensions of common programming languages like \python or \texttt{Julia}. Nevertheless, stand-alone implementations exist as well: for example, \texttt{DiffTaichi} \citep{Hu2019-xc} is an independent new differentiable programming language tailored for building physics simulators.

Physics simulators that incorporate AD can be seamlessly coupled to neural networks and thus benefit directly from advances in deep learning \citep{Cranmer2019-gj}. Still, the use of AD in the physical sciences is in its infancy: examples can be found in computational quantum physics \citep{Torlai2020-qg}, while \citet{Baydin2020-df} discusses the advantages and challenges of using AD in the context of high-energy physics. In a previous work \citep{Chianese_2020} we presented the first auto-differentiable strong-lensing image analysis pipeline.

\subparagraph{Probabilistic programming} elevates probability densities and random sampling to first-class constructs of the programming language. In this paradigm, probability density estimation (required for inference) and sampling (required for simulation) are usually done with the \emph{same} code. This enables practitioners to focus on writing simulator code, and the subsequent statistical analysis is then performed automatically. Various probabilistic programming languages (PPLs) exist, often as libraries and extensions, e.g.\ \texttt{Stan} \citep{STAN} and \pyro \citep{pyro}. The latter integrates with AD and deep learning frameworks and thus combines the advantages of deep learning, AD, and variational inference.

\paragraph*{Variational inference} (VI) allows the approximation of extremely high-dimensional Bayesian posteriors with simple proposal distributions by solving an optimisation problem \citep[for a review see][]{Zhang2017-lt}. The idea of VI is not new, but an efficient implementation requires a differentiable simulator. Accordingly, it regained traction in the context of deep neural networks when \citet{vae} introduced the very well-known variational auto-encoders (VAEs), which are now increasingly used also in physics both for parameter inference and as generative models \citep[e.g.][]{Green2020-it, Otten2019-lt}. With the advent of powerful general AD frameworks that can differentiate through physics simulators, VI has the potential to drastically improve the analysis of high-dimensional (in particular, image) data. It allows one, often with little overhead above simple maximum a posteriori estimation, to derive meaningful posteriors for thousands or millions of parameters in situations where sampling methods would utterly fail.

In the case of strong-lensing image analysis with pixellated sources, the number of parameters can be extremely large, $\order{\num{e5}}$ and more. Due to the non-linearities introduced by lensing, marginal likelihoods are intractable and require approximate inference. The commonly adopted sampling techniques do not scale well to extremely large numbers of parameters, and indeed existing pixellated source reconstruction pipelines only derive MAP estimators for source parameters. VI enables us to go significantly beyond this limitation.

\paragraph*{Gaussian processes} (GPs) are excellent non-parametric models for uncertain functional dependencies \citep{Murphy2021-yo}. While the source regularisation schemes discussed above are difficult to interpret and control since the various penalty terms and their hyperparameters define the correlation structure only implicitly, in GP models it is directly expressed through the covariance kernel, which effectively connects flux, gradient, curvature, and higher mode regularisation. In the astronomy and physics communities GPs have been used for instance in the contexts of component separation for \SI{21}{\centi\metre} lines \citep{Mertens2017-lu}, Ly~$\alpha$ absorber detection \citep{Garnett2016-ji}, light-curve classification \citep{staccato}, particle physics bump hunting \citep{Frate2017-ut}, radio image analysis \citep{Arras2020-yt}, and time series analyses \citep{Foreman-Mackey2017-zf}.

Numerous machine learning libraries for GP regression exist. However, they usually focus on stationary kernels and simple correlation structures, which makes them unsuitable for modelling the correlations in strong-lensing images.

\medskip

\paragraph*{In this paper} we present a fully end-to-end differentiable GPU-accelerated strong-lensing image analysis pipeline embedded in the probabilistic programming language \pyro. We propose a new method for modelling the sources in strong-lensing images inspired by GP regression. However, we formulate it as an equivalent mixture model, which allows us to replace time-consuming GP optimisation with fast approximate VI. Our approach accounts both for correlations induced by overlapping pixels in the source plane and for the intrinsic correlation structure of the source flux. The resulting pipeline enables us to quickly infer and constrain the parameters of the main lens, as well as the amount of intrinsic source inhomogeneity at various length scales, and obtain estimates for the uncertainties of the reconstructed source. We believe that the proposed methods can be beneficial for a large variety of image analysis problems in astronomy. Yet, our main motivation is to use the approximate posterior predictive distribution of the fitted model to draw examples for the targeted training of inference networks for dark matter substructure. We briefly demonstrated this approach in \citet{Coogan:2020yux} and will present a more thorough exposition in a companion paper \citep{upcoming}.

The rest of this paper is structured as follows. The overall framework is introduced, and a brief description of strong gravitational lensing is given in \cref{section:lensing}. The source model is elaborated and related to Gaussian process regression in \cref{section:source}. Variational inference and variational GP optimisation are discussed in \cref{section:elbo,section:vigp}. Finally, the application of the framework to mock observations is presented and discussed in \cref{section:mock-data-analysis}, with more examples shown in \cref{apx:fit-gallery}.

\section{Overview of the model and gravitational lensing}\label{section:lensing}

\begin{figure}
	\centering
	\tikzsetnextfilename{\currfilebase}%
\begin{tikzpicture}[
	semithick,
	every label/.append style={font=\small,},
	pale group/.style={draw, dotted, subgraph text none},
	lens params/.style={pale group,
		prefix after command={\pgfextra{\tikzset{every label/.append style={
		label position=north east, anchor=south east}}}}}
]
	\graph[no layout] {
		// [layered layout, grow=down, component sep=7em, components go down absolute left aligned] {
			LENSparams/"$\params{lens}$" [label={[rotate=90, anchor=south]left:lens model}] // [tree layout, components go right top aligned, component sep=1em, nodes={param}] {
				SPLEparams [lens params, "SPLE"]
				// [tree layout, component sep=0.5em] {"$\comp{\ximg}_{0, x}$", "$\comp{\ximg}_{0, y}$", "\ellangle", "\ellq", "\splerein", "\spleslope"};
				EXTSHEARparams [lens params, "ext. shear"]
				// [tree layout, component sep=0.5em] {"$\extshear_1$", "$\extshear_2$"}
			};
			P/"\P" [input] -!- disp/"$\variable{\disp}$" [calc] -> p/"$\p$" [calc] -> varKernel/"\varKernel" [calc, inner sep=1pt, label={left:pixellation}];
			{
				GPparams/"\hparamsGP" [not target, subgraph text top=text centered, graphnode]
				// {ovar/"\ovar" [param], kernelsize/"\kernelsize" [input, save path=\kernelsizepath]};
				kernelsize -> T/"\T" [calc],
				ovar
				-- prior/"$\Normal(\variable{0}, \ovar^2\varmatrix{\Identity})$" [dist]
				-> y/"\y" [param]
			} -> f/"\flux" [calc] -- { [same layer]
				likelihood/"$\Normal(\flux, \dataerr^2 \Identity)$"
					[dist, label={left:instrument}]
				-!- data/"\anydata"
					[output, sibling pre sep=1em, label={right:observation}]
			};
			{[same layer] P [sibling post sep=2em], disp, kernelsize, ovar
				-!- GUIDE/"$\proposal_{\gparam}(\param, \hparam)$"
					[sibling pre sep=3em, guide, label={[align=right, anchor=south east, name={vp label}]north east:variational\\posterior}]};
			{[same layer] varKernel -!- T [sibling pre sep=2em], y};
		};
		LENSparams -> [tail anchor=south, to path={-| (\tikztotarget)}] disp;
		P -> {disp, p};
		likelihood -> data;
		{p, varKernel} -> T;
		
		GUIDE -> ovar;
		GUIDE -> [to path={|- (\tikztotarget)}] y;
		
		"layers plate" [subgraph text none, draw, label={[align=center]Gaussian process\\source model}] // {GPparams, prior, y, T, f};
		"lensing physics" [pale group, label={[anchor=south west]north west:ray tracing}] // {P, disp, p};
	};

	\path[name path=isect horiz] (LENSparams.south east) -- (LENSparams.south west);
	\path[name path=isect diag] (LENSparams) -- (vp label);
	\draw[name intersections={of={isect horiz and isect diag}}, guide edge, ->] (vp label) -- (intersection-1);

	\node at (layers plate.south east) [above left] {$\nlayers$};

	\begin{scope}[on background layer]
		\fill[filled, even odd rule] let \p1=(kernelsize.north), \p2=(kernelsize), \n1={\y2-\y1} in (kernelsize) circle (\n1) (GPparams.north west) rectangle (GPparams.south east);
		\drawnode[filled]{LENSparams}
	\end{scope}
\end{tikzpicture}%
	\pcaption{Graphical representation of the model. Nodes without borders (only $\P$) are (non-stochastic) input, while the shaded circular node ($\anydata$) is the observed data. Parameters (collectively labelled $\param$, or $\hparam$ for \emph{hyper}parameters, and depicted as shaded boxes) are inferred variationally using a proposal posterior ($\proposal$, see \cref{section:guide}) and have constant priors determined outside the model (except $\y$, as indicated). In this work we fix the kernel size hyperparameter $\kernelsize$, so it is depicted as an input. The source model, represented as a plate, is replicated (independently) $\nlayers$ times, with the outputs of those \enquote{layers} then summed to produce the \enquote{true} modelled flux ($\flux$). See \cref{tab:graph} for descriptions of the variables and \cref{tab:gparams} for the parameters of the proposal distribution. \\ \captionnotes{The solid lines indicate deterministic relations, except when they \enquote{flow} through a distribution (for $\y$ and $\flux$). The direction of the arrows indicates the order of operations in a forward (generative) pass through the model.}\label{fig:graph}}
\end{figure}

\begin{table}
	\centering
	\pcaption{Summary of the variables involved in the model. We use small bold letters to denote \enquote{arrays} (ordered sets) and an arrow to indicate spatial vectors. A bold variable with an arrow indicates an array of vectors, and indexing it returns, naturally, a single vector. Bold capital letters are used for matrices operating on arrays, and a double underscore denotes a matrix that operates on spatial vectors. The variable $\varKernel$ is an array of spatial matrices.\label{tab:graph}}
	\begin{tabularx}{\linewidth}{c X}
		\toprule
		$\ximg_0$, \ellangle, \ellq & main lens location, position angle, and axis ratio\\
		\splerein, \spleslope & Einstein radius and power law slope of SPLE main lens \\
		$\extshear_1$, $\extshear_2$ & external shear components \\
		\P, \p & \multirow{2}{\linewidth}{pixel centre positions in the image- and source-plane and the lensing displacement field relating them through \cref{eqn:lens}} \\
		\disp & \\
		\varKernel & source-plane pixel shape covariance matrices (\cref{section:windowing}) \\
		\T & \enquote{transmission matrix} (\cref{section:gp-gen,section:gp-factorisation}) \\
		\kernelsize, \ovar & hyperparameters of source: kernel size and (root) variance \\
		\y & \enquote{source parameters} of GP (\cref{section:gp-gen}) \\
		\flux & true \enquote{flux} (predicted by the source) \\
		\data, \dataerr & observed data and instrumental uncertainty \\
		\bottomrule
	\end{tabularx}
\end{table}

We model observational data (a single-channel telescope image) that represents surface brightness\footnote{technically called \enquote{radiance}. Since we are also given a noise level for the same quantity to act as scale, the data could also represent radiant intensity. We will, however, call it simply \enquote{flux}.} using a pipeline of three components: a source, which will be discussed in the next \namecref{section:source}, a lens, and an instrument (i.e.\ telescope). In this work we model the instrumental effects as simply imbuing the observation with Gaussian noise $\dataerr$ that is constant across all pixels and independent for each of them, but the framework can seamlessly be extended to account for a point-spread function (PSF), a varying observational uncertainty, and for correlations between different colour channels in an image, all of which will be important for analysing telescope data. We present a graphical overview of the model in \cref{fig:graph} and summarise the involved quantities in \cref{tab:graph}.

\subsection{The physics of strong lensing}\label{section:lensing-theory}

\begin{figure}
	\centering
 	\usetikzlibrary{decorations.text}
\pgfmathparse{scalar(\linewidth / 2.5cm)}
\edef\lensingscale{\pgfmathresult}

\tikzsetnextfilename{\currfilebase}%
\begin{tikzpicture}[scale=\lensingscale]
	\coordinate (O) at (0, 0);
	\coordinate (Oimg) at (1, 0); \coordinate (Osrc) at (2, 0);
	
	\coordinate (ximg) at ($(Oimg) + (-0.15, 0.2)$);
	\coordinate (x at img) at ($(Oimg) + (0.1, 0.08)$);
	
	\coordinate (ximg at src) at ($(O)!2!(ximg)$);
	\coordinate (x) at ($(O)!2!(x at img)$);
	
	\coordinate (mass) at ($(ximg)!1.4!(x at img)$);
	
	\newcommand{\plane}[6][0.8]{
		\path[draw, postaction={decorate}, fill=white, draw opacity=1, fill opacity=#1] (#2, #3) -- (#2, #5) -- (#4, #5) -- (#4, #3) -- cycle;
		\path[
			decorate,
			decoration={
				raise=0.5pt,
				reverse path,
				text effects along path, text={#6},
				text align=center,
			},
			text effects={
				characters={anchor=base},
				character widths={inner xsep=0pt}
			},
		] (#2, #5) -- (#4, #5);
	}
	
	\begin{scope}[
		x={(1, -0.25)}, y={(0, 1)}, >={Stealth[round,sep=2pt]},
		oax/.style={dashed},
		ray/.style={ultra thick},
		inv/.style={dotted}
	]
		\begin{scope}[shift={(Osrc)}]
			\plane{0.3}{-0.1}{-0.5}{0.4}{source plane}
			
			\tkzDrawPoint(Osrc);
			\tkzDrawPoint(x); \tkzDrawPoint(ximg at src);
			
			\node[below] at (x) {$\x*$};
			\draw[->] (x) -- (ximg at src) node[midway, above right=-3pt and -1.5pt] {$\disp^{\prime}$};
			
			\draw[oax] (Oimg) -- (Osrc);
			\draw[ray] (ximg) -- (x);
			\draw[inv] (x at img) -- (x) (ximg) -- (ximg at src);
		\end{scope}
		
		\begin{scope}[shift={(Oimg)}]
			\plane{0.3}{-0.1}{-0.4}{0.35}{image plane}
			
			\tkzDrawPoint(Oimg);
			\tkzDrawPoint(ximg); \tkzDrawPoint(x at img);
			\node[above] at (ximg) {$\ximg$};
			\draw[->] (x at img) -- (ximg) node[midway, above right=-3pt and -1.5pt] {$\disp$};
			
			\draw[oax] (O) -- (Oimg);
			\draw[ray] (O) -- (ximg);
			\draw[inv] (O) -- (x at img);
			
			\draw[fill=black!80, decorate, decoration={random steps,segment length=0.3pt,amplitude=0.2pt}] (mass) circle(0.05);
			\node at ($(mass) + (0, +0.1)$) {$\ximg^\prime$};
			\node at ($(mass) + (0, -0.1)$) {$M$};
			\tkzDrawPoint(mass)
		\end{scope}
	\end{scope}
	
	\tkzDrawPoint(O)
	\node at (O) [below] {observer};
 
	\coordinate [below=of O] (Ob); \coordinate [below=1em of Ob] (Obb);
	\coordinate [below=of Oimg] (Oimgb);
	\coordinate [below=of Osrc] (Osrcb); \coordinate [below=1em of Osrcb] (Osrcbb);
	\path[|-|, midway]
		(Ob) edge ["$\dl$", above] (Oimgb)
		(Oimgb) edge ["$\dls$", above] (Osrcb)
		(Obb) edge node[below] {$\ds$} (Osrcbb);
\end{tikzpicture}%
	\pcaption{Thin lens geometry. The lensing mass $M$ located at $\ximg^\prime$ bends the light ray (thick line) emanating from the point $\x$, so that to the observer it looks like it is coming from the direction of $\ximg$. General relativity predicts the deflection angle $\disp^\prime$ \emph{as viewed from the image plane} based on the mass $M$ and the impact parameter $\vector{b} = \ximg - \ximg^\prime$. It then has to be rescaled by $\dls / \ds$ to obtain the displacement field $\disp$ (as viewed by the observer) for use in \cref{eqn:lens}. Angles are all assumed to be small enough that they can be used for Euclidean calculations. The dashed line is the optical axis perpendicular to the planes and connects the origins of the coordinate systems for each plane.\label{fig:lensing}}
\end{figure}

The general theory of relativity (GR) allows for solving for the propagation of light through arbitrary spacetimes, but this can, accordingly, be arbitrarily difficult. Usually, therefore, when considering \enquote{ordinary} non-extreme systems like galaxies and clusters one adopts a weak-gravity approximation which assumes that mass densities are low enough and that relativistic effects can be ignored, so that the Newtonian approximation to GR can be used; i.e.\ one assumes the metric can be fully described by a scalar gravitational potential $\potfull$. Furthermore, usually there is a clear separation of scales in the system, with the lensing mass being located much further from the observer and the source of light than the size of the gravitating object. This setup lends itself to the \emph{thin lens approximation} in which all the mass is assumed to lie in a single plane (the \emph{image plane}), while all the light is assumed to be coming from a \emph{source plane} (see \cref{fig:lensing}).

We will be using $\x$ and $\ximg = \qty(\comp{\ximg}_x, \comp{\ximg}_y)$ as general angular coordinates in the source and image plane, respectively, labelling with $z$ the coordinate perpendicular to the planes. Since the deflections involved in a weak gravity setting are small, we can use the angular measures as a Cartesian coordinate system in the two planes.

The configuration, then, is determined by two fields: the distribution of surface brightness in the source plane, $\sbr(\x)$, and the distribution of mass in the image plane, described by the projected potential:
\begin{equation}
	\pot(\ximg) \equiv \int\limits_{-\infty}^{\infty} \potfull(\comp{\ximg}_x, \comp{\ximg}_y, z) \dd{z}.
\end{equation}

Under these assumptions, the image-plane and source-plane coordinates of a light ray are related by the simple \emph{raytracing equation} \citep{lensing}:
\begin{equation}\label{eqn:lens}
	\x = \ximg - \disp\qty(\ximg).
\end{equation}
The (reduced) displacement field $\disp$ is determined by the projected potential:\footnote{The factor of two here is the only difference between the Newtonian treatment of centuries past and GR.}
\begin{equation}\label{eqn:disp-pot}
	\disp(\ximg) = \frac{\dls}{\ds} \frac{2}{\speedoflight^2} \frac{\vectgrad[\ximg] \pot(\ximg)}{\dl},
\end{equation}
where the gradient is taken in the image plane and should have dimensions of inverse length, whence the introduction of the observer--lens distance\footnote{All distances considered in lensing are, naturally, \emph{angular diameter distances}. We use the routines from the \astropy package \citep{astropy_2013, astropy_2018} and the flat cosmology implied by \citet{Planck2015} to calculate them.} $\dl$. The additional factor $\dls/\ds$ is geometric (as illustrated in \cref{fig:lensing}) and motivates the \enquote{reduced} qualification.

From a computational standpoint it is more convenient to represent \cref{eqn:disp-pot} as an integral. This is accomplished through the use of the Poisson equation $\vectlap \potfull = 4\pi\gravG\rho$ and setting appropriate boundary conditions \citep{lensing}:
\begin{equation}\label{eqn:disp-sdens}
	\disp(\ximg) = \frac{\dls}{\ds} \frac{4\gravG}{\speedoflight^2 \dl} \int \sdens(\ximg^\prime) \frac{\ximg - \ximg^\prime}{\abs{\ximg-\ximg^\prime}^2} \dd[2](\dl \ximg^\prime),
\end{equation}
where the integral is performed over the image plane. Here $\sdens$ is the projected (surface) mass density related to the 3D mass density by an integration along $z$:
\begin{equation}
	\sdens(\ximg) \equiv \int_{-\infty}^{\infty} \rho(\comp{\ximg}_x, \comp{\ximg}_y, z) \dd{z},
\end{equation}
similarly to the expression for the projected potential.

In fact, \cref{eqn:disp-sdens} can be interpreted as a properly rescaled sum of the well-known contribution $4 \gravG M / b \speedoflight^2$ (which \citet{Dyson_1920} confirmed) from infinitesimal point masses $M = \sdens \dd[2](\dl\ximg^\prime)$ to the deflection of a ray with impact parameter $\vector{b} = \ximg-\ximg^\prime$. This is an expression of the linearity of thin weak gravity\footnote{The weak gravity regime is not to be confused with weak lensing. Even in the Newtonian limit (weak gravity) the deflections can be large in comparison to the system size (strong lensing).} lensing, which allows the lensing effects of arbitrary mass distributions to be trivially superposed by summing their respective contributions to the displacement field.

The expression in \cref{eqn:disp-sdens} is simplified by introducing the \emph{critical surface density} for lensing and the closely related \emph{convergence}:
\begin{equation}
	\label{eqn:sdenscrit}
	\sdenscrit \equiv \frac{\speedoflight^2}{4\pi\gravG} \frac{\ds}{\dl\dls},
	\qquad \convergence(\ximg) = \frac{\sdens(\ximg)}{\sdenscrit}.
\end{equation}
The former is a bound above which a lens can form multiple images. It only depends on the relative strength of gravity ($\gravG$) to the speed of light ($\speedoflight$) and dictates that further sources are more easily lensed (for a fixed distance to the lens) since the deflections required are smaller. The multiple images condition is equivalently stated as $\convergence > 1$ (see \citet[section 2.6]{lensing} for discussion and illustration).

The lens equation can be viewed as a coordinate transformation and then conveniently characterised through its Jacobian, which can be shown to take the form
\begin{eq}\label{eqn:lens-jacobian}
	\dv{\x}{\ximg\unsmashhere} = \mqty(\dmat[0]{1-\convergence, 1-\convergence}) - \mqty(\shear_1 & \shear_2 \\ \shear_2 & -\shear_1),
\end{eq}
conveniently split into an isotropic and a shear part. The (inverse) Jacobian defines the magnification
\begin{eq}\label{eqn:magnification}
	\magnification \equiv \vectdet{\dv{\ximg}{\x}} = \qty[(1-\convergence)^2 - \qty(\shear_1^2 + \shear_2^2)]^{-1},
\end{eq}
which can be interpreted as the overall scaling of the area of small patches of the source plane after lensing. The magnification can take any real value, with negative magnification signifying a flipped image. It can also become extremely large (positive or negative) where the argument in brackets is close to zero. The regions where this occurs are called caustics and critical lines in the source and image planes, respectively. The magnitude of magnification differentiates \emph{strong} from \emph{weak} lensing: for the former $\convergence$ and $\shear$ are comparable to (and/or even exceed) unity, and so sources are \emph{strongly} magnified.

It is important to note, however, that lensing does not create or destroy photons (and thus, conserves energy\footnote{at least for stationary (non-rotating) spacetimes. The interested reader is referred to the effect of \emph{superradiance} \citep{superradiance} as an example of extraction of energy by \enquote{very strong gravitational lensing}.}). Thus, the amount of light arriving at the observer is proportional to the area \emph{in the image plane} that the source subtends, which effectively acts as a collecting area for light that is then simply rerouted just as with conventional lenses. This means that \emph{lensing conserves surface brightness} around infinitesimal points, and thus the surface brightness registered by the observer (in the absence of other sources) is
\begin{eq}
	\sbr(\ximg) = \sbr(\x\qty(\ximg)).
\end{eq}

It is obvious from this expression that the lens and the source are entirely degenerate: for any observation there is for every deflection field (that does not form multiple images!) a surface brightness configuration that can reproduce the observation. The reverse (that one can find deflection fields that map distinct sources to the same image) is also sometimes true, giving rise to the mass-sheet \citep{masssheetdegeneracy} and source-position degeneracies \citep{sourcepositiontransformation} (the latter being more general but approximate). As a consequence, the spatial scale of the source cannot be conclusively determined (in the absence of time-delay data) since a sheet of constant mass density could be magnifying it uniformly by an arbitrary amount \citep{lensing-short}.

\subsection{Lens model}\label{section:lensing-model}

Our framework admits the use of an arbitrary combination of mass distributions as the lens model due to the aforementioned linearity of thin weak gravity lensing. Ultimately, we are interested in investigating the statistical properties of the overall distribution on different size (and mass) scales. As a first step, it is usual to split the lensing mass into a macroscopic smooth component, which is constrained by the need to converge the multiple observed images of the source onto the same location in the source plane, and a \enquote{substructure} component. For simplicity, at present we do not consider substructure and defer all investigation into and discussion about it to upcoming works.

Almost exclusively, the smooth component is realised using an analytical model giving the displacement as a (relatively) easily computable expression. While in computation-driven studies the singular isothermal sphere or ellipsoid (SIS/SIE) model seems to dominate due to its simplicity \citep{Hezaveh_2017, Brehmer_2019}, analyses of observational data \citep[e.g.][]{SLACS-5} require further flexibility and hence often resort to the singular power law ellipsoid (SPLE) model, which is an extension of the SIS/SIE. Furthermore, the SPLE has been shown to be a more adequate representation of the combined DM and baryon mass distribution in the inner regions of galaxies, to which strong lensing is most sensitive \citep{Suyu:2008zp}.

A (spherical) power law lens is parametrised by its 3D density exponent $\spleslope$ (or \emph{slope}) and an Einstein radius $\splerein$. In terms of a general (2D) radial coordinate $\ellR(\ximg)$ its surface density is:
\begin{equation}
	\frac{\sdens(\ximg)}{\sdenscrit} = \convergence(\ximg) = \frac{3-\spleslope}{2} \qty(\frac{\splerein}{\ellR(\ximg)\unsmashhere})^{\spleslope - 1}.
\end{equation}
An ellipticity can be induced by first transforming from Cartesian, $\qty(\comp{\ximg}_{x}, \comp{\ximg}_{y})$, to elliptical coordinates, $\qty(\ellR, \ellarg)$. The transformation is controlled by a scale factor $\ellq$, and, for generality, a rotation through an angle $\ellangle$, and an overall translation, $\ximg_0$:
\begin{align}
	\mqty(\ellR_x \\ \ellR_y) & = \mqty(\dmat[0]{\sqrt{\ellq}, 1 / \sqrt{\ellq}}) \mqty(\cos{\ellangle} & \sin{\ellangle} \\ -\sin{\ellangle} & \cos{\ellangle}) \mqty(\comp{\ximg}_x - \comp{\ximg}_{0, x} \\ \comp{\ximg}_y - \comp{\ximg}_{0, y})
	\\ \tan{\ellarg} & = \ellR_y / \ellR_x.
\end{align}

The SPLE displacement field has a closed form expression in the complex notation, in which $\comp{\disp} = \comp{\disp}_x + \i \comp{\disp}_y$, \citep{ORiordan_2020, Tessore_2015}:
\begin{eq}\label{eqn:dispsple}
	\comp{\disp}^{\text{SPLE}}\qty(R, \ellarg)
	= &\ \splerein \frac{2 \sqrt{\ellq}}{1+\ellq} \qty(\frac{\splerein}{\ellR(\ximg)\unsmashhere})^{\spleslope-2}
	\\ & \times \prescript{}{2}{F_1}\qty(1, \tfrac{\spleslope-1}{2}; \tfrac{5-\spleslope}{2}; -\tfrac{1-\ellq}{1+\ellq} \e^{2\i\ellarg}) \e^{\i \ellarg}.
\end{eq}
Here $\ellarg$ is the elliptical angular coordinate (and not the ellipse orientation $\ellangle$!), and $\prescript{}{2}{F_1}$ is the hypergeometric function\footnote{See \href{https://mathworld.wolfram.com/HypergeometricFunction.html}{https://mathworld.wolfram.com/HypergeometricFunction.html}}. In the circular ($\ellq=1$) isothermal ($\spleslope=2$) case this reduces to $\comp{\disp} = \splerein \e^{\i\ellarg}$, i.e.\ is an isotropic constant, as is well-known. The function $\prescript{}{2}{F_1}$, however, is not implemented in \torch\footnote{as of version 1.8}, let alone in an auto-differentiable fashion, so we instead use the procedure described in full in \citet[Appendix A]{Chianese_2020}, which relies on pre-tabulating the angular part of \cref{eqn:dispsple} (through an alternative formulation) and interpolating at runtime. This results in very fast code with negligible output degradation.

Additionally, we include a displacement field component due to an externally caused shear:
\begin{equation}
	\disp^{\text{ext. shear}}(\ximg) = \mqty(\extshear_1 & \extshear_2 \\ \extshear_2 & -\extshear_1)\, \ximg
\end{equation}
in analogy with \cref{eqn:lens-jacobian}. It represents the combined weak lensing effect of all unmodelled lenses along the light path (assumed constant across the image). As discussed in \cref{section:lensing-theory}, an overall magnification is degenerate with scaling the source plane, so it is not modelled explicitly.

Overall, the lens model has eight trainable parameters that we collect in a vector $\params{lens}$: $\qty{\comp{\ximg}_{0, x}, \comp{\ximg}_{0, y}, \ellq, \ellangle, \splerein, \spleslope}$ from the SPLE and $\qty{\extshear_1, \extshear_2}$ for external shear.

\section{Modelling the source light}\label{section:source}

The goal of this section is to describe a source model flexible enough to enable fits to the lensed image at the level of image noise but also statistically rigorous enough to not overfit in areas where the observed details could be due to lensing substructure. Inspired by common non-parametric Gaussian process (GP) regression, whose usual application is precisely statistically justified interpolation of arbitrary data, and motivated by the need to avoid costly operations like matrix inversions, which would preclude application to future high-resolution datasets, we replace the usual regression strategy with optimisation via variational inference. Furthermore, because of the peculiarities of the lensing transformation and the possible overlap between non-adjacent pixels when projected into the source plane, we need to introduce a method for modelling the thus-induced correlations, which goes well beyond typical applications of GP. Still, in certain limits, our model does reduce to conventional GP regression. We expect that in the field of gravitational lensing it outperforms existing methods in terms of accuracy, precision, speed and flexibility.

\subsection{Overview of Gaussian processes}

From a mathematical perspective a (Gaussian or not) \emph{process} is a distribution over functions. In the simplest case, the functions may be the possible surface brightness profiles of the source galaxy (but see \cref{section:windowing} for how this is modified by pixellation). The assumption of Gaussianity then directly provides the likelihood of observing some data $\data$ with observational covariance\footnote{
	As the name suggests, $\datavar$ is only determined by instrumental effects, which in the present work we restrict to simple uncorrelated noise, i.e.\ $\datavar \rightarrow \dataerr^2 \varmatrix{\Identity}$. We will, however, where possible, keep to the general notation so as to include the possibility of, e.g.\ a point-spread function.
} $\datavar$, marginalised over the possible \enquote{true} values of the flux \citep[Chapters~2~\&~5]{gp}:
\begin{equation}\label{eqn:gp-mll}
	2 \ln \prob(\data \given \hparamsGP) = \contract[\qty(\K_{\hparamsGP} + \datavar)^{-1}]{\data} + \ln\abs{2\pi \qty(\K_{\hparamsGP} + \datavar)}.
\end{equation}
The process is entirely defined\footnote{
	We assume a zero-mean process. A prior belief that the data should have some dependence on position can be encoded into a mean function $\mu(\x)$, which would also require a trivial modification to the marginal likelihood equation: $\data \rightarrow \data - \mu(\p)$.
} by the prior covariance matrix of the true values, $\K$, whose entries are assumed to be given by a covariance function applied to the positions\footnote{
	These are the positions in the source plane as calculated via \cref{eqn:lens}.
} $\p$ associated to the observations:
\begin{equation}
	\varcomp{\K}_{ij} = \cov(\varcomp{\flux}_i, \varcomp{\flux}_j) \equiv \kfunc_{\hparamsGP}(\varcomp{\p}_1, \varcomp{\p}_2).
\end{equation}
With $\hparamsGP$ we indicate the \emph{hyperparameters} that the covariance function (and hence also the evidence in \cref{eqn:gp-mll}) depends on. 

Although a variety of covariance functions can be used\footnote{
	See, e.g., \citet{gp-web} for a demonstration of the effect of different covariance functions.
}, we will restrict ourselves to the most common one, the Gaussian radial basis function (RBF) with hyperparameters $\ovar^2$ for the overall variance and $\Kernel$ as a covariance \emph{kernel}. It is given by
\begin{equation}\label{eqn:grbf}
	\krbf(\x_1, \x_2) = \ovar^2 \exp(- \frac{\contract[\Kernel^{-1}]{\qty(\x_1 - \x_2)}}{2}).
\end{equation}

Often one aims to infer a single kernel best fitting the data. However, in galaxy images there may be a number of different relevant scales, like the scale of the whole galaxy and that of dust lanes and individual stellar clumps, and so a single kernel is usually not optimal for modelling them. Furthermore, one usually can get an estimate of the appropriate kernel size by e.g.\ first fitting a smooth analytical source model. Hence, we model the covariance using a sum of $\nlayers$ Gaussian RBFs as in \cref{eqn:grbf}:
\begin{equation}\label{eqn:grbf-layers}
	\kfunc_{\hparamsGP}(\x_1, \x_2) = \sum_{k=1}^{\nlayers} \layerindex{\krbf}{k}(\x_1, \x_2),
\end{equation}
assigning to each an isotropic kernel $\layerindex{\Kernel}{k} \rightarrow \layerindex{\kernelsize^2}{k} \vectmatrix{\Identity}$, where $\Identity$ is the identity, and an overall variance $\layerindex{\ovar^2}{k}$, which together form the set of GP hyperparameters. Due to the linearity of Gaussian processes, this is equivalent to considering a number of linearly superimposed independent (a priori) GP sources, which we call \emph{layers}.

We do not vary the length scales $\layerindex{\kernelsize}{k}$ but fix them, along with the total number of layers, based on an initial fit so that all relevant correlation scales are covered (see \cref{section:mock-data-analysis}). On the other hand, the variance hyperparameters are optimised variationally by maximising the model evidence (see \cref{section:elbo}). In a multilayer GP they act as weights for the layers and can suppress and enhance the modelled source variations on the relevant scales as determined from the data. In the following we will usually omit the layer subscripts and instead imply that the descriptions apply to each layer individually and independently.

\subsection{Approximate pixellated Gaussian processes}\label{section:windowing}

\begin{figure}
	\centering
	\newcommand{\analysePixel}[5]{
	\foreach \A/\B in {#1/#2,#2/#3,#3/#4,#4/#1} {
		\tkzDefMidPoint(\A,\B) \tkzGetPoint{c\A\B}
	}
	\tkzDefMidPoint(c#1#2,c#3#4) \tkzGetPoint{c#5}
}

\newcommand{\definePoints}{
	\useasboundingbox (-1.5, -1.5) rectangle (2.5, 2);
	
	\tkzDefPoint(-0.5, -1){A1}
	\tkzDefPoint(0, -1){A2}
	\tkzDefPoint(1.5, 0.5){A3}
	\tkzDefPoint(-1, 1.5){A4}
	
	\analysePixel{A1}{A2}{A3}{A4}{A}
	\tkzDrawPolygon[red, thick](A1,A2,A3,A4)
	
	\tkzDefPoint(1, -1){B1}
	\tkzDefPoint(2, -0.5){B2}
	\tkzDefPoint(2, 1.5){B3}
	\tkzDefPoint(-0.5, 1){B4}
	
	\analysePixel{B1}{B2}{B3}{B4}{B}
	\tkzDrawPolygon[blue, thick](B1,B2,B3,B4)
}

\newlength{\windowingpanelwidth}
\setlength{\windowingpanelwidth}{0.32\linewidth}
\pgfmathparse{scalar(\windowingpanelwidth / 4cm)}
\edef\windowingpanelscale{\pgfmathresult}

\begin{subfigure}{\windowingpanelwidth}
	\centering
	\tikzsetnextfilename{windowing-a}%
	\begin{tikzpicture}[scale={\windowingpanelscale}]
		\definePoints
		\begin{scope}[blend group=screen]
			\tkzFillPolygon[red, opacity=0.5](A1,A2,A3,A4)
			\tkzFillPolygon[blue, opacity=0.5](B1,B2,B3,B4)
		\end{scope}
	\end{tikzpicture}%
	\pcaption{}
\end{subfigure}
\hfill
\begin{subfigure}{\windowingpanelwidth}
	\centering
	\tikzsetnextfilename{windowing-b}%
	\begin{tikzpicture}[scale={\windowingpanelscale}]
		\definePoints
		
		\begin{scope}[blend mode=screen]
			\fill[rotate around={172.027:(cA)}, red, opacity=0.5] (cA) ellipse (0.785 and 1.035);
			\fill[rotate around={-148.283:(cB)}, blue, opacity=0.5] (cB) ellipse (0.900 and 1.077);
		\end{scope}
		
		\tkzDrawSegment[red](cA1A2,cA3A4)
		\tkzDrawSegment[red](cA2A3,cA4A1)
		\labelPoint[below left](cA){$O$}
		\labelPoint[below](cA1A2){$A$}
		\labelPoint[above](cA3A4){$A^\prime$}
		\labelPoint[right](cA2A3){$B$}
		\labelPoint[left](cA4A1){$B^\prime$}
		
		\tkzDrawSegment[blue](cB1B2,cB3B4)
		\tkzDrawSegment[blue](cB2B3,cB4B1)
	\end{tikzpicture}%
	\pcaption{}
\end{subfigure}
\hfill
\begin{subfigure}{\windowingpanelwidth}
	\centering
	\tikzsetnextfilename{windowing-c}%
	\begin{tikzpicture}[scale={\windowingpanelscale}]
		\begin{axis}[
			xmin=-1.5, xmax=2.5, ymin=-1.5, ymax=2,
			domain=-1.5:2.5, y domain=-1.5:2, samples=21, shader=interp,
			view={0}{90},
			width=4cm, height=3.5cm, at={(-1.5cm, -1.5cm)},
			scale only axis, hide axis, axis lines=none,
			colormap default colorspace=rgb,
			colormap={red}{color=(\globalwhitecolor) color=(red)},
			colormap={blue}{color=(\globalwhitecolor) color=(blue)},
			point meta min=0, point meta max=1,
		]
			\begin{scope}[blend group=multiply]
				\addplot3[surf, colormap name=red]{
					exp(-2.52814556738587*x^2 + 0.29742889028069*x*y - 1.48714445140345*y^2)
				};
				\addplot3[surf, colormap name=blue]{
					0.0826881584134624*exp(-1.77838335437341*x^2 - 0.523053927756885*x*y + 4.13212602927939*x - 1.51685639049497*y^2 + 1.34686386397398*y)
				};
			\end{scope}
		\end{axis}
		
		\definePoints
	\end{tikzpicture}%
	\pcaption{}
\end{subfigure}
	\pcaption{The projection of two pixels onto the source plane (a). A full treatment should consider their exact shapes and use an indicator function that is nonzero only inside the shaded areas. Instead, we approximate the pixels as ellipses defined by the (projection of) the pixel centres $O_i$ and the vectors $\vector{a} = \vv{AA^\prime}$ and $\vector{b} = \vv{BB^\prime}$, as shown in (b). The indicator functions are Gaussian with covariances derived from those ellipses (see the text for the exact definition) and are depicted in (c). \label{fig:windowing}}
\end{figure}


A particular complication arising in the modelling of multiply lensed systems is that it is not enough to assign the observed data to zero-sized points in the source plane, as this will disregard the correlations arising from the overlap of the light-collecting areas of different pixels (see \cref{fig:windowing}). To fully account for the (projected) pixel shapes and sizes, one must introduce a set of indicator functions $\gpwindow_i(\x)$, each being non-zero only within the source-plane light-collecting area for the respective pixel, so that
\begin{equation}\label{eqn:window-definition}
	\varcomp{\flux}_i \equiv \int\limits_{\mathclap{\text{source plane}}} \dd{\x} \gpwindow_i(\x)\, \sbr(\x),
\end{equation}
where $\varcomp{\flux}_i$ are the noiseless observations (GP true values), and $\sbr(\x)$ is the surface brightness, which we treat as the function being modelled by the Gaussian process. Hence, we assume that the covariance from \cref{eqn:grbf-layers} in fact applies to the underlying surface brightness:
\begin{equation}
	\cov(\beta(\x_1), \beta(\x_2)) = \kfunc_{\hparamsGP}(\x_1, \x_2)
\end{equation}
and consequently modify the covariance matrix of the observed fluxes to take into account the correlations from all pairs of points within the light-collecting areas:
\begin{equation}\label{eqn:windowing}
	\varcomp{\K}_{ij} = \iint \dd{\x_1} \dd{\x_2} \gpwindow_i\qty(\x_1)\, \kfunc(\x_1, \x_2)\, \gpwindow_j\qty(\x_2).
\end{equation}
Note that this covariance is not generally stationary since it is not just a function of the positions $\varcomp{\p}_i$ and $\varcomp{\p}_j$ of the observations.

Calculating this integral exactly is infeasible even if the projected pixels are assumed to be polygonal (which is not necessarily true), since it would involve numerous intersection checks and code branching points, which are extremely inefficient on GPUs. Instead, we opt for a simplified treatment that approximates the indicator functions to Gaussians located at the pixel centres:
\begin{eq}\label{eqn:window-function}
	\gpwindow_i\qty(\x) & = \Gaussian(\x-\varcomp{\p}_i, \varcomp{\varKernel}_i)
	\\ & = \vectdet{2\pi \varcomp{\varKernel}_i}^{-\frac{1}{2}} \exp[-\frac{\contract[\varcomp{\varKernel}_i^{-1}]{\qty(\x - \varcomp{\p}_i)}}{2}].
\end{eq}

The conservation of surface brightness (as described in \cref{apx:windowing-norm}) demands that
\begin{alignat}{2}
	\label{eqn:window-cond1}
	& \int \dd{\x} \gpwindow_i\qty(\x) && = 1,
	\\ \label{eqn:window-cond2}
	& \int \dd{\x} \gpwindow_i^2 \qty(\x) && = A_i^{-1}.
\end{alignat}
The first requirement simply means that the indicator function acts as a \enquote{contribution density} and is satisfied by any properly normalised Gaussian function. The second condition, where $A_i$ is the projected pixel area, or an estimate thereof, demands that the indicator function is similar in size to the pixel and ensures that the variance of a pixel's brightness scales inversely with its light-collecting area. In the same \namecref{apx:windowing-norm} we show that this is satisfied when the \enquote{pixel shape covariance matrix} is
\begin{equation}\label{eqn:window-kernel}
	\varcomp{\varKernel}_i = \frac{\vector{a}_i \vector{a}_i^\transpose + \vector{b}_i \vector{b}_i^\transpose}{4\pi}
\end{equation}
with $\vector{a}$ and $\vector{b}$ the axes of the pixel\footnote{
	In the actual implementation only the pixel centres are projected to the source plane. This makes it impossible to use the midsegments as illustrated. Instead, the axes $\protect\vector{a}$ and $\protect\vector{b}$ are calculated as half the distance between the centres of the two adjacent pixels along the respective axes of the grid.
} (depicted as $\vv{AA^\prime}$ and $\vv{BB^\prime}$ in \cref{fig:windowing}).

The indicator function from \cref{eqn:window-function} allows us to easily solve the integral in \cref{eqn:windowing}. Assuming that the surface brightness $\beta\qty(\x)$ has the Gaussian RBF covariance from \cref{eqn:grbf}, the prior covariance matrix of the data is
\begin{flalign}
	\label{eqn:covariance}
	\begin{aligned}[b]
		\varcomp{\K}_{ij} & = \ovar^2 \sqrt{\vectdet{2\pi \Kernel}}
		\\ & \mathrlap{
			\mkern12mu \times \mkern-6mu\int\mkern-12mu\int
			\! \Gaussian(\x_1-\varcomp{\p}_i, \varcomp{\varKernel}_i)
			\, \Gaussian(\x_1 - \x_2, \Kernel)
			\, \Gaussian(\x_2-\varcomp{\p}_j, \varcomp{\varKernel}_j)
			\dd{\x_1} \dd{\x_2}
		}
		\\ & = \ovar^2 \sqrt{\vectdet{2\pi \Kernel}}
		\, \Gaussian(\varcomp{\p}_i-\varcomp{\p}_j, \varcomp{\varKernel}_i + \Kernel + \varcomp{\varKernel}_j).
	\end{aligned}
	&&
\end{flalign}
Reassuringly, in the limit $\varcomp{\varKernel}_k \rightarrow 0$, in which pixels are treated as points, this covariance reduces to the usual expression. A similar construction for nonstationary covariance functions has been considered in the Gaussian process literature \citep{10.5555/2981345.2981380}.

\begin{figure}
	\centering
	\pgfplotsset{
	scale only axis, width=0.7\linewidth, colormap name=traffic,
}%
\tikzsetnextfilename{\currfilebase-hist}%
\begin{tikzpicture}[trim axis left, trim axis right]
	\sisetup{round-mode=places,round-precision=1, zero-decimal-to-integer}
	\begin{loglogaxis}[
		height=0.4\linewidth,
		xmin=0.9, xmax=12, ymin=4.2e-3, ymax=1.2,
		xtick={1, 2, 3, 5, 10},
		ytick={1e-3, 2e-3, 5e-3, 1e-2, 2e-2, 5e-2, 10e-2, 20e-2, 50e-2, 100e-2},
		yticklabel={\pgfmathparse{100 * exp(\tick)}\SI{\pgfmathresult}{\percent}},
		xticklabel={\pgfmathparse{exp(\tick)}$\divideontimes \num{\pgfmathresult}$},
		xlabel={approximate / exact},
		ylabel={$p(\text{ratio} > \text{factor of } \cdot)$},
		no markers, cycle list={[colors of colormap={16,263,838}]},
		cycle list shift=-2, filled legend
	]
		\addplot[black, dashed] table[x=all, y=y-all-only] {tikz/windowing-check/hist.txt}; \addlegendentry{all}
		\addplot[black] table[x=masked, y=y-masked-only] {tikz/windowing-check/hist.txt}; \addlegendentry{masked}
		\pgfplotsinvokeforeach{0.05,0.10,0.50}{
			\addplot table[x=thr-#1, y=y-thr-#1-only] {tikz/windowing-check/hist.txt}; \addlegendentry{$s_{ij} \ge \num[round-mode=figures]{#1}$}
		}
	\end{loglogaxis}
\end{tikzpicture}%
\\[1em]%
\tikzsetnextfilename{\currfilebase}%
\begin{tikzpicture}[trim axis left, trim axis right]
	\begin{loglogaxis}[
		clip marker paths=true,
		colorbar right, colorbar style={
			ylabel={$s_{ij} = \sqrt{\qty(\text{shape index})_i \times \qty(\text{shape index})_j}$},
			ylabel style={at={(ticklabel* cs:0.5,1.8em)}, rotate=180},
			ytick={-1.30103, -1. , -0.69897, -0.30103, {ln(1)}},
			yticklabels={$0.05$, $0.1$, $0.2$, $0.5$, $1$},
			minor ytick={-1.22184875, -1.15490196, -1.09691001, -1.04575749, -0.52287875, -0.39794001, -0.22184875, -0.15490196, -0.09691001, -0.04575749}, xminorticks=true,
			colormap name=traffic,
		},
		height=\pgfkeysvalueof{/pgfplots/width},
		xmin=3e-4, xmax=0.5, ymin=3e-4, ymax=0.5,
		xlabel={exact $\varcomp{\K}_{ij}$}, ylabel={approximate $\varcomp{\K}_{ij}$},
	]
		\addplot[
			scatter, only marks, point meta=explicit,
			scatter/use mapped color={fill=mapped color, draw=mapped color},
			mark size=1pt, fill opacity=0.5, line width=0pt, draw opacity=0,
			point meta min=-1.32,
			restrict expr to domain={sqrt(\thisrow{shape})}{0.05:inf},
			restrict expr to domain={\thisrow{mask}}{0.5:1.5}
		] table [x=cov0, y=cov, meta expr={log10(sqrt(\thisrow{shape}))}] {tikz/windowing-check/data-min.txt};
		\addplot[name path=less2, no markers, black, ultra thin] coordinates {(1e-6, 0.5e-6) (1, 0.5)};
		\addplot[name path=more2, no markers, black, ultra thin] coordinates {(1e-6, 2e-6) (1, 2)};
		\addplot[fill=black!50, fill opacity=0.5] fill between[of=less2 and more2];
		\addplot[no markers, green!60!black, thin] coordinates {(1e-6, 1e-6) (1, 1)};
	\end{loglogaxis}
\end{tikzpicture}%
%
	\pcaption{
		Bottom: comparison of pixel--pixel covariance calculated exactly using projected pixel geometry (horizontal axis) to the approximation used in this work (\cref{eqn:window-function,eqn:window-kernel,eqn:covariance}). The colour scale depicts a \enquote{shape index} $= 4\pi\times\text{area}/\text{perimeter}^2$, which is an indicator of elongation, being \num{\ll 1} for very elongated polygons. The shaded area indicates the region a factor of 2 away from the exact result. Top: cumulative fraction of pixel pairs (subsets as indicated in the legend) for which the approximation is off (either smaller or bigger) from the exact calculation by more than a given factor.
		\\
		The setup is an example lensed pixel grid of size \num{100x100} for a typical lensing configuration from \cref{section:mock-data-analysis}, and a single layer of kernel size comparable to the typical (projected) pixel sizes. The \enquote{exact} calculation was performed by averaging \cref{eqn:grbf} over \num{1000} uniformly sampled points in each pixel. Only pixels that would be considered for fitting (masked\footref{foot:mask} as in \cref{section:mock-data-analysis}) are included in the bottom panel, and a further cut was enacted to remove abnormally elongated pixels (located e.g.\ near critical lines).
		\label{fig:windowing-check}}
\end{figure}

To test the approximation, we apply it to a pixel grid deformed by a realistic lens configuration as in \cref{section:mock-data-analysis} and compare it to the exact calculation according to \cref{eqn:windowing} with \enquote{hard} indicator functions, i.e.\ non-negative only within the polygon formed by the projected pixel corners\footnote{This definition sometimes leads to \enquote{invalid} geometry along the critical curves/caustics where parts of the pixel may be flipped and/or sent to infinity. We exclude such cases from this test.}. We directly compare the approximated covariance matrix elements in \cref{fig:windowing-check} and verify that for the majority of pixel pairs that would be considered for fitting (see \cref{section:mock-data-analysis} for the \enquote{masking} procedure) the approximation is adequate, with \SI{95}{\percent} of the entries reproduced to within a factor of \num{2}. As evident from the figure, where significant deviations occur, either one or both of the pixels in the pair have a very elongated shape, as quantified by a \emph{shape index} related to the area-to-perimeter ratio of the pixel projection.

Lastly, we note that the Gaussian approximation leads to a biased estimate of the pixel variances, $\varcomp{\K}_{ii}$, since in \cref{eqn:window-cond2} we chose to normalise for the case of a vanishing kernel size. Thus, Gaussian indicator functions are more concentrated (\enquote{smaller}) than the polygons they approximate and hence end up having a slightly bigger variance. This effect is on the order of \SI{<10}{\percent} in typical cases, and we do not expect it to have a noticeable effect on the final results.

\subsection{Formulation as a generative model}\label{section:gp-gen}

In general, the positions $\p$ at which observations are made are a function of other parameters in the overall model. In the case of lensing studies they depend on the lens configuration. Thus, a way to optimise the lens parameters is to perform gradient descent on the marginal likelihood of \cref{eqn:gp-mll}, for which they act as hyperparameters. However, this would require performing expensive matrix inversion and determinant calculations, both of which have a complexity scaling as $\order{\n^3}$ with the number of observations. Even though software like \gpytorch \citep{gpytorch} improve on na\"{\i}ve implementations by using the conjugate gradient method \citep[Section~10.6]{numrecipes}, it still needs to be performed at each step of parameter optimisation. Furthermore, the method of calculating the determinant presented in \citet{gpytorch} is stochastic and can sometimes introduce noise comparable with the effects of substructure (\numrange[range-phrase={--}]{1}{2}\si{\percent}). This renders typical GP regression suboptimal for lensing studies and particularly so for substructure searches.

Instead, we propose a GP-\emph{inspired} source model that reduces to a GP in ideal situations but circumvents expensive matrix operations and allows other model parameters to be fit via gradient descent simultaneously with the optimisation of the source hyperparameters. Furthermore, the fitting procedure produces a simple analytic posterior for the true values which can be used to marginalise them out directly or to obtain training samples for subsequent analysis using inference networks \citep{Coogan:2020yux, upcoming}.

The source model relies on the framework of variational inference (VI), which will be discussed shortly. A key component of it is a generative model that produces mock data according to the prior. A generative description of a GP as above is\footnote{assuming a single GP layer. A multilayer model generates \enquote{true values} $\flux = \sum_k \layerindex{\flux}{k}$ with covariances $\layerindex{\K}{k}$.}
\begin{equation}\label{eqn:gp-gen1}
	\data \sim \Normal\qty(\flux, \datavar) \qc \flux \sim \Normal(\variable{0}, \K),
\end{equation}
which upon marginalisation of $\flux$ indeed gives \cref{eqn:gp-mll}. Sampling $\flux$ can further be expressed as
\begin{equation}\label{eqn:gp-gen2}
	\flux = \T\y \qc \y \sim \Normal(\variable{0}, \ovar^2 \varmatrix{\Identity})
\end{equation}
as long as $\T$ satisfies
\begin{equation}\label{eqn:decomposition}
	\K = \alpha^2 \T \T^\transpose.
\end{equation}
Here we introduced the set of \emph{source parameters} $\y$, which we optimise by ELBO maximisation as presented in \cref{section:vigp} in order to approximate the exact GP solution.

This procedure shifts the complexity of the GP into calculating $\T$ appropriately and so requires only sampling with diagonal (or, otherwise, constant) covariance and one matrix--vector multiplication. Crucially, all operations\footnote{
	barring, in the general case, the final data sampling $\data \given \flux, \datavar$, which, however, is trivial for uncorrelated noise or requires one matrix inversion per dataset for a general covariance matrix. In some cases (as for a PSF), this final step can even be split into a deterministic \enquote{postprocessing} of the fluxes (a convolution with the PSF) and an uncorrelated sampling step.
} have complexity $\order{\n^2}$ at worst and efficient standard GPU implementations. A slight caveat is that, if there as many source parameters as pixels in the observation, the matrix $\T$ becomes too large to store at once in memory, without necessarily being sparse. To circumvent this, we instead implement directly the product $\T\y$ via the \keops library \citep{charlier2020kernel}, which enables on-demand compilation of symbolically defined (e.g.\ as in \cref{eqn:transmission}) matrices to extremely efficient reduction routines optimised for graphical processing units. Critically, \keops fully integrates into the \torch ecosystem by also supporting automatic differentiation of these operations.

Our source model can also be understood in terms of Bayesian linear regression onto a set of basis functions defined by the transmission matrix $\T$ and shifted to the positions $\p$ associated with the observations. What we term \enquote{source parameters} play the role of the weights, over which a normal prior is placed. Our expressions can be translated into those found in \citet[Section~2.1]{gp} by substituting $\y \to \bm{w}$, $\ovar^2 \varmatrix{\Identity} \to \Sigma_p$, $\T \to \Phi^\top$, and $\data \to \bm{y}$.

\subsection{Factorisation of the covariance matrix}\label{section:gp-factorisation}

\shortcut{\Afunc}{\mathcal{A}}
\shortcut{\A}{\variable{\Afunc}}
\shortcut{\tKernel}{\vectmatrix{\tilde{\KernelSymbol}}}
\shortcut{\tvarKernel}{\vectmatrix{\tilde{\variable{\varKernelSymbol}}}}
\shortcut{\vq}{\x}

Given a covariance matrix $\K$, linear algebraic methods like the Cholesky decomposition \citep[Section~2.9]{numrecipes} can produce the required factorisation, generally in $\order{\n^3}$ time, which, as discussed, is to be avoided. Instead, motivated by the fact that a matrix product is analogous to a convolution, and the convolution of two Gaussian functions is again a Gaussian, we approximate the (Gaussian) covariance matrix from \cref{eqn:covariance} as the product of two other (\enquote{transmission}) matrices of the same form
\begin{equation}\label{eqn:transmission}
	\varcomp{\T}_{ik} = \varcomp{\A}_k \Gaussian(\varcomp{\p}_i - \varcomp{\q}_k, \varcomp{\tvarKernel}_i + \tKernel),
\end{equation}
where $\q$ are a set of \enquote{inducing} points, and $\varcomp{\A}_k$ are a set of normalisation constants which, along with $\tKernel$ and $\{\varcomp{\tvarKernel}_i\}$, are determined by requiring that
\begin{flalign}
	\begin{aligned}[b]
		\frac{\varcomp{\K}_{ij}}{\ovar^2} & = \sum_k \varcomp{\T}_{ik} \varcomp{\T}_{kj}
		\\ & = \mathrlap{
			\sum_k \varcomp{\A}_k^2
			\, \Gaussian(\varcomp{\p}_i - \varcomp{\q}_k, \varcomp{\tvarKernel}_i + \tKernel)
			\, \Gaussian(\varcomp{\p}_j - \varcomp{\q}_k, \varcomp{\tvarKernel}_j + \tKernel)
		}
		\\ & \approx \mathrlap{
			\int \dd{\vq} \density_{\q}(\vq) \Afunc^2(\vq)
			\, \Gaussian(\varcomp{\p}_i - \vq, \varcomp{\tvarKernel}_i + \tKernel)
			\, \Gaussian(\varcomp{\p}_j - \vq, \varcomp{\tvarKernel}_j + \tKernel)
		}
		\\ & = \tilde{\Afunc}^2
		\, \Gaussian(\varcomp{\p}_i-\varcomp{\p}_j, \varcomp{\tvarKernel}_i + 2\tKernel + \varcomp{\tvarKernel}_j),
	\end{aligned}
	&&
\end{flalign}
where $\density_{\q}(\vq)$ is the density\footnote{
	We use a simple density estimator based (in 2D) on the area of a circle reaching the $k$\textsuperscript{th} closest to $\vq$ point in $\q$; we use $k=20$.
} of inducing points $\q$ around $\vq$, and $\tilde{\Afunc}^2 = \density_{\q}(\vq) \Afunc^2(\vq)$ is set to be independent of $\vq$. The approximation becomes an equality as the inducing points become infinitely dense and cover the whole space. Comparing with \cref{eqn:covariance}, we can identify
\begin{align}
	\tKernel & = \Kernel / 2,
	\\ \varcomp{\tvarKernel}_i & = \varcomp{\varKernel}_i,
	\\ 
	\Afunc\qty(\varcomp{\q}_k) = \varcomp{\A}_k & = \frac{\vectdet{2\pi\Kernel}^{1/4}}{\sqrt{\density_{\q}\qty(\varcomp{\q}_k)}}.
\end{align}

This derivation is akin to the one presented by \citet[p.~84]{gp} and relates to the well-known fact that the squared-exponential covariance function corresponds to Bayesian linear regression onto an infinite linear combination of Gaussian basis functions.

\subsection{Aside: Source reconstruction and inducing points}\label{section:inducing-points}

As discussed, our source model can be interpreted as a regression onto a finite set of Gaussian basis functions each with kernel $\tKernel = \Kernel / 2$ located at the inducing points and weighted by the source parameters. Intuitively, this corresponds to treating the source as a collection of (a finite number of) Gaussian \enquote{blobs} attached to the inducing points and free to scale in intensity and with size fixed by the kernel. Thus, once the source parameters / weights (and hyperparameters) have been determined, the source surface brightness can be reconstructed at arbitrary locations $\x$ by evaluating all the basis functions at $\x$ and summing their contributions. This is succinctly expressed in matrix notation by simply replacing $\varcomp{\p}_i \rightarrow \x$ in \cref{eqn:transmission} and then evaluating \cref{eqn:gp-gen2}:
\begin{eq}\label{eqn:gp-reconstruction}
	\sbr(\x) = \sum_k \varcomp{\T}_{k}(\x) \varcomp{\y}_k = \sum_k \varcomp{\A}_k \Gaussian(\x - \varcomp{\q}_k,\, \Kernel / 2)\, \varcomp{\y}_k.
\end{eq}
Note that here we do not include the \enquote{pixel shapes} $\varcomp{\tvarKernel}_i$ since we are directly evaluating the underlying surface brightness. Alternatively, we can assume that, in contrast to observational data, the source reconstruction is evaluated on a high-resolution regular grid for which the point approximation is adequate. This equation in fact defines a posterior distribution over the functions $\sbr(\x)$, arising from the inferred posterior distribution of the source parameters\footnote{
	and inducing points, if they are varied in the fit! This is a manifestation of the true VI spirit in which all the parameters are drawn at once, and then the model is run forward to produce a posterior sample. The variability of the inducing locations arises in our model since we \enquote{entangle} them with the image-plane pixels, and those are differently de-projected to the source plane based on the concrete sample of lens parameters.
}, as appropriate for a \emph{process} model. Furthermore, for fixed inducing points and covariance hyperparameters, this distribution will be Gaussian.

\subsubsection{Reconstructed uncertainty}\label{section:rec-uncertainty}
A well-known problem of basis function regression, and one of the main advantages of the full GP treatment, is that it is not well-suited for making predictions away from the inducing points, especially in the case of decaying bases like the squared exponential. It is plain to see from \cref{eqn:gp-reconstruction} that sufficiently far away from the data a source model as we describe, however trained, predicts unfailingly $\varcomp{\flux}_* = 0$, i.e.\ $\varcomp{\flux} \given \emptyset \sim \delta(\varcomp{\flux})$, where $\emptyset$ denotes the absence of data. Even though this is not unreasonable when it comes to the mean---indeed, the GP predictive mean also reverts to the prior (which is null in our case) away from data---it also paradoxically predicts with vanishing uncertainty, which is in stark contrast to the expected regress to the prior covariance of a true GP\@.

This is not a problem either for our stated goal of training a model that is then able to produce further training samples for substructure inference or for plotting the reconstruction in the image plane since these will only require evaluations at the same points as the model was trained.\footnote{
	Even though the locations at which the source is evaluated will vary as the lens parameters are drawn from the posterior, in practice they will do so by much less than the kernel size of the source-plane source, while the inducing points of the image-plane source are always \enquote{dragged} along.
} However, in plotting high-resolution source reconstructions (in the source plane), the uncertainty will be concentrated \enquote{unnaturally} around the inducing points, an artefact further concealed by the integration over pixel areas during training (and when generating image-plane reconstructions). One solution might be to only evaluate the posterior at the inducing point locations and then perform mesh interpolation or kernel interpolation using the Gaussian indicators as kernels. However, this would defeat the philosophy of using \emph{process}es for interpolation, and hence in figures like \cref{fig:results-img-simg} (and \cref{fig:vigp}, although there it is less prominent) we still present the mean of the source reconstruction as derived from \cref{eqn:gp-reconstruction} while mesh interpolating only the predicted variance.

\subsubsection{Choice of inducing points}\label{section:ips-sps}

Note also that the basis functions are entirely defined at the inducing points, without reference to the observational data: $\varcomp{\tvarKernel}_i$ appear in \cref{eqn:transmission} only to account for the integration over the pixel indicator functions and should in fact be regarded as part of the \emph{instrumental} component of the model. This means that, even layer-by-layer, the inducing points, and crucially, their number, can be chosen freely. If, for example, only a small number (fewer than the image pixels) of inducing points is used, the model may be solvable exactly because of the reduced complexity of the inversion and determinant calculations: this is the so-called \emph{low-rank approximation}, which we will explore in further work.

We employ two strategies of choosing the inducing points, which are useful for modelling source variations at different scales. The first is to fix $\layerindex{\q}{\text{ip}} = \p$, i.e.\ attach the source-plane Gaussians to the image pixels and have them move with the change of lens parameters. The advantages of this prescription are that it automatically results in a higher density of inducing points in highly magnified regions of the source plane, akin to triangulation schemes, but also reduces the correlations between the lens and source parameters by \enquote{dragging} the observed flux along the source plane as the lens changes. We dub this GP configuration the \emph{image-plane GP} (hence the subscript) and use it to model small-scale image variations.

However, this strategy is unsuitable for modelling variations on scales larger than the separation between the projected pixel locations, which can be very small in highly magnified source regions. Indeed, in a situation in which the inducing points are significantly closer than the kernel size $\kernelsize$ of the GP layer, the Gaussians in the sum \cref{eqn:gp-reconstruction} overlap substantially, resulting in significant degeneracies between source parameters. While one option for mitigating this problem would be to use only a subset of the image pixels as inducing points, we instead introduce a large\=/$\kernelsize$ layer with inducing points fixed to a predetermined grid in the source plane with separation comparable to $\kernelsize$: $\layerindex{\q}{\text{sp}} = \p_{\text{grid}}$. We call this source setup the \emph{source-plane GP}. Since we only use it to model large-scale variations, for this layer we drop the pixel shape contribution in \cref{eqn:transmission}, which corresponds to treating each basis function as approximately constant over the projected size of a single pixel.

\subsection{Variational inference by ELBO maximisation}\label{section:elbo}

The method of maximisation of the evidence lower bound (ELBO) \citep{Saul_1996, Jordan_1999, Hoffman_2013} is particularly suitable for fulfilling the two main goals of our source modelling: to provide an estimate of the posterior and optimise the source hyperparameters (which, in some sense, include the lens parameters). Traditionally, the latter is performed by maximising the model evidence, and this is a natural outcome of ELBO maximisation as the name suggests. The former is achieved by using a proposal distribution, the so-called \emph{variational posterior}, $\proposal_{\gparam}(\param)$, to approximate the true posterior $\posterior(\param\given\data)$ of the model parameters $\param$ and optimising its parameters $\gparam$ using gradient ascent on the function
\begin{multline}\label{eqn:elbo}
	\elbo[\prob(\param, \data), \proposal_{\gparam}(\param)]
	\\ = \expectation_{\proposal_{\gparam}(\param)}[\ln{\prob(\data, \param)} - \ln{\proposal_{\gparam}(\param)}],
\end{multline}
where $\prob(\data, \param) = \prob(\data\given\param) \prob(\param) = \posterior(\param\given\data) \prob(\data)$ is the joint likelihood of the generative model. The ELBO is thus composed of two terms: the overlap between the proposal and posterior (rescaled by the evidence) and the entropy of the proposal, both of which we, naturally, would like to maximise. More formally, it can be shown that the difference between the (logarithm of the) model evidence and the ELBO is the Kullback–-Leibler (KL) divergence between the variational and true posteriors \citep{vi}:
\begin{multline}
	\ln \prob(\data) - \elbo[\prob(\param, \data), \proposal_{\gparam}(\param)]
	\\ \begin{aligned}[b]
		& = \expectation_{\proposal_{\gparam}(\param)} \qty[\ln{\proposal_{\gparam}(\param)} - \ln{\posterior(\param\given\data)}]
		\\ & = \kl{\proposal_{\gparam}(\param)}{\posterior(\param\given\data)},
	\end{aligned}
\end{multline}
which is strictly non-negative for any two distributions and zero only if they coincide. This not only justifies the ELBO's name, but also means that ELBO maximisation produces the optimal approximation of the posterior within the confines of the parametrisation of the proposal.

\subsubsection{Stochastic gradient descent (SGD)}
Note that, unlike typical marginalisation where integration is performed over the model parameters and can often be intractable, the ELBO only requires samples from the proposal, which is usually much simpler and often represented by analytical distributions. Given a model $\prob(\data, \param)$ and a proposal $\proposal_{\gparam}(\param)$, however, it is often impossible to calculate the expectation value in the ELBO analytically, let alone its gradient with respect to $\gparam$. Even if a Monte Carlo estimate is available, taking its gradient na\"{\i}vely does not capture the dependence of the sampling function on $\gparam$ and hence does not give the correct result.

There are two strategies to circumvent this, both of which express the gradient of the ELBO as an expectation of a \enquote{surrogate function}. The first approach, called \emph{black box variational inference} \citep{bbvi}, uses
\begin{multline}
	\vargrad[\gparam] \elbo(\gparam)
	\\ = \expectation_{\proposal_{\gparam}(\param)}[(\vargrad[\gparam] \ln{\proposal_{\gparam}}) (\ln{\prob(\param, \data)} - \ln{\proposal_{\gparam}})],
\end{multline}
which is an unbiased estimate for general proposal and model and can be sampled using Monte Carlo. However, such estimates often suffer from large variance, especially when the proposal does not cover the majority of the posterior mass (e.g.\ in the early stages of a fit).

An alternative is to \emph{re-parametrise} the proposal distribution so that its stochasticity is independent of the parameters we are taking a gradient with respect to \citep{vae}. This \enquote{re-parametrisation trick} is applicable to the majority of common analytical distributions, for which sampling is implemented as a series of deterministic manipulations on a \enquote{unit} random variable $\epsilon$ that is usually uniform on $[0, 1)$ or $\epsilon \sim \Normal(0, 1)$. Crucially, the distribution $\prob(\epsilon)$ is not parametrised, and the gradient needs to \enquote{flow} only through the deterministic transformations. Formally, this can be expressed as
\begin{eq}\label{eqn:reparam}
	\vargrad[\gparam](\expectation_{\proposal_{\gparam}(\param)}[f(\param, \gparam)])
	& = \vargrad[\gparam](\expectation_{\prob(\epsilon)}[g(\param, \gparam, \epsilon)])
	\\ & = \expectation_{\prob(\epsilon)}[\vargrad[\gparam] g(\param, \gparam, \epsilon)]
	\\ & = \expectation_{\prob(\epsilon)}[h(\param, \gparam, \epsilon)],
\end{eq}
whence we can again obtain a stochastic estimate of the required derivative via standard Monte Carlo methods. Of course, automatic differentiation is still practically indispensable in computing the gradient for any non-trivial model.

Both methods of SGD are implemented in the package \pyro with re-parametrised samples used where possible. Indeed, all the distributions that we use are re-parametrisable.

\subsection{Variational inference of a Gaussian process}\label{section:vigp}

\begin{figure*}
	\centering
	\begin{minipage}{\linewidth}\setlength\linewidth{0.9\linewidth}
	\tikzsetnextfilename{\currfilebase}%
\begin{tikzpicture}[vigp defs, smooth]
	\newcommand{\prefix}{tikz/vigp/400-}
	\newcommand{\plotdata}[1][]{\addplot[data] table {\prefix vigp.txt};}
	\newcommand{\percentiles}[3][]{
		\foreach \perc/\opacity in {95/0.4} {
			\foreach \ul in {l, u} {
				\edef\temp{
					\noexpand\addplot[percentile, name path global={#3_\ul\perc}] table [y={#3_\ul\perc}] {#2};
				}\temp
			}
			\edef\temp{
				\noexpand\addplot[percentile, opacity=\opacity, #1] fill between[of={{#3_l\perc} and {#3_u\perc}}];
			}\temp
		}
		\addlegendimage{area legend, draw opacity=0, fill opacity=0.4, fill, #1}
	}
	\newcommand{\labelexact}{exact GP, $\ovartrue$}
	\newcommand{\labelfull}{full VGP, $\ovartrue$}
	\newcommand{\labeldiagfixed}{diag. VGP, $\ovartrue$}
	\newcommand{\labeldiagfree}{diag. VGP, $\ovarvi$}

	\begin{groupplot}[
		group style={
			group size=1 by 2, vertical sep=1em,
			xlabels at=edge bottom, xticklabels at=edge bottom,
		},
		tight layout=1, scale only axis,
		xlabel={$x$}, xmin=-0.05, xmax=1.93,
		every axis y label/.append style={at={(ticklabel* cs:0.5, 2em)}},
		table/x=x,
		legend style={/tikz/every even column/.append style={column sep=1em}},
		every axis plot/.append style={thick},
		data/.style={
			table/y=y,
			only marks, mark size=0.75pt, mark options={draw opacity=0},
		},
		percentile/.style={
			no markers, draw opacity=0, forget plot
		},
		clip marker paths=true,
	]
	
		\nextgroupplot[
			height=0.26\linewidth,
			ylabel={$\varcomp{\flux}$}, ymin=-1.1, ymax=1.3,
			y label style={rotate=-90},
			legend pos=south east, legend columns=2,
			every axis legend/.append style={at={(1, 0)}, xshift=-0.2em, yshift=0.2em},
			legend image post style={xscale=0.75}
		]
		\addlegendimage{empty legend}\addlegendentry{}
		
		\percentiles[exact]{\prefix vigp-grid.txt}{exact-true_nonoise}
		\addlegendentry{\labelexact}
		
		\percentiles[diag]{\prefix vigp.txt}{diag_nonoise}
		\addlegendentry{\labeldiagfree}
		
		\percentiles[full]{\prefix vigp.txt}{full_nonoise}
		\addlegendentry{\labelfull}
		
		\addplot[exact, forget plot] table[y=exact-true_nonoise_med] {\prefix vigp-grid.txt};
		\addplot[full, forget plot] table[y=full_nonoise_med] {\prefix vigp-grid.txt};
		\addplot[diag, forget plot] table[y=diag_nonoise_med] {\prefix vigp-grid.txt};
		
		\plotdata
		
		\begin{scope}[shift={(axis direction cs:0, -1)}, <->, >=|, below, yshift=3em, xshift=1em, every node/.append style={anchor=base, pos=0}]
			\begin{scope}[every node/.append style={yshift=-1em}]
				\draw[postaction={decoration={markings, mark={at position 0.5 with {\fill circle (0.75pt);}}}, decorate}] (0, 0) -- (0, 0.3) node {$2\dataerr$};
				\draw[xshift=2em] (0, 0) -- (0, 0.5) node {$\layerindex{\ovar}{1}$};
				\draw[xshift=4em] (0, 0) -- (0, 0.2) node {$\layerindex{\ovar}{2}$};
			\end{scope}
			\begin{scope}[yshift=-2em, left, every node/.append style={xshift=-1em}]
				\draw (0, 0) -- (0.2, 0) node {$\layerindex{\kernelsize}{1}$};
				\draw[yshift=-0.8em] (0, 0) -- (0.02, 0) node {$\layerindex{\kernelsize}{2}$};
			\end{scope}
		\end{scope}
	
		\begin{scope}[/pgfplots/.cd, only marks, mark=|, mark repeat=4, table/y expr={1.15}]
			\addplot[black] table[x=x1] {\prefix xs.txt};
			\addplot[black, yshift=0.5em] table[x=x2] {\prefix xs.txt};
		\end{scope}
		
		\nextgroupplot[
			height=0.2\linewidth,
			ymin=-2, ylabel={$(\varcomp{\flux}-\varcomp{\mpred}_{\varcomp{\flux}}) / \varcomp{\epred}_{\varcomp{\flux}}$},
			legend pos=north east, legend columns=2,
			every axis legend/.append style={at={(1, 1)}, xshift=-0.2em, yshift=-0.2em, name=legend},
			legend image post style={xscale=0.75}
		]
		\addplot[exact, forget plot, name path=exact null uncertainty] coordinates {(\pgfkeysvalueof{/pgfplots/xmin}, 0) (\pgfkeysvalueof{/pgfplots/xmax}, 0)};
		\addplot[percentile, name path=exact uncertainty] coordinates {(\pgfkeysvalueof{/pgfplots/xmin}, 1) (\pgfkeysvalueof{/pgfplots/xmax}, 1)};
		\addplot[exact, percentile, opacity=0.4] fill between[of={exact null uncertainty and exact uncertainty}];
		
		\addplot[percentile, name path=null uncertainty] coordinates {(0, 0) (1.88, 0)};
		
		\addlegendimage{empty legend}
		\addlegendentry{$(\hat{\mpred}_{\flux} - \meanpred) / \errpred$}
		\addlegendimage{empty legend}
		\addlegendentry{$\errpredhat / \errpred$}
		
		\newcommand{\listofthings}
		
		\foreach \col/\fname in {diag-fixed/vigp-scaled.txt, diag/vigp-scaled-cross.txt, full/vigp-scaled.txt} {
			\edef\temp{
				\noexpand\addplot[\col] table[y=\col _nonoise_med_e] {\prefix \fname};
				
				\noexpand\addplot[percentile, name path=\col uncertainty] table[y=\col _nonoise_u68] {\prefix \fname};
				\noexpand\addplot[\col, opacity=0.4, area legend, draw opacity=0] fill between[of={null uncertainty and \col uncertainty}];
			}\temp
		}
		\addlegendentry{\labeldiagfixed}\addlegendentry{\labeldiagfixed}
		\addlegendentry{\labeldiagfree}\addlegendentry{\labeldiagfree}
		\addlegendentry{\labelfull}\addlegendentry{\labelfull}
		
		\begin{scope}[densely dashed, no markers]
			\addplot[diag-fixed] table[y expr={0.1872 / \thisrow{yscale_true}}] {\prefix vigp.txt};\label{vigp-plot-theory-true}
			\addplot[diag] table[y expr={0.1127 / \thisrow{yscale_true}}] {\prefix vigp.txt};\label{vigp-plot-theory-vi}
		\end{scope}
		\addplot[diag, densely dotted] table[y=cross_median] {\prefix vigp-scaled-cross.txt};\label{vigp-plot-theory-bias}
	\end{groupplot}

	\node[xshift=-0.2em, below left, draw] (uncertainty legend) at (legend.north west) {
		\shortstack[c]{
			\cref{eqn:paramvariance-layer-main} \\
			\shortstack[l]{
				\ref{vigp-plot-theory-true} $\ovar = \ovartrue$ \\
				\ref{vigp-plot-theory-vi} $\ovar = \ovarvi$
			}
		}
	};
	\node[xshift=-0.2em, below left, draw] at (uncertainty legend.north west) {
		\ref{vigp-plot-theory-bias} \cref{eqn:bias-main}
	};
\end{tikzpicture}%
	\end{minipage}
	\pcaption{Comparison of the exact GP solution of a 1\=/dimensional problem to variational approximations. The observed data (black dots) were drawn from a GP with two layers of sizes $\layerindex{\kernelsize}{1} = \num{0.2}$ and $\layerindex{\kernelsize}{2} = \num{0.02}$ and root variances $\layerindex{\ovar}{1}=0.5$ and $\layerindex{\ovar}{2}={0.2}$ with noise $\dataerr = \num{0.15}$ (shown as segments). The locations of the observations (every fourth one shown as a dash along the top of the top plot) imitate two overlapping magnified grids as in a lensing system. The central \SI{95}{\percent} of the posteriors for the true values $f(x)$ (marginalised over $f(x^\prime \neq x)$) are shown as shaded areas in gray (the exact solution \cref{eqn:gp-exact}), green (a variational fit with full covariance matrix), and red (with diagonal covariance). Solid lines trace the posterior means of the respective fits. All but the diagonal fit have the hyperparameters fixed at the true values. The bottom panel depicts the 1\=/sigma intervals normalised to the location and size of the exact solution. It also includes a diagonal fit with fixed true hyperparameters (yellow) and depicts the expected standard deviations of the diagonal fits from \cref{eqn:paramvariance-layer-main} as dashed lines. Finally, the red dotted line is the bias of the mean predicted by \cref{eqn:bias-main}.\label{fig:vigp}}
\end{figure*}

According to the generative model \cref{eqn:gp-gen1}, the posterior for the true values $\flux$, given an observation $\data$, has the form of a multivariate normal distribution (MVN) \citep[Section~2.2]{gp}:
\begin{align}
	\label{eqn:gp-exact}
	\posterior(\flux\given\qty{\data, \datavar}) & = \Normal(\meanpred, \varpred)
	\\ \label{eqn:gp-exact-mean} \qq{with} \meanpred & = \K\qty(\K + \datavar)^{-1} \data
	\\ \label{eqn:gp-exact-var} \qq{and} \varpred & = \K - \K\qty(\K+\datavar)^{-1}\K.
\end{align}
Since the source parameters $\y$ are related to $\flux$ by a linear transformation, their posterior, too, will be multivariate normal. Hence, an MVN is the correct form to assume for the proposal distribution.

However, an MVN proposal requires fitting $\n (\n+1) / 2$ parameters of the covariance matrix, which cannot be stored in reasonable memory, let alone optimised in reasonable time. Therefore, we are forced to use a mean field approximation in which we treat the posterior covariance as diagonal and use a univariate normal proposal distribution for each source parameter independently of the rest:
\begin{eq}
	\proposal_{\meany, \erry^2}(\y) = \prod_i \Normal(\varcomp{\y}_i \given \meanyi, \erryi^2), \label{eqn:gp-guide}
\end{eq}
where $\meany$ and $\erry^2$ are, respectively, the variational mean and variance vectors. A full analytical treatment of a variationally optimised GP, including detailed account of the ramifications of this choice, is presented in \cref{apx:vigp-analytical}.

\subsubsection{Hyperparameter optimisation}

Hyperparameter optimisation is a special case of Bayesian model selection, which is usually performed by maximising the model evidence, thus accounting for the posterior uncertainties of the latent variables (in regular GP regression these are the \enquote{true values}, which we have reformulated in terms of the \enquote{source parameters}). However, variational inference does not have access to the true evidence, but rather only to the ELBO, which contains the additional KL divergence term. As discussed, the latter is null if and only if the variational posterior matches the true one, and in our case this cannot be achieved since we are forcing the proposal covariance to be diagonal. Hence, the KL divergence may (and indeed does) bias the model selection procedure.

As described in \cref{apx:vigp-analytical}, using a diagonal approximation to the covariance overestimates the posterior variance, and therefore the fit will attempt to diminish it, which it can achieve by decreasing the prior variance. Hence, we anticipate the recovered hyperparameter values, $\ovarvi^2$, to scale inversely with the average number of correlated observations, $\nc$, with respect to the \enquote{true} hyperparameters, $\ovartrue^2$:
\begin{eq}\label{eqn:hyper-scaling}
	\ovarvi^2 \approx \ovartrue^2 / \nc.
\end{eq}
In \cref{apx:vigp-analytical-hyper} we show that this is indeed so.

\subsubsection{Example}
An illustration of the validity of the variational approach to solving a Gaussian process and comparison with the analytical solution is presented in \cref{fig:vigp} for the case of a (relatively) small 1\=/dimensional dataset designed to model the main aspects of lensing images in the source plane. Given the true hyperparameters, a variational optimisation with a full covariance matrix reproduces exactly the analytical solution within the domain of the data (i.e.\ not too far away from the observation), whereas a diagonal fit locates the mean but attains a constant uncertainty
\begin{equation}\label{eqn:paramvariance-layer-main}
	\layerindex[, i]{\qty(\erry^2)}{k} \rightarrow \frac{\layerindex{\ovar^2}{k} \dataerr^2}{\layerindex[]{\ovar^2}{k} + \dataerr^2},
\end{equation}
as derived in \cref{apx:vigp-analytical}. There we also justify why this usually leads to an overestimation of the posterior variance of the true values, as can also be observed in \cref{fig:vigp}; that is, our fitting strategy is usually conservative.

The uncertainties are somewhat reduced if the hyperparameters are also optimised. As discussed, this leads to a reduction of the prior variance, and hence, also of the posterior. However, this also means that predictions away from zero are more strongly penalised, which biases the mean with respect to the analytical treatment (observe the offset between the red and green lines in \cref{fig:vigp}). As derived in \cref{apx:vigp-analytical-bias}, the variational mean is a factor
\begin{equation}\label{eqn:bias-main}
	\frac{\meanpred \given \ovarvi^2}{\meanpred \given \ovartrue^2} \approx \frac{\ovartrue^2}{\dataerr^2 + \ovartrue^2}
\end{equation}
smaller (in absolute value) than the analytical result. Evidently, the offset depends inversely on the signal-to-noise ratio, and hence we do not expect it to significantly bias the parameters we are actually interested in while fitting lensing images. Note, furthermore, that variational inference still recovers the exact posterior mean \emph{at the optimised values of the hyperparameters}, so the bias is not due to the fitting procedure directly, but rather is a side effect of the different hyperparameters, themselves influenced by the additional KL term in the ELBO\@.

\section{Mock data analysis}\label{section:mock-data-analysis}

While there are currently on the order of a hundred available observations of galaxy--galaxy lenses suitable for substructure detection (mostly contained within the SLACS \citep{SLACS-1, SLACS-13} and BELLS \citep{BELLS-1, BELLS-3} optical surveys\footnote{
	These surveys identify candidates in large spectroscopic surveys (the SDSS and BOSS), by looking for objects that seem to combine light with distinct redshifts. Follow-up observations have been performed with the Hubble Space Telescope to obtain the final deep high-resolution images. JVA/CLASS \citep{CLASS-1} is an older radio survey that also provides suitable images. Other collections have also been produced \citep{More_2012, Huang_2020}, for which high enough quality imagery is generally still not available.
}), instruments in the near future are expected to deliver \emph{hundreds of thousands} more \citep{Collett_2015}. The main contributors are expected to be the Vera Rubin Observatory's Legacy Survey of Space and Time \citep[LSST,][]{LSST} and the Euclid space telescope \citep{Euclid}. The latter will in fact provide imagery of the necessary resolution and, more importantly, seeing to be directly usable for substructure modelling. On the other hand, the expected point spread function (PSF) of the LSST will be on the order of the Einstein radius for the majority of the discovered lenses, and so follow up will still be required. In that respect, the James Webb Space Telescope and the Extremely Large Telescope (ELT) will improve on the imaging capabilities of the Hubble Space Telescope (HST), with the latter expected to produce $\ang{;;10} \times \ang{;;10}$ diffraction-limited images (resolution is expected to be at least \SI{2}{mas} (\href{https://www.eso.org/sci/facilities/eelt/instrumentation/phaseA.html}{ELT Instrumentation}), compared to Hubble's \SI{50}{mas} (\href{https://hst-docs.stsci.edu/acsihb/chapter-5-imaging/5-3-wide-field-optical-ccd-imaging}{ACS Handbook}) and Euclid's \SI{100}{mas} (\href{https://sci.esa.int/web/euclid/-/euclid-vis-instrument}{Euclid VIS Instrument})).

In this section we demonstrate that our pipeline is capable of rapidly analyzing mock observations representative of what upcoming high-resolution telescopes such as ELT will produce. After describing our mock data setup we study a benchmark system in detail, showing that we can accurately fit the observation, recover the source light distribution, and precisely infer lens parameters with reasonable variational uncertainty estimates. We also comment on several key technical points, most importantly the independence of the lens parameters' variational posteriors from the number of GP source layers. Fits for several other mock lensing systems are presented in \cref{apx:fit-gallery}.

\subsection{Data generation and source model}\label{section:data-generation}

\begin{table}
	\centering
	\pcaption{List of lens parameters, their priors, true values in the mock image of Hoag's object, and results of the final fit with \num{3} image-plane GP layers. The \enquote{fit} column quotes the means and standard deviations derived from the fitted multivariate proposal described in the text. The full variational posteriors from all fits (to this mock observation) are depicted in \cref{fig:indepth-corner}.\label{tab:lens-params}}
	\sisetup{
		table-sign-mantissa, table-figures-integer=1, table-figures-decimal=4,
		table-number-alignment=center
	}
	\begin{tabular}{@{\hskip 0.5\tabcolsep} r l l S[table-figures-uncertainty=1] S s@{\hskip 0.5\tabcolsep}}
		\toprule
		& & {Prior} & {Fit} & {True} & \\
		\midrule
		\vertmultirow{6}{6em}{SPLE}
		& $\comp{\ximg}_{0, x}$ & $\uniform\qty(\num{-0.2}, \num{0.2})$ & \fitresult{main/x} & \truth{main/x} & \arcsecond \\
		& $\comp{\ximg}_{0, y}$ & $\uniform\qty(\num{-0.2}, \num{0.2})$ & \fitresult{main/y} & \truth{main/y} & \arcsecond \\
		& $\ellq$ & $\uniform\qty(\num{0.1}, \num{1})$ & \fitresult{main/q} & \truth{main/q} & \\
		& $\ellangle$ & $\uniform\qty(\num{0}, \num{2\pi})$ & \fitresult{main/phi} & \truth{main/phi} & \radian \\
		& $\splerein$ & $\uniform\qty(\num{1}, \num{1.7})$ & \fitresult{main/r_ein} & \truth{main/r_ein} & \arcsecond \\
		& $\spleslope$ & {---} & {---} & 2.2193 & \\
		\midrule
		\vertmultirow{2}{6em}{Ext.\\shear}
		& $\extshear_1$ & $\Normal\qty(\num{0}, \num{0.05})$ & \fitresult{ext/gamma_1} & \truth{ext/gamma_1} & \\
		& $\extshear_2$ & $\Normal\qty(\num{0}, \num{0.05})$ & \fitresult{ext/gamma_2} & \truth{ext/gamma_2} & \\
		\bottomrule
	\end{tabular}
\end{table}

\begin{figure}
	\centering
	\showofffigure{hoags_object}%
	\pcaption{Mock ELT observation (left panel) and the original HST image used as a source (right panel).	The red curves depict the tangential critical curve and the corresponding caustic line, respectively in the image and source planes, along which the magnification is (formally) infinite. While we show the images in colour, the analysis was performed on a grayscale version.\label{fig:showoff}}
\end{figure}

For our mock data we assume a resolution of \SI{12.5}{mas}, which is significantly larger than the expected size of ELT's point spread function \citep{MACADO}, allowing us to neglect it at this stage. We generate images of size \num{400 x 400} pixels, spanning an area of $\ang{;;5} \times \ang{;;5}$ on the sky. In order to simulate integration of the light across the pixel areas and to avoid pixellation artefacts, we initially generate the mock observation at 10\=/times higher resolution and then downsample to the target size by local averaging. We set the noise level so that (after downsampling) the brightest pixel has a signal-to-noise ratio (SNR) of \num{\sim 30}, in line with HST observations of the SLACS lenses.\footnote{
	Our mock observation has an overall SNR of about \num{700} summed over the pixels that contain significant light from the source (as defined by the masking procedure described in this \namecref{section:data-generation}).
}

The properties of the lensing system are representative of the expected distribution of strong galaxy--galaxy lenses detectable by next generation instruments, as derived by \citet{Collett_2015}. Specifically, we set an Einstein radius $\splerein \sim \ang{;;1.5}$ and the size of the unlensed source to \ang{;;0.4}.\footnote{
	At source and lens redshifts of $\zsrc \sim \num{0.6}$ and $\zlens \sim \num{2.5}$, corresponding to the modes of the respective distributions for all galaxy--galaxy lenses in the universe, these values imply a lens mass (within the Einstein radius) of \SI{\sim 6e11}{\Msun} and physical radius of the source galaxy of \SI{\sim 30}{\kilo\parsec}.
} The other parameters of the SPLE lens and the external shear are drawn at random from the priors listed in \cref{tab:lens-params}. As a source we use high-resolution public domain galaxy images. For the main example we interpolate over a post-processed \num{1185 x 1185} HST image \stscirelease{2002}{21} of Hoag's object \citep{Hoag_original}, converted to grayscale. We choose this object because it combines a largely featureless \enquote{early-type} core with a featureful \enquote{late-type} star-forming ring, which allow us to test the model simultaneously in both relevant regimes of image complexity. Lastly, we do not model the lens galaxy's light at this stage\footnote{
    In our framework foreground light could be accounted for by a GP source located in the image plane, possibly in combination with an analytical model. In future work we plan to implement this, while modelling simultaneously multi-wavelength observations to facilitate the separation of foreground from source light.
}, assuming it has been perfectly accounted for in a preprocessing step as in \citet{SLACS-1}. \Cref{fig:showoff} displays the mock observation, source, critical curve, and caustic.

Finally, we create a mask\footnote{Everywhere \enquote{masked} refers to the region of high SNR, which we include in the analysis, and \enquote{masked out} refers to the remainder of the image.\label{foot:mask}} for our observation by first applying a \SI{\sim 35}{mas} Gaussian blur to the observation and keeping only the pixels with SNRs larger that \num{1}.

{\sisetup{round-mode=figures, round-precision=2, zero-decimal-to-integer}
In the fits we model the source via a multilayer image-plane GP and a single source-plane GP (cf.\ \cref{section:inducing-points}). We set the kernel size of the latter to
\pgfmathparse{1000*\getdata{hoags_object/ngpl-3/src/sigma}}$\spindex{\kernelsize} = \SI{\pgfmathresult}{mas}$,
while for the former we use a geometric progression of spatial scales ranging between
\pgfmathparse{1000*\getdata{hoags_object/ngpl-3/gp/sigma/0}}\SI{\pgfmathresult}{mas} and
\pgfmathparse{1000*\getdata{hoags_object/ngpl-3/gp/sigma/2}}\SI{\pgfmathresult}{mas}.
}
The layer sizes span from \num{\sim 1/6} to \num{\sim 3} times the image pixel size and thus allow simultaneously modelling highly magnified areas in detail and the large-scale appearance of the source. We test from one (either on the small or the large end of the size range, indicated by subscripts \enquote{s} and \enquote{b}, respectively) up to eight image-plane layers. The image-plane GP is set up to only model the masked pixels, i.e.\ the ones with a significant source-light contribution, which are also used as inducing points. In contrast, the source-plane source is evaluated across the whole image, and its grid of inducing points has a size of \num{40 x 40} with a separation of \SI{30}{mas}, spanning an area about \SI{20}{\percent} larger than the images used for mock data generation. The variance hyperparameter $\spindex{\ovar}$ is initialised to \num{10} times the noise level (similar to the highest SNR) for the source-plane GP and to the level of the observational noise for the image-plane GP layers.

\subsection{Variational posterior}\label{section:guide}
\shortcut{\L}{\varmatrix{L}}
\shortcut{\C}{\varmatrix{C}}
\newcommand{\Errysp}[1][]{\spindex[\y,]{\Err\ifthenelse{\equal{#1}{}}{}{^{#1}}}}
\newcommand{\errysp}[1][]{\spindex[\y,]{\err\ifthenelse{\equal{#1}{}}{}{^{#1}}}}
\shortcut{\ngrid}{\n_{\text{grid}}}
\shortcut{\nlens}{\n_{\text{lens}}}

\begin{table*}
	\centering
	\pcaption{Parameters of the variational posterior, $\gparam$.\label{tab:gparams}}
	\begin{tabular}{l l r S[table-parse-only, table-number-alignment=left, table-figures-integer=5, table-figures-decimal=0, table-text-alignment=left] l l}
		\toprule
		param. & initial & size & {size} & constraint & description \\
		\midrule
		$\meanparams{lens}$ & within \SI{\sim 10}{\percent} of truth & $\nlens$ & 7 & prior domain & mean lens parameters (see \cref{tab:lens-params}) \\
		$\L_{\text{lens}}$ & $\num{0.01}\, \Identity$ & $\nlens (\nlens+1) / {2}$ &  28 & lower Cholesky & scale matrix of lens parameters \\
		$\spindex{\meany}$ & \variable{0} & $\ngrid$ & 1600 & --- & mean for source-plane GP \\
		$\errysp$ & \variable{0.01} & $\ngrid$ & 1600 & positive & st.\ dev.\ for source-plane GP \\
		$\C$ & \varmatrix{0} & $\ngrid \times \nlens$ & 11200 & --- & lens-source covariance \\
		$\layerindex{\meany}{k}$ & \variable{0} & $\nlayers \n$ & $\order{\num{e5}}$ & --- & means for image-plane GP layers \\
		$\layerindex{\err_{\y,}{}}{k}$ & \variable{0.001} & $\nlayers \n$ & $\order{\num{e5}}$ & positive & st.\ dev.\ for image-plane GP layers \\
		$\layerindex{\mean{\ovar}}{k}$ & $\sim\!\max(\data);\enspace \dataerr$ & $1 + \nlayers$ & {\numrange[range-phrase={ -- }]{2}{9}} & positive & variance hyperparameters of GP \\
		\bottomrule
	\end{tabular}
\end{table*}

The slope $\spleslope$ has a complex degeneracy with the source, related to the mass-sheet and source-position transformation degeneracies \citep{masssheetdegeneracy,sourcepositiontransformation}. Breaking it requires additional information or assumptions on e.g.\ the dynamics of the lens galaxy \citep{Barnabe:2007ns,Barnab__2009,2020MNRAS.496.1718E}. For simplicity in the current work we do not optimise $\spleslope$ but instead fix it to the true value from \cref{tab:lens-params}. For the remaining seven lens parameters we use a multivariate normal (MVN) proposal distribution, while we optimise the GP source parameters $\layerindex{\y}{k}$ via a diagonal normal proposal, as discussed in \cref{section:source}. However, since the source-plane GP layer has relatively few ($\order{1}$ with the number of image pixels) source parameters, it is computationally feasible to at least partly model their off-diagonal covariance and their correlation with the lens parameters. We achieve this by using a \enquote{partial} MVN proposal:
\begin{equation}
	\params{lens},\, \spindex{\y} \overset{\proposal}{\sim} \Normal\qty[\mqty(\meanparams{lens} \\ \spindex{\meany}), \L\L^\transpose],
\end{equation}
where
\begin{equation}
	\L \equiv \mqty(
		\L_{\text{lens}} & \varmatrix{0} \\
		\C & \Errysp[1/2]
	)
\end{equation}
is the scale matrix of the MVN with $\L_{\text{lens}}$ an $\nlens \times \nlens$ lower-triangular matrix with positive diagonal entries (\enquote{lower Cholesky}), $\Errysp$ a diagonal matrix with $\errysp[2]$ on the diagonal, and $\C$ the $\ngrid \times \nlens$ matrix correlating the lens and the source-plane source. Under this proposal the covariance of $\params{lens}$ is $\L_{\text{lens}}\L_{\text{lens}}^\transpose$, while that of $\spindex{\y}$ is $\Errysp + \C\C^\transpose$, so apart from the lens-source correlations, $\C$ also accounts for a part of the full posterior GP covariance. Exploring more flexible variational posteriors capable of better accounting for the most important correlations between source parameters is left for future work.

Thus, our full variational posterior can be written as
\begin{eq}\label{eqn:guide}
	\proposal_{\gparam}(\param, \hparam) =&\
	\proposal_{\text{lens, sp}}(\params{lens}, \spindex{\y})
	\\ & \times \prod_{\mathclap{k \neq \text{sp}}} \Normal(\layerindex{\y}{k} \given \layerindex{\meany{}}{k},\, \layerindex[, k]{\err^2}{\y})
	\\ & \times \prod_{k} \delta(\layerindex{\ovar}{k} \given \layerindex{\mean{\ovar}}{k}).
\end{eq}
Here the last line is a technical way to write the optimisation of hyperparameters, for which a posterior is not derived. The complete set of variational parameters $\gparam$ are listed in \cref{tab:gparams}.

 Finally, we note that the parameter constraining functionality of \torch\ /\ \pyro was used to ensure that samples for the lens parameters always fall within the domains of the respective priors (listed in \cref{tab:lens-params}). This is implemented by transforming the values \emph{after they have been sampled from the proposal}, which means that in fact the samples \emph{do not} follow the distribution as written in \cref{eqn:guide}. However, since $\proposal$ is Gaussian, and the posterior uncertainties we obtain are very small in comparison with the prior ranges, the transformation amounts to no more than a linear rescaling of the uncertainty parameters (and a transformation of the means), which we take into account when plotting and citing the posteriors. Furthermore, we account for the transformation Jacobian while fitting by explicitly adding it to the loss.

\subsection{Optimisation and initial fit problem}

\begin{figure}
	\tikzsetnextfilename{\currfilebase}%
\begin{tikzpicture}[trim axis group right]
	\newcommand{\nullloss}{237200}
	\pgfkeys{/pgf/fpu=true}
	\pgfmathsetmacro{\failloss}{270000 - \nullloss}
	\pgfkeys{/pgf/fpu=false}
	\begin{groupplot}[
		group style={group size=2 by 1, horizontal sep=0pt, y descriptions at=edge left},
		height={0.7\linewidth},
		every axis title shift={0em},
		ylabel={relative loss}, ylabel near ticks,
		scaled x ticks={base 10:-3}, xtick scale label code/.code={},
		ymode=log, ymin={100}, ymax={4e5},
		y filter/.expression={ln(y-\nullloss)},
		every axis plot/.append style={line width=1pt},
		legend image post style={opacity=1, line width=2pt},
	]
		\nextgroupplot[
			title={source-plane source},
			width={0.60\linewidth}, xmin=-1e3, xmax=20e3,
			xticklabel style={xshift={- (\tick == 20) * 0.5em}},
			xlabel={step (thousands)}, x label style={at=(ticklabel cs: 1)},
			failed/.style={red, opacity=0.2},
			lr/.style={opacity=0.07},
			colormap name=viridis,
			lr1/.style={lr, color of colormap=0},
			lr2/.style={lr, color of colormap=1000},
			lr3/.style={lr, color of colormap=500},
			legend image post style={scale=0.5}, filled legend
		]
			\addlegendimage{empty legend}\addlegendentry{\hspace{-0.2cm}$\mathtt{lr}$:}
			\addlegendimage{lr1}\addlegendentry{\ttfamily 1e-2};
			\addlegendimage{lr2}\addlegendentry{\ttfamily 1e-3};
			\addlegendimage{lr3}\addlegendentry{\ttfamily 1e-4};
			
			\axhline[/pgfplots/failed, opacity=1, dashed]{\failloss};
			\node[right] at (2e3, 1e5) {\color{red} failed};
			
			\input{scripts/indepth/tikz-indepth-losses.tex}
		
		\nextgroupplot[
			title={image-plane source},
			width={0.60\linewidth},
			xmin=20e3, xmax=85e3, xtick={25e3,35e3,...,\pgfkeysvalueof{/pgfplots/xmax}},
			ytick pos=right, ylabel style={rotate=180},
			table/x=step, every axis plot/.append style={very thick},
			legend columns=2, legend image post style={scale=0.75},
			legend style={/tikz/every even column/.append style={column sep=0.5em}}
		]
			\draw[fill=black, fill opacity=0.2, draw=none] ({{axis cs:20e3, 1}} |- {{rel axis cs:0, 0}}) rectangle ({{axis cs:25e3, 1}} |- {{rel axis cs:0, 1}});
		
			\addlegendimage{empty legend}\addlegendentry{\hspace{-0.5cm}$\nlayers$:}
			\pgfplotsinvokeforeach{1s,1b,2,3,4,5,6,7,8}{
				\addplot[ngpl-#1, opacity=0.7, each nth point=10] table {tikz/indepth-losses-final.txt};
				\addlegendentry{$\ngplformat{#1}$}
			}
	\end{groupplot}
\end{tikzpicture}%
%
	\pcaption{Loss history of our fits to the image of Hoag's object. The left panel is the initial fit stage and shows \num{100} trials with lenses initialised at random from the prior (but see text). The lines are colour-coded according to the current learning rate. At step \num{2000} we discard any fit that has not achieved a minimum loss indicated by the dashed red line. The right panel shows the subsequent fitting of the full variational posterior including image-plane source parameters (in the shaded area it was trained without varying the rest of the model) for different numbers of layers. We reset the training to $\mathtt{lr} = \num{e-3}$ every \num{10000} steps, hence the spikes. All losses have been smoothed over \num{100} iterations and shifted by an arbitrary constant for presentation.\label{fig:indepth-losses}}
\end{figure}

We optimise our model in stages using Adam \citep{adam} with the default \torch settings. Initially, we remove the image-plane GP layers and fit only the variational posterior parameters for the lens and source-plane GP in order to obtain an initial estimate of the lens parameters and the appearance of the source. We find that generally the true parameters are discovered by fits initialised at random locations covering the whole prior ranges for the Einstein radius and position of the lens, as well as starting from a null external shear. However, it is necessary to have a prior estimate of the axis ratio $\ellq$ within \num{\sim 0.1} and the position angle $\ellangle$ within \SI{\sim 0.5}{\radian}, otherwise the fit usually gets stuck in a local minimum. These could potentially be obtained with neural networks (using e.g.\ the network from \citet{Hezaveh_2017}) or by using the lens galaxy's light. Instead, here we point out that since the local minima have a substantially higher loss value, and it can be determined with certainty whether the fit has entered one of them or has discovered the global (true) values within a few thousand steps (see \cref{fig:indepth-losses}), which usually take about a minute, a brute force initial search is a viable simple strategy.

Moreover, we examine the robustness and convergence properties of initial fits, when they indeed locate the global optimum. Starting at a learning rate of $\mathtt{lr} = \num{e-2}$, we let \torch's \verb|ReduceLROnPlateau| reduce it when the loss does not change significantly down to $\mathtt{lr} = \num{e-4}$ over the span of \num{20000} fitting steps. These fits usually take around \SI{10}{\minute} on a single Titan RTX GPU. As seen in the left panel of \cref{fig:indepth-losses}, the final loss is consistent within a scatter of \num{\sim 100} despite the random initialisation.\footnote{
	Even though the absolute scale of the loss is arbitrary (in plots it is presented relative to an arbitrary zero point), a useful basis of comparison is the number of pixels in the observation since random Gaussian noise contributes a loss of $\order{1}$ per pixel, and so \num{100} is indeed a small scatter for an image of $\order{\num{e5}}$ pixels.
} Furthermore, for our mock observation, the approximate posteriors derived from these initial fits (depicted as pale red to green ellipses in the corner plot of \cref{fig:indepth-corner}) are consistent and unbiased, having uncertainties in the third decimal\footnote{which is only a couple of orders of magnitude bigger than the machine precision of single-precision floats and certainly smaller than the accuracy of the SPLE interpolation tables used in our lens model as well as the deviations from ideal lenses found in nature}.

In the next stage of fitting we reintroduce the image-plane source and optimise its variational parameters for \num{5000} steps at a $\mathtt{lr} = \num{e-2}$, while keeping the parameters of the proposal related to the rest of the model fixed. This stage is represented in the shaded area of \cref{fig:indepth-losses}. After that we train all parameters, starting at $\mathtt{lr} = \num{e-3}$ and again reducing it automatically upon encountering a plateau in the loss. In this stage we use \num{4} samples (instead of one for the previous stages) to estimate the gradient of the ELBO in \cref{eqn:reparam} in order to constrain better the posterior (co)variances. We find that in general the loss stabilises after about \num{10000} steps and starts decreasing exponentially slowly. After \num{\sim 50000} more steps all fits reached a rate of decreasing loss of $\order{10}$ per \num{10000} steps, and so we stopped them arbitrarily at this point.

As expected, introducing more GP layers leads to a lower overall loss and a better data reconstruction loss in particular (see \cref{fig:ngplayers-recloss}). However, if a few layers are already present, the improvements are minimal.

\subsection{Final results}\label{section:results}

\begin{figure*}
	\resultsfigure[hoags_object]{3}
	\pcaption{Mock ELT image reconstruction in the image plane (top row) and source plane (bottom row). The middle two columns depict the mean and standard deviation of images derived from \num{100} samples from the trained variational posterior with three small-scale GP layers. All the fluxes, as well as the residuals in the rightmost column, have been normalised by dividing by the observational noise $\dataerr$. See \cref{section:rec-uncertainty} for a note and caveat about the source-plane reconstruction uncertainty.\label{fig:results-img-simg}}
\end{figure*}

\begin{figure*}
	\centering
	\hlayersfigure[hoags_object]{3}
	\pcaption{Source reconstruction split across the different layers: the \enquote{source-plane source}, evaluated on a regular grid as shown, and three \enquote{image-plane} layers, induced at the projected pixel centres. The bottom row shows the posterior mean of each layer, derived from samples as in \cref{fig:results-img-simg}, with the standard deviation above it. A circle in the upper right corner indicates the kernel size of the respective layer. Note that the smaller-scale layers are less prominent in terms of both mean flux (see the diminishing colourbar range), and noise contribution (the dimming of the top row), which means that the hyperparameter optimisation correctly regularises the additional irrelevant parameters (indeed, we see that this is the case with up to eight layers).\\ \captionnotes{All fluxes (colour scales) are in units of the observational noise. See \cref{section:rec-uncertainty} for a note and caveat about the source-plane reconstruction uncertainty.}\label{fig:results-layers}}
\end{figure*}

\begin{figure}
	\resids[hoags_object][3][true]
	\pcaption{Data reconstruction for the mock image of Hoag's object with three small-scale GP layers. Only the masked\footref{foot:mask} pixels are shown, and they are split according to whether they originate from the core (featureless part) or the ring (detailed region) of Hoag's object. The horizontal axis is the true flux (mock observation before adding instrumental noise). On the vertical axis in the middle panel is the mean of the posterior (panel~b.\ of \cref{fig:results-img-simg}), while the bottom panel shows the difference between the true and the reconstructed flux normalised by the posterior standard deviation (around half the noise level, as seen in panel~c.\ of the same \namecref{fig:results-img-simg}). Note that despite the name, these \enquote{residuals} are not with respect to the noisy data as usual but rather with the underlying truth. Finally, the true flux distributions for the two sets of pixels are shown in the top panel. (This figure uses the fit with $\nlayers = \ngplformat{3}$, but the residuals are only slightly improved by adding up to eight layers.)\label{fig:resids}}
\end{figure}

\begin{figure}
	\tikzsetnextfilename{\currfilebase}%
\begin{tikzpicture}[trim axis group right]
	\newcommand{\dohists}[1]{
		\pgfplotsinvokeforeach{1s,1b,2,3,4,5,6,7,8}{
			\addplot[histogram, ngpl-##1] table[y=resid-true] {tikz/resids/hoags_object/resids-hoags_object-ngpl-##1-#1.txt};
			\addlegendentry{$\nlayers = \ngplformat{##1}$}
		}
		\addplot[samples=201, no markers, black] expression {exp(-x^2/2) / sqrt(2*pi)};
		\addlegendentry{$\Gaussian(0, 1)$};
		
		\node[labelnode, #1] {#1};
	}
	
	\begin{groupplot}[
		group style={group size=1 by 2, x descriptions at=edge bottom, vertical sep=0.75em},
		width=\linewidth, height=0.7\linewidth,
		histogram/.append style={
			hist={bins=40, density},
			fill, fill opacity=0.8, draw opacity=0, area legend,
			legend image post style={fill opacity=1, draw opacity=0}
		},
		xmin=-3, xmax=5, ymin=0, xlabel style={at={(ticklabel cs:0.375,0)}},
		xlabel={$(\flux_0 - \meanpred) / \errpred$}, ylabel={PDF}, yticklabels={},
		/tikz/labelnode/.style={at={(axis description cs:0, 1)}, anchor=north west, xshift=0.5em, yshift=-0.5em, font=\bfseries},
	]
		\nextgroupplot[legend pos=north east, xlabel style={at={(ticklabel cs: 0.375,0)}}]
		\dohists{core}
		
		\nextgroupplot
		\dohists{ring}
		\legend{}
	\end{groupplot}
\end{tikzpicture}%
	\pcaption{Distributions of the residuals (corresponding to the bottom panel of \cref{fig:resids}) for the fits with different number of small-scale layers, compared to a standard Gaussian. All histograms, except for $\ngplformat{1s}$, approximately coincide, indicating the approximate independence of the result from the number of layers used. Moreover, they are narrower than a standard Gaussian (black lines), indicating a conservative estimate of the uncertainty. Pixels have been split into \enquote{core} (top) and \enquote{ring} (bottom) as in \cref{fig:resids}.\label{fig:resids-hist}}
\end{figure}

The final fit to the mock image of Hoag's object is depicted in \cref{fig:results-img-simg} (with more examples presented in \cref{apx:fit-gallery}). The mean and standard deviation images were computed using \num{100} samples of the lens and source parameters from the approximate posterior. For the image-plane reconstruction (top row) these were processed by the whole model in order to calculate the corresponding noiseless lensed image. We find that the posterior mean reproduces the observation with high fidelity and residuals dictated by the observational uncertainty. Furthermore, we find no evidence for an \emph{overall} systematic bias of the reconstructed flux (see \cref{fig:resids,fig:resids-hist}).

Using the same samples we apply \cref{eqn:gp-reconstruction} over a fine grid in the source plane in order to obtain high-resolution source reconstructions, which can be thought of as samples from the approximate posterior for the flux in the source plane. Its statistics are presented in the second row of \cref{fig:results-img-simg}. As in the image plane, the mean is an unbiased estimator of the true source flux, and the standard deviation is below the observational noise level, further decreasing in areas of higher projected pixel density, especially near caustics. Still, we refer the reader to \cref{section:rec-uncertainty} for the caveats relating to source-plane flux reconstruction.

We observe that our model does not reproduce the details of the source to arbitrarily small spatial scales, as evidenced by the presence of abnormally negative residuals in \cref{fig:resids} (see also panel~h.\ in \cref{fig:results-img-simg}), irrespective of the fact that we do include GP layers with sufficiently small kernel sizes. There are multiple reasons for this, the first one being that the resolution of the observation, projected onto the source plane, imposes a resolution limit for the reconstruction. This limit, furthermore, varies across the source plane, depending on the multiplicity of the projection and the magnification. Secondly, the uncertainty in the lens parameters smears details in the source plane over the scale of the induced variations of the projected locations of the pixels. We attempt to account for these by comparing the source reconstruction not to the original high-resolution image, but with a Voronoi tessellation\footnote{in fact, with the mean of Voronoi tessellations for samples of the approximate posterior lens} of the observed fluxes projected onto the source plane. Still, some detail is not recovered, and this is due to the fact that the source model imposes a global variance, $\layerindex{\ovar^2}{k}$, for each spatial scale $\layerindex{\kernelsize}{k}$, while the power of the relevant fluctuations in the actual image varies with position. Since the average power even of the smallest scales of fluctuations is low, the correspondingly low $\layerindex{\ovar^2}{k}$ penalises the reconstruction of small isolated details, even when the data requires them. In essence, this is a representation of the fact that real sources are not, in fact, samples from a multilayer GP model. We reserve exploring possible remedies, like allowing a spatially varying $\ovar^2(\x)$ with regularisation, e.g.\ by interpolating over a coarse grid, or replacing the prior distribution for the source parameters in the forward model with one with more probability mass in the tails for future work.

Crucially for the purposes of subsequent substructure analysis, the fits also produce an estimate of the uncertainty of the reconstruction (panels~c.\ and~g.\ of \cref{fig:results-img-simg}). We find that the standard deviation of the reconstructions is usually below the observational noise level and that, furthermore, it is a conservative estimate: as seen in the histograms of \cref{fig:resids-hist}, the true (noiseless) observation is more tightly distributed around the posterior mean than expected from the respective normal distribution.

We also examine the effect of having a different number of image-plane GP layers and find that the reconstructions are all almost equally good, except for the case of a single small-scale layer of $\kernelsize = \SI{2}{mas}$ (labelled $\ngplformat{1s}$), owing to the fact that the most relevant variations are captured by the source-plane and largest image-plane layers. We observe that in general all layers are assigned non-negligible power, with the total (corresponding to the total amount of detail in the image) roughly preserved (see \cref{fig:ngplayers-ovar}).

Similarly, all fits (except for $\ngplformat{1s}$) have comparable uncertainty in the reconstruction. Even though each image-plane GP layer contributes a number of free parameters comparable to the size of the data, the model does not overfit the data because it is optimised variationally, and even though the small-scale layers are uncorrelated and weakly constrained, optimising their variance hyperparameters prevents them from contributing excessive noise to the reconstruction (see \cref{fig:results-layers}).

Finally, we present the full approximate posterior for the lens parameters from all fits in \cref{fig:indepth-corner}. In general, adding a small-scale GP source lowers the uncertainties with respect to the initial (source-plane source only) fit by a factor \numrange[range-phrase={--}]{\sim 3}{5}, while remaining unbiased. The final posteriors are similar across the fits with different number of small-scale layers with only slight improvement noticeable with the increase of layer count. Again the exception is $\nlayers = \ngplformat{1s}$, which still improves upon the initial fit but by a smaller factor, while remaining unbiased.

\begin{figure*}
	\centering
	\begin{minipage}{\linewidth}
	\setlength\parindent{0.07\linewidth}
	\setlength\linewidth{0.86\linewidth}
 	\tikzset{
	hist1d/.append style={opacity=0.1, line width=0.5pt},
	cntr/.append style={fill=none, opacity=0.15, line width=0.5pt},
	final/.style={opacity=1, line width=0.75pt},
	final hist1d/.style={hist1d, final, /pgfplots/update limits=false},
	final cntr/.style={cntr, final}
}

\newsavebox\legendbox
\begin{lrbox}{\legendbox}
	\tikzset{external/optimize=false}
	\begin{tikzpicture}
		\matrix[inner sep=0.2em, column sep=0.2em] {
			\pgfplotsforeachungrouped \ngpl in {1s,1b,2,3,4,5,6,7,8}{\edef\tmp{
					\noexpand\draw[final, ngpl-\ngpl] (0,0) -- (0.5,0); \noexpand&
					\noexpand\node[right] {$\noexpand\nlayers = \noexpand\ngplformat{\ngpl}$};
				}\tmp\\}\\
		};
	\end{tikzpicture}%
\end{lrbox}

\tikzsetnextfilename{\currfilebase}
\begin{tikzpicture}[trim axis group left, trim axis group right]
	\begin{groupplot}[
		tight layout={7}, corner, group style={group size=7 by 7},
		label style={at={(ticklabel* cs:0.5, 1cm)}, anchor=base},
		colormap name=traffic reverse,
	]
		\input{scripts/indepth/tikz-indepth-corner.tex}
	\end{groupplot}

	\node[above] (legend) at (group c6r5.south) {\usebox\legendbox};
	\draw (group c6r5.south west) rectangle ({{group c6r5.north east}} |- {{legend.north east}});
\end{tikzpicture}%
 	\end{minipage}
	\pcaption{Variational posteriors for the lens parameters of the mock ELT observation of Hoag's object. The paler lines come from the (successful) \enquote{initial fits} and are colour coded from red to green according to the average loss value of the last \num{100} steps (green is lower loss), while the thicker lines come from the final posteriors with number of small-scale GP layers as indicated in the legend. Contours in off-diagonal panels encompass \SI{95}{\percent} of the approximate posterior probability, while the diagonal plots depict the marginal PDFs (in different arbitrary units for each panel). Black lines are placed at the true values.\label{fig:indepth-corner}}
\end{figure*}

\section{Conclusion}\label{section:conclusion}



Our aim with this work was three-fold: (1) to present a fully differentiable pipeline for the analysis of strong lensing images, (2) to introduce Gaussian processes (GPs) as a statistical model for the surface brightness of the source, and (3) to lay the foundations for targeted generation of training data for neural inference models.

Based on the deep learning framework \torch, our simulator code allows gradients to be calculated automatically for arbitrary lensing and instrumental setups, while benefitting from trivial parallelisation on graphical processors. This enables powerful statistical methods like variational inference, which optimises a lower bound on the Bayesian evidence (ELBO) and thus enables posteriors to be derived for all lens and source parameters simultaneously with the optimisation of any relevant hyperparameters. The probabilistic programming approach of \pyro integrates tightly with the \torch ecosystem and enables the physics simulator and inference machinery to use the same code, resulting in fast analyses that scale to the high-resolution, large datasets that will be produced by upcoming telescopes.

At the same time Gaussian processes provide a principled non-parametric Bayesian framework for learning correlation structures directly from data, while quantifying the reconstruction uncertainty. The hyperparameters we have used admit straightforward interpretation as the spatial scales on which the source light varies and the amplitudes of these variations. Computing the exact GP covariance \emph{matrix}, however, is computationally expensive since beyond the intrinsic correlations of the source on different spacial scales, it must also account for the overlaps between pixel projections in the source plane. Furthermore, the full GP likelihood is accessible only through a matrix inversion and determinant calculation, both of which have prohibitive complexity. Hence, we have resorted to a number of approximation techniques, which we have introduced and validated:
\begin{itemize}
    \item By rephrasing exact GP as basis function regression defined over a set of inducing points, we have constructed a source model whose (hyper\=/)parameters can be optimised via gradient descent, circumventing inversion and determinants.
    \item We have introduced an accurate approximation scheme for pixel overlaps to stand in for the GP covariance. A \keops-based implementation produces optimised GPU routines that fit seamlessly in the overall auto-differentiable framework.
    \item We have refined the standard Gaussian covariance function into a set of layers to allow for a flexible effective GP kernel, modelling multiple relevant scales simultaneously and automatically adjusting their relative contributions.
    \item Lastly, for the different layers we have used two strategies for choosing inducing points: for fast reconstruction of the source's large-scale appearance we have defined a fixed grid, while for recovering fine details we have allowed inducing points to move with the fitting of the lens. The latter has the effect of focusing the inducing points onto the relevant highly magnified regions while also reducing correlations between the source parameters and the lens. We have also used a proposal posterior for VI that partially captures these correlations.
\end{itemize}

Combining GPs and VI, we have been able to reconstruct observations down to the noise level (\cref{fig:results-img-simg}) while making unbiased inferences for the main lens parameters (\cref{fig:indepth-corner}) from high-resolution mock observations. We have shown that the results depend only weakly on how the source layers are configured and that adding many small-scale layers has only a marginal impact on the residuals. Meanwhile, optimising their variance hyperparameters effectively prevents them from contributing excessive noise. Through analytic arguments and numerical one-dimensional tests, we have demonstrated that the uncertainties of the variational posteriors for the source parameters are conservative. While recognising that replacing the GP marginal likelihood with the ELBO biases the GP hyperparameters and flux reconstructions towards zero, we have shown that the offset scales inversely with the observation's signal-to-noise ratio, and so we do not expect it to significantly impact our analysis. We have found our modelling sufficient to obtain unbiased lens parameter estimates with resolution-limited precision. More complex lens models and proposal posteriors could be employed in future work if necessary.

The computational efficiency of the pipeline allows obtaining reasonable lens parameter posteriors in \SI{\sim 10}{\minute} with full convergence in a few hours. Importantly, the runtime of the pipeline scales linearly with model complexity, making it imminently applicable to more realistic analysis configurations. We will extend fits to multi-band observations with non-trivial instrumental effects (a realistic point-spread function, for example), explore more complex main lens models, and include the lens galaxy's light in follow-up work. Additionally, we plan to improve modelling of the smallest-scale source details, which are difficult to fully resolve at present.

Lastly, our main scientific goal is to make inferences about dark matter substructure and ultimately the fundamental features of dark matter. A simultaneously accurate and precise reconstruction of the source and main lens in all their complexity is crucial for detections of the percent-level fluctuations introduced by substructure. Furthermore, due to the statistical challenges of the substructure inference problem itself, the approaches best suited for tackling it require a wealth of well-targeted but also realistic training data, which our variational source analysis is well-positioned to deliver. In upcoming work, therefore, we will use the pipeline presented here as a targeted simulator \citep{Coogan:2020yux} to generate training data to teach neural networks to make substructure inferences, enabling heretofore intractable analyses.

\section*{Acknowledgements}

We thank Marco Chianese, Simona Vegetti, Sam Witte, Noemi Anau Montel, Camila Correa, and Rajat Thomas for helpful discussions.

This work was carried out on the DAS\=/5 cluster \citep{das5} (funded by the Netherlands Organization for
Scientific Research (NWO/NCF)) and Lisa Compute Cluster at SURFsara. Lisa runs on \SI{100}{\percent} wind energy.

Software used: \python, \numpy \citep{harris2020array}, \scipy \citep{2020SciPy-NMeth}, \astropy \citep{astropy_2013,astropy_2018}, \torch \citep{pytorch}, \pyro \citep{pyro}, \keops \citep{charlier2020kernel}, \matplotlib \citep{Hunter:2007}, \tqdm \citep{casper_da_costa_luis_2021_4663456}.

\newcommand{\jmlr}{JMLR}
\bibliographystyle{mnras}
\bibliography{mlens}

\begin{appendices}

\section{Gallery of mock analyses}\label{apx:fit-gallery}

\begin{figure}
	\centering
	\showofffigure{antennae}\\[0.5em]
	\showofffigure{arp142}\\[0.5em]
	\showofffigure{arp148}\\[0.5em]
	\showofffigure{ngc4414}
	\pcaption{Mock observations of, top to bottom: the Antennae galaxies \stscirelease{2006}{46}, Arp142 \stscirelease{2013}{23}, Arp148 \stscirelease{2008}{16}, and NGC4414 \stscirelease{1999}{25}.\label{fig:showoff-all}}
\end{figure}

\setlength{\floatsep}{0pt}\setlength{\dblfloatsep}{\floatsep}

\begin{figure*}
	\pgfplotsset{sourceplane/.append style={extent=0.6}}
	\resultsfigure[antennae]{3}
	\pcaption{Fit to the mock observation of the Antennae galaxies. See \cref{fig:results-img-simg} for details.\label{fig:results-antennae}}
\end{figure*}
\begin{figure*}
	\resultsfigure[arp142]{3}
	\pcaption{Fit to the mock observation of Arp142. See \cref{fig:results-img-simg} for details.\label{fig:results-arp142}}
\end{figure*}
\begin{figure*}
	\resultsfigure[arp148]{3}
	\pcaption{Fit to the mock observation of Arp148. See \cref{fig:results-img-simg} for details.\label{fig:results-arp148}}
\end{figure*}
 \begin{figure*}
 	\pgfplotsset{imageplane/.append style={extent=2.1}}
 	\resultsfigure[ngc4414]{3}
 	\pcaption{Fit to the mock observation of NGC4414. See \cref{fig:results-img-simg} for details.\label{fig:results-ngc4414}}
 \end{figure*}

We present a gallery of mock strong-lensing observations based on various high-resolution galaxy images (\cref{fig:showoff-all}) with arbitrary lens parameters from the priors in \cref{tab:lens-params}, to which we perform fits with the same methodology as described in \cref{section:mock-data-analysis}, using $\nlayers = \ngplformat{3}$ image-plane layers and a single source-plane layer. Once again we are able to quickly obtain very precise and accurate lens parameter posteriors, starting from reasonable guesses for $\ellq$ and $\ellangle$ and random guesses for the other parameters. We are able to fit the observation at the noise level and derive uncertainties. In the case of NGC4414, the \enquote{raw} uncertainties approach the original resolution at which the mock image was created, and so we put a lower bound on them by artificially introducing \enquote{sub-pixel}-localisation noise on the sampled lens parameters. In all cases we can reproduce small-scale details of the source galaxies (\cref{fig:results-antennae,fig:results-arp142,fig:results-arp148,fig:results-ngc4414}) with the same caveats as discussed in \cref{section:mock-data-analysis}.

\section{On the number of image-plane GP layers}\label{axp:ngplayers}

\begin{figure}
	\centering
	\tikzsetnextfilename{\currfilebase}%
\begin{tikzpicture}
	\newcommand{\getovarcoord}[2]{(\getdata{hoags_object/#1/gp/sigma/#2}, \getdata{hoags_object/#1/gp/alpha/#2})}
	
	\begin{semilogxaxis}[
		width=\linewidth, height={0.75*\pgfkeysvalueof{/pgfplots/width}},
		x dir=reverse,
		xtick={0.002,0.003,...,0.011},
		xticklabel={\pgfmathparse{1000*exp(\tick)}\pgfmathprintnumber[fixed, precision=0]{\pgfmathresult}},
		yticklabel={\pgfmathprintnumber[fixed, precision=2]{\tick}},
		xlabel={kernel size, $\layerindex{\kernelsize}{k} / \si{mas}$},
		ylabel={fitted root variance, $\layerindex{\mean{\ovar}}{k}$},
		legend columns=2,
		clip mode=individual,
	]
			\addlegendimage{empty legend}\addlegendentry{}
			\pgfplotsinvokeforeach{1s,1b,2,3,4,5,6,7,8}{
				\addplot[ngpl-#1, mark=*, thick] coordinates {\directlua{ovarcoords(\ngplnum{#1}, "#1")}};
			\addlegendentry{$\nlayers = \ngplformat{#1}$}
		}
	\end{semilogxaxis}
\end{tikzpicture}%
	\pcaption{Final inferred overall variances of the image-plane GP layers in fits with different number of layers. For comparison, the root overall variance of the source-plane layer, $\spindex{\ovar}$, for the different fits is between \num{1.8} and \num{2}. We observe a \enquote{redistribution} of the \enquote{power} of the modelled features. \label{fig:ngplayers-ovar}}
\end{figure}

\begin{figure}
	\centering
	\tikzsetnextfilename{\currfilebase}%
\begin{tikzpicture}
	\newcommand{\ngpllist}{1b,2,3,4,5,6,7,8}
	\newcommand{\getreclosscoord}[1]{(\ngplnum{#1}, \getdata{hoags_object/ngpl-#1/recloss/\reclosskey})}
	\newcommand{\drawline}[1][]{
		\addplot[no markers, #1] coordinates {
			\luadirect{for i, ngpl in ipairs({"1b","2","3","4","5","6","7","8"}) do tex.print("\\getreclosscoord{" .. ngpl .. "}") end}
		};
	}
	\newcommand{\drawmarks}[1][]{
		\expandafter\pgfplotsinvokeforeach\expandafter{\ngpllist}{\edef\tmp{
			\noexpand\draw[ngpl-##1, mark=*, #1] plot coordinates {\getreclosscoord{##1}};
		}\tmp}
	}
	\newcommand{\drawlabels}{
		\expandafter\pgfplotsinvokeforeach\expandafter{\ngpllist}{\edef\tmp{
			\noexpand\node[below, yshift=-0.5em] at \getreclosscoord{##1} {$\ngplformat{##1}$};
		}\tmp}
	}
	
	\begin{axis}[
		width=\linewidth, height={0.5*\pgfkeysvalueof{/pgfplots/width}},
		xtick={1,2,...,8},
		xlabel={number of image-plane layers, $\nlayers$},
		ylabel={data MSE, $\abs{\qty(\data - \meanpred) / \dataerr}^2 / \n$},
		ymin=1, yticklabel={\pgfmathprintnumber[fixed, fixed zerofill, precision=2]{\tick}},
		every axis plot/.append style={thick}
	]
		\providecommand{\reclosskey}{}
		\renewcommand{\reclosskey}{obs/masked}
		\drawline\drawmarks\drawlabels
		\addlegendentry{masked}
		
		\renewcommand{\reclosskey}{obs/all} \drawline[dashed]\drawmarks[mark=square*]
		\addlegendentry{all}
	\end{axis}
\end{tikzpicture}%
	\pcaption{
		\sisetup{round-precision=2, round-mode=places}
		Final image reconstruction losses, defined as the mean squared normalised errors between the observation and the mean of the PPD, for the fits with different numbers of image-plane GP layers. For the circles and solid line only the masked pixels are considered, while the squares and dashed line include all pixels. The data MSEs for the $\ngplformat{1s}$ fit (not shown) are \num{\getdata{hoags_object/ngpl-1s/recloss/obs/masked}} (masked) and \num{\getdata{hoags_object/ngpl-1s/recloss/obs/all}} (all).\label{fig:ngplayers-recloss}}
\end{figure}

See \cref{fig:ngplayers-ovar,fig:ngplayers-recloss} for further details on the reconstruction of the image of Hoag's object with different numbers of image-plane GP layers. In general we find that the improvement in reconstruction fidelity is minimal after \numrange[range-phrase={--}]{\sim 3}{5} layers. Meanwhile, due to the automatic regularisation via the optimisation of the variance hyperparameters, the overall \enquote{power} accounted for by the source model (related to the area under the curve in \cref{fig:ngplayers-ovar}) is kept roughly constant and is simply redistributed across the layers. Furthermore, we reiterate that the additional many parameters, which for the smallest layers are independent of one another and largely unconstrained by data, neither lead to overfitting, nor contribute excessive predictive variance, and so the reconstructions and the associated losses remain similar.

\section{Windowing normalisation derivation}\label{apx:windowing-norm}

Due to conservation of surface brightness, the brightness received in pixel $i$ is the integral over the pixel area \emph{in the image plane} of the source surface brightness at the projected locations in the source plane; i.e.
\begin{eq}
	\varcomp{\flux}_i \propto \int \dd{\ximg} \gpwindowimg\qty(\ximg) \sbr\qty(\x\qty(\ximg)),
\end{eq}
where $\ximg$ is an image-plane and $\x$ a source-plane coordinate, $\beta$ is the source surface brightness, and $\gpwindowimg$ is the pixel's image-plane indicator function, which is simply $1$ inside the pixel and zero outside. The coordinate transformation\footnote{assuming we are away from critical lines/caustics and the transformation is well-defined} $\ximg \rightarrow \x$ introduces a factor equal to the magnification from \cref{eqn:magnification}, which can be approximated by the ratio of image-plane to source-plane pixel areas:
\begin{eq}
	\varcomp{\flux}_i \propto \int \dd{\x} \abs{\dv{\ximg}{\x}} \gpwindowimg\qty(\ximg\qty(\x)) \sbr(\x)
	\propto \int \dd{\x} \frac{\gpwindowimg\qty(\ximg\qty(\x))}{A_i} \sbr(\x).
\end{eq}
From here we can identify $\gpwindow\qty(\x) \propto \gpwindowimg\qty(\ximg\qty(\x)) / A_i$ with $A_i$ the projected (source-plane) pixel area. This is a function that is still zero outside the pixel but has a value $A_i^{-1}$ inside, which leads to the normalisations
\begin{eq}
	\int \dd{\x} \gpwindow\qty(\x) \propto 1 \qand \int \dd{\x} \gpwindow^2\qty(\x) \propto A_i^{-1}
\end{eq}
with the same coefficient of proportionality for which it is only important not to depend on the \emph{projected} pixel properties, i.e.\ on the kernel.

The first condition is trivially satisfied by a properly normalised Gaussian. Plugging \cref{eqn:window-function} and \cref{eqn:window-kernel} into the second condition, we can verify:
\begin{multline}
	\int\dd{\x} \gpwindow^2\qty(\x) = \int\dd{\x} \qty[\Gaussian(\x - \vector{p}_i, \varKernel_i)]^2
	\\ \begin{aligned}[b]
		& = \qty(\frac{1}{\sqrt{\vectdet{2\pi \varKernel_i}}})^2 \int\dd{\vector{r}} \exp(-\frac{\vector{r}^\transpose \qty(2\varKernel_i^{-1}) \vector{r}}{2})
		\\ & = \frac{1}{\vectdet{2\pi\varKernel_i}} \sqrt{\vectdet{2\pi \frac{\varKernel_i}{2}}}
		= \frac{1}{\sqrt{\vectdet{4\pi \varKernel_i}}}
		= \vectdet{\vector{a}_i \vector{a}_i^\transpose + \vector{b}_i \vector{b}_i^\transpose}^{-\frac{1}{2}}
		\\ & = \qty[\qty(a_{i,x}^2 + b_{i,x}^2)\qty(a_{i,y}^2 + b_{i,y}^2) - \qty(a_{i,x} a_{i,y} + b_{i,x} b_{i,y})^2]^{-\frac{1}{2}}
		\\ & = \qty[\qty(a_{i,x} b_{i,y} - a_{i,y} b_{i,x})^2]^{-\frac{1}{2}}
		= \matrixdeterminant{\vector{a}_i & \vector{b}_i} \approx A_i, \qq{QED.}
	\end{aligned}
\end{multline}
In the last line it is presumed that the pixel is a parallelogram with sides $\vector{a}_i$ and $\vector{b}_i$.

\section{Analytical treatment of a variationally optimised Gaussian process}\label{apx:vigp-analytical}

An exact multilayer GP has a (prior) covariance matrix that is the sum of the covariances for the different layers: $\K = \sum_k \layerindex{\ovar^2}{k} \layerindex{\K}{k}$. We assume that each covariance layer can be factorised into the outer product of a transfer matrix with itself: $\layerindex{\K}{k} = \layerindex{\ovar^2}{k} \layerindex{\T}{k} \layerindex{\T}{k}^\transpose$, and instead define a generative model under which $\flux = \sum_k \layerindex{\T}{k} \layerindex{\y}{k}$. These equations can be conveniently recast in matrix form by defining $\T$ and $\y$ to be the concatenation of the transfer matrices (row-wise) and the source parameters of all layers:
\begin{align}
	\T & \equiv \mqty(\layerindex{\T}{1} & \layerindex{\T}{2} & ... & \layerindex{\T}{\nlayers}),
	\\ \y^\transpose & \equiv \mqty(\layerindex{\y}{1}^\transpose & \layerindex{\y}{2}^\transpose & ... & \layerindex{\y}{l}^\transpose)^\transpose.
\end{align}
The transformation equation is then the familiar $\flux = \T\y$. On the other hand, the total covariance can be written as $\K = \T \Ovar^2 \T^\transpose$ by defining $\Ovar^2$ to be an $\n\nlayers \times \n\nlayers$ block diagonal matrix with $\layerindex{\ovar^2}{k} \varmatrix{\Identity}_n$ on the diagonal. Given that the transformation is linear, it is also applicable to the posterior means\footnote{In our case, there are as many parameters \emph{in each layer} as there are observed pixels, so this transformation is degenerate, and therefore it is \enquote{easy} to find parameters that reconstruct the true mean, but this is nothing to say of their uncertainty and correlations}, i.e.
\begin{eq}
	\meanpred = \T\meanpredy.
\end{eq}
Similarly, for the posterior variance, which for the true values is given in \cref{eqn:gp-exact},
\begin{eq}\label{eqn:paramvariance-exact}
	\variance_{\posterior}(\T\y) & = \T \varpredy \T^\transpose
	\\ \variance_{\posterior}(\flux) & = \K - \K\qty(\K+\datavar)^{-1}\K
	\\ & = \T \qty[\Ovar^2 - \Ovar^2 \T^\transpose \qty(\K+\datavar)^{-1} \T \Ovar^2 ] \T^\transpose
	\\ \implies \varpredy & = \Ovar \qty[\varmatrix{\Identity}_{\n\nlayers} - \Ovar\T^\transpose\qty(\K+\datavar)^{-1}\T\Ovar] \Ovar
	\\ & = \qty(\Ovar^{-2} + \T^\transpose \datavar^{-1} \T)^{-1}.
\end{eq}

ELBO maximisation in this case can be considered analytically by examining the KL divergence between the MVN true posterior and a diagonal approximation. It is given by \citep{gaussiankl}:
\begin{multline}\label{eqn:kl-diag-full}
	\kl{\Normal(\meany, \Erry)}{\Normal(\meanpredy, \varpredy)} =
	\\ = \frac{1}{2} \qty{\tr(\varpredy^{-1} \Erry) + \contract[\varpredy^{-1}]{\qty(\meany - \meanpredy)} - \n + \ln\frac{\vardet{\varpredy}}{\vardet{\Erry}}},
\end{multline}
where for notational convenience we use $\Erry$ to denote a matrix with $\erryi^2$ as diagonal entries\footnote{In the multilayer case, $\meany$ and $\erry$ are concatenations of the parameters for the different layers, as we had before.}. Evidently, this KL divergence is minimised when the true mean is recovered, i.e.\ $\meany = \meanpredy$. Minimising with respect to the variational (co)variance leads to
\begin{eq}\label{eqn:paramvariance}
	\erryi^2 = \frac{1}{\vpred_{\y, ii}^{-1}} = \frac{1}{\qty(\Ovar^{-2} + \T^\transpose \datavar^{-1} \T)_{ii}} \rightarrow \frac{1}{\Ovar^{-2}_{ii} + \dataerr^{-2}},
\end{eq}
where the last expression holds for a diagonal data covariance $\datavar = \dataerr^2 \varmatrix{\Identity}$ because $\layerindex{\T}{k}\layerindex{\T}{k}^\transpose$ are normalised to a unit diagonal and $\layerindex{\T}{k}$ are symmetric. Thus, we should expect a fit to produce a constant, albeit differing per layer, variance of the source parameters
\begin{eq}\label{eqn:paramvariance-layer}
	\layerindex[, i]{\qty(\erry^2)}{k} \rightarrow \frac{\layerindex{\ovar^2}{k} \dataerr^2}{\layerindex[]{\ovar^2}{k} + \dataerr^2}.
\end{eq}

We note that, by the Schur complement formula\footnote{Briefly, $\erryi^2 = 1 / \vpred_{\y, ii}^{-1} = \vpred_{\y, ii} - \mathscr{P}_i$, where $\mathscr{P}_i$ is given by $\mathscr{P}_i = \sum_{j, k \neq i} \vpred_{\y, ij} \vpred^{-1}_{\y, jk} \vpred_{\y, ki}$, which is an inner product with the inverse of a positive definite matrix, which is also positive definite. Hence, $\mathscr{P}_i$ is non-negative, and $\erryi^2 \leq \vpred_{\y, ii}$.} \citep{schur}, this is smaller than the marginal variance of the true posterior ($\vpred_{\y, ii}$), and so as expected of a mean-field approximation, the per-parameter variance is underestimated.

However, we are interested instead in the reproduced posterior covariance of the true values, i.e.
\begin{eq}
	\variance_{\proposal}(\flux) = \T \variable{\Sigma_{\y}} \T^\transpose,
\end{eq}
in which the transfer matrix combines the variances of a number of apriori independent source parameters that are, in the exact case, a posteriori anti-correlated (due to the second term in \cref{eqn:paramvariance-exact}): intuitively, the information that nearby source parameters produce similar effects in the observed data is lost in a diagonal approximation. This leads to an overestimation of the posterior variance of $\flux$ that scales roughly with the number of a priori correlated observations\footnote{This can be estimated, for a stationary covariance function, as $\int \kfunc(\vector{0}, \x) \dd{\x}$ times the local point density.}: see \cref{fig:vigp}.

\subsection{Hyperparameter optimisation}\label{apx:vigp-analytical-hyper}
To verify \cref{eqn:hyper-scaling}, we will directly compute the ELBO already minimised with respect to the variational mean and variance at a fixed set of hyperparameters $\layerindex{\ovar^2}{k}$ (from which we can construct the matrix $\Ovar$ as above). Substituting $\meany = \meanpredy$ and \cref{eqn:paramvariance} into \cref{eqn:kl-diag-full}, we get
\begin{equation}
	\min_{\meany, \Erry} \kl{\Normal(\meany, \Erry)}{\Normal(\meanpredy, \varpredy)} = \ln{\vardet{\varpredy}} - \ln{\vardet{\Erry}} - \n.
\end{equation}
Using this and the evidence from \cref{eqn:gp-mll} we can write the loss function, which is the negative of the ELBO, up to constant terms and a proportionality, as
\begin{equation}\label{eqn:hyper-loss-full}
	\Loss \simeq \contract[\qty(\K + \datavar)^{-1}]{\data} + \ln{\vardet{\K + \datavar}} + \ln{\vardet{\varpredy}} - \ln{\vardet{\Erry}}.
\end{equation}

The log-determinant of $\varpredy$ can be written, via \cref{eqn:paramvariance-exact}, as
\begin{equation}
	\ln{\vardet{\varpredy}} = \ln{\abs{\Ovar^2}} + \ln{\vardet{\datavar}} - \ln{\vardet{\T\Ovar^2\T^\transpose + \datavar}},
\end{equation}
whence we see that the last term cancels the log-determinant of the evidence, and we are left with a loss
\begin{eq}\label{eqn:hyper-loss}
	\Loss & \simeq \contract[\qty(\K + \datavar)^{-1}]{\data} + \ln{\abs{\Ovar^2}} - \ln{\vardet{\Erry}}
	\\ & \simeq \contract[\qty(\K + \datavar)^{-1}]{\data} + \sum_k \n \ln(\layerindex{\ovar^2}{k} + \dataerr^2)
\end{eq}
by \cref{eqn:paramvariance-layer}.

To proceed, we will look into the case of fitting a single-layer GP to data that consists of clumps of observations, where the points within each clump are perfectly correlated with each other but perfectly independent of the points in any other clump. We can assume each clump consists of $\nc$ observations, given, for a stationary covariance function, by
\begin{equation}
	\nc = n \int \frac{\kfunc(\vector{0}, \x)}{\kfunc(\vector{0}, \vector{0})} \dd{\x} \approx \frac{1}{N} \sum_{ij} \frac{\varcomp{\K}_{ij}}{\varcomp{\K}_{ii}},
\end{equation}
with $n$ the average point density. Then the covariance matrix $\K$ is block diagonal, and each $\nc \times \nc$ block is filled with the value $\ovar^2$, i.e.\ we can write a block as
$\block{\K} = \ovar^2 \u\u^\transpose$, where $\u$ is a vector of $\nc$ ones.

Crucially, now we can compute the inner product in \cref{eqn:hyper-loss} (for (block\=/)diagonal data covariance) by inverting $\K + \datavar$ block-wise via the Sherman-Morrison formula \citep{Sherman_Morrison_1950}:
\begin{eq}
	\qty(\blockdatavar + \block{\K})^{-1} & = \blockdatavar^{-1} - \blockdatavar^{-1} \frac{\ovar^2 \u\u^\transpose}{1 + \ovar^2 \u^\transpose \blockdatavar^{-1} \u} \blockdatavar^{-1}
	\\ & \rightarrow \frac{1}{\dataerr^2} \qty[\varmatrix{\Identity}_{\nc} - \frac{\ovar^2}{\dataerr^2 + \nc \ovar^2} \u\u^\transpose].
\end{eq}
Naturally, all blocks are equal, so applying this in the inner product yields
\begin{equation}\label{eqn:hyper-loss-withsums}
	\contract[\qty(\K + \datavar)^{-1}]{\data} \rightarrow \frac{1}{\dataerr^2} \qty[\sum_i \varcomp{\data}_i^2 - \frac{\ovar^2}{\dataerr^2 + \nc \ovar^2} \sum_{ij} \varcomp{\data}_i \varcomp{\data}_j].
\end{equation}

Supposing that the data was generated from a GP with the true prior variance $\ovartrue^2$ and correlation structure as we assumed above, on expectation the two sums can be written as
\begin{align}
	\sum_i \varcomp{\data}_i^2 & \approx \trace(\Ktrue + \datavar) = \n \qty(\ovartrue^2 + \dataerr^2),
	\\ \sum_{ij} \varcomp{\data}_i \varcomp{\data}_{j} & \approx \norm{\Ktrue + \datavar}_1 = \n \qty(\nc \ovartrue^2 + \dataerr^2).
\end{align}

After substitution into \cref{eqn:hyper-loss-withsums} and then into \cref{eqn:hyper-loss} and some manipulation we arrive, finally, at
\begin{equation}\label{eqn:hyper-loss-approx}
	\Loss \simeq \frac{\ovartrue^2 + \dataerr^2 + \qty(\nc - 1) \ovar^2}{\nc \ovar^2 + \dataerr^2} + \ln(\ovar^2 + \dataerr^2).
\end{equation}
Indeed, it can be shown that this function is extremised, in the limit $\dataerr \rightarrow 0$, by $\ovarvi^2 = \ovartrue^2 / \nc$ as anticipated. It is compared to the exact calculation of \cref{eqn:hyper-loss-full} and the exact evidence of \cref{eqn:gp-mll} in \cref{fig:vigp-hyper}.

\subsection{Bias of the mean}\label{apx:vigp-analytical-bias}
\shortcut{\nci}{\nc{}_{,i}}

Using different hyperparameters of the prior naturally modifies the posterior. The effect on the variance was derived in \cref{eqn:paramvariance-layer}: using smaller hyperparameter values diminishes the posterior uncertainty. Whereas that result depends on the particular fitting strategy (using a diagonal covariance), the effect on the posterior mean can be studied by considering only the exact solution since the variational mean follows it exactly, as stated previously.

Starting from \cref{eqn:gp-exact-mean} and using the same form of the covariance matrix as when deriving the optimal hyperparameters, we can obtain a simple estimate for the posterior mean:
\begin{equation}\label{eqn:bias}
	\meanpred \approx \frac{\nc \ovar^2}{\dataerr^2 + \nc \ovar^2} \bar{\data},
\end{equation}
where $\bar{\data}$ is the local average of the data (weighted by the covariance function, i.e.\ $\varcomp{\bar{\data}}_i \propto \sum_j \varcomp{\K}_{ij} \varcomp{\data}_j$). Thus, for the optimal (in terms of ELBO) hyperparameters, the mean scales in the same way as the variance from \cref{eqn:paramvariance-layer}, and for strongly correlated setups ($\nc \gg 1$) is closer to zero than the exact solution by a factor $\ovartrue^2 / \qty(\dataerr^2 + \ovartrue^2)$, but in high-signal-to-noise scenarios, in which $\ovartrue^2 \gg \dataerr^2$, this bias is expected to be small.

If the number density of points is not uniform, we can further let $\nc \rightarrow \nci$ be the \emph{local} number of correlated points, i.e.\ $\nci = \sum_j \varcomp{\K}_{ij} / \varcomp{\K}_{ii}$. Finally, to take into account multiple layers, we first calculate the number of correlated points per layer and then replace $\nci \ovar^2$ by its mean over the layers. We plot this solution in \cref{fig:vigp} to confirm that it is indeed a good approximation.

\begin{figure}
	\tikzsetnextfilename{\currfilebase}%
\begin{scaletowidth}{\linewidth}
\begin{tikzpicture}[vigp defs]
	\newcommand{\prefix}{tikz/vigp/400-}
	\newcommand{\plotcontours}{
		\addplot[no markers, exact, smooth] table {\prefix contours-exact.txt} -- cycle;
		\addplot [only marks, exact] coordinates {(0.5, 0.2)};
	}
	\newcommand{\plotalphavi}{
		\addplot [only marks, diag] coordinates {(0.0675, 0.1215)};
	}
	
	\begin{groupplot}[
		group style={
			group size=1 by 3, vertical sep=1em,
			xlabels at=edge bottom, xticklabels at=edge bottom,
			every plot/.style={scale=\thescale}
		},
		xmin=0, xmax=1, ymin=0, ymax=0.5, axis equal image, view={0}{90},
		xlabel={$\layerindex{\ovar}{1}$}, ylabel={$\layerindex{\ovar}{2}$},
		colorbar right, colormap name={viridis},
		colorbar style={y label style={at={(ticklabel* cs:0.5, 2.2em)}}},
		/tikz/mark=x, /tikz/mark size=0.5em,
		colormesh/.style={surf, shader=interp, patch type=bilinear,},
		table/x=alpha_1, table/y=alpha_2,
	]
		\nextgroupplot[colorbar style={ylabel={$-\Delta\ln\prob(\data\given\ovar^2)$}}]
			\addplot3[colormesh, point meta max=100] table [z=exact] {\prefix mll.txt};
			\plotcontours
		
		\nextgroupplot[colorbar style={ylabel={$\min\limits_{\mathclap{\meany, \Erry \given \alpha}}\Loss$}}]
			\addplot3[colormesh, point meta max=1000] table [z=theoretical] {\prefix mll.txt};
			\plotcontours
			\plotalphavi

		\nextgroupplot[colorbar style={ylabel={$\min\limits_{\mathclap{\meany, \Erry \given \alpha}}\Loss$}}]
			\addplot3[colormesh, point meta max=1000] table [z=approximate] {\prefix mll.txt};
			\plotcontours
			\plotalphavi
			\addplot [only marks, orange] coordinates {(0.08297999, 0.16666989)};
			
			\draw[red] (0.0114, 0) -- (0.0114, 0.5); 
			\draw[red] (0, 0.0360) -- (1, 0.0360);
	\end{groupplot}
\end{tikzpicture}%
\end{scaletowidth}
	\pcaption{Hyperparameter optimisation of the Gaussian process in \cref{fig:vigp}. The top panel is (minus) the exact model evidence \cref{eqn:gp-mll}, which serves as a \enquote{posterior} for the hyperparameters (assuming flat priors). The contours (replicated in the other panels for comparison) are \num{1}- and \num{2}-$\sigma$ credible regions. On the other hand, the middle and bottom panels show the ELBO for an optimised diagonal variational posterior (the exact calculation of \cref{eqn:hyper-loss} and the approximation \cref{eqn:hyper-loss-approx}, respectively) with a red and orange cross marking the positions of the respective optima. The thin red lines are the locations beyond which the approximation is expected to deviate significantly ($\nc \ovar^2 < \dataerr^2$).\label{fig:vigp-hyper}}
\end{figure}

\end{appendices}

\bsp	
\label{lastpage}
\end{document}